\documentclass[twocolumn,superscriptaddress,floatfix,preprintnumbers,prd,nofootinbib]{revtex4-2}
\usepackage{bm}
\usepackage{bbm}
\usepackage{amsmath}
\usepackage{etoolbox}
\usepackage{float}
\usepackage{needspace}
\newcommand{\eg}{{\emph{e.g.~}}}
\newcommand{\ie}{{\emph{i.e.~}}}

\usepackage[utf8]{inputenc}
\usepackage{verbatim}
\usepackage{amsmath,amsfonts,amssymb}
\usepackage[breaklinks,colorlinks]{hyperref}
\usepackage{float}
\usepackage{natbib}
\usepackage{tikz}
\usepackage{float}
\usepackage{graphicx}
\usepackage{adjustbox}
\usepackage{tabularx}

\usepackage{amsbsy}
\usepackage{orcidlink}
\hypersetup{
    colorlinks=true,
    allcolors=black,
}

\begin{document}

\preprint{DESY-25-111}
\preprint{CERN-TH-2025-251}

\title{Axion Hair and Pulsar Electrodynamics: modeling, discharge dynamics, and particle-in-cell simulations}

\author{Samuel J. Witte\orcidlink{0000-0003-4649-3085}}\email{Samuel.Witte@physics.ox.ac.uk}
\affiliation{Rudolf Peierls Centre for Theoretical Physics, University of Oxford, UK}
\affiliation{Deutsches Elektronen-Synchrotron DESY, Notkestra\ss{}e 85, 22607 Hamburg, Germany}
\affiliation{
II. Institute of Theoretical Physics, Universit{\"a}t Hamburg, 22761, Hamburg, Germany}
\author{Andrea Caputo\orcidlink{0000-0003-3516-8332}}\email{Andrea.Caputo@cern.ch} 
\affiliation{Department of Theoretical Physics, CERN, Esplanade des Particules 1, P.O. Box 1211, Geneva 23, Switzerland}
\affiliation{Dipartimento di Fisica, ``Sapienza'' Universit\`a di Roma \& Sezione INFN Roma1, Piazzale Aldo Moro
5, 00185, Roma, Italy}
\affiliation{Department of Particle Physics and Astrophysics, Weizmann Institute of Science, Rehovot 7610001, Israel}

\author{Stefan Stelzl\orcidlink{0000-0001-5964-1054}}\email{SStelzl@ifae.es}
\affiliation{Institute of Physics,Theoretical Particle Physics Laboratory, Ecole Polytechnique Federale de Lausanne,  Switzerland}
\affiliation{Institut de F\'{i}sica d'Altes Energies (IFAE) and
Barcelona Institute of Science and Technology (BIST), 
Campus UAB, 08193 Bellaterra (Barcelona), Spain
}

\author{Alexander Chernoglazov}
\affiliation{Institute for Advanced Study, Princeton, NJ 08540, USA}

\author{Alexander A. Philippov} 
\affiliation{Department of Physics, University of Maryland, College Park, MD 20742, USA}

\author{Surjeet Rajendran}
\affiliation{Department of Physics and Astronomy, Johns Hopkins University, Baltimore, MD 21218, USA}

\begin{abstract}
In a companion paper, we demonstrated that static axion field gradients sourced by dense nuclear matter (\emph{axion hair}) can dominate the near-field electrodynamics of old rotation-powered pulsars, leading to new constraints on light QCD axions and on CP-violating axion-nucleon interactions.
This article provides the extended theoretical and numerical framework underlying those results. We begin by providing a detailed description of the sourcing of axion hair from dense nuclear matter, computing self-consistent field profiles for each interaction across the relevant parameter space. We then study the modification to the electrodynamics induced in the polar gap region by these axion gradients; this is done at the analytic level by studying the modification induced by axion field gradients on the effective discharge parameter (computed in the force-free limit of the split monopole magnetic field configuration, and looking at leading deviations from the force-free limit for dipolar field configurations), and numerically by developing dedicated 1D particle-in-cell simulations which capture the leading order dynamical behavior near the star. Our results demonstrate that axion hair serves to either enhance acceleration, or enhance screening, where the relevant effect changes between the northern and southern hemispheres of the star, and between the field lines which support out-flowing and return currents.  
\end{abstract}

\maketitle

\section{Introduction}

Pulsars provide some of the most extreme environments in the Universe, characterized by super-strong electromagnetic fields, strong gravity, and rapid rotation. These conditions establish neutron stars as premier laboratories for studying high-energy plasma physics and quantum electrodynamics (QED) effects that are inaccessible in terrestrial experiments. 

There has been immense progress over the last few decades in understanding how these objects radiate and dissipate their energy into the surrounding medium. Broadly speaking, the goal of these studies is to identify self-consistent solutions for the electromagnetic fields, and the charge and current distributions, across a wide range of dynamical scales, ranging from the plasma scale (often on the order of centimeters) to the scales well-beyond the light cylinder (which can extend to thousands of stellar radii away from the star). Particle-in-cell (PIC) simulations, capable of resolving both plasma production and the plasma phase space distribution, have played a major role in recent years in bridging the small-scale and large-scale physics, and connecting dynamical processes with the observed emission~\cite[see, e.g.,][for reviews]{Cerutti:2016ttn, 2022ARA&A..60..495P}. %

As progress in our understanding of these systems have advanced, so too have the prospects for turning these objects into laboratories for new fundamental physics. Typically, one would expect that any new  beyond the Standard Model physics must interact with known particles so feebly that, to leading order, one can treat their presence and interactions as tiny perturbations operating on top of the background dynamics. This assumption, however, is not always valid. For example, axions (\ie light pseudoscalars with a approximate shift symmetry)~\cite{WilczekAxion, Peccei:1977hh, Weinberg:1977ma} couple to  electromagnetism~\cite{Wilczek:1987mv}, and can induce non-trivial modifications to the acceleration and pair production processes operating in pulsar magnetospheres; in order for such effects to be non-negligible, however, the axion field values must be exceptionally large.

In a companion paper~\cite{Witte:CompanionPRL}, we have demonstrated that large axion field gradients (which we refer to as ``axion hair'') can be directly sourced from the dense nuclear matter found in neutron stars\footnote{It was also shown in ~\cite{Noordhuis:2023wid,Witte:2024akb,Caputo:2023cpv}, that large field gradients can be sourced by the continued accumulation of gravitationally-bound axions sourced from small scale dynamical electromagnetic fields operating in the polar caps of neutron stars. In this case, however, the axion field gradients are expected to saturate when effect of the axion becomes comparable to that of the background.}. These gradients can be so large that their contribution to Maxwells equations (in particular, Gauss' law) dominate over the leading order background contribution; in effect, we have demonstrated this axion hair serves to  decouple near-field particle acceleration and plasma production processes from the pulsar rotational period, shifting the location of pulsar death in a manner which is seemingly incompatible with the observed pulsar population~\cite{Witte:CompanionPRL}. This work contains the details of the modeling of the electrodynamic processes underpinning those results. In particular, this article: presents a detailed derivation of the axion field profiles sourced from the neutron star, derives  an analytic understanding of how axion hair modifies the electrodynamics operating on the open field line bundle near the surface of the star, and develops dedicated 1D numerical particle-in-cell (PIC) simulations capable of capturing pair discharge processes in the presence of axion hair. These results lay the foundation for the work in ~\cite{Witte:CompanionPRL}, which demonstrates how modifications of pulsar electrodynamics can be used to constrain the axion parameter space. The framework described in this work, however, is more general and can be extended to any static field configuration around neutron stars.

This paper is organized as follows.
In Sec.~\ref{sec:sourcing} we review how dense nuclear matter alters the structure of the axion potential, and derive self-consistent neutron-star structure and the scalar-field solutions for the couplings and parameter space of interest. In Sec.~\ref{sec:pairdis} we derive a leading order estimate for how pulsar electrodynamics operates in the presence of axion hair, focusing specifically on how these axion gradients modify acceleration in the open field lines near the surface of the star. In Sec.~\ref{sec:pic} we outline the development of the 1D PIC simulations with external axion gradients, and demonstrate how the formation of vacuum gap and pair discharge is altered when such axion gradients are present. We conclude in Sec.~\ref{sec:summary}.

\section{The sourcing of axion hair by neutron stars}\label{sec:sourcing}

We start by reviewing how light QCD axions are sourced in nuclear matter, and derive the axion field profiles that we will use throughout this work (and our companion paper). Again, as already stressed in the introduction, our framework can be applied to different models producing static field gradient around neutron stars. 

We consider the axion, a pseudo-scalar field $a$ coupled to gluons
\begin{equation}
\mathcal{L} \supset \frac{1}{2}(\partial a)^2 + \frac{g_s^2}{32\pi^2} \, \frac{a}{f_a} G \tilde{G},
\end{equation}
where $f_a$ the decay constant, $g_s$ the strong coupling, and $G_{\mu\nu}$ the gluon field strength. Below the QCD scale, the axion obtains the potential~\cite{DiVecchia:1980yfw, GrillidiCortona:2015jxo}:
\begin{equation}\label{eq:lightQCDpotential}
V(a) = -\epsilon m_\pi^2 f_\pi^2 \left( \sqrt{1 - z_{\rm ud} \sin^2 \left(\frac{a}{2 f_a} \right)}  - 1\right),
\end{equation}
where $z_{\rm ud} = 4 m_u m_d / (m_u + m_d)^2$, and $\epsilon \le 1$ is a parameter to tune the axion lighter than naively expected. For symmetry-based models producing lighter QCD axions with $\epsilon < 1$ see e.g.~\cite{Hook:2017psm,Hook:2018jle, DiLuzio:2021pxd, DiLuzio:2021gos, Banerjee:2022wzk,Banerjee:2025kov}. Although the mass of these models is well described by  Eq. \ref{eq:lightQCDpotential}, some of them differ in their vacuum structure away from the origin, see  Appendix, Sec.~\ref{sec:Zn}. Similarly, a shift-symmetry breaking interaction between axions and nucleons is generated,
\begin{equation}\label{eq:axionnucleoninteraction}
\mathcal{L} \supset  \sigma_{\pi N} \bar{N} N \left( \sqrt{1- z_{\rm ud} \sin^2 \left( \frac{a}{2 f_a}\right)} - 1 \right),
\end{equation}
where $\sigma_{\pi N} \simeq 50$ MeV is the pion-nucleon sigma term. The interaction in Eq. \ref{eq:axionnucleoninteraction} can be interpreted as an axion-dependent nucleon mass. 

\begin{figure}
    \centering
    \includegraphics[width=\linewidth]{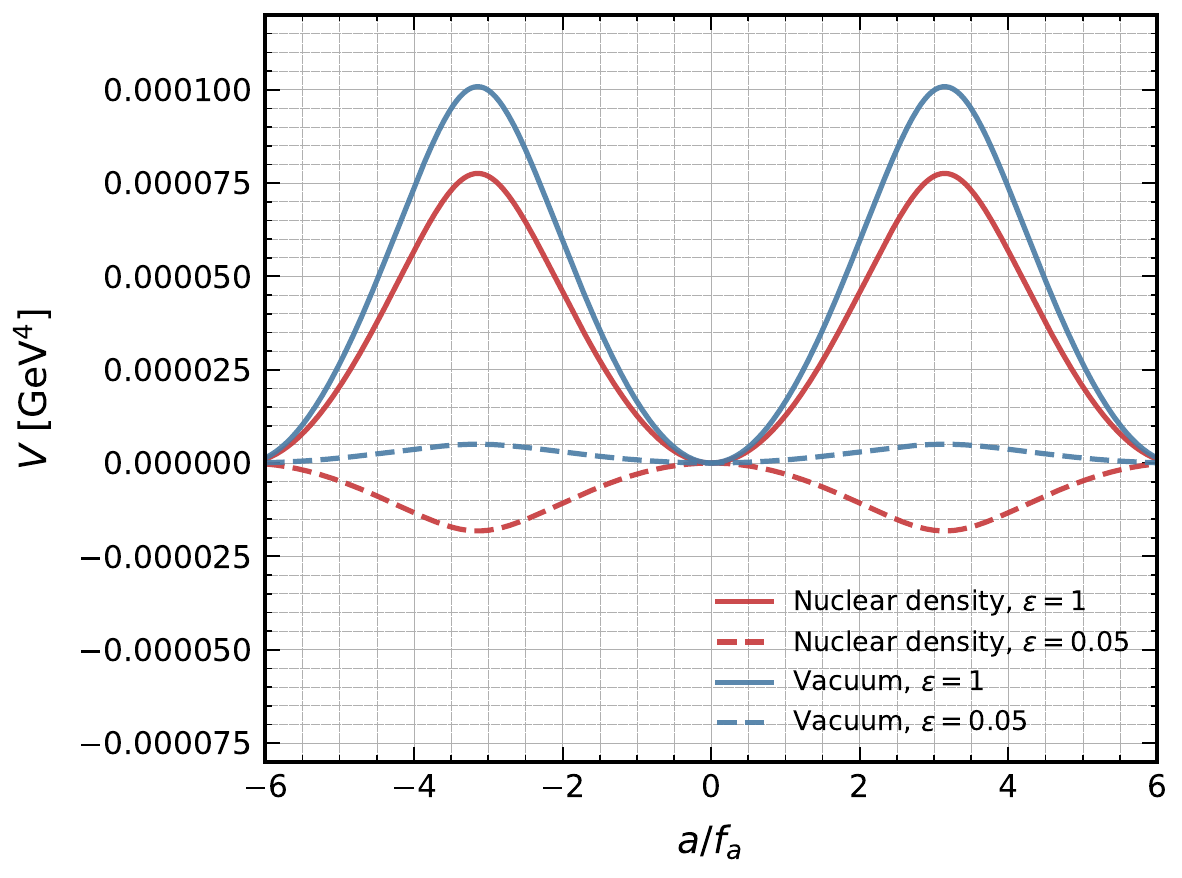}
    \caption{Potential of the QCD axion in vacuum (blue) and at nuclear density (red), for $\epsilon = 1$ (solid) and $\epsilon = 0.05$. For QCD axions that have been tuned sufficiently light $\epsilon \lesssim \mathcal{O}(0.1)$, the presence of dense matter flips the potential and causes the axion to settle at $a / f_a = \pm \pi$.}
    \label{fig:V_lqcd}
\end{figure}

\begin{figure}
    \centering
    \includegraphics[width=0.95\linewidth]{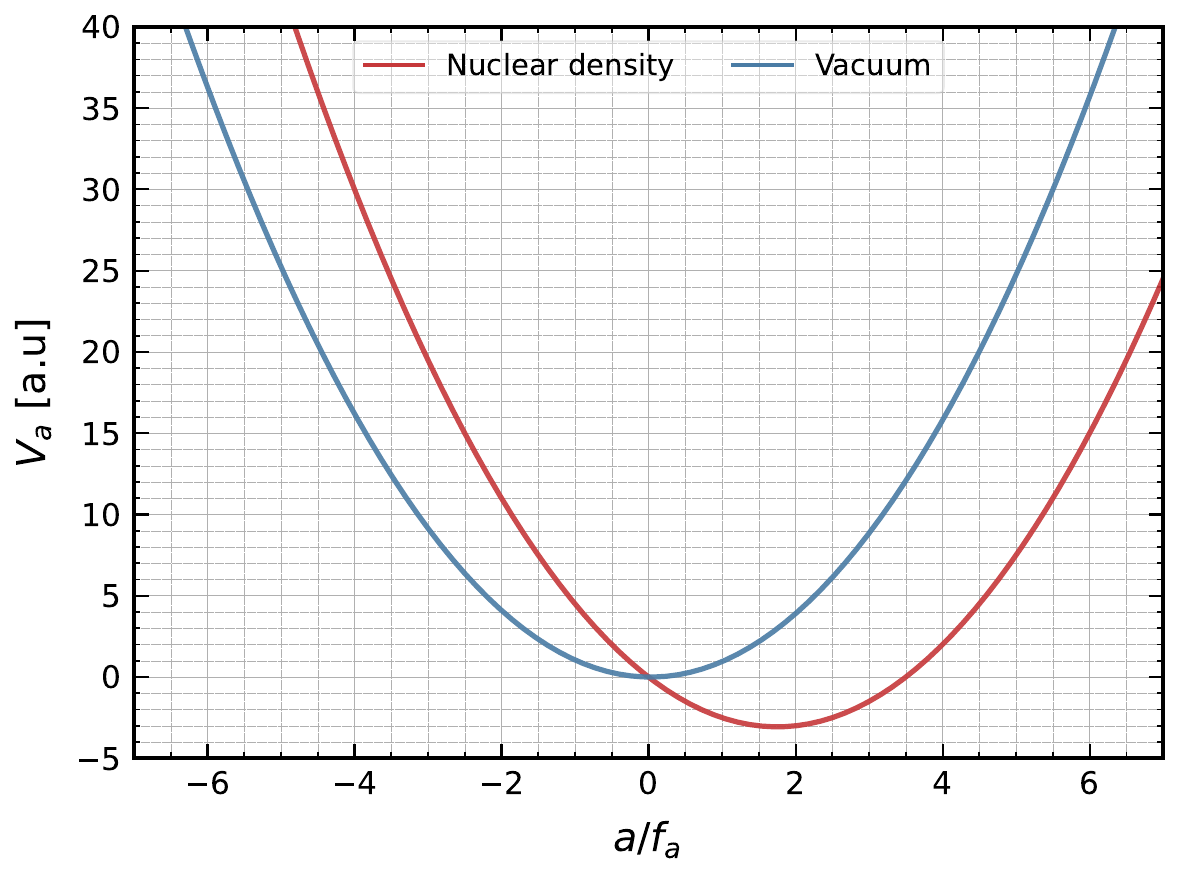}
    \caption{Potential (in arbitrary units) of linearly coupled axion in vacuum (blue) and at nuclear density (red).}
    \label{fig:LinPot}
\end{figure}

In a dense enough environment, with number density  $n \equiv \langle \bar{N} \gamma^0 N \rangle$, it can be energetically preferable for the axion field to be displaced from its vacuum minimum.
More quantitatively, the main task in order to find the axion field profile is to solve the coupled equations of gravity (TOV equations) and the static equation of motion of the scalar,
 \begin{widetext}
\begin{subequations}
    \label{eq:coupledTOV}
\begin{align}
    &a''\bigg[1-\frac{2GM}{r}\bigg]+\frac{2}{r}a'\left[1-\frac{GM}{r}-2\pi Gr^2\left(\varepsilon-p\right)\right]
    =\frac{\partial V(a)}{\partial a} + \rho_s \frac{\partial m_N(a)}{\partial a}, \\
    &p'=-\frac{GM\varepsilon}{r^2}\bigg[1+\frac{p}{\varepsilon}\bigg]\left[1-\frac{2GM}{r}\right]^{-1}\left[1+\frac{4\pi r^3}{M}\left(p+\frac{(a')^2}{2}\left\{1-\frac{2GM}{r}\right\}\right)\right]
    -a'\left[ \frac{\partial V(a,\rho_s)}{\partial a} + \rho_s \frac{\partial m_N(a)}{\partial a} \right],\label{eq:coupledTOV1}\\
    &M'=\,\,4\pi r^2\left[\varepsilon+\frac{1}{2}\left(1-\frac{2GM}{r}\right)\left(a'\right)^2\right]\label{eq:coupledTOV2},
\end{align}
\end{subequations}
\end{widetext}
where $m_N(a)$ is the effective field-dependent nucleon mass, i.e. the $a$-dependent prefactor of the $\bar{N} N$ term in the Lagrangian, and we impose
 boundary conditions for the axion satisfying $\lim_{r \to \infty} a(r,t) = 0$ and $ a'(0,t)=0 $. 
 The fact that there is no time dependence in the scalar EOM comes from a separation of time-scales: the axion settles in its ground state quickly compared to the life-time of the NS.
To gain some intuition for sourcing, we show the effective scalar potential, $V(a) + \rho_s m_N(a)$ for a given $\rho_s$, for both the light QCD axion in Fig. \ref{fig:V_lqcd} and the linearly coupled scalar in Fig. \ref{fig:LinPot}.
 
 These equations must be supplemented with an equation of state (EOS), which can, for example, be defined by the energy density $\epsilon$, pressure $p$, and scalar density $\rho_s$ as functions of number density $\rho$ and $a$,
 $\epsilon\left(\rho, a\right)$, $p\left(\rho, a\right)$, and $\rho_s\left(\rho, a\right)$.
 Even without a sourced axion, the equation of state at densities in neutron stars is unknown. We briefly describe how the presence of a sourced axion modifies the EOS by looking at the toy model of a free Fermi gas of neutrons. With a sourced axion, the energy density and pressure are given by
  \begin{subequations}
\begin{align}
\label{eq::epsandptot}
\varepsilon  &= \varepsilon_{\psi}(m_N(a),\rho)+V(a) \\ p &=p_{\psi}(m_N(a),\rho)-V(a) \,,
\end{align}
\end{subequations}
with
 \begin{subequations}
 \label{eq:eosffg}
\begin{align}
\varepsilon_{\psi}(m_N(a),\rho) &= 2\int^{k_F(\rho)}\frac{\mathrm{d}^3k}{(2\pi)^3} \sqrt{\mathbf{k}^2+m_N^{2}(a)}\,,
\\
p_{\psi}(m_N(a),\rho) &= \frac{2}{3}\int^{k_F(\rho)}\frac{\mathrm{d}^3k}{(2\pi)^3} \frac{k^2}{\sqrt{\mathbf{k}^2+m_N^{2}(a)}}\,,
\\
\rho_s(m_N(a),\rho) &= (\varepsilon_\psi-3p_\psi)/m_N(a)\,.
\end{align}
\end{subequations}
This toy model neglects all interactions, and in the presence of interactions the modifications of the EOS are more subtle: In particular for light QCD axions, also the pion gets lighter, thereby making the nuclear force more attractive.

Let us now focus on the scalar field profiles for light QCD axions.
The axion dependent nucleon mass is given by
\begin{equation}
 m_N(a) = m_N +   \sigma_{\pi N} \left( \sqrt{1- z_{\rm ud} \sin^2 \left( \frac{a}{2 f_a}\right)} - 1 \right).
\end{equation}
The task of solving the combined equations was performed in Ref.~\cite{Balkin:2023xtr}, where NS were described by a degenerate free Fermi gas coupled to gravity. More recently in Ref.~\cite{Kumamoto:2024wjd} it was studied in the limit where axion gradients are negligible, including effects of nuclear interactions using relativistic mean field theory.
Depending on the value of $\epsilon$, there are two types of solutions that one finds. In the case of the free Fermi gas, for $\epsilon\lesssim 0.07$, the axion is sourced all the way to the end of the star. This is because the axion profile has strong a back-reaction on the star, and leads to a new ground state of matter, where the axion is sourced and the energy-per-particle is less than that of separated neutrons. The star then ends abruptly with a minimal density where the axion is still sourced. 
On the other hand, for $\epsilon\gtrsim 0.07$, there is a phase transition from the sourced phase to the unsourced phase inside the star. However, the axion field cannot go to zero instantly, and there is some axion gradient leaking outside of the star. This can be especially important for large values of the axion decay constant. When including interactions, the critical value of $\epsilon$ at which the axion is sourced until the boundary of the star is expected to increase. This comes from the fact that the pion is lighter with the axion sourced, making nuclear interactions more attractive. For the largest values, $\epsilon \simeq 1$, it remains a point of speculation if the axion gets sourced inside NSs (see e.g. \cite{Balkin:2020dsr}), as calculability breaks down at densities before the sourcing happens. To be conservative, in this work we restrict ourselves to $\epsilon\lesssim 0.07$, in which case the axion is certainly sourced until the boundary of the star. 

In this case the field configurations found in Ref.~\cite{Balkin:2023xtr} are well approximated by
\begin{equation}\label{eq:axionprofile}
    a(r) = \begin{cases}
    \pm \pi \, f_a ,& \text{if } r < R_\text{star}\\
    \pm \pi \, f_a \, \frac{R_\text{star}}{r} \, e^{-\frac{\sqrt{\epsilon z_{ud}}\,m_\pi f_\pi}{2f_a}(r-R_\text{star})},              & r > R_\text{star},
\end{cases}
\end{equation}
where the characteristic length scale over which the axion field decays is set by the inverse in-vacuum mass,
\begin{equation}
    \lambda \equiv m_a^{-1} = \frac{2 f_a}{\sqrt{\epsilon z_{ud}}\,m_\pi f_\pi} \simeq 16 \, {\rm cm} \, \left(\frac{f_a}{10^{12} \, {\rm GeV}} \right) \left(\frac{10^{-2}}{\epsilon} \right)^{1/2} \, .
\end{equation}

In all our death line analysis we use the simplified field profiles of
Eq.~\ref{eq:axionprofile} which fit the full numerical solutions found in \cite{Balkin:2023xtr} well, up to $O(1)$ pre-factors. In particular, modifications to this form have a small impact on the observables we study, and are subdominant with respect to astrophysical uncertainties in NS population modeling. Note that there are two qualitatively different regimes depending on the axion mass (or equivalently wavelengths). For small axion masses the axion profile drops as $1/r$ in the region of interest, while for larger axion masses it drops exponentially fast. The transition between the regimes is when the axion wavelength is of order of the neutron star radius, $\lambda \simeq R_\text{NS} \simeq 10\text{km}$.

We now turn to the case of the linear interaction
\begin{eqnarray}
    \mathcal{L} \supset 
    - g_N a \, \bar{N} \, N \, ,
\end{eqnarray}
with $g_N \equiv \frac{m_N}{f_a}$ the linear scalar nucleon coupling. The linear coupling is tightly constrained by fifth force experiments. While the EOS of Eq. \ref{eq:eosffg} still applies, the modifications compared to the EOS without sourcing are small. Due to the small coupling, the back-reaction on the star is negligible, see discussion in \cite{Balkin:2023xtr}, and the TOV equations decouple from the scalar field equation.  In this case, to find the scalar field profile, one can find the time-independent solution of the scalar EOM in the high density background.  Here, the potential of the scalar field is more sensitive to the shape of the density profile in the neutron star, as the minimum appearing is continuously shifting from $\phi=0$. In order to compute the scalar field profiles we consider a realistic neutron star density profile, in particular we adopt the APR equation of state~\cite{Akmal:1998cf} and consider a NS with mass $1 M_\odot$. We also considered  different accretion histories on the NS crust, but our results are not sensitive to these changes.

With a density profile at hand we can solve the time-independent equation of motion for the field, which in spherical coordinates reads
\begin{equation}
    a'' + \frac{2}{r}a'  = m_a^2 a + g_N n_N,
\end{equation}
with boundary conditions $\lim_{r \to \infty} a(r,t) = 0$ and $ a'(0,t)=0 $. 
While this equation can be solved using the shooting method, it is most easily done by solving the time-dependent EOM and adding a small friction term, such that the field settles down in its ground state.
In general, \textit{outside the star}, we expect a solution of the form
$a \propto \frac{R_{\rm{star}}}{r}e^{-m_a (r-R_{\rm{star}})}$.
One can identify a few limits for this equation of motion.
For very small scalar field masses, the scalar field is sensitive to the density in the entire star. The prefactor of the scalar field solution is independent of the axion mass, as it is set by the axion gradient; for $m_a \lesssim 10^{-11} \, \rm eV$ the numerically found solution reads
\begin{equation}
a \simeq  1.5\times 10^{13} \, \rm{GeV}\,\left(\frac{g_N}{10^{-23}}\right) \frac{R_\text{star}}{ \, r} \exp[-m_a (r - R_\text{star})].
\label{eq:lin1}
\end{equation}

\begin{figure}
    \includegraphics[width = 0.45\textwidth]{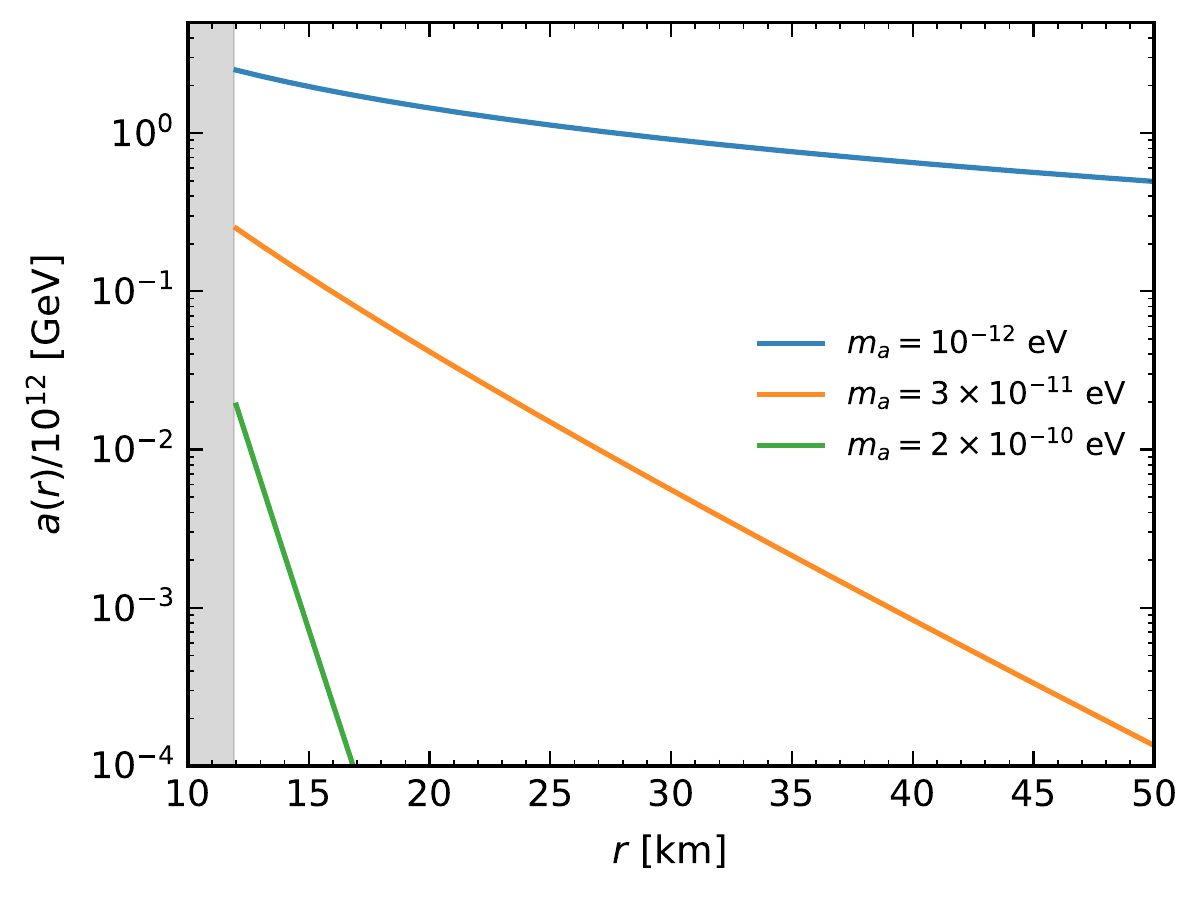}
    \caption{Axion profile for $g_N=10^{-23}$, and three different masses $m_a = 10^{-12},\, 3\times10^{-11}, 2\times10^{-10} \, \rm eV$, for the three different fitting formulas in Eq.~\ref{eq:lin1}-~\ref{eq:lin3}.}
    \label{fig:LinCouplProfiles}
\end{figure}

In the opposite limit, when the axion mass is large, $1/m_a \lesssim \rm km$, then the field starts to be very sensitive to the very external part of the NS profile, when the density is much smaller than in the inner few kilometers. The field profile outside the star then drops more quickly and we find that for $m_a \gtrsim 2 \times 10^{-10} \rm eV$ a better fit to the numeric solution outside the star is

\begin{eqnarray}
a &\simeq& 10^{11} \, \rm{GeV} \, \left(\frac{g_N}{10^{-23}}\right)\left(\frac{2 \times 10^{-10}\rm eV}{m_a}\right)^2 \nonumber \\ &\times & \frac{R_c}{r}\exp[-m_a (r - R_c)]
\label{eq:lin2}
\end{eqnarray}
with $R_c \simeq 10.5 \, \rm km $. We numerically verified this fit for masses up to $10^{-9} \, \rm eV$, where the field value outside the star is already quite small.

\begin{figure}
    \includegraphics[width = 0.5\textwidth]{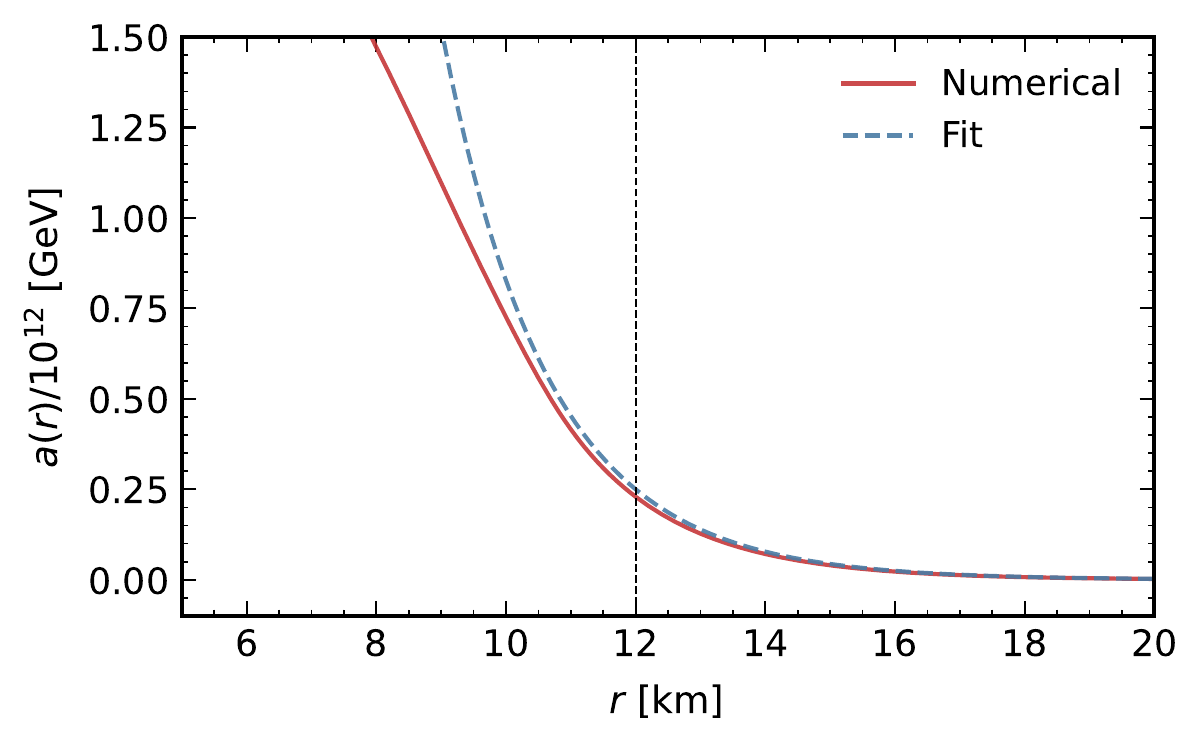}
    \caption{Axion profile for $m_a = 10^{-10} \, \rm eV$, $g_N = 10^{-23}$. The black solid line is the full numerical solution, while the dashed black line is the approximation in Eq.~\ref{Eq:SolIntermediate}.}
    \label{fig:test}
\end{figure}

When the inverse of the scalar mass is comparable to the size of the entire star, that is to say $m_a \sim \mathcal{O}(\text{few})10^{-11}-10^{-10} \, \rm eV$, the solution is more sensitive to the density profile of the star.
We find that the solution outside the star is well approximated by the analytical form 
\begin{eqnarray}\label{Eq:SolIntermediate}
    a &\simeq& 2.5 \times 10^{11} \, \rm{GeV} \, \left(\frac{g_N}{10^{-23}}\right)\left(\frac{10^{-10} \, \rm eV}{m_a}\right)^2  \nonumber \\ &\times &  \frac{R_\text{star}}{r} \exp[-m_a (r - R_\text{star})],
\label{eq:lin3}
\end{eqnarray}
where $R_\text{star} = 12 \, \rm km$ and the numerical prefactor have been fixed using the APR density profile. In Fig.\ref{fig:LinCouplProfiles}, we also explicitly show, for reference, the axion profiles corresponding to the three different fitting formulas. Furthermore, in Fig.~\ref{fig:test}, we show the numerical solution (black solid line) and the analytical approximation for the profile outside the star (black dashed line) for an axion mass of $10^{-10},\mathrm{eV}$. A similarly good agreement is found for the other fitting formulas we provide. 

In all the field profiles, the coupling $g_N$ has been taken to roughly saturate the bound from the weak equivalence principle (WEP), where we used Fig.~1 of Ref.~\cite{OHare:2020wah}. %

\subsection{The special case of the $\mathbb{Z}_N$ axion}\label{sec:Zn}
So far we have focused on the minimal light QCD axion, for which the vacuum potential has a unique minimum (modulo the $2\pi$ periodicity) and finite-density effects inside a neutron star can displace the field from that vacuum value, sourcing a nonzero external gradient.
However, it is useful to briefly comment on models in which the axion remains parametrically light because its potential is multi-branched, as in the $\mathbb{Z}_N$ axion constructions discussed in Refs.~\cite{Hook:2018jle,DiLuzio:2021pxd}.
In such scenarios the zero-density potential contains a tower of local minima between $a=0$ and $a=\pi f_a$, which we can label by $k=1,\dots,(N+1)/2$.
At finite density, the ordering of these minima can change, so that a metastable branch ($k>1$) is energetically preferred over part of the stellar interior, leading to an internal phase boundary where the field interpolates between branches.
Outside the star the profile is again controlled by the radius at which the preferred minimum reverts to the vacuum branch (set by the local nucleon density), so the near-zone gradient entering the magnetospheric electrodynamics is well captured by the sourcing estimates used in the main text.
A major difference compared to the case of the potential of Eq. \ref{eq:lightQCDpotential} is that for the $\mathbb{Z}_N$ axion for $N < 33$, the structure of the star is only significantly modified at the densities around the phase transition, while the outer layers of the star remain mostly unaltered, see also \cite{Gomez-Banon:2024oux}. Thus, the bounds derived there and in \cite{Kumamoto:2024wjd} are not valid for the $\mathbb{Z}_N$ model.
However, this is only true in the neglible gradient limit, and we will now work out the position of the axion brane, and it's effect on the envelope. 
To estimate this, a simple energetics argument suffices. The axion brane tension scales as
\begin{equation}
    \sigma \simeq \frac{\sqrt{\epsilon(N)} m_\pi f_\pi }{N} f_a,
\end{equation}
and thus the energy required to change the position at which the brane is located from $R$ to $R+\Delta R$ is
\begin{eqnarray}
    \Delta E_\sigma =  8 \pi R \sigma \Delta R \simeq 8\pi \frac{\sqrt{\epsilon} m_\pi f_\pi }{N} f_a R_{NS} \Delta R.
\end{eqnarray}
This change in energy needs to be provided by the mass change in nucleons, given by
\begin{eqnarray}
    \Delta E_N \simeq 4 \pi R_{NS}^2 \Delta m_N n \Delta R .
\end{eqnarray}
Equating both changes in energy gives us the density at which the axion brane is located,
\begin{eqnarray}
   n \simeq \frac{2 \sqrt{\epsilon} m_\pi f_\pi f_a}{\Delta m_N N R_{NS}} \sim 10^{34} \text{cm}^{-3} \Big(\frac{f_a}{10^{13}\text{GeV}}\Big)\frac{\sqrt{\epsilon}}{N}
\end{eqnarray}
where we used 
\begin{equation}
\Delta m_N(a=\pi) = - \sigma_{\pi N} \left( \sqrt{1- z_{\rm ud} \sin^2 \left( \frac{\pi}{2}\right)} - 1 \right) \simeq 33 \, \text{MeV},
\nonumber
\end{equation}
and we normalised to typical number densities in the envelope region of neutron stars~\cite{1983ApJ...272..286G}.
The axion brane pushes nucleons to this density, and the star ends abruptly. Thus if this density is of the order of envelope densities, the backreaction on the envelope is large, and cooling bounds in \cite{Gomez-Banon:2024oux} should apply. As a naive estimate, we assume that if this number density is larger than typical number densities in the envelope, the backreaction on the envelope is large enough such that the cooling bounds are actually valid. 

Let us now comment on the bounds derived in our companion paper. There are two scenarios, depending on the values of $f_a$ and $N$. For small values of the brane tension, it has negligible backreaction on the envelope. In this case, the axion brane starts \textit{outside} the envelope, and thus the bound derived in this work is also valid for the $Z_N$ case. On the other hand, for large values of the brane tension, the brane starts well within the envelope. However, it has a strong backreaction on the form of the envelope, and in particular pushed nucleons inside. Even in this case, an $O(1)$ fraction of the brane will be \textit{outside} the envelope, and thus our bounds are unaffected by this vacuum structure (although, as mentioned in the main text, subtlties nonetheless arise in the heavy axion limit).
Since most of the transition of axion from $a = \pi ((N-1)/2)$ towards zero in any case lies outside of the star, the bounds derived in this work coming from modifying the electrodynamics outside the star, and not the structure of the star itself, apply for all values of $N$ where the axion is sourced inside the star. Assuming a free fermi gas of nucleons for neutrons, and trusting the description until nuclear density, we find that for all $N\ge3$ sourcing happens inside NSs, and our bound applies. If the vanilla QCD axion with $N=1$ is sourced inside NSs is an open question, see e.g. \cite{Balkin:2020dsr,Balkin:2023xtr,Kumamoto:2024wjd}. However, in this case the vacuum structure is different from the $Z_N$ case with $N\ge3$, and so is the backreaction on the star. This will be discussed in the following.

\section{Pulsar Electrodynamics}
\label{sec:pairdis}

Having derived the scalar profiles we will be using and described the models of interest, we now start outlining the conventional picture of pulsar electrodynamics and providing a brief analytic example to illustrate the effect of axion hair. 

For pulsars with a sufficient supply of charges (corresponding to pulsars sufficiently far from the death line), the magnetosphere should approach the so-called force-free solution, which corresponds to the stable plasma configuration in which charges experience no net force, \ie
\begin{eqnarray}\label{eq:ffe}
 m n \frac{d(\gamma \vec{v})}{dt} = \rho \vec{E} + \vec{j} \times \vec{B} \simeq 0 \, .
\end{eqnarray}
This condition is equivalent to demanding $\vec{E}\cdot \vec{B} = 0$ together with 
$\partial_t (\vec{E}\cdot \vec{B}) = 0$. In other words, charges (which are confined to flow along magnetic 
field lines) have fully screened the component of the 
electric field capable of driving particle acceleration. Force-free electrodynamics (FFE), however, cannot be a perfect description of the full magnetosphere. From an observational perspective one can conclude that acceleration (and therefore nonzero values of $E \cdot B$) must be present in order to generate radiation. 
Nevertheless, deviations from the FFE regime must be rather limited, as even modest deviations in the near-field regime become unstable to pair cascades which quickly re-drive the system towards the FFE limit; consequently, FFE dynamics serve as a natural starting point for understanding the charge distribution and current flows in pulsar magnetospheres.

In order to be concrete, let us begin with an illustrative example. We will focus on the well-known case of Michel's split monopole~\cite{michel1973rotating} (note that this is one of the few configurations that admits analytic solutions) which describes an axi-symmetric aligned monopolar field configuration with a sign flip about the equatorial plane (this cannot represent the near-field magnetospheric geometry of realistic pulsars, but is expected to be qualitatively similar to the far-field configuration near, and beyond, the light cylinder), \ie
\begin{eqnarray}\label{eq:B_sm}
    B_r &=& B_0 \left(\frac{r_{\rm NS}}{r} \right)^2 {\rm sign}(\pi/2 - \theta) \\
    B_\theta &=& 0 \\
    B_\phi &=& - \Omega \, r \, \sin\theta \, B_r \, .  
\end{eqnarray}
In the FFE limit, the electric field can be determined by setting the field to zero in the co-rotating reference frame, and transforming back to the reference frame of the pulsar, \ie $\vec{E} = - (\vec{\Omega} \times \vec{r}) \times \vec{B}$, which for the case of the split monopole in Eqns.~\ref{eq:B_sm} yields $E_r = E_\phi = 0, E_\theta = -\Omega r \sin\theta B_r$. The current density in the FFE limit can be derived either by combining Eq.~\ref{eq:ffe} with Maxwells' equations, or by deriving the necessary condition for $\vec{E} \cdot \vec{B} = 0$ and $\partial_t (\vec{E} \cdot \vec{B}) = 0$. Writing the current in terms of the parallel and perpendicular components, $\vec{j} = \frac{(\vec{j} \cdot \vec{B})}{|B|^2}\vec{B} + (\vec{j} - \frac{(\vec{j} \cdot \vec{B})}{|B|^2}\vec{B})  \equiv j_{||} \frac{\vec{B}}{|B|} + \vec{j}_\perp$, one finds the FFE constraint demands
\begin{eqnarray}\label{eq:jffe}
    j_{||} &=& \frac{1}{B} \left[ (\nabla \times \vec{B}) \cdot \vec{B}
    \;-\; \vec{E} \cdot (\nabla \times \vec{E}) \right] \\
    \vec{j}_{\perp} &=& \frac{ - (\vec{j} \times \vec{B}) \times \vec{B} }{B^2}
    \;=\; \frac{ (\nabla \cdot \vec{E}) \, \vec{E} \times \vec{B} }{B^2}
    \;=\; \frac{ \rho \, (\vec{E} \times \vec{B}) }{B^2} \, \nonumber .
\end{eqnarray}

Inserting the split monopole field configuration leads to a current configuration (away from the equatorial plane, where the sign flip yields a discontinuity) with an amplitude $j_r = -2 \Omega \cos\theta B_r = \rho_{\rm GJ}$, where $\rho_{\rm GJ}=\nabla\cdot \vec{E}$ is the co-rotation charge density.

Thus far we have worked in the flat space limit. This result can be generalized to include relativistic effects (see \eg~\cite{Gralla:2016fix,Gralla:2017nbw}), where the metric outside the star $r > r_{\rm NS}$ is given by
\begin{eqnarray}
    ds^2 &=& \left(1 - \frac{2 M}{r} \right) dt^2 + \left(1 - \frac{2 M}{r} \right)^{-1} dr^2  \nonumber \\[10pt] &+& r^2 \left[ d\theta^2 + \sin^2\theta (d\phi - \Omega_Z dt)^2 \right]  \, .
\end{eqnarray}
Here, $\Omega_z \equiv 2 \hat{I} \Omega / r^3$ is the frame-drag frequency, with $\hat{I}$ being the moment of inertia. For the split monopole, the ratio of the radial current to the GJ charge density is now given by
\begin{equation}
   \frac{j_r}{\rho_{\rm GJ}} = \frac{\left(1 - \frac{2 M}{r} \right)}{  (1 - \frac{\Omega_z}{\Omega})} \, . 
\end{equation}
For acceptable values of $M$, $r_{\rm NS}$, and $\hat{I}$, one finds that $j_r / \rho_{\rm GJ} > 1$ (in contrast to the flat space limit in which $j_r / \rho_{\rm GJ}  = 1$).

A useful ratio for understanding the electrodynamics of pair discharges is the so-called ``discharge parameter" (see \eg~\cite{TimokhinArons2013}, and also~\cite{Caputo:2023cpv} for a discussion on how dense, dynamic, axion field configurations can alter this behavior), given by:
\begin{equation}
  \alpha_0 \equiv {\rm sign}(j_r) \frac{|\vec{j} \cdot \vec{b}| }{\rho_{\rm GJ}} \equiv \frac{j_{||}}{\rho_{\rm GJ}} \, ,
\end{equation}
where $\vec{b}$ is a unit vector along the B field. The importance of this parameter emerges when considering the charge distribution along a field line near the neutron star.  It determines whether the required current $j_{||}$ can be supplied while 
also screening the local electric field: only for $0 \le \alpha_0 < 1$ is this simultaneously achievable. For $\alpha_0 \ge 1$, a potential drop develops, leading to particle acceleration and pair discharges.

\begin{figure*}
    \includegraphics[width=0.45\textwidth]{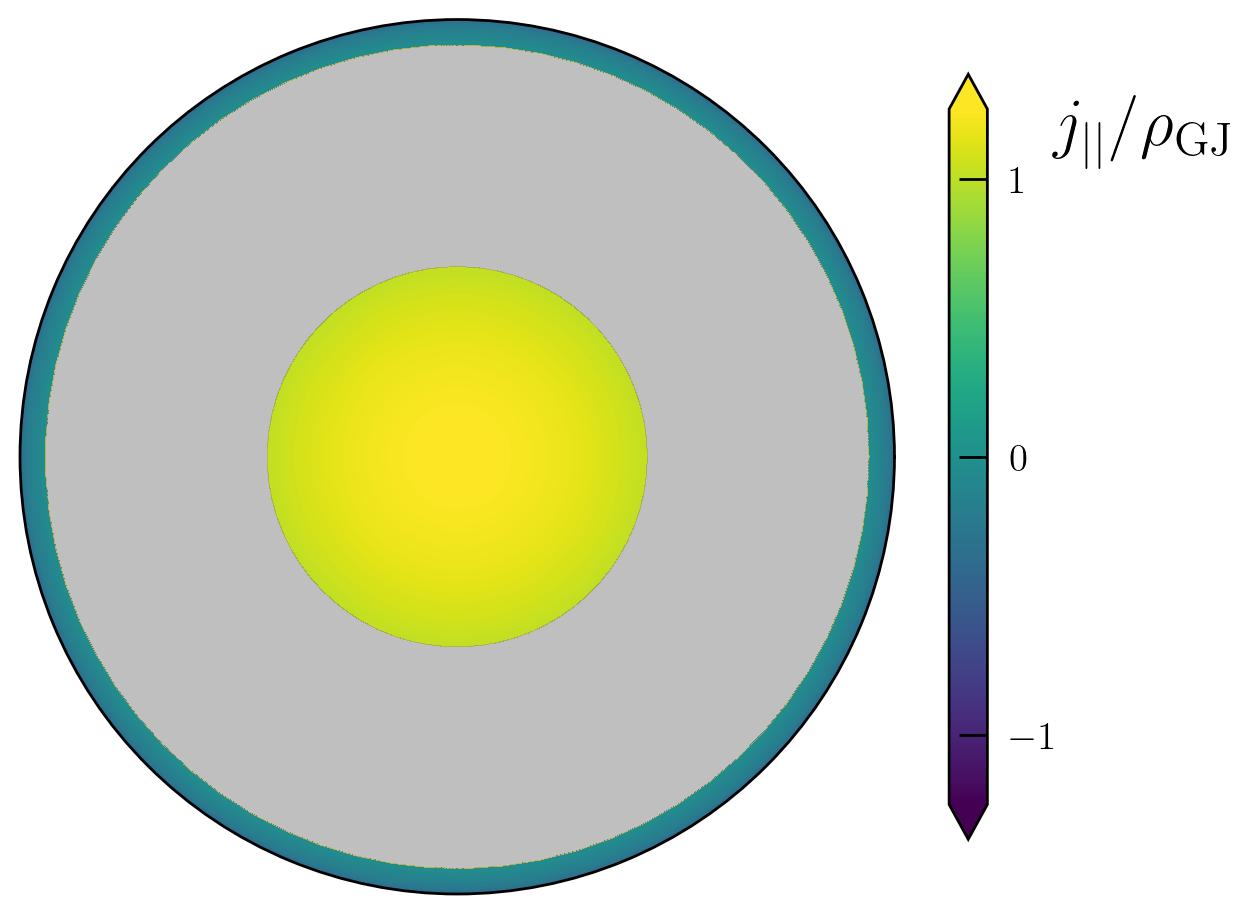}
    \includegraphics[width=0.45\textwidth]{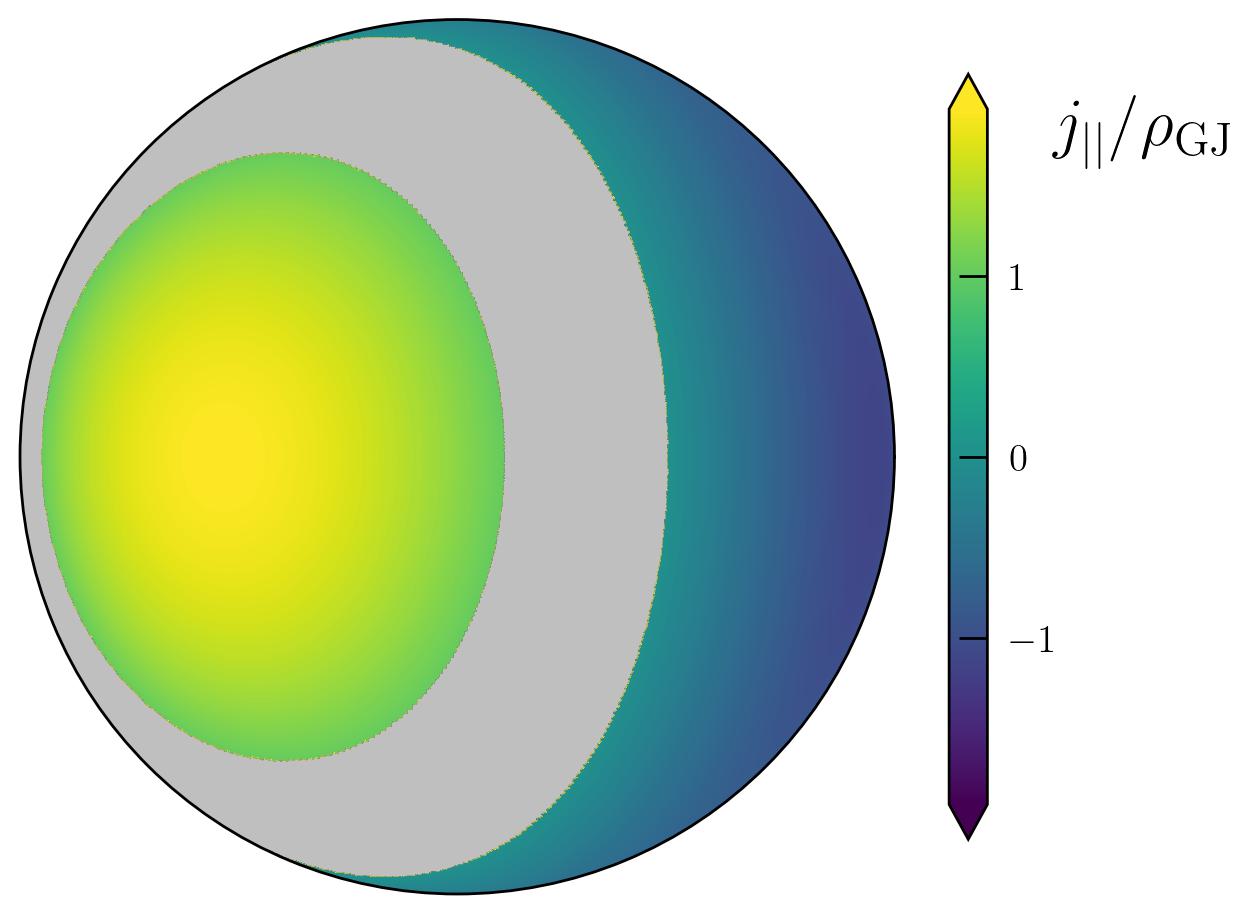}
    \caption{Discharge parameter across the polar cap for a dipolar field with a misalignment angle $\chi = 0^\circ$ (left) and $\chi = 60^\circ$ (right), obtained using the fitting formulas of~\cite{Gralla:2017nbw} (computed in the force-free limit). We have set the color scheme to gray for values of the discharge parameter between $0 \leq \alpha_0 < 1$ in order to emphasize that these field lines do not support discharges.}\label{fig:current}
\end{figure*}

\begin{figure*}
    \centering
     \includegraphics[width=.8\textwidth]{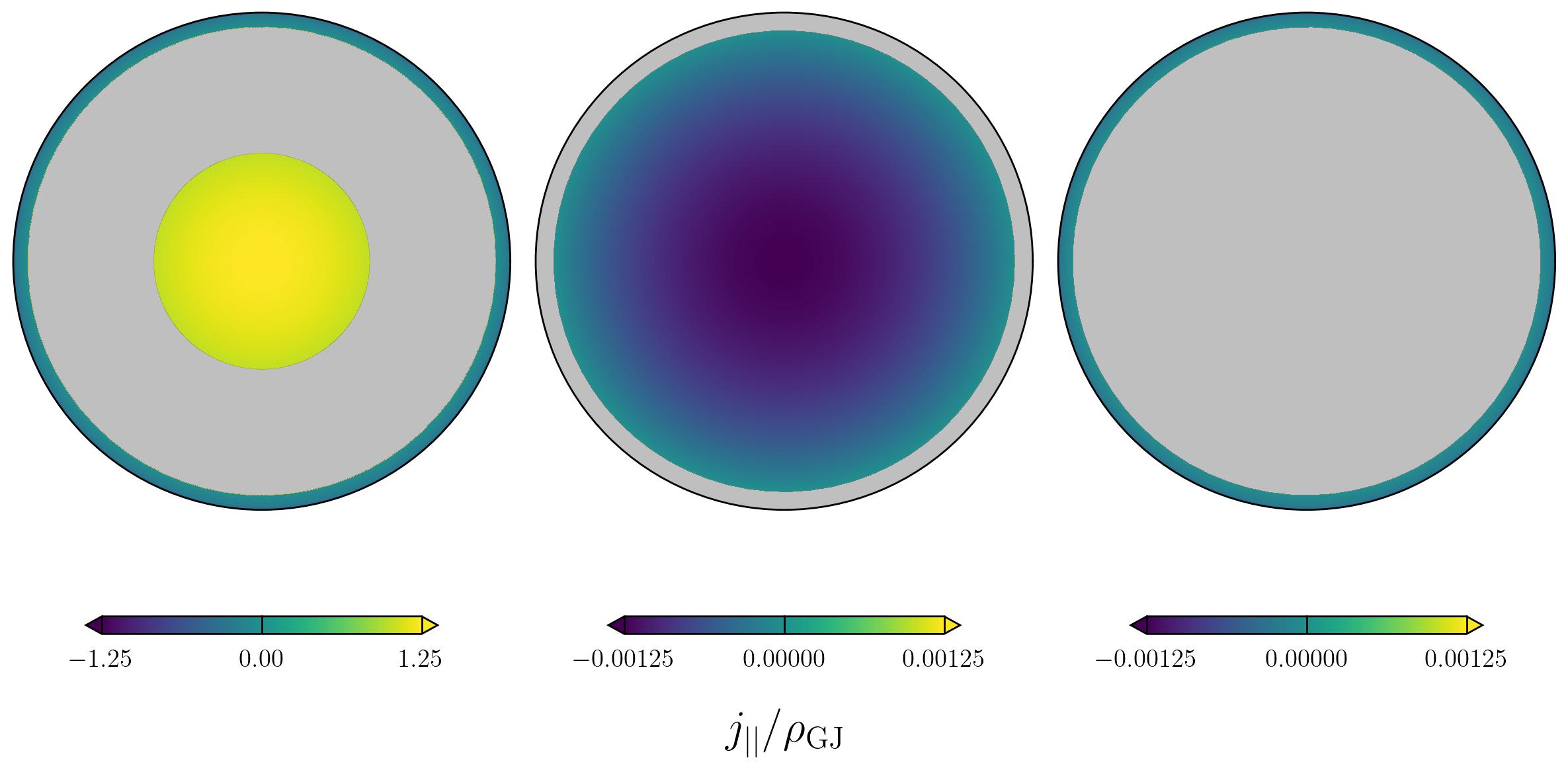}
     \includegraphics[width=.8\textwidth]{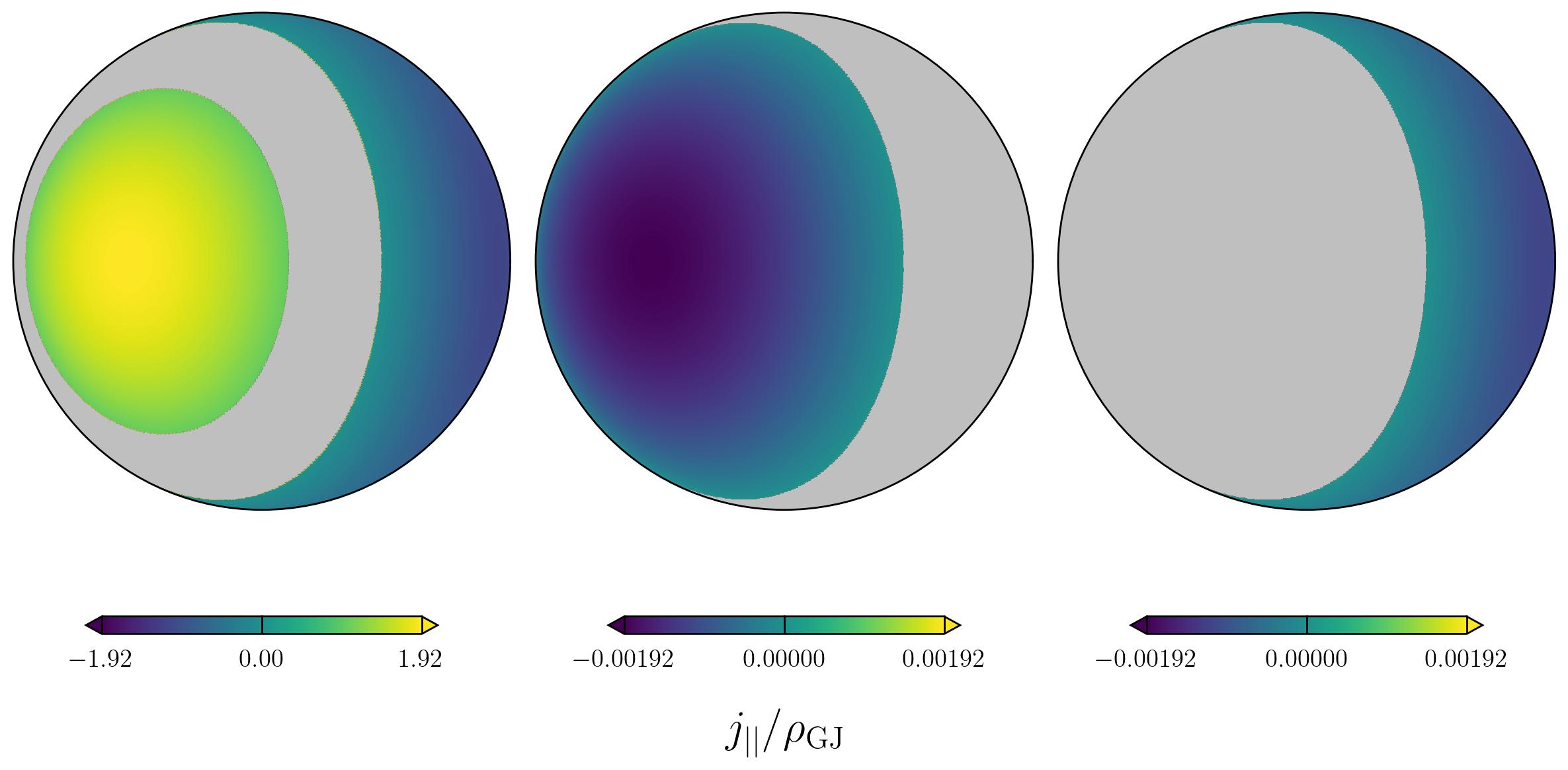}
    \caption{Value of discharge parameter for an aligned force-free dipole without an axion gradient (left), with an axion charge density $\rho_a / \rho_{\rm GJ} = -10^3$ (center), and with $\rho_a / \rho_{\rm GJ} = 10^3$ (right).  Top panel corresponds to $\chi =0^\circ$, bottom panel corresponds to $\chi = 60^\circ$.}
    \label{fig:current_asym}
\end{figure*}

Returning to the split monopole, we see that relativistic corrections drive 
$\alpha_0 > 1$, implying that pair discharges must appear. The split monopole is, however, 
somewhat unphysical, as the non-zero divergence of the magnetic field requires a current 
sheet to extend all the way to the stellar surface. Force-free solutions with dipolar field 
configurations must therefore be obtained numerically, and many groups have presented 
consistent results over the years (see, \eg, \cite{timokhin2006force,Spitkovsky2006,
kalapotharakos2009three,petri2012pulsar}). Dipolar fields differ fundamentally from monopolar configurations in that a large fraction 
of the field lines emerging from the stellar surface are closed, reconnecting with the star 
at both ends. These closed field lines do not carry any twist, $\nabla\times \vec{B} = 0$, and 
thus support stable configurations where there is no parallel current $j_{||}=0$, and with $E_{||}=0$. The open field lines 
originating in the polar-cap region, however, share many qualitative similarities with the 
split-monopole case. In particular, the discharge parameter, $\alpha_0$, is of order unity, with relativistic frame dragging producing an overall upward shift. Recall that the distribution of parallel currents $j_{||}$ is set by the \emph{global} 
geometry of the magnetosphere---that is, by the twist of the magnetic field lines near the light cylinder. References~\cite{Gralla:2016fix,Gralla:2017nbw} provide numerical fits to the 
spatial distribution of the discharge parameter for dipolar configurations, given by

\begin{multline}
    \frac{j_{||}}{\rho_{\rm GJ}} \simeq \frac{1}{1-\Omega_Z / \Omega} \left[J_0(2 \arcsin(r / \sqrt{\zeta_0})) \right. \nonumber \\[10pt] \left. + J_1(2 \arcsin(r / \sqrt{\zeta_0}))\tan\chi \cos\phi \right] \, , \vspace{-1cm}
\end{multline}
where $J_{0,1}$ are Bessel functions, $\Omega_Z / \Omega \simeq 2 C / 5$ captures the Lense-Thirring effect (with compactness $C = 2 M / R_{\rm NS} \sim 0.5$), $\zeta_0$ defines the size of the polar cap, $r \equiv r_\perp / R_{\rm NS}$ gives the distance from the center of the polar cap, and $\phi$ gives the azimuthal angle around the magnetic polar axis, and $\chi$ is the inclination angle between the rotational and magnetic axes. The distribution of the discharge parameter in the polar cap region of a dipolar
magnetosphere for two representative inclination angles is shown in Fig.~\ref{fig:current}. For completeness, in Fig.~\ref{fig:current_asym} we show the shift in the discharge parameter induced by axion hair.

Let us now return to the question of how a large, static, axion field gradient modifies this picture. Let us start by noting that Maxwell's equations in the presence of axions take the generalized form:
\begin{align}
\nabla \cdot \vec{E} &= \rho - g_{a\gamma\gamma} \, \vec{B} \cdot \nabla a , \label{eq:GaussApp} \\
\nabla \times \vec{B} - \partial_t \vec{E} &= \vec{j} + g_{a\gamma\gamma} \, \dot{a} \, \vec{B} + g_{a\gamma\gamma} \, (\nabla a \times \vec{E}) \, . \label{eq:curlBApp}
\end{align}
In the case of axion hair, one use spherical symmetry to freely set $\partial_t a \simeq 0$. We can then use these modified equations to  generalize the result of Eq.~\ref{eq:jffe}; the result is the following:
\begin{eqnarray}
    j_{||} = j_{||}^{\rm std} - \frac{1}{B} \left[g \vec{B} \cdot (\nabla a \times \vec{E})\right]\\
    \vec{j}_{\perp} = j_{\perp}^{\rm std} + \frac{(g \vec{B} \cdot \nabla a) \vec{E} \times \vec{B}}{B^2} \, .
\end{eqnarray}
Let us note that this result could have been read directly from the modified form of Amp\`{e}re's law. Similarly, the modification to $\rho_{\rm GJ}$ can be directly inferred by transforming Gauss' law to the co-rotating reference frame -- this procedure yields $\nabla \cdot \vec{E} = \rho - \rho_{\rm GJ} - g_{a\gamma\gamma} \vec{B} \cdot \nabla a$, implying one can introduce a re-definition of the GJ charge density as $\rho_{\rm GJ}^{\rm eff} \simeq \rho_{\rm GJ} + g_{a\gamma\gamma} \vec{B} \cdot \nabla a$. One can immediately see that the modification of the current density is of higher order, since 
$|E| \ll |B|$ within the light cylinder (while at large distances the axion field is exponentially 
suppressed\footnote{For axion masses $m_a \lesssim 10^{-15}\,\mathrm{eV}$, the exponential 
suppression sets in beyond the light cylinder, and one might wonder whether such light axions 
could induce a non-negligible modification to the current density. The short answer is no: the axion field also carries a $(R_{\rm NS}/r)$ pre-factor, ensuring that its contribution becomes negligible at 
large distances.})---this makes clear that the dominant effect of the axion configuration is a direct modification of the potential drop near the stellar surface.

\begin{figure}
    \centering
    \includegraphics[width=0.9\linewidth]{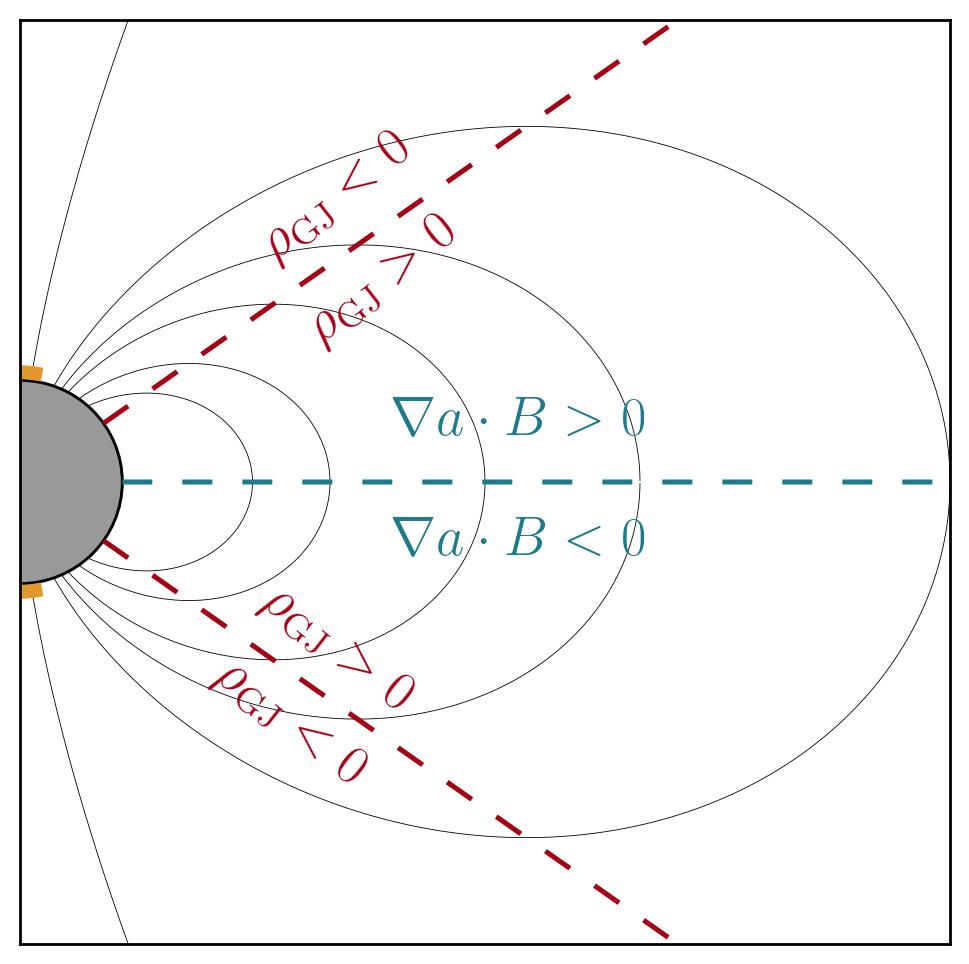}
    \caption{ 
    Side view of aligned pulsar, with polar caps highlighted in yellow (their sizes have been enlarged in order to make them visible). Magnetic field lines are shown with thin black lines. Null surfaces, corresponding to surfaces with $\rho_{\rm GJ} = 0$ (corresponding to the boundaries which separate positively and negatively charged regions), are shown with dashed red lines\footnote{In the example shown, we have assumed $\vec{B} \cdot \vec{\Omega} > 0$, corresponding to an aligned magnetic field and rotational axis. One could alternatively consider an anti-aligned system, and the characteristic charge densities would flip.}. The parity asymmetry of the axion contribution in Gauss' law is highlighted with the blue dashed line. }
    \label{fig:parity}
\end{figure}

\begin{figure}
    \centering
    \includegraphics[width=\linewidth]{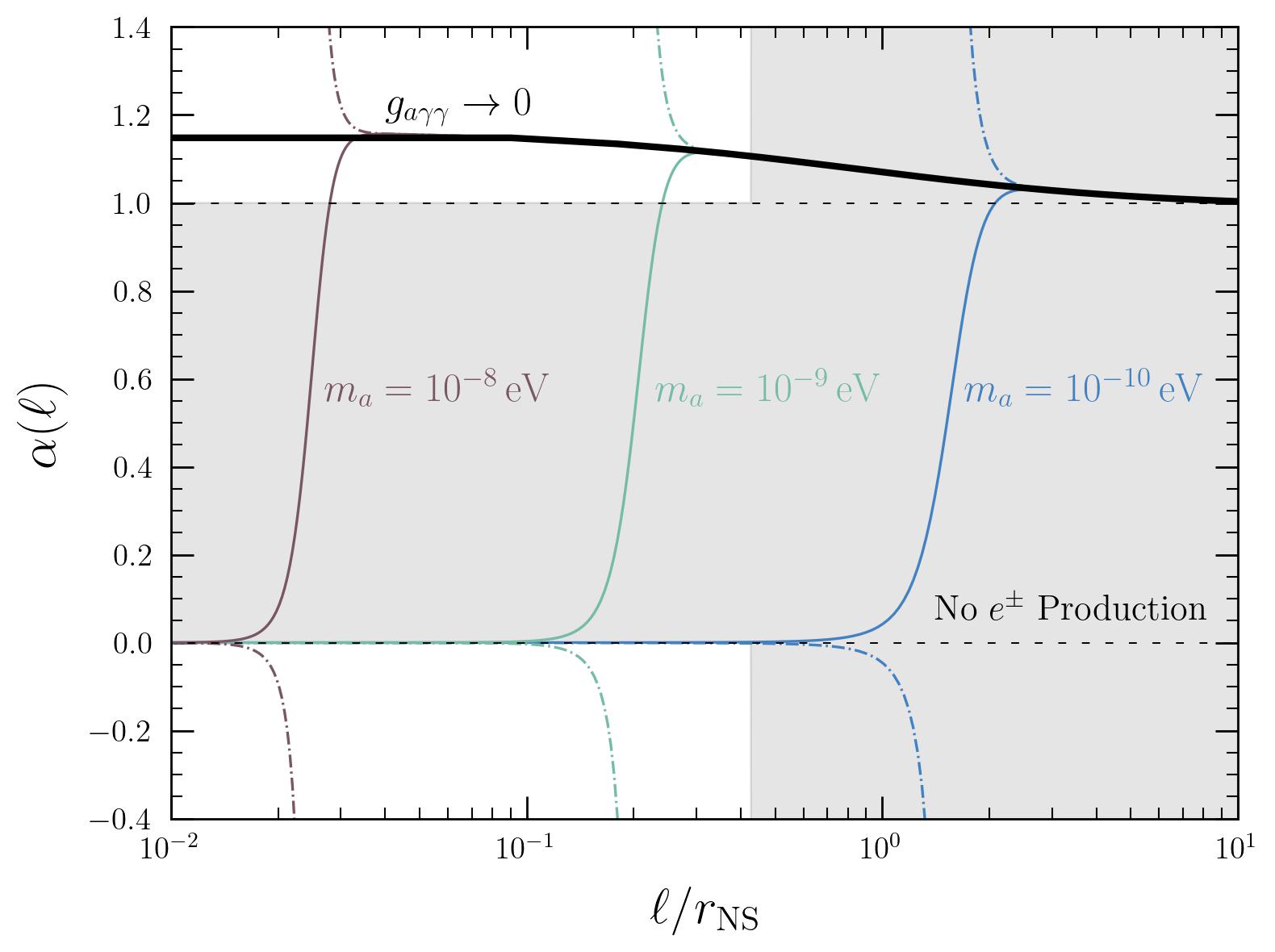}
    \caption{Discharge parameter, $\alpha_0$, as a function of distance $\ell$ along a field line (defined such that $\ell \rightarrow 0$ corresponds to the magnetic field footprint on the stellar surface), for a field line carrying out-flowing current (in the absence of axion), $\alpha_0>0$. Results are shown for a surface magnetic field value $B_0 = 10^{12}$ G, a rotational frequency $P = 2 \, {\rm s}$, an alignment angle $\chi = 20^\circ$, a neutron star mass and radius $M_{\rm NS} = 1.4 \, M_\odot$ and $r_{\rm NS} = 11$ km, and a magnetic field footprint situated at an angle $\theta_m = 0.1 \times \theta_{\rm pc}$ with respect to the magnetic axis, where $\theta_{\rm pc}$ is the angle defining the boundary of the polar cap.  In the standard scenario (\ie the limit in which $g_{a\gamma\gamma} = 0$, shown with a thick black line), $\alpha_0 > 1$ at the stellar surface on both magnetic poles, implying an unscreened electric field will appear and pair production will ensue. When an axion is included, an asymmetry is induced between the two poles, and $\alpha_0$ is driven to $0^\pm$ near the surface. For the pole with $\alpha_0 < 0$ (dot-dashed), a null surface forms close to the star. Close to the star, the current-carrying charges are unable to screen the electric field, and pair production will ensue. In the opposite pole (solid, colored), the axion enhances the GJ charge density, leading to a dense cloud flowing at non-relativistic velocities that efficiently screens the electric field. This screening remains effective until $\ell \gtrsim m_a^{-1}$, beyond which the axion gradient becomes exponentially suppressed and particle acceleration can occur, as in the axion-free case. Note that in this regime the accelerating potential is comparable to that without axions, and thus pair production is not triggered in old pulsars located below the death line. Should $\alpha_0$ transition above unity when the magnetic field is still strong, pair production will take place (in this case, at a displaced distance from the stellar surface); for sufficiently light axion masses, however, this transition takes place far away from the star, where the magnetic field is weak, which would not allow pair production. The region where $e^\pm$ pair production no longer takes place is roughly highlighted in gray (for the large radial region, we estimate this threshold by defining a new `effective surface magnetic field' $B(r)$, and identifying the radial distance for which $\dot{P}\left(B(r)\right)$ becomes equivalent to the standard death line in Fig. 1 of the companion paper~\cite{Witte:CompanionPRL}).   }
    \label{fig:alpha_radial}
\end{figure}

In case of the split monopole, the leading order contribution from the axion gradient to the discharge parameter can be computed analytically; in the near-field regime, $r \Omega \ll 1$, and away from the equatorial plane, one finds (in the flat space limit)
\begin{eqnarray}\label{eq:splitM_alp}
    \alpha_0 \simeq \frac{2 \Omega}{2\Omega - \sec\theta g_{a\gamma\gamma} \partial_r a } \, .
\end{eqnarray}
In the limit where the axion field gradient dominates ($g_{a\gamma\gamma} \partial_r a  \gg 2 \Omega \cos\theta$), we see that $\alpha_0 \rightarrow \pm \epsilon$, with $\epsilon > 0$. Here, the relative sign depends on whether one is in the upper or lower hemisphere -- this asymmetry, discussed in the main text, is shown more clearly for an aligned dipolar field configuration in Fig.~\ref{fig:parity}. In one of the poles, the axion-induced modification yields $\alpha_0 < 0$, enabling particle acceleration and subsequent pair discharges. At the opposite pole, screening remains efficient, and particles are not accelerated. The assumption, however, that the axion field gradient is dominant, is a local statement -- $\partial_r a$ scales as $e^{-m_a r} (1 + m_a r)/ r^2$, and thus will necessarily become sub-dominant at some radii $r > r_{\rm NS}$. Thus, one expects $\alpha_0$ along a field line to asymptote to the standard scenario as one moves away from the surface. We show this evolution along a field line for a specific example in Fig.~\ref{fig:alpha_radial}. Here, the standard scenario of a pair-producing field line is shown in black, while the solid and dashed curves show the evolution of the discharge parameter along a field line for different axion masses (differentiated by color) and for two different polar regions (solid vs dashed). Pair production does not occur for $0 \leq \alpha_0 < 1$, or if the transition to $\alpha_0 > 1$ occurs at sufficiently large radii, since the magnetic field will be significantly weakened in this region. For large axion masses, this transition must occur near the stellar surface since the axion field value is exponentially damped at $r \gtrsim m_a^{-1} \sim 20 \, {\rm m} \, (10^{-8} \, {\rm eV} / m_a)$ (note that since $|\rho_a|$ can be much larger than $|\rho_{\rm GJ}|$, one may require as many as $\mathcal{O}(10)-$e folds of suppression before this transition occurs), while for very light axions this transition may only occur very far from the star itself. This latter point is particularly interesting, since the appearance of a gap at very large radial distances does not necessarily imply pair production, since the magnetic field in far-field region is significantly smaller, and thus requires higher energy primary photons.

\begin{figure*}
    \centering
    \includegraphics[width=0.45\linewidth]{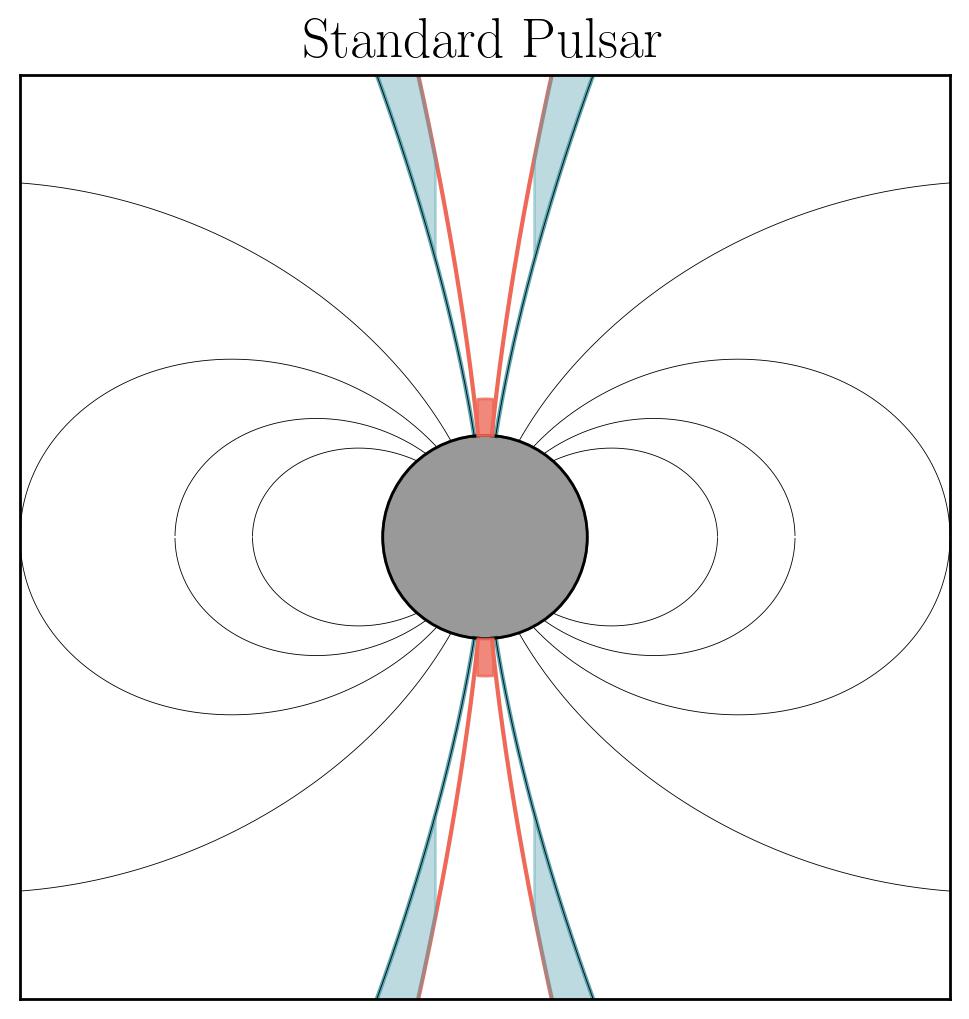}
    \includegraphics[width=0.45\linewidth]{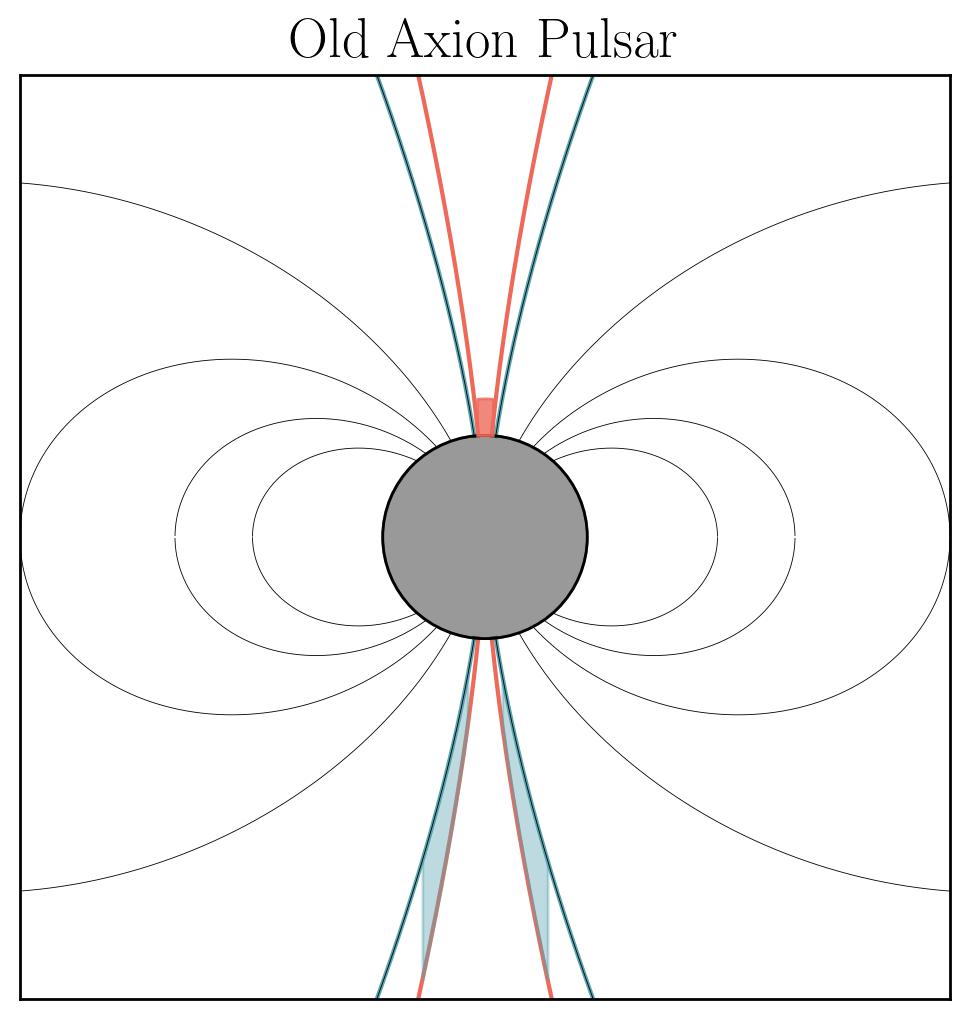}
    \caption{ Rough illustration of near-field pair producing regions (here, we do not discuss pair production occurring in the outer magnetosphere -- see~\cite{SashaReview} for a more general discussion). Left: for a standard pulsar, pair production occurs near the star in the region hosting super-GJ currents (red), and along return currents ($\alpha_0 < 0$) at distances on the order $\mathcal{O}(R_{\rm NS})$ away from the star (blue). Right: for an old pulsar (located below the conventional death line on the $P-\dot{P}$ diagram) with axion hair, one does not expect pair production in the out-flowing current zone in the southern pole, and in the northern  pole's return current region. Meanwhile, pair production in the northern pole's out-flowing current and in the southern pole's return-current regions, which would otherwise be absent in such old pulsars, is triggered by the axion-induced voltage. Note that pair production in the return-current region shifts closer to the stellar surface. We have verified using global PIC magnetospheric simulations that suppressing pair production at only one pole does not alter the global magnetospheric structure, since active pair production in the southern pole's return current ensures the formation of the current sheet and global current closure.
     }
    \label{fig:pp_pairity}
\end{figure*}

Eq.~\ref{eq:splitM_alp} applies only to the split monopole, however the impact of a static axion configuration in a dipolar (or, e.g., quadrudipolar) magnetosphere is similar -- the axion-induced modification to the current density will be heavily suppressed relative to the modification to $\rho_{\rm GJ}$ (this follows merely from $|E| \ll |B|$), with large axion field gradients serving to amplify the potential drop, and driving $\alpha_0 \rightarrow 0^\pm$ across the open field lines. As discussed in the main text, one important subtlety arises from the fact that one must consider pair discharges on field lines supporting both out-flowing, $\alpha_0 > 0$, and return,  $\alpha_0 < 0$, currents. The geometry in this case is slightly more subtle; for comparison, we show in Fig.~\ref{fig:pp_pairity} where pair discharges typically occur for active pulsars without axions (left), and for slowly rotating pulsars with axions (right). Here, in the northern pole, axion effects activate pair discharge along field lines with out-flowing current $\alpha_0>0$, but do not affect field lines supporting return current, $\alpha_0<0$; conversely, in the southern pole, pair discharge is not sustained on field lines with out-flowing current, but is activated on field lines with return current (note also that the pair discharge region in the return current has been shifted closer to the stellar surface, illustrating that the axion-induced voltage drop is needed to ignite pair cascades). More generally, the presence of axion hair modifies the near-field electrodynamics in a variety of different ways, including:
\begin{itemize}
    \item there will be an intrinsic asymmetry in how pair discharges operate in the northern and southern polar cap regions, which could be searched for using interpulse pulsars;
    \item closed field lines will host an amplified charge density, although only near the surface of the star (regions which are difficult to probe directly) 
    \item there will likely be a modification to the luminosity and spectrum of radio emission produced in these systems as a result of the modified electric potential. Here, one can estimate the characteristic radio luminosity generated from the discharge process as roughly $L \sim (\overline{E} \sin\alpha)^2 (\eta r_{\rm pc}^2) / (4\pi)$~\cite{Tolman:2022unu}, where $\overline{E}$ is the average electric field, $\alpha$ is the typical angle between the background magnetic field and wave propagation, and $\eta$ characterizes the fraction of pair producing field lines. The frequency spectrum is instead set by the characteristic plasma frequency, which is intrinsically related to the scale of $\rho_{\rm GJ}$. The axion modifies $\overline{E}$, $\rho_{\rm GJ}$, and $\eta$ in a non-trivial way, and thus is likely to impact the radio flux. 
\end{itemize}
Unfortunately, some of these effects are observationally difficult to test, while others are still lacking firm theoretical calculations which would be necessary in order to differentiate standard physics from exotic. In the future, it may be interesting to return and investigate these more subtle signatures.

\subsection*{A comment on the continuity equation}
\label{sec:ChargeCont}

In the proceeding derivation, we argued that the the correction from the axion to the current density was a higher order effect that could be neglected at leading order. Here, we briefly digress to show that this hierarchy is consistent with charge continuity, and thereby does not introduce any conceptual problems in our approach. 

Let us start by taking the divergence of the Amp\`{e}re-Maxwell law, Eq.~\ref{eq:curlBApp}. Here, one obtains
\begin{equation}
    \partial_t \Big( \nabla \cdot \vec{E} + g_{a\gamma\gamma} \, \vec{B} \cdot \nabla a \Big) = - \nabla \cdot \vec{j} ,
\label{eq:cont}
\end{equation}
where we used the vector identities 
$\nabla \cdot (\vec{E} \times \nabla a) = (\nabla a) \cdot (\nabla \times \vec{E}) - \vec{E} \cdot (\nabla \times \nabla a)$, 
$\nabla \times (\nabla a) = 0$, 
and $\nabla \cdot \vec{B} = 0$. Then, using Gauss' law, Eq.~\ref{eq:cont} simply reproduces the continuity equation, $\partial_t \rho + \nabla \cdot \vec{j} = 0$. It is important to note that the continuity equation is only preserved in this form when time variations in both the axion field and the magnetic field are included (prematurely dropping $\partial_t a$ would have lead to a modified continuity equation).  In developing our numerical simulations of the pair discharge process (described in detail in the following section),  we only modify Gauss' law by introducing the axion-induced charge density, leaving Amp\`{e}re's law unchanged -- let us emphasize that this leads to a minor inconsistency which amounts to neglecting corrections of order $\mathcal{O}(E/B) \lesssim 10^{-5}$ in Amp\`{e}re's law (and the continuity equation).

\section{Particle-in-cell simulations of pair discharge}\label{sec:pic}

Having outlined in the proceeding section the fundamental physics determining how and when pair discharge occurs, we now demonstrate the pair discharge process using kinetic particle-in-cell simulations, and how it is modified in the presence of axion hair. 

We perform 1D time-dependent simulations using the code {\tt{Tristan v2}}~\cite{hayk_hakobyan_2023_7566725}, largely following the setup outlined in~\cite{Chernoglazov:2024rvo}. 

The surface of the neutron star is covered by a thin gravitationally-supported atmosphere which serves to provide a reservoir of charged particles. As in Ref.~\cite{Chernoglazov:2024rvo}, we simulate the atmosphere by placing a thermal plasma layer on the left edge of the simulation domain, which has a spatial Boltzmann distribution $n = n_{\rm peak} {\rm exp}(-x / h)$, where $n_{\rm peak} = 10 \, n_{\rm GJ}$, and $h$ is the gravitational scale height. Non-neutral plasma is initialized across the active domain to ensure that Gauss' law in the co-rotating reference frame, which acts as a constraint equation, is satisfied consistently with the initialization $E = 0 $. These conditions imply that 
\begin{eqnarray}
    \rho_{\rm init} = \rho_{\rm GJ} + \rho_a \, ,
\end{eqnarray}
where we adopt co-rotation charge density profile $\rho_{\rm GJ} = \rho_{\rm GJ}^0 (1 + 0.8 x / L)$, which contains a non-zero gradient in order to imitate the spatial dependence induced by the Lense-Thirring effect. We additionally include a dense neutral plasma with density $n_{e} \approx 10\,\rho_{\rm GJ}/e$. These initial conditions are intended to mimic the plasma state left behind by a previous discharge episode. All charges are initialized with non-relativistic velocities. The baseline charge density, $\rho_{\rm GJ}^0$, is fixed such that the corresponding skin depth is resolved by a few numerical cells.

We adopt an axion charge density profile $\rho_a = N_a \, {\rm exp}(- m_a \, x)$, where we take the norm to be $N_a = \xi \rho_{\rm GJ}^0$. In practice, we perform three simulations, taking $\xi = +10, -10,$ and $0$, and we fix $m_a = 5 \times 10^{-3}$ in units of the inverse grid spacing.

The electromagnetic fields are decomposed into the sum of the FFE solution and a deviation, \ie $\vec{B} = \vec{B}_{\rm FFE} + \delta \vec{B},$ $\vec{E} = \vec{E}_{\rm FFE} + \delta \vec{E}$, where $\vec{B}_{\rm FFE} = \vec{B}_0 + \vec{B}_\phi$ and $\vec{E}_{\rm FFE} = - \vec{\Omega} \times \vec{r} \times \vec{B}_0$. Here, $\vec{B}_0$ is the background dipolar magnetic field, the magnetospheric current is determined as $\vec{j}_{\rm mag} = \nabla \times \vec{B}_\phi$, the fields then satisfy\footnote{Note that these equations support two stationary solutions: the force-free solution ($\delta \vec{E} = \delta \vec{B} = 0$), and an un-twisted inactive solution with $j =0$, $\delta B = - B_\phi$~\cite{Chernoglazov:2024rvo}.}
\begin{eqnarray}
    \partial_t \delta \vec{E} &=& \nabla \times \delta \vec{B} - (\vec{j} - \vec{j}_{\rm mag}) \\[7pt]
    \partial_t \delta \vec{B} &=& -\nabla \times \delta \vec{E} \, \\[7pt] 
    \nabla \cdot \delta \vec{E} &=& \rho - \rho_{\rm GJ} \, .
\end{eqnarray}
In 1D, the first two equations reduce to one evolution equation, $\partial_t (\delta E_{||}) = -j_{||} + j_{\rm mag}$, and the final equation serves as a constraint equation. For simulations of out-flowing current, we fix $j_{\rm mag} = 2 \rho_{\rm GJ}$, and for those of the return current, we adopt $j_{\rm mag} = -2 \rho_{\rm GJ}$.

In the simulations, we set $B_0 / B_Q = 1$ (note that this choice is somewhat arbitrary, as is merely chosen to ensure that the gap collapse process occurs on scales smaller than the box size, for rescaled parameters), we fix the Lorentz factor achievable through acceleration in the vacuum electric field, $\gamma_{\rm PC} = 3.2 \times 10^7$ \footnote{This factor is typically of the order of $\gamma_{\rm PC} \sim 0.5 (r_{\rm pc} / d_e^{\rm GJ})^2$, where $r_{\rm pc}$ and $d_e^{\rm GJ}$ are the polar cap size and the GJ skin depth, correspondingly, and, thus, effectively amounts to a rescaling of the polar cap. }, set the Lorentz factor at the radiation reaction limit $\gamma_{\rm rad} = 8 \times 10^5$ (\ie, the Lorentz factor where acceleration by the vacuum electric field,  $\rho_{\rm GJ} r_{\rm PC}$, is balanced by radiative losses), and fix the parameter $\gamma_{\rm emit} = 10^4$, which sets the Lorentz factor of the particle that emits a curvature-radiation photon whose characteristic energy is equal to $m_e$. The scales of these parameters have been reduced to ease the calculation, but the hierarchy of energy scales $\gamma_{\rm pc} \gg \gamma_{\rm rad} \gg \gamma_{\rm emit}$ has been maintained. Additional details on the parameter and the simulation setup can be found in Ref.~\cite{Chernoglazov:2024rvo}.

 In our fiducial analyses, we initialize a one-dimensional spatial domain that resolves $L=3.2 \times 10^4$ cells, corresponding to a resolution of 16,000 times the GJ skin depth $d_e^{\rm GJ}$. This box size is not always sufficient, however, and can influence the gap dynamics for larger gaps, or for gaps opening in the center of the domain; in these cases, we extend the box size to $L=5.6 \times 10^4$ or $L=6.4 \times 10^4$ cells (at fixed skin depth), depending on the simulation of interest, to ensure reliable results.

\begin{figure*}
    \centering
    \begin{adjustbox}{max width=0.76\textwidth, max height=0.76\textheight}
        \begin{tabular}{ccc}
            \includegraphics[trim={0cm 1.7cm 0cm 0cm},clip]{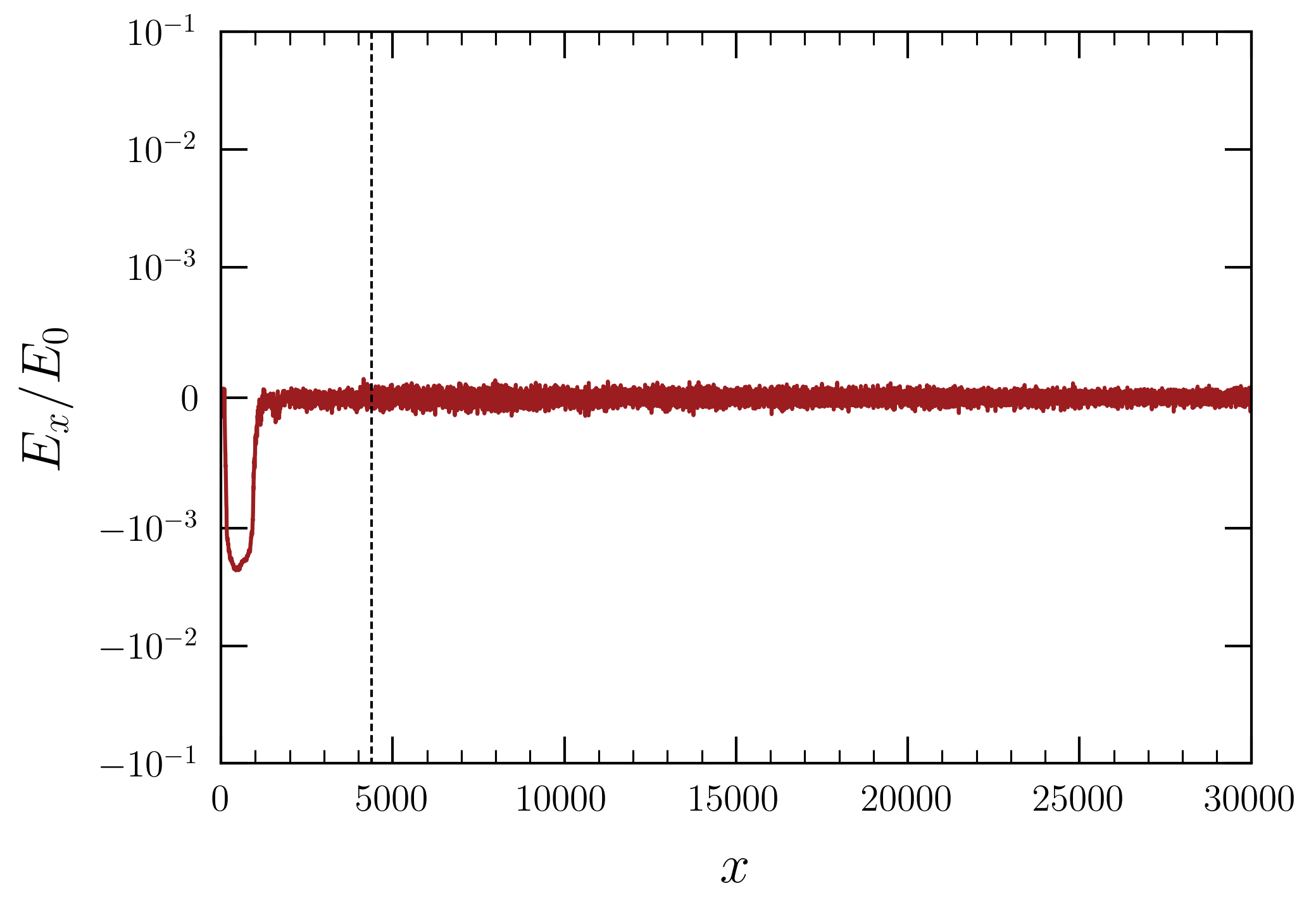} &
            \includegraphics[trim={2.7cm 1.7cm 0cm 0cm},clip]{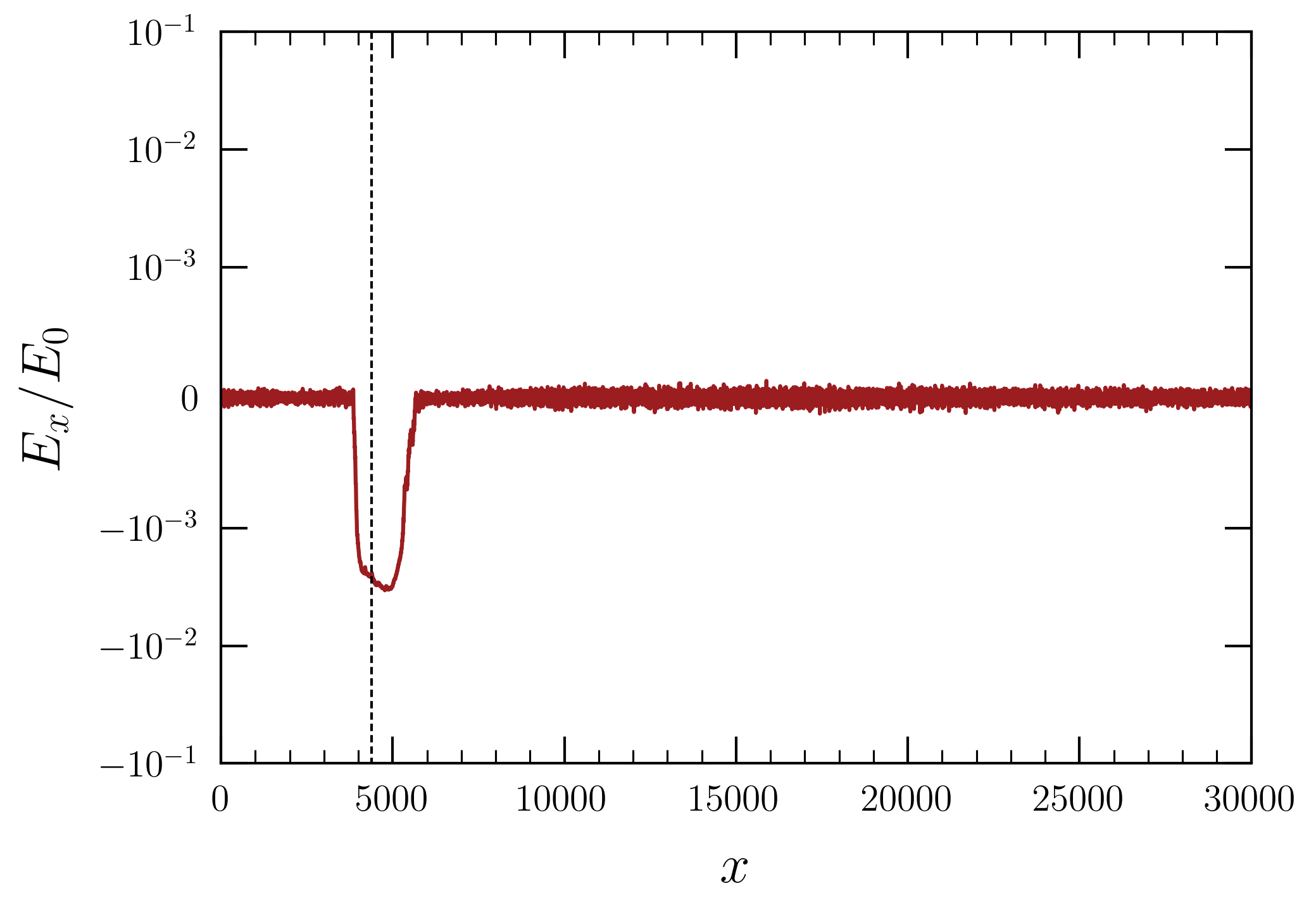} &
            \includegraphics[trim={2.7cm 1.7cm 0cm 0cm},clip]{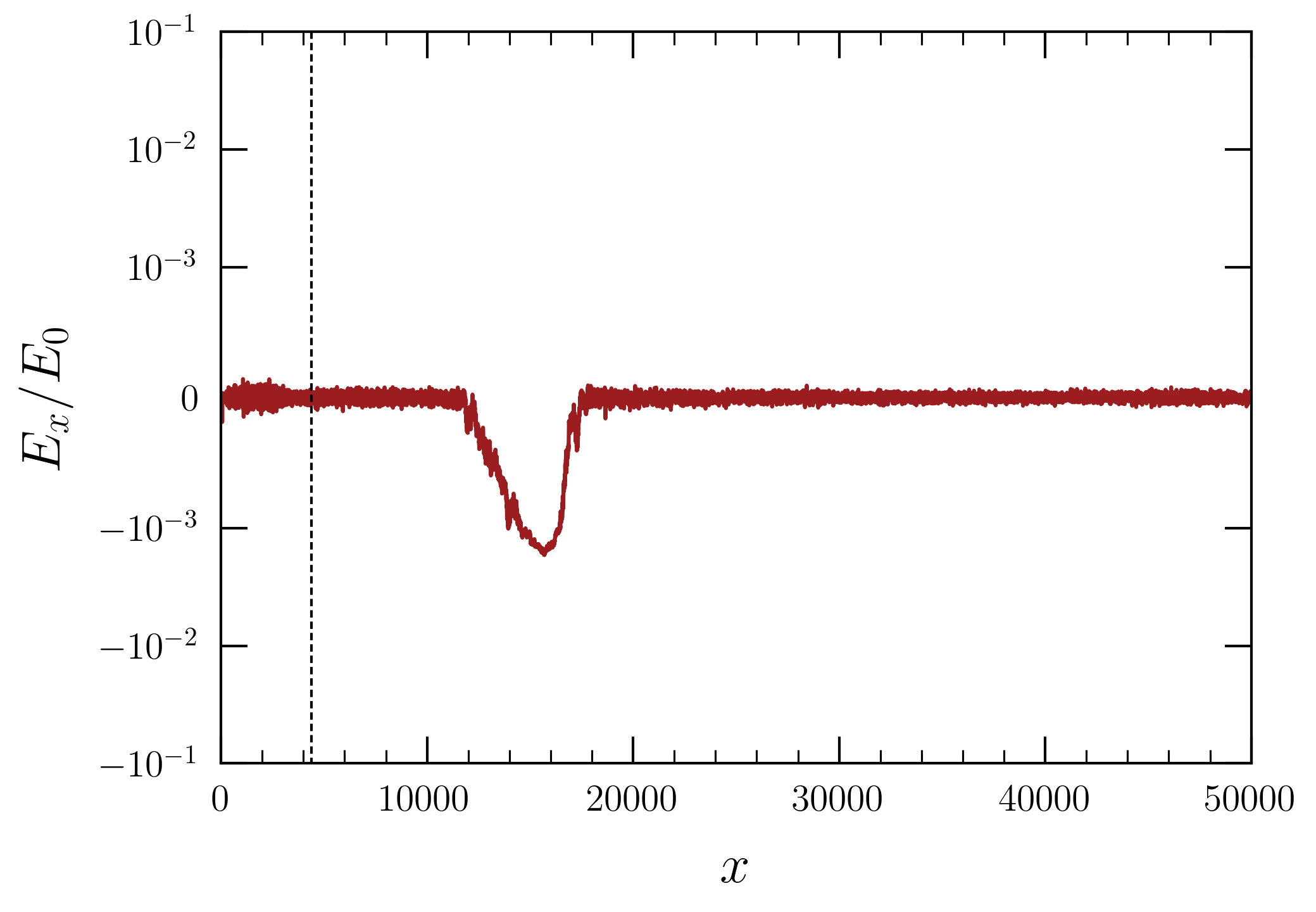} \\
            
            \includegraphics[trim={0cm 1.7cm 0cm 0cm},clip]{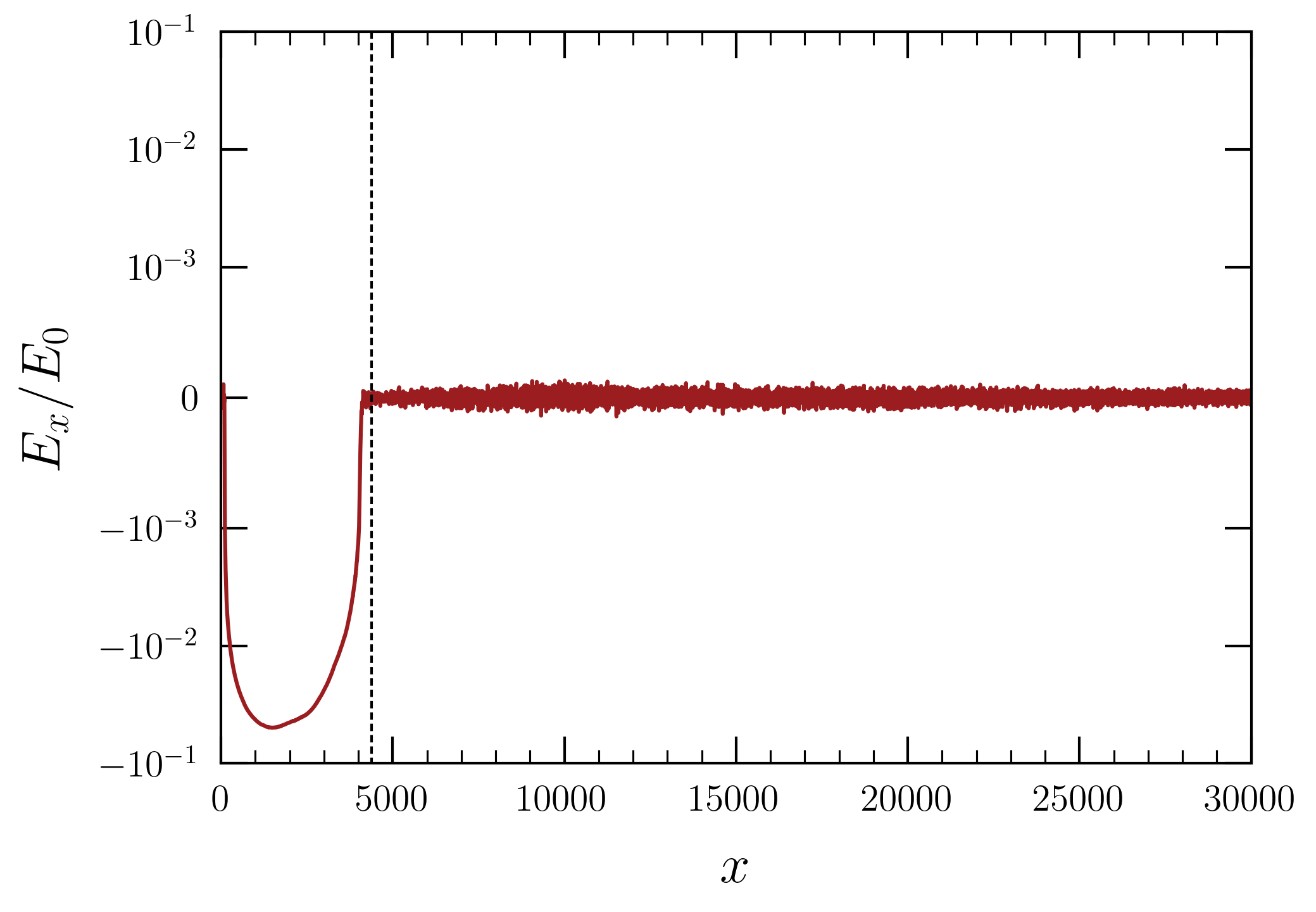} &
            \includegraphics[trim={2.7cm 1.7cm 0cm 0cm},clip]{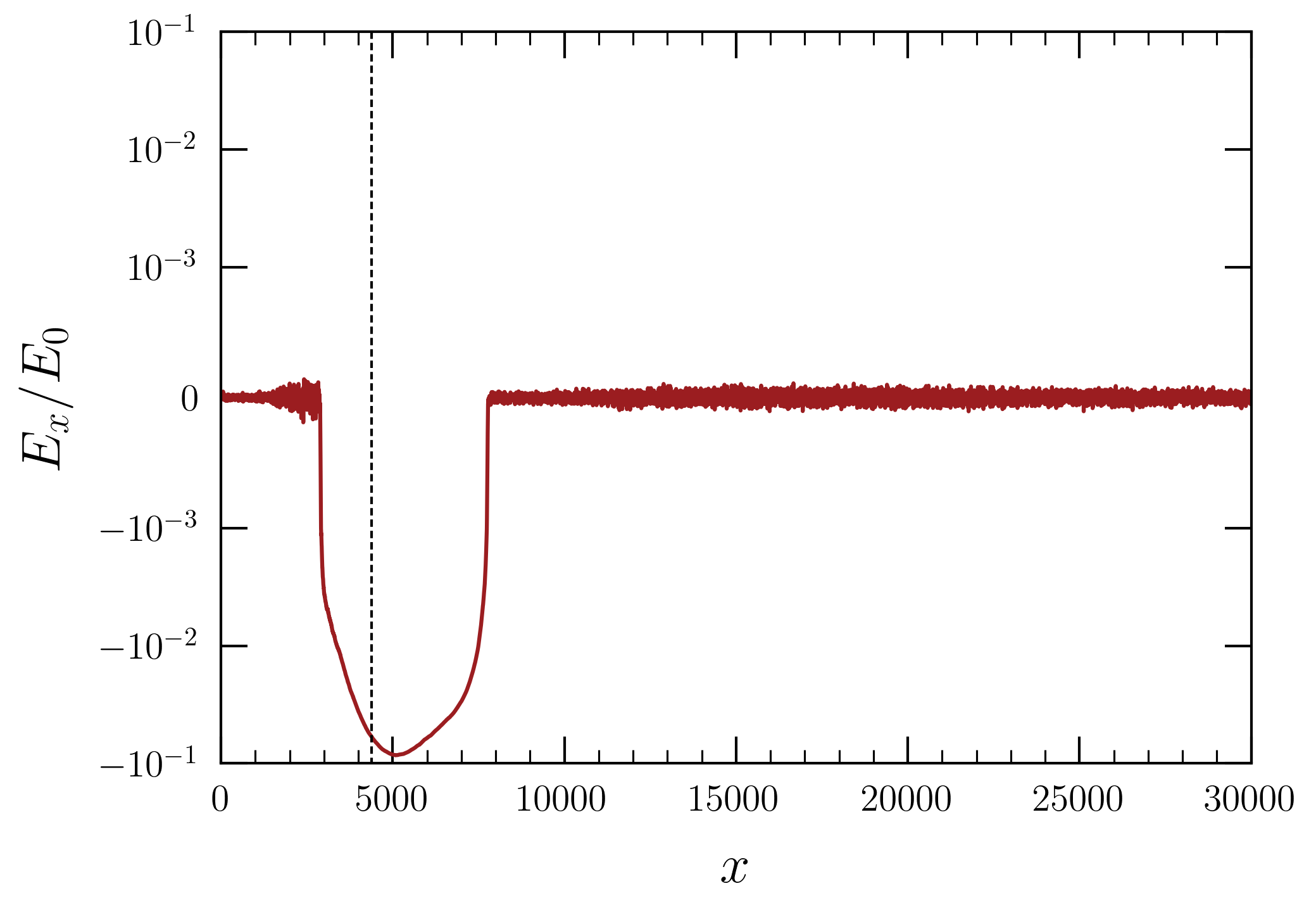} &
            \includegraphics[trim={2.7cm 1.7cm 0cm 0cm},clip]{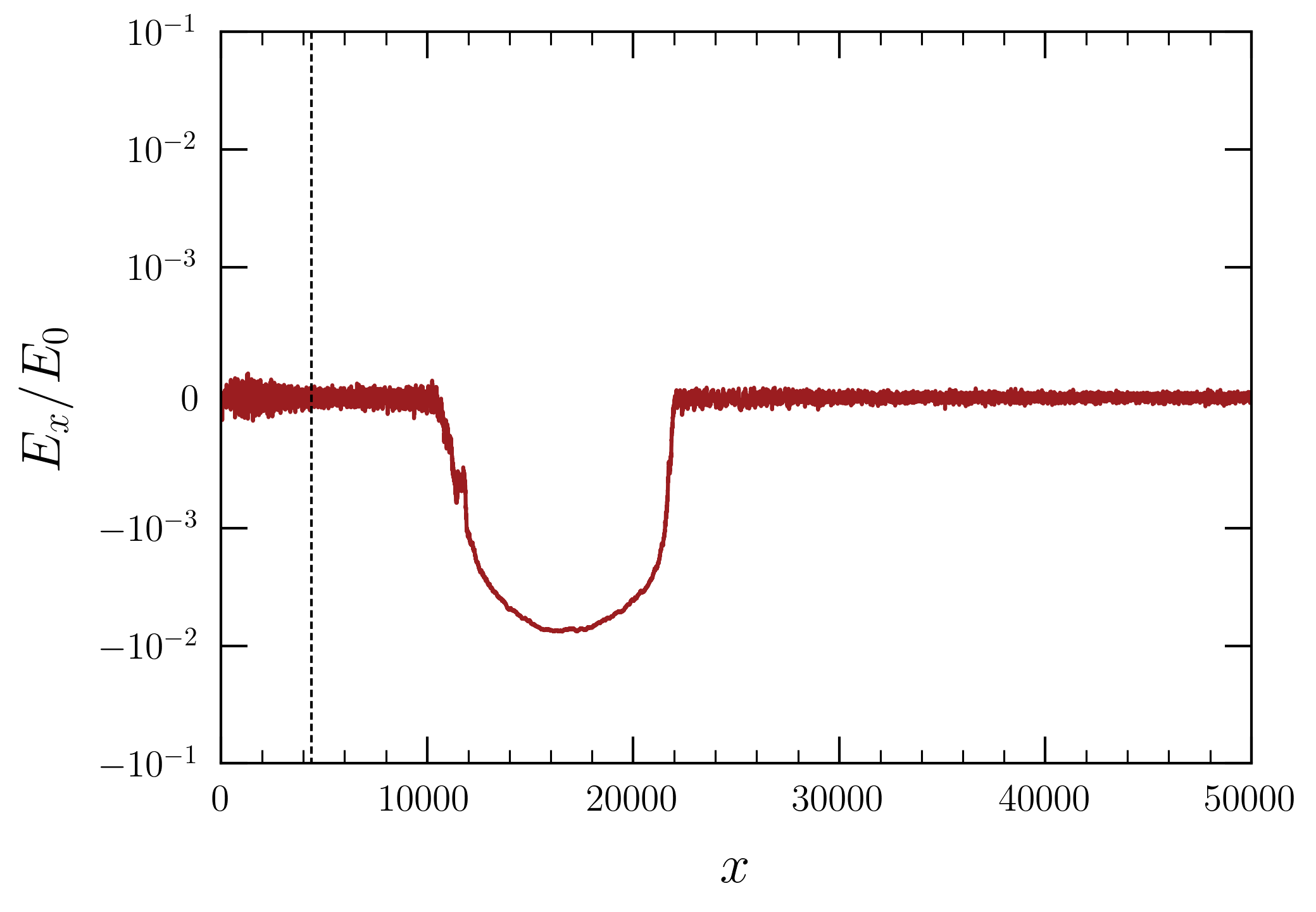} \\
            
            \includegraphics[trim={0cm 1.7cm 0cm 0cm},clip]{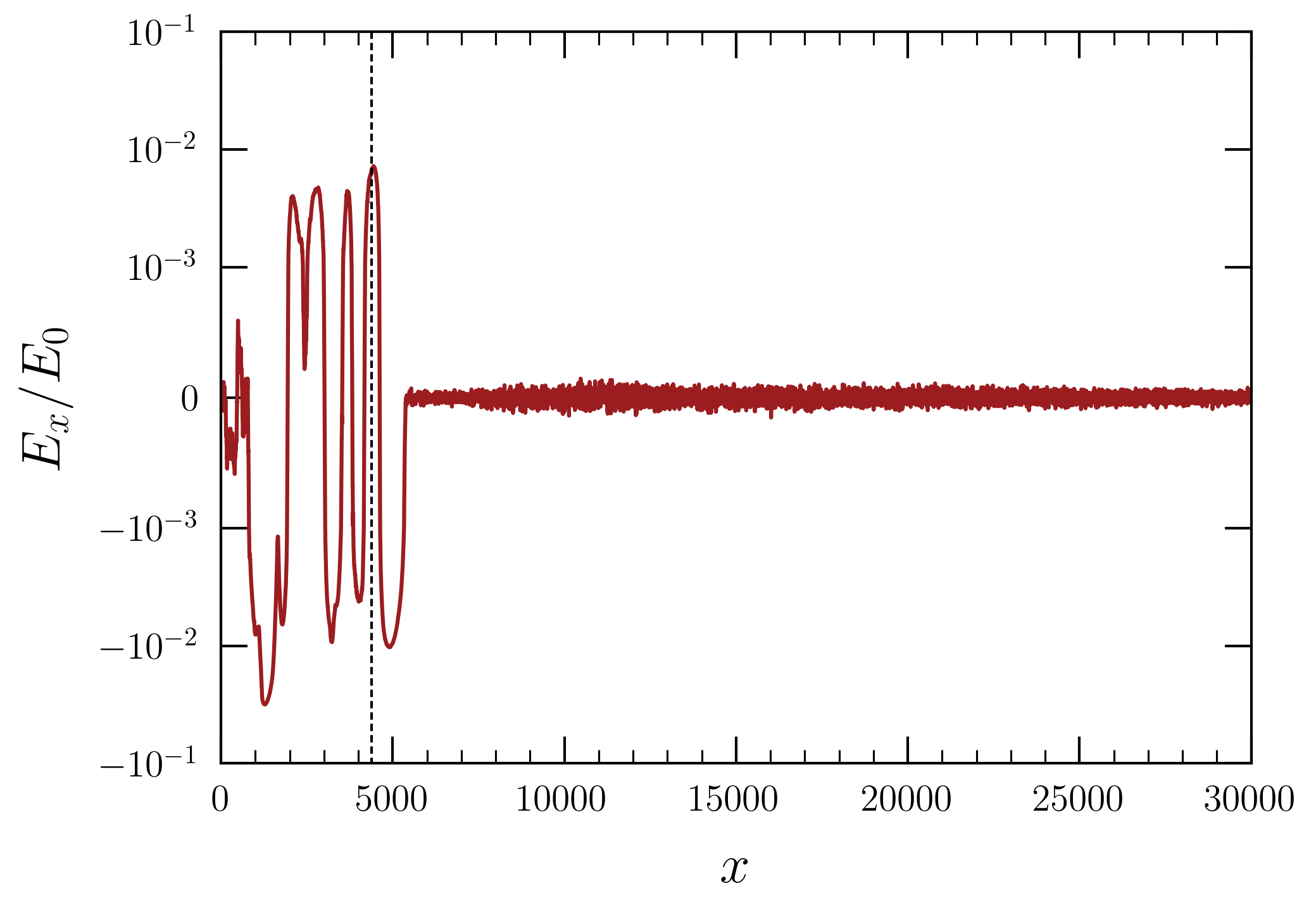} &
            \includegraphics[trim={2.7cm 1.7cm 0cm 0cm},clip]{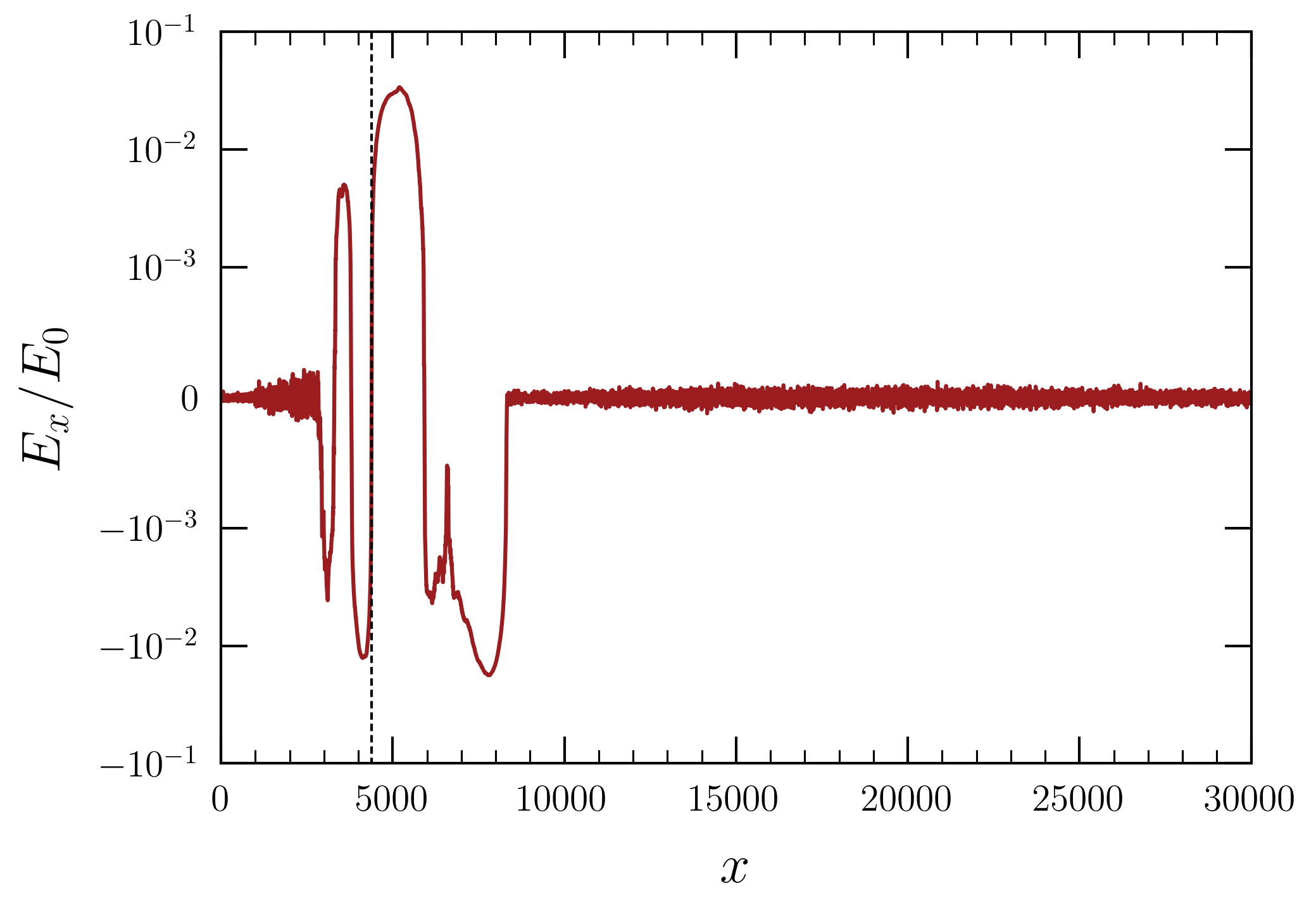} &
            \includegraphics[trim={2.7cm 1.7cm 0cm 0cm},clip]{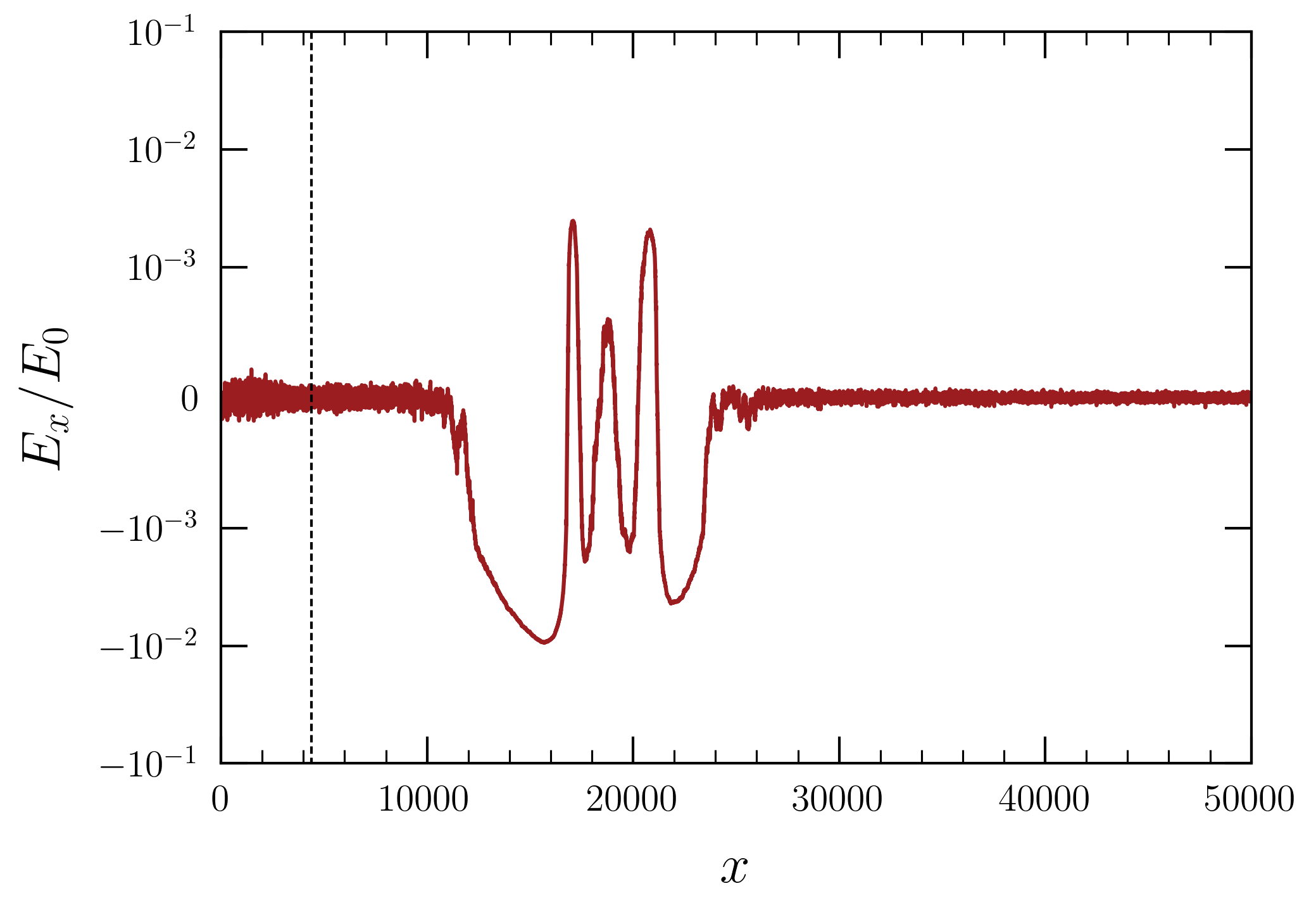} \\
            
            \includegraphics[trim={0cm 1.7cm 0cm 0cm},clip]{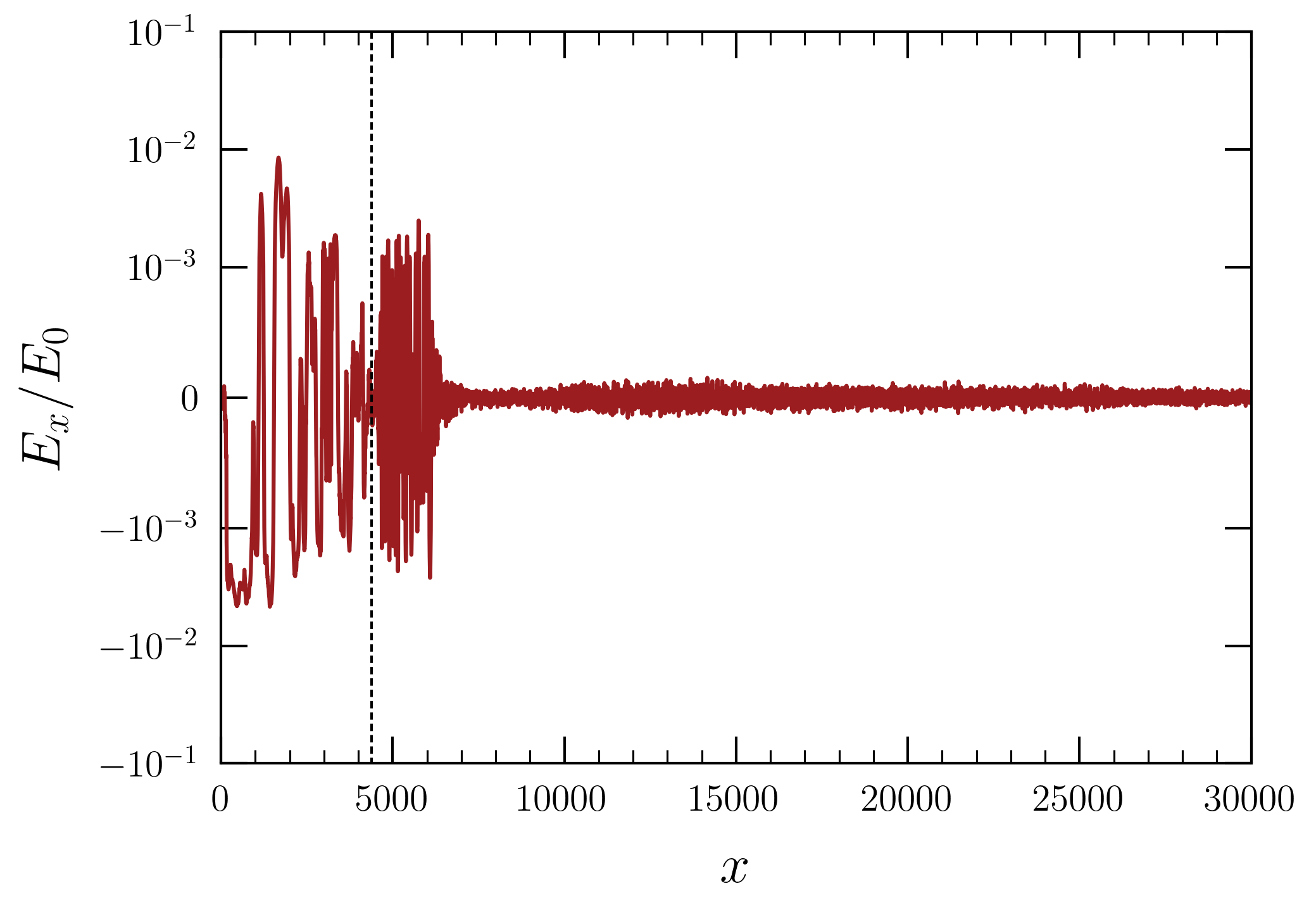} &
            \includegraphics[trim={2.7cm 1.7cm 0cm 0cm},clip]{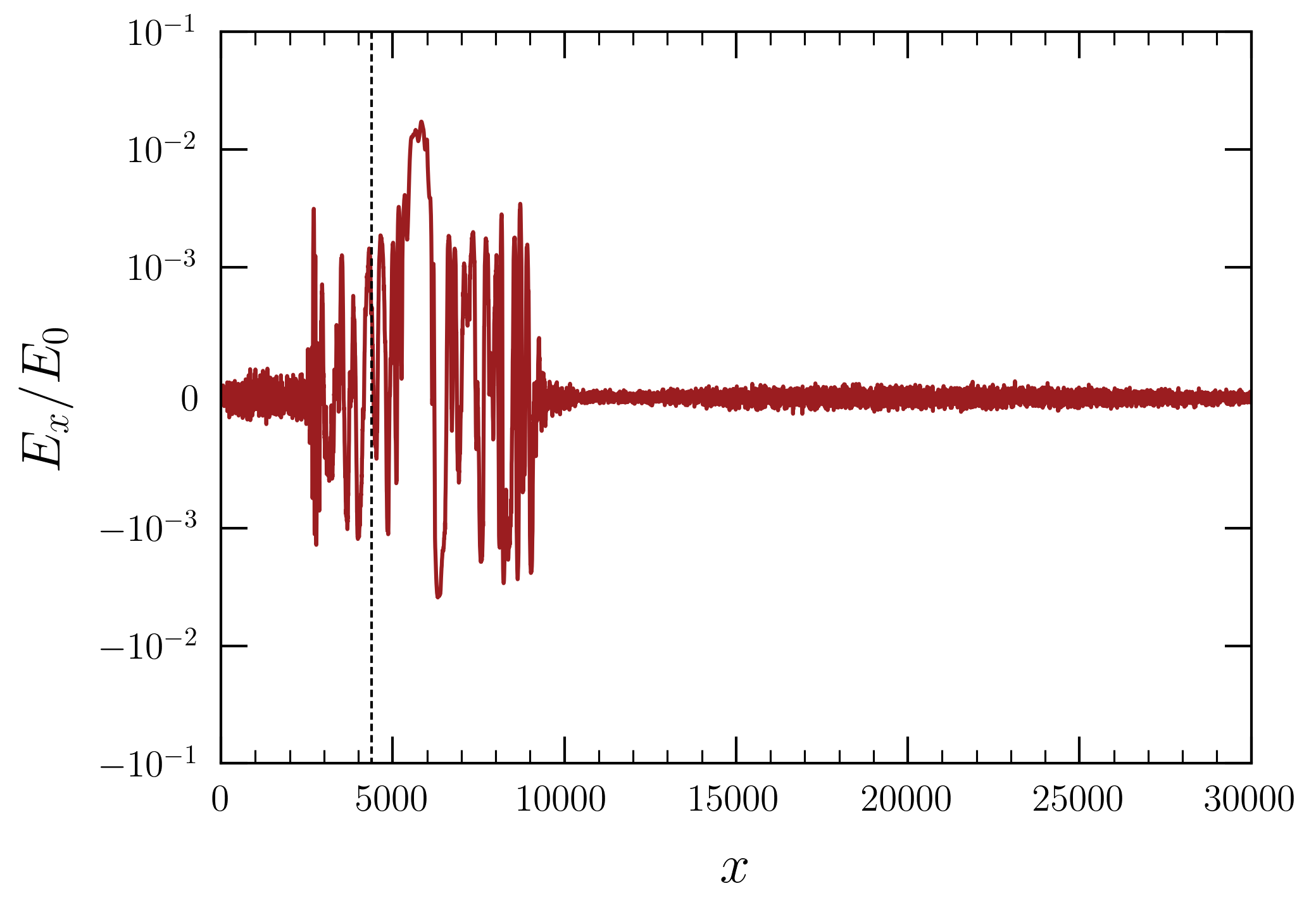} &
            \includegraphics[trim={2.7cm 1.7cm 0cm 0cm},clip]{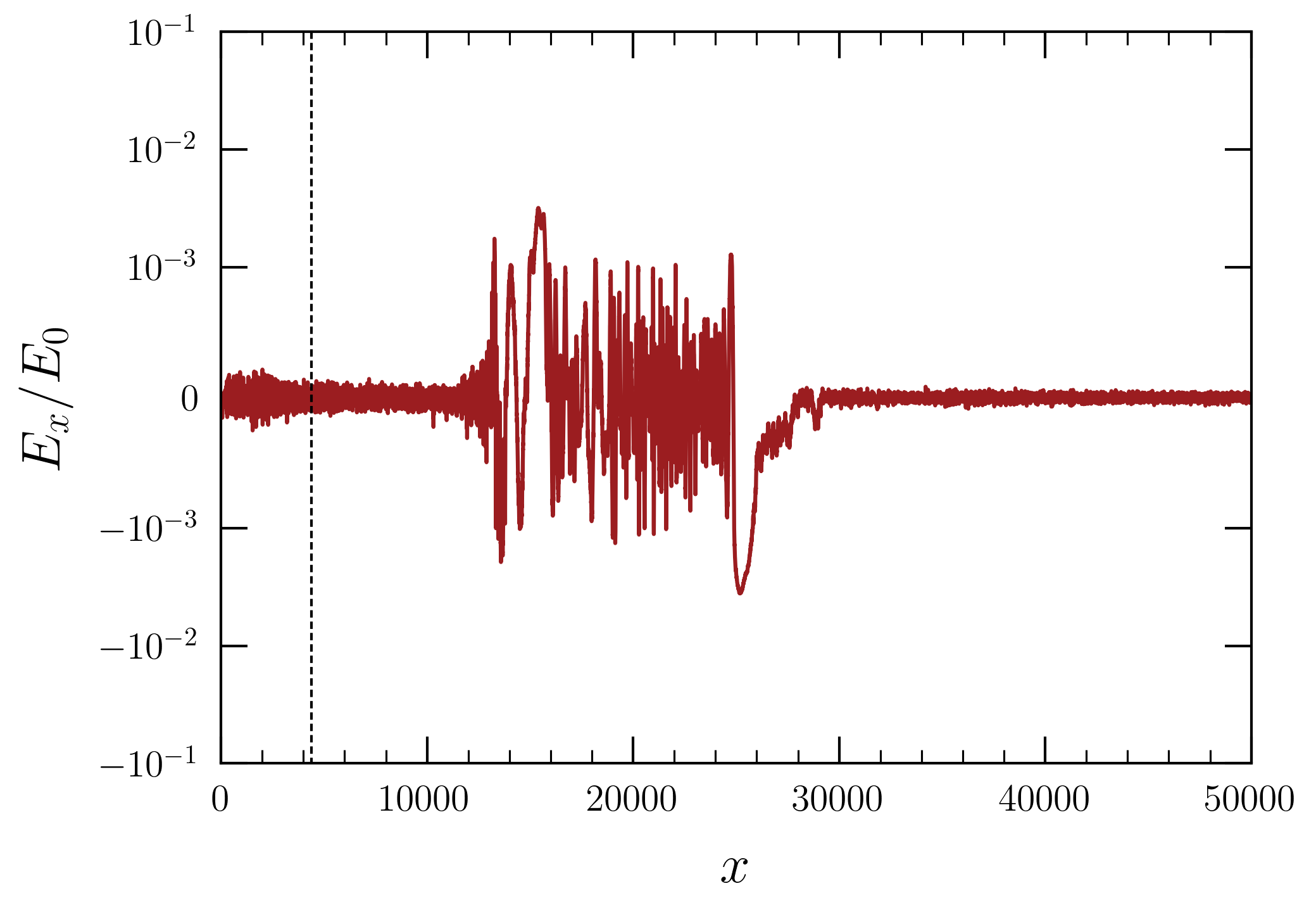} \\
            
            \includegraphics[trim={0cm 1.7cm 0cm 0cm},clip]{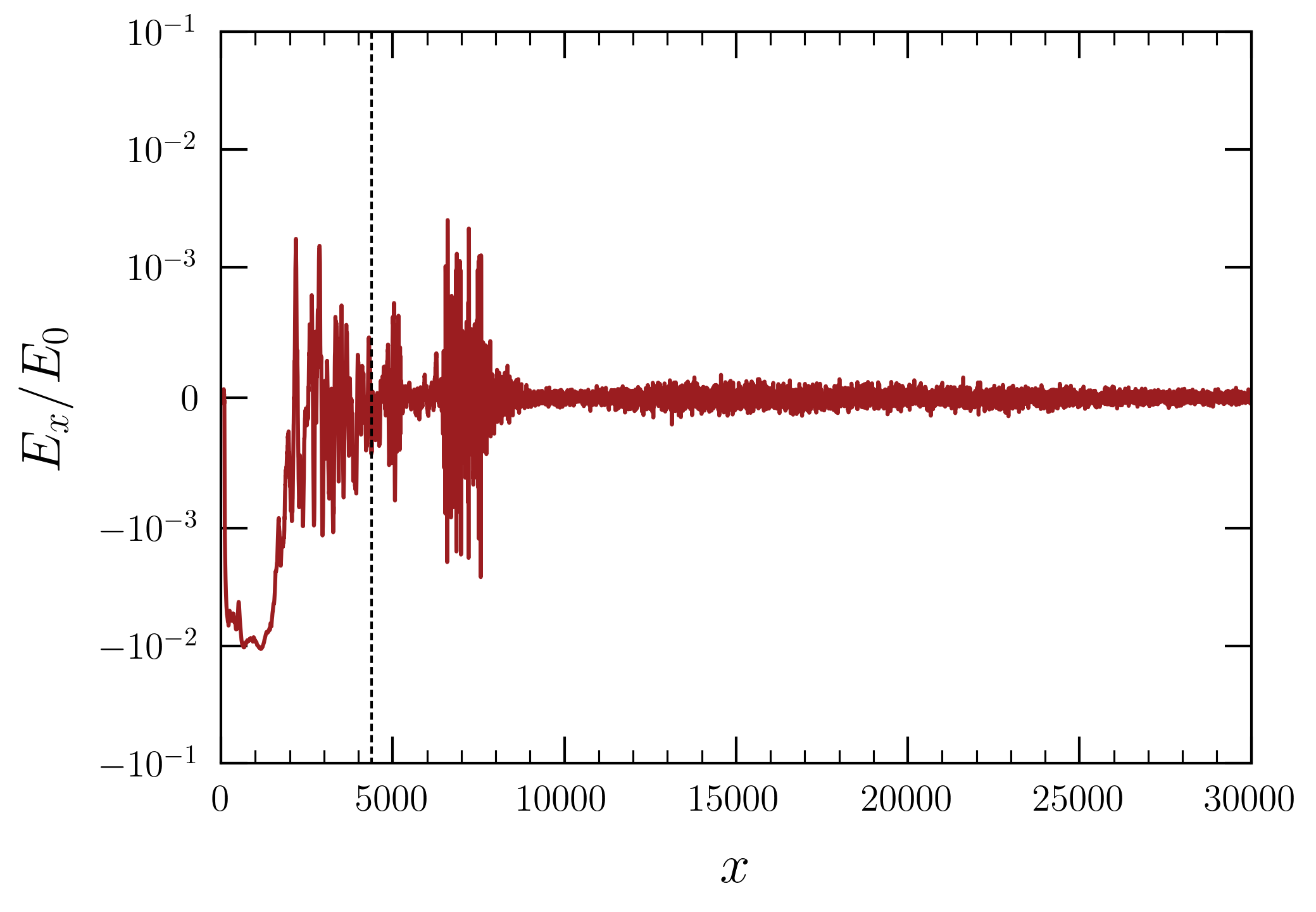} &
            \includegraphics[trim={2.7cm 1.7cm 0cm 0cm},clip]{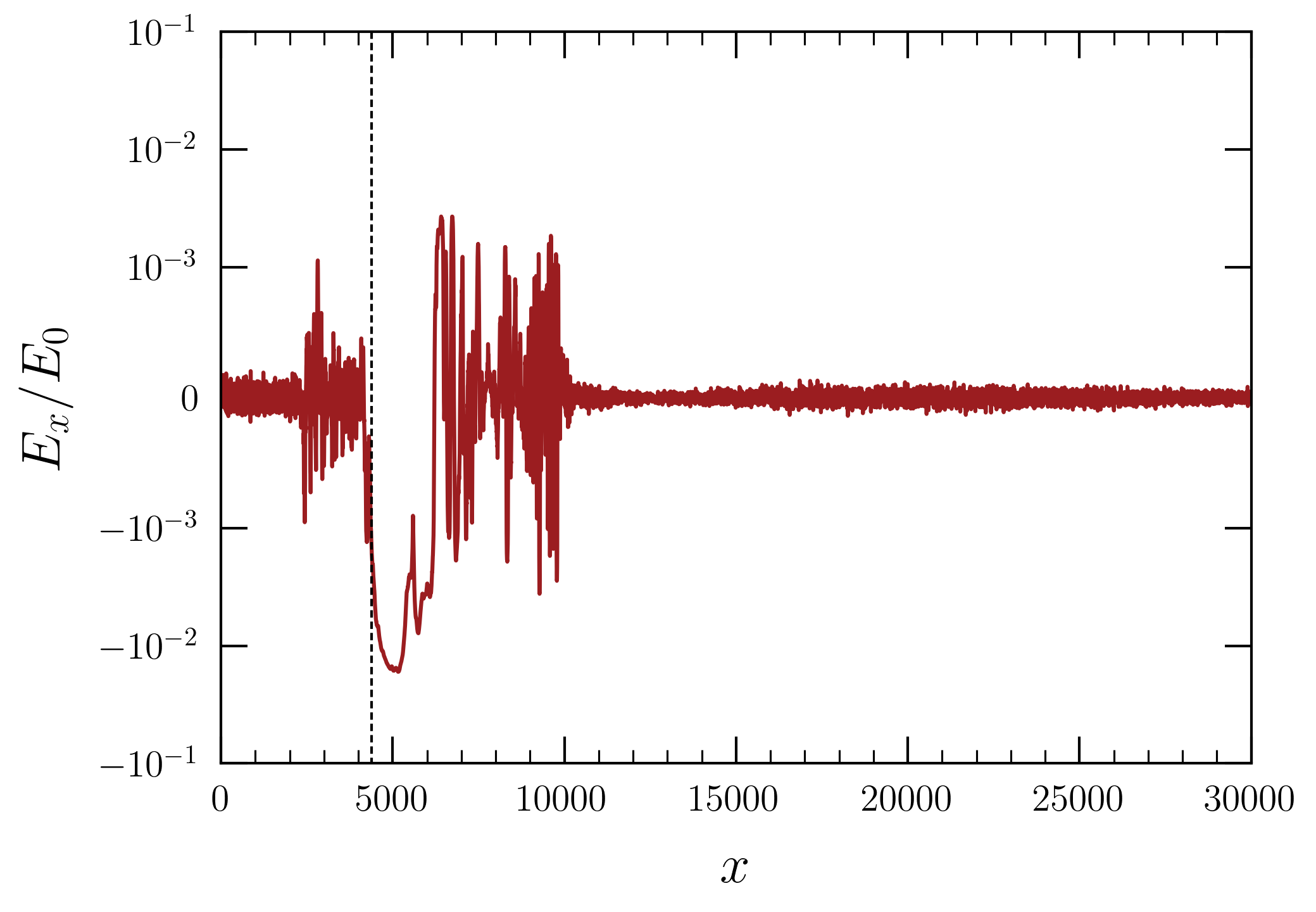} &
            \includegraphics[trim={2.7cm 1.7cm 0cm 0cm},clip]{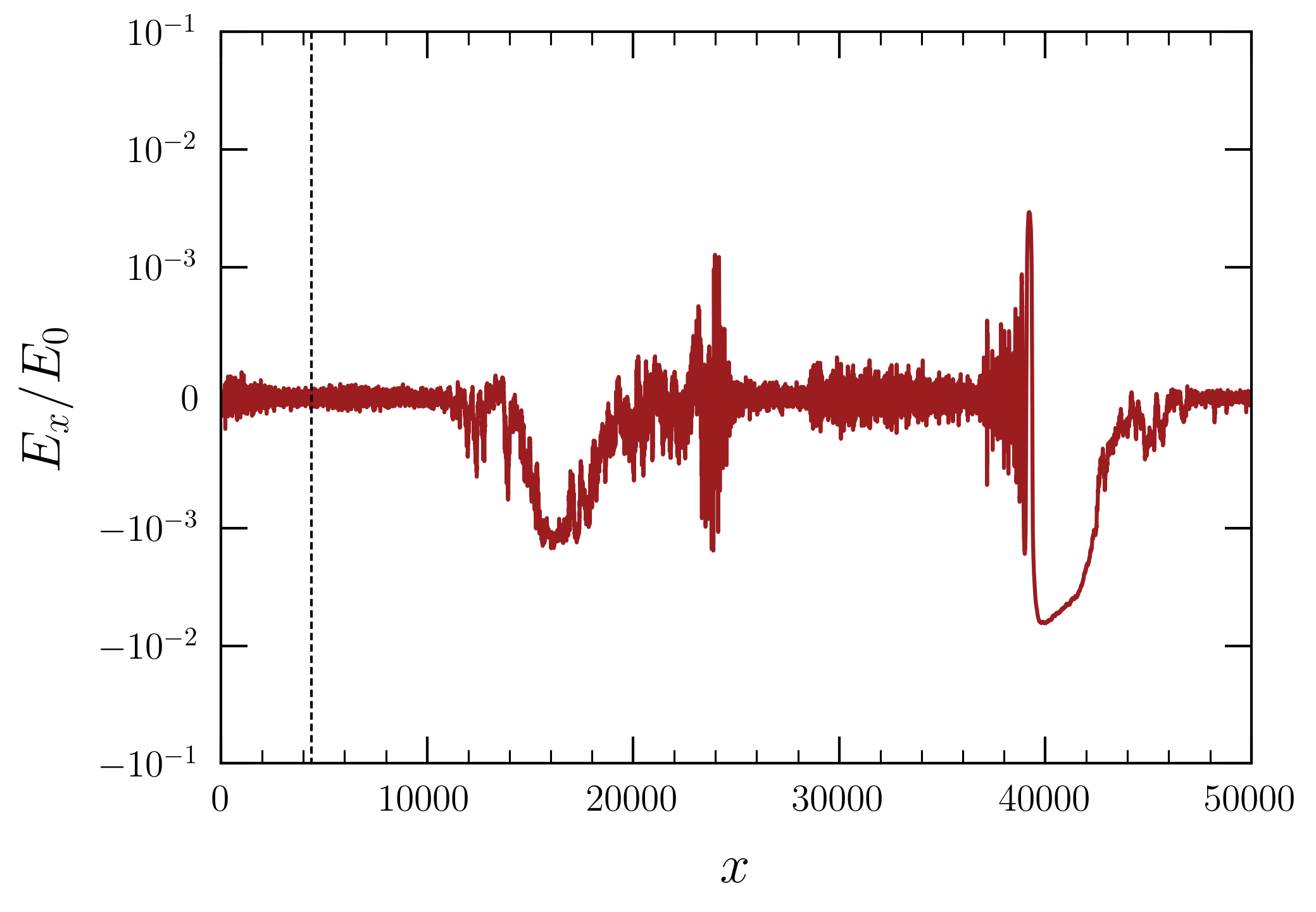} \\
            
            \includegraphics[trim={0cm 0cm 0cm 0cm},clip]{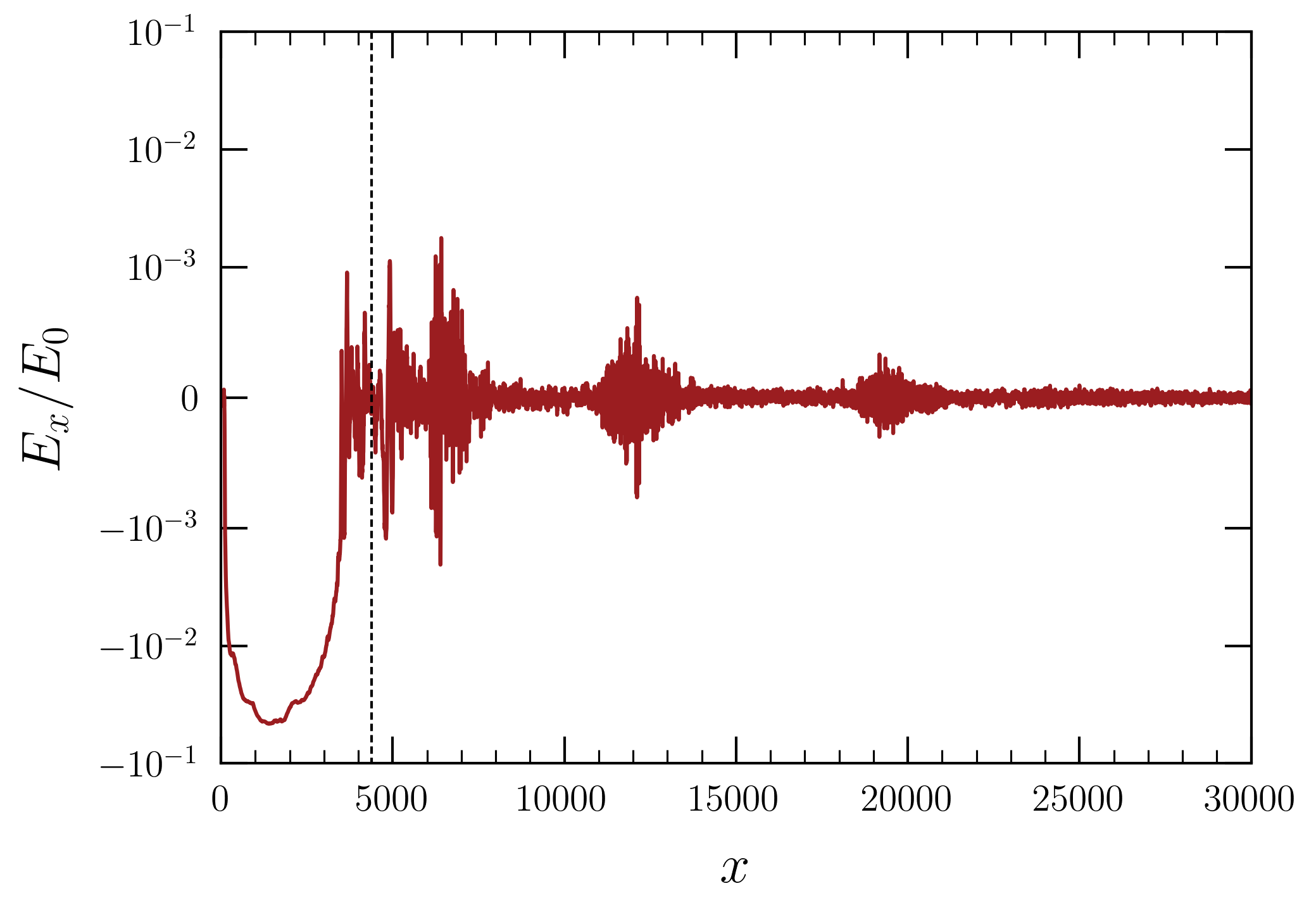} &
            \includegraphics[trim={2.7cm 0cm 0cm 0cm},clip]{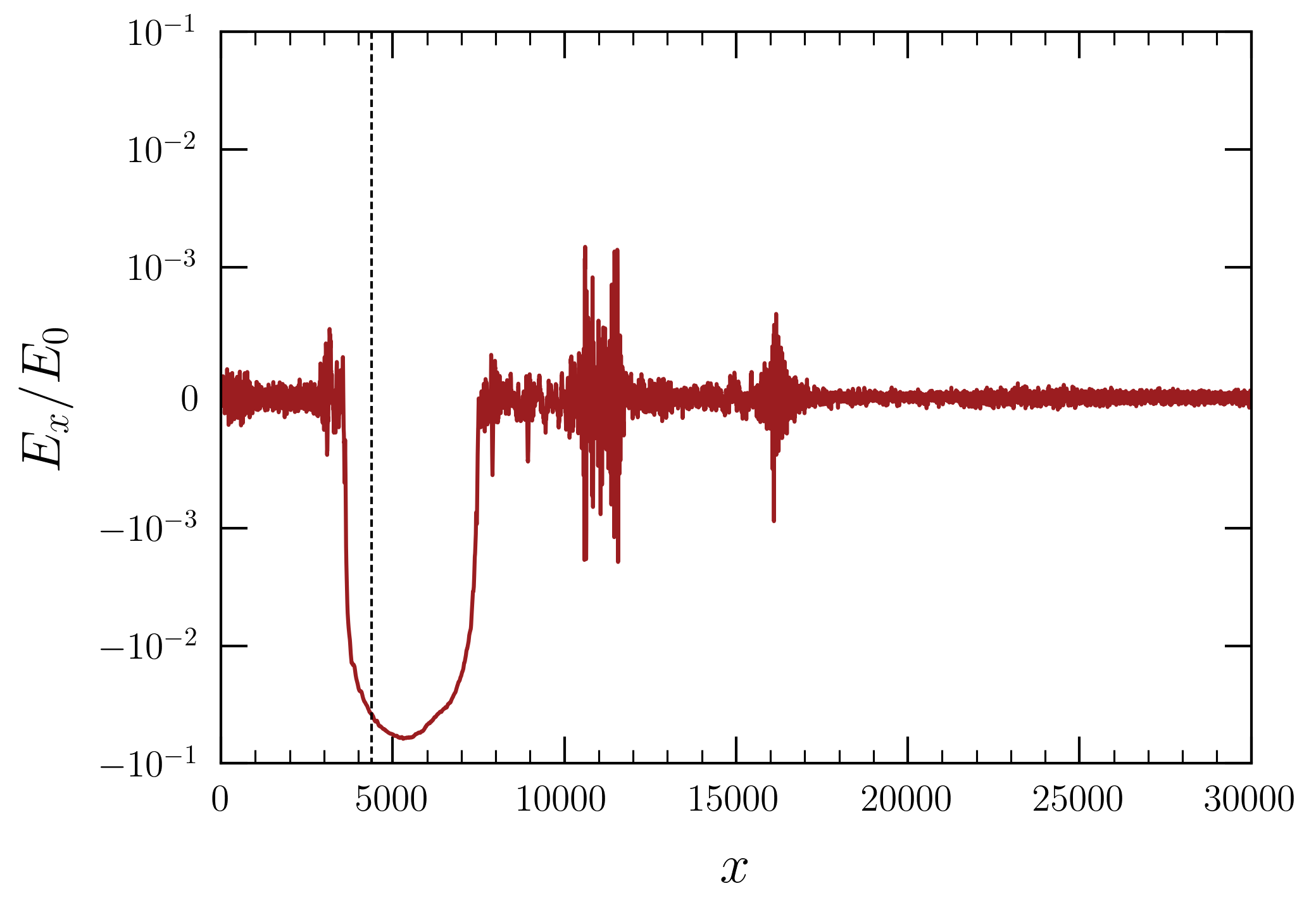} &
            \includegraphics[trim={2.7cm 0cm 0cm 0cm},clip]{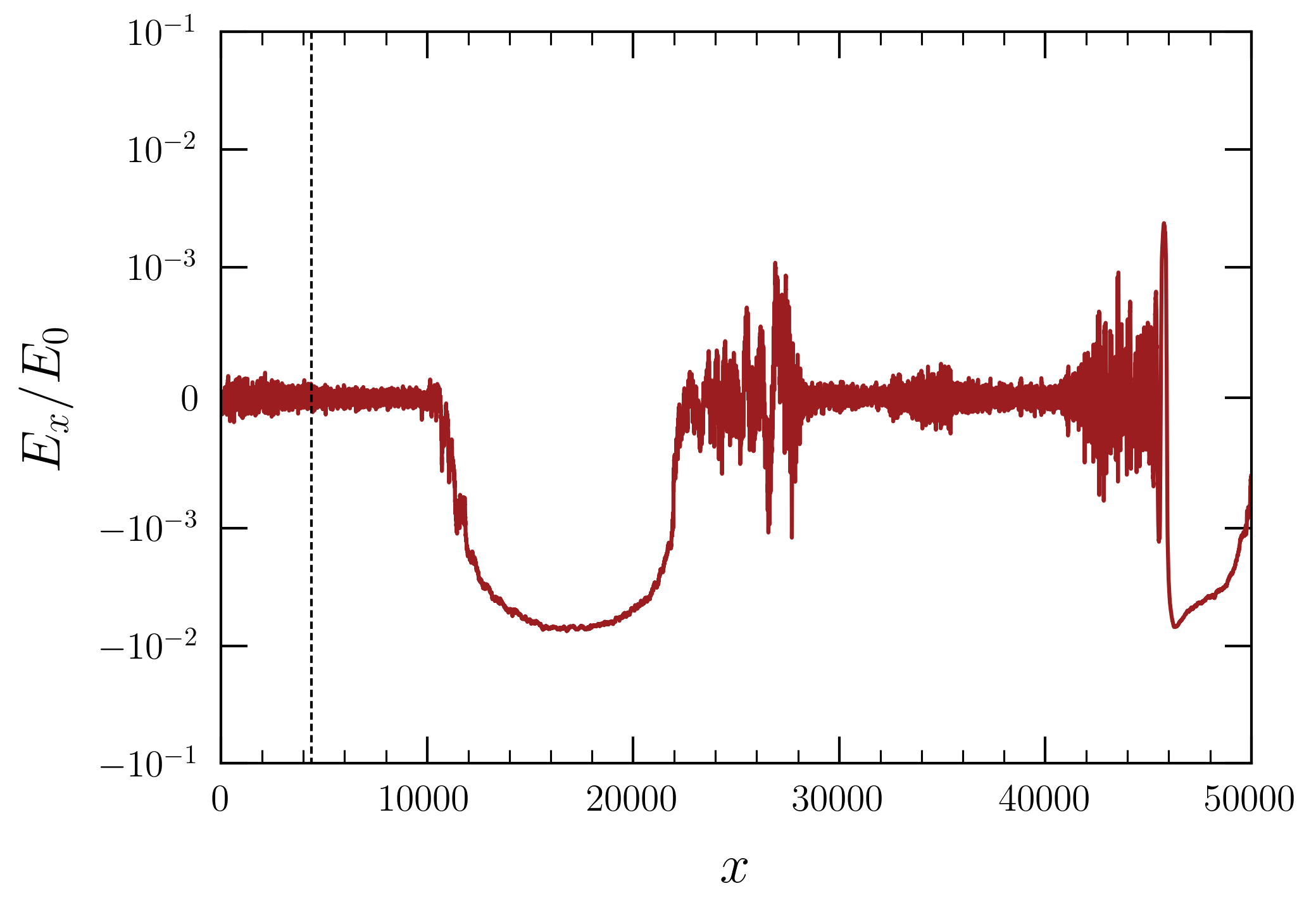} \\
        \end{tabular}
    \end{adjustbox}
    \caption{Evolution of $E_{||}$ (in normalized units) computed using particle-in-cell simulations. Left column is computed without axions, center column with $\rho_a / \rho_{\rm GJ} = -10$ , and right column with $\rho_a / \rho_{\rm GJ} = 10$. Each row corresponds to a fixed time snapshot, and the full evolution tracks the opening, collapse, and dynamical damping of the gap. A vertical dashed line has been placed at the position $x$ where $|\rho_a(x)| = |\rho_{\rm GJ}(x)|$, taking $|\rho_a(x=0)| = 10 |\rho_{\rm GJ}(x=0)| $. Note that the simulations in the right panel have been performed with a larger box size in order to avoid spurious effects arising when the gap is not sufficiently isolated from the boundary. The snapshots in the left panel correspond to time step: 4500, 9500, 13000, 17500, 22000, 50000; in the center panel to time step: 6000, 11000, 12500, 17500, 20500, 38000; and in the right panel to time step: 88500, 99500, 103500, 108500, 135000, 144000.}
    \label{fig:picE}
\end{figure*}

\begin{figure*}
    \centering
    \begin{adjustbox}{max width=0.76\textwidth, max height=0.76\textheight}
        \begin{tabular}{ccc}
            \includegraphics[trim={0cm 1.7cm 0cm 0cm},clip]{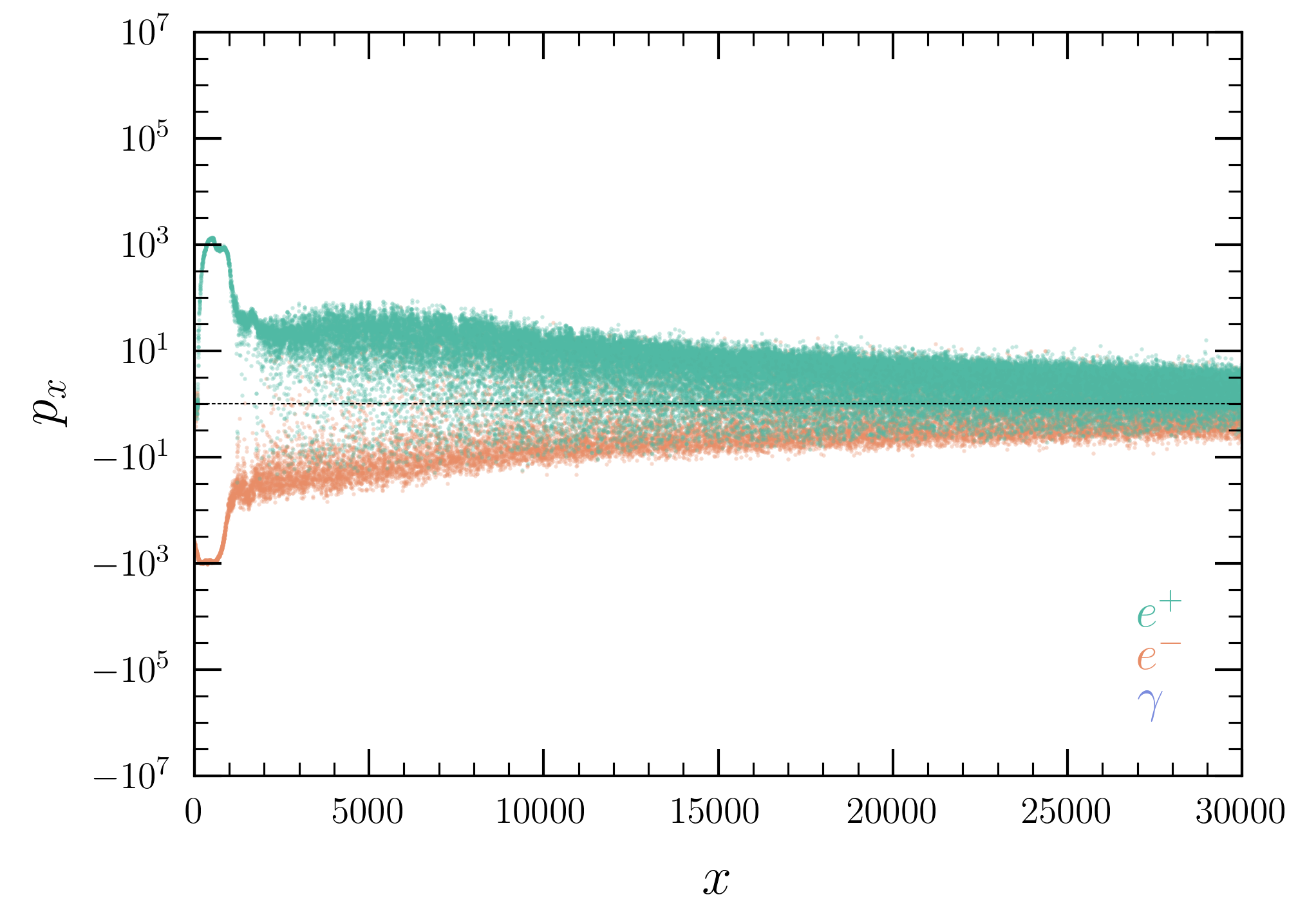} &
            \includegraphics[trim={2.7cm 1.7cm 0cm 0cm},clip]{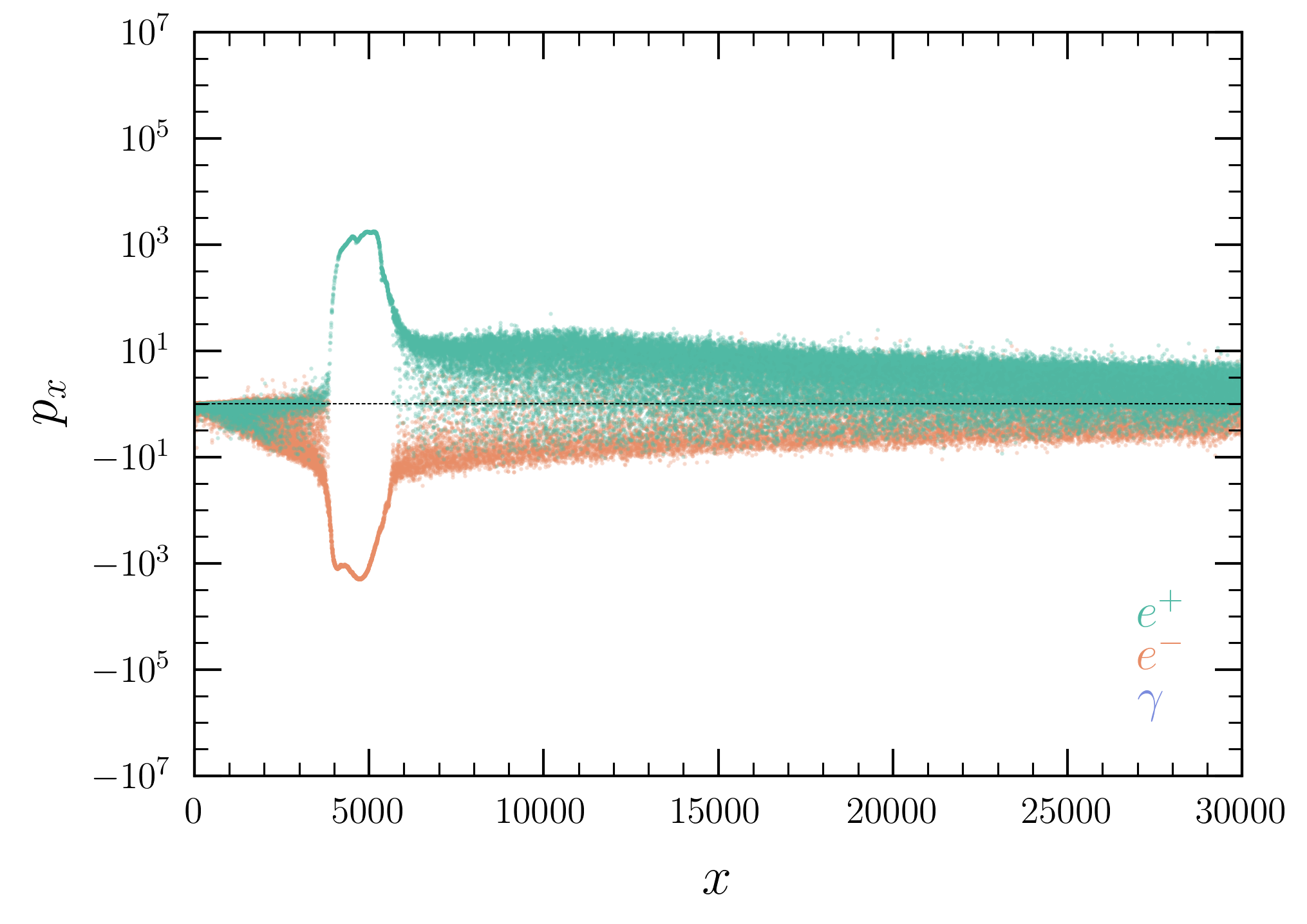} &
            \includegraphics[trim={2.7cm 1.7cm 0cm 0cm},clip]{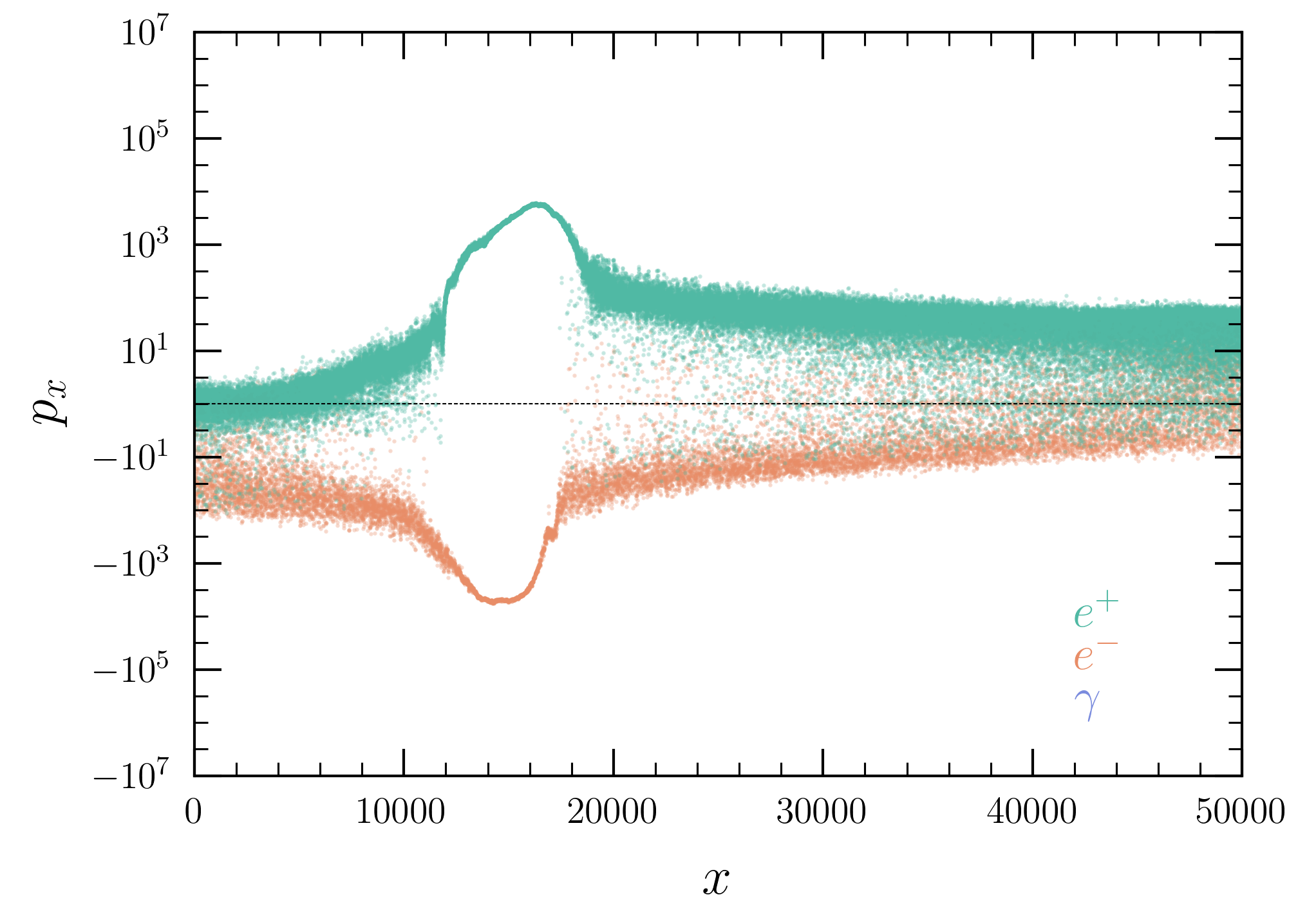} \\
            
            \includegraphics[trim={0cm 1.7cm 0cm 0cm},clip]{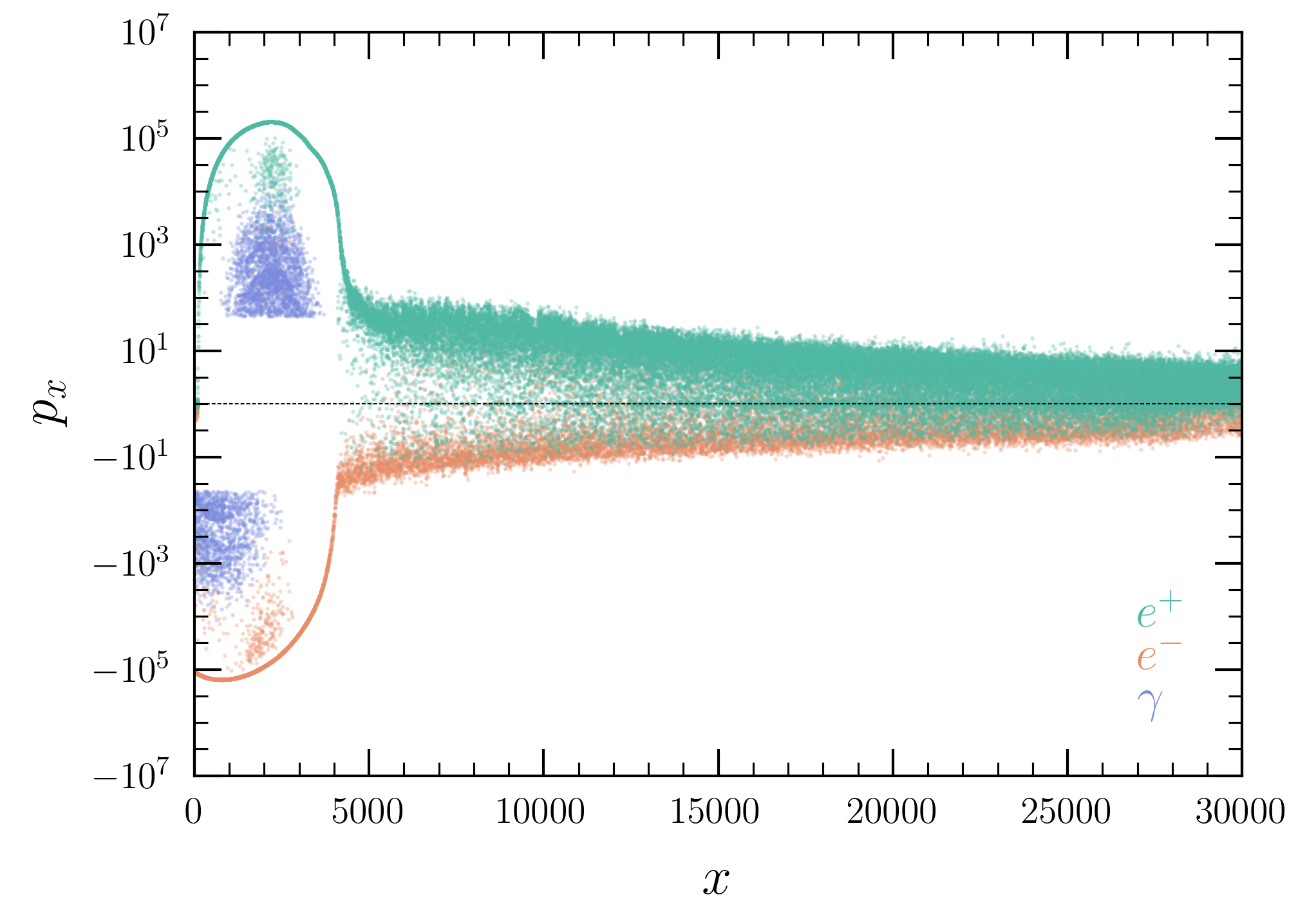} &
            \includegraphics[trim={2.7cm 1.7cm 0cm 0cm},clip]{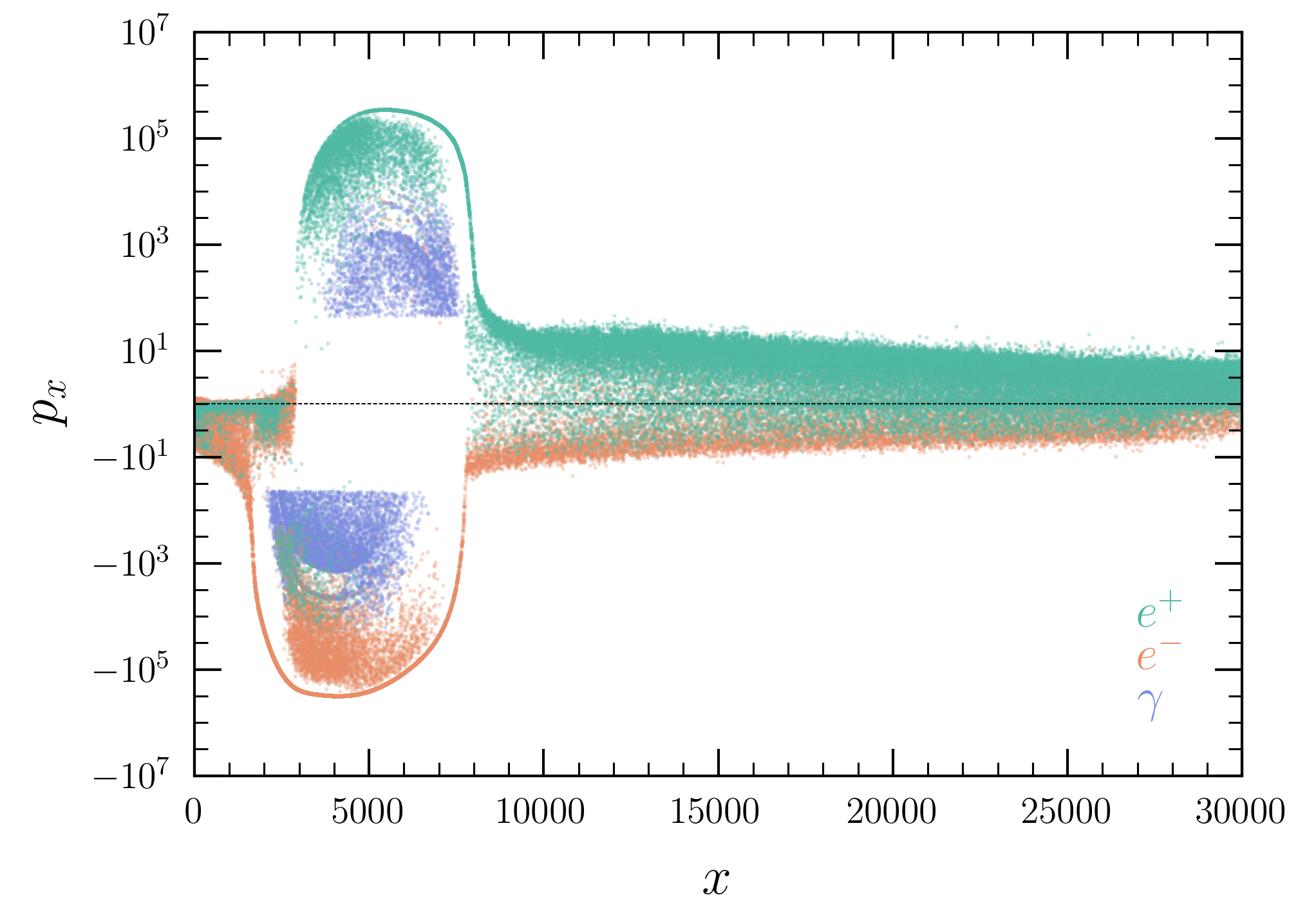} &
            \includegraphics[trim={2.7cm 1.7cm 0cm 0cm},clip]{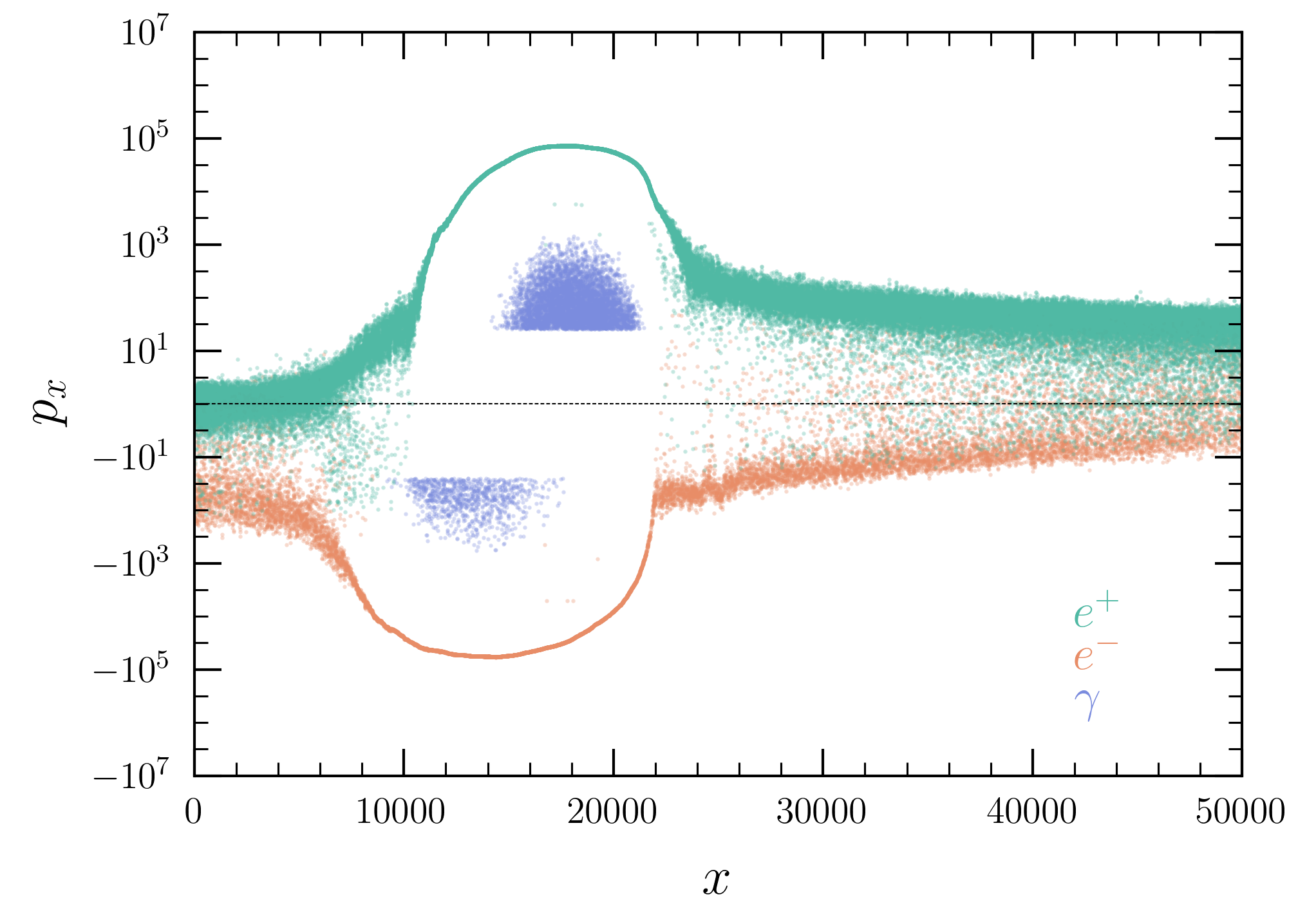} \\
            
            \includegraphics[trim={0cm 1.7cm 0cm 0cm},clip]{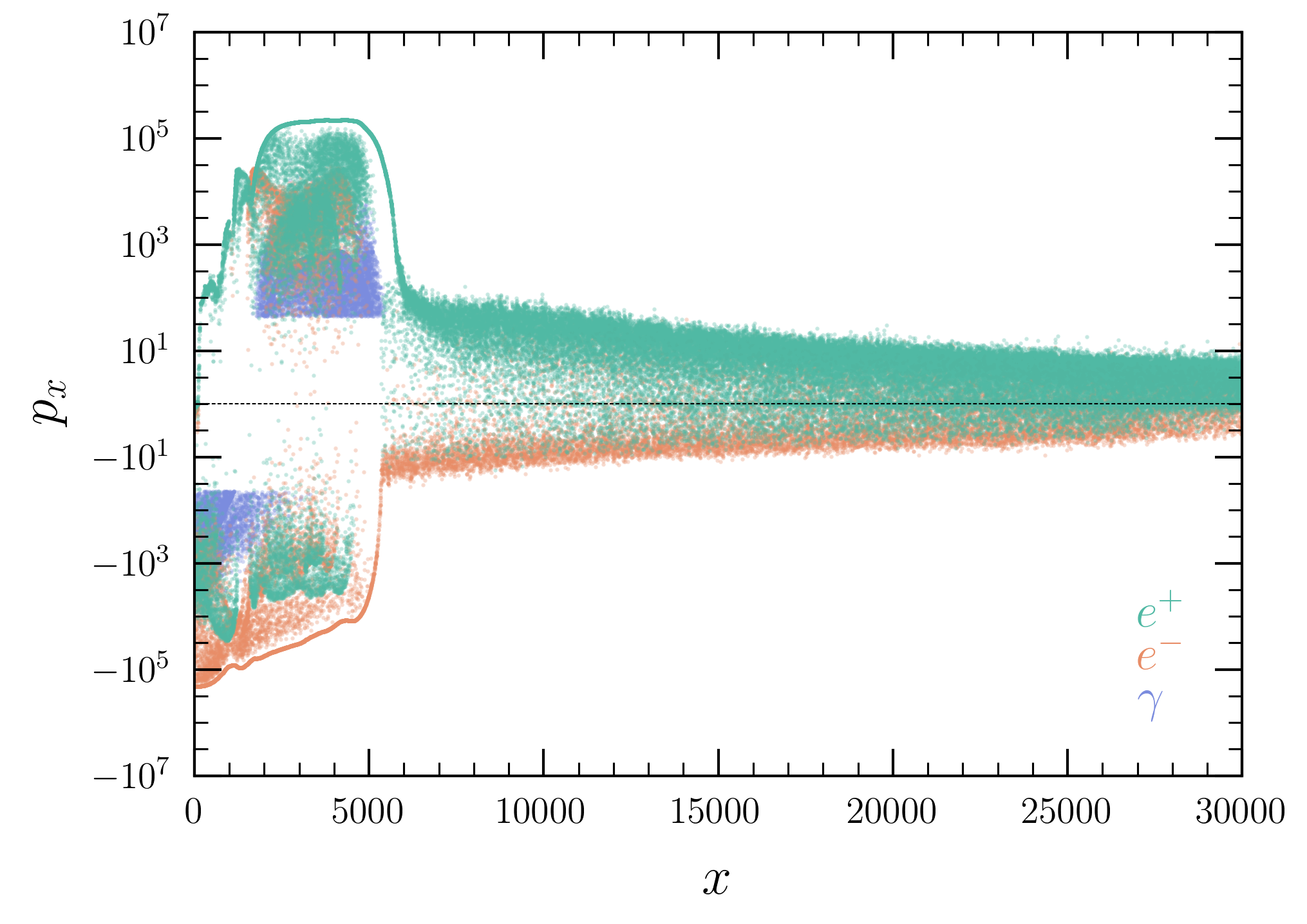} &
            \includegraphics[trim={2.7cm 1.7cm 0cm 0cm},clip]{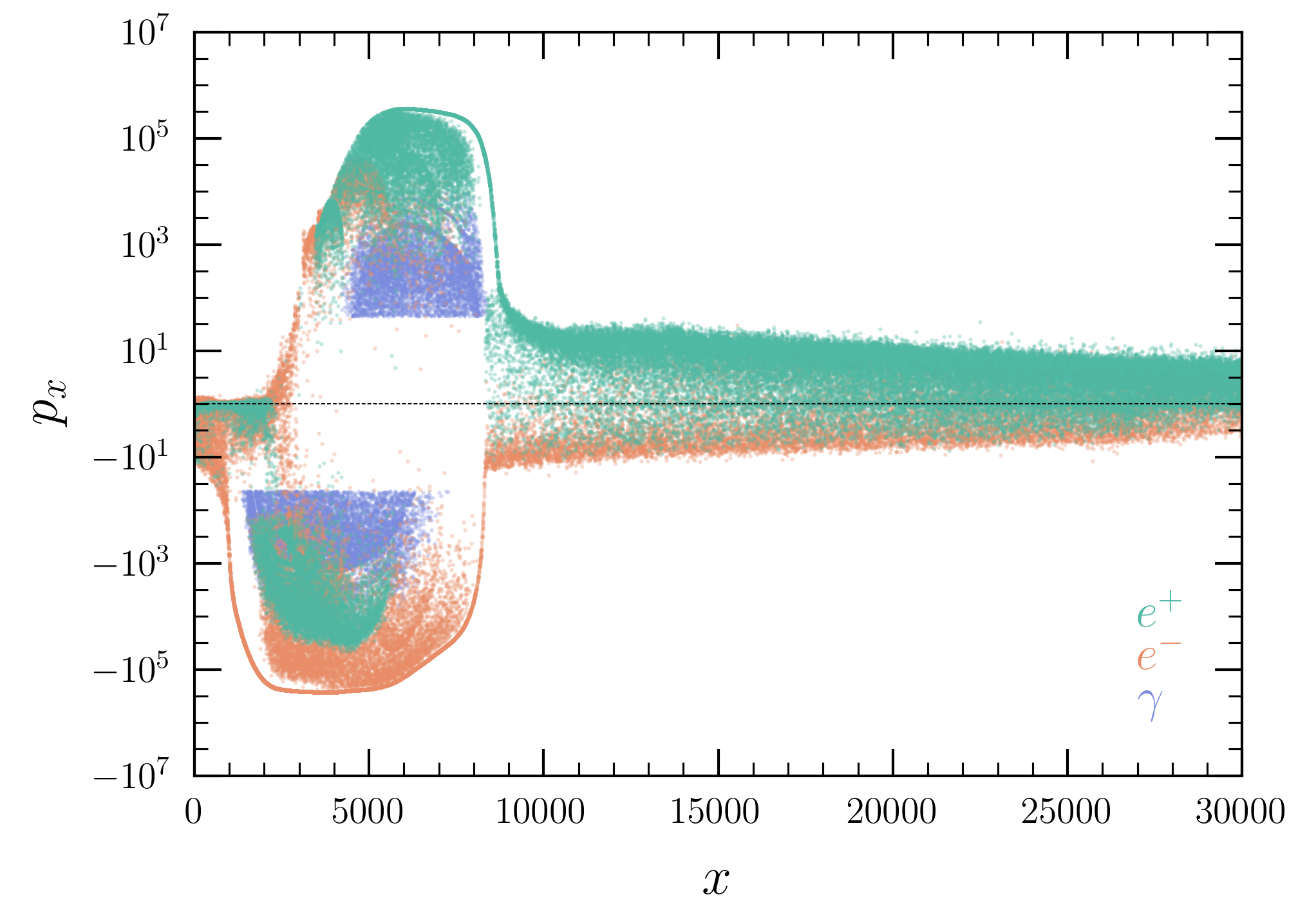} &
            \includegraphics[trim={2.7cm 1.7cm 0cm 0cm},clip]{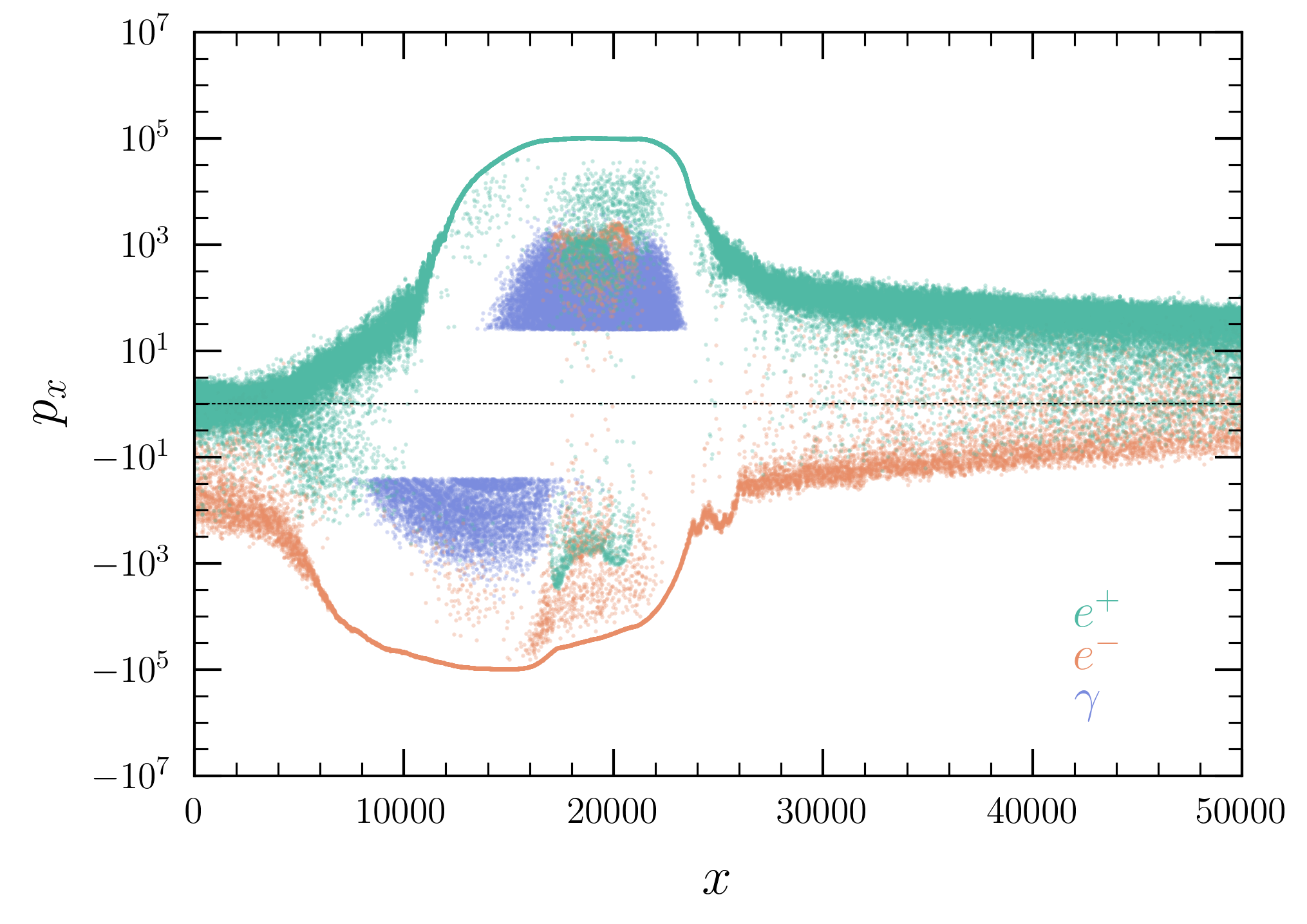} \\
            
            \includegraphics[trim={0cm 1.7cm 0cm 0cm},clip]{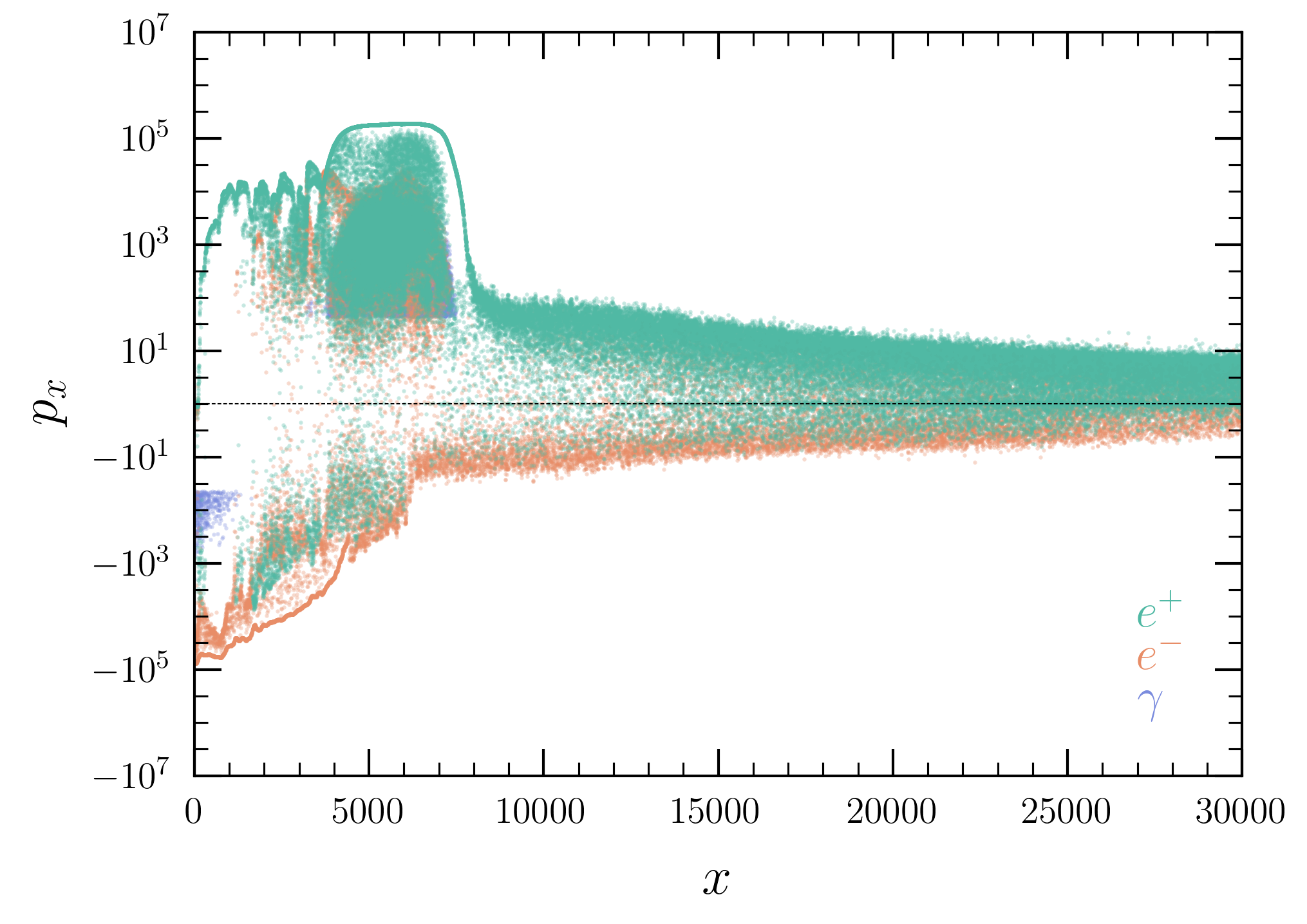} &
            \includegraphics[trim={2.7cm 1.7cm 0cm 0cm},clip]{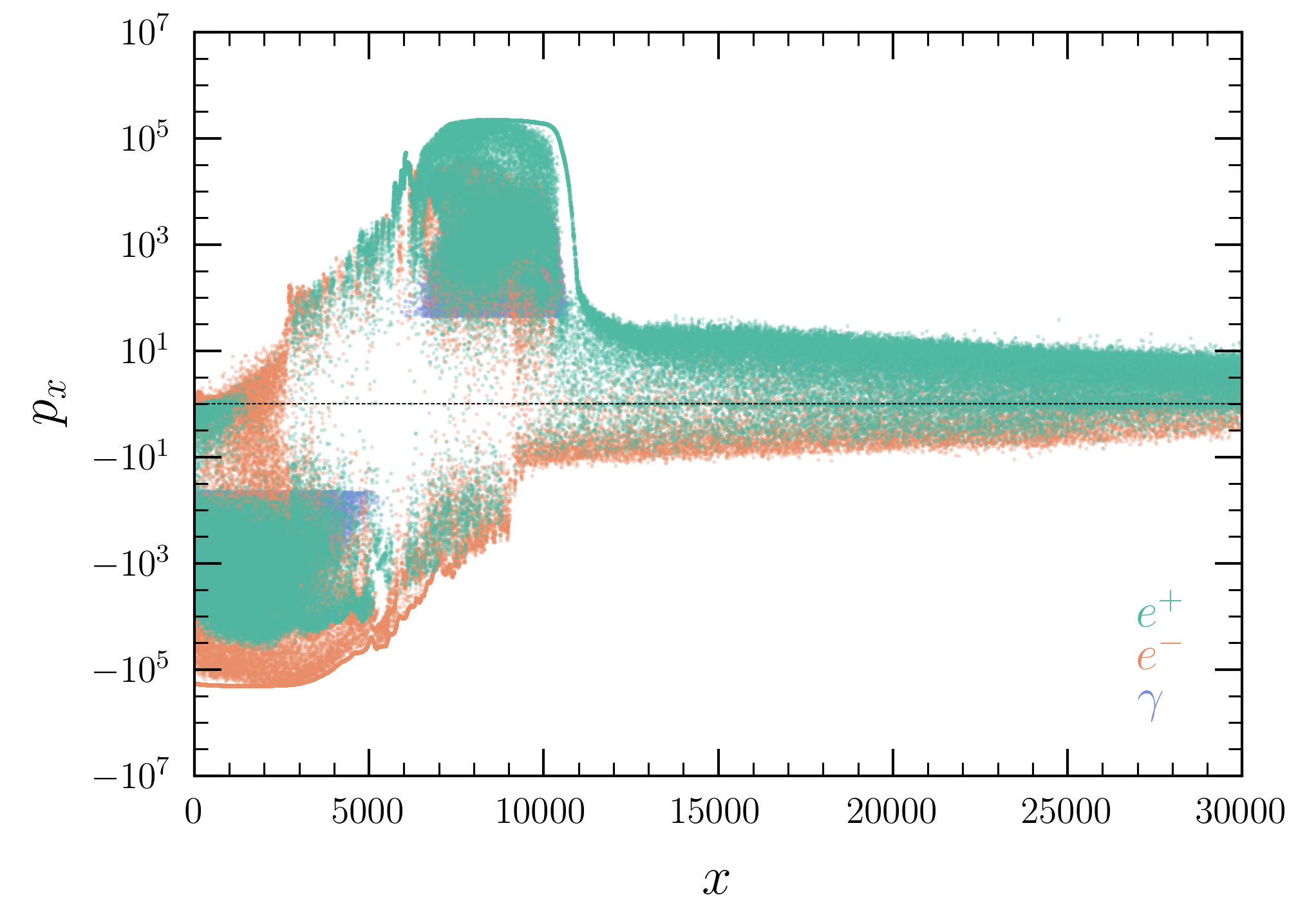} &
            \includegraphics[trim={2.7cm 1.7cm 0cm 0cm},clip]{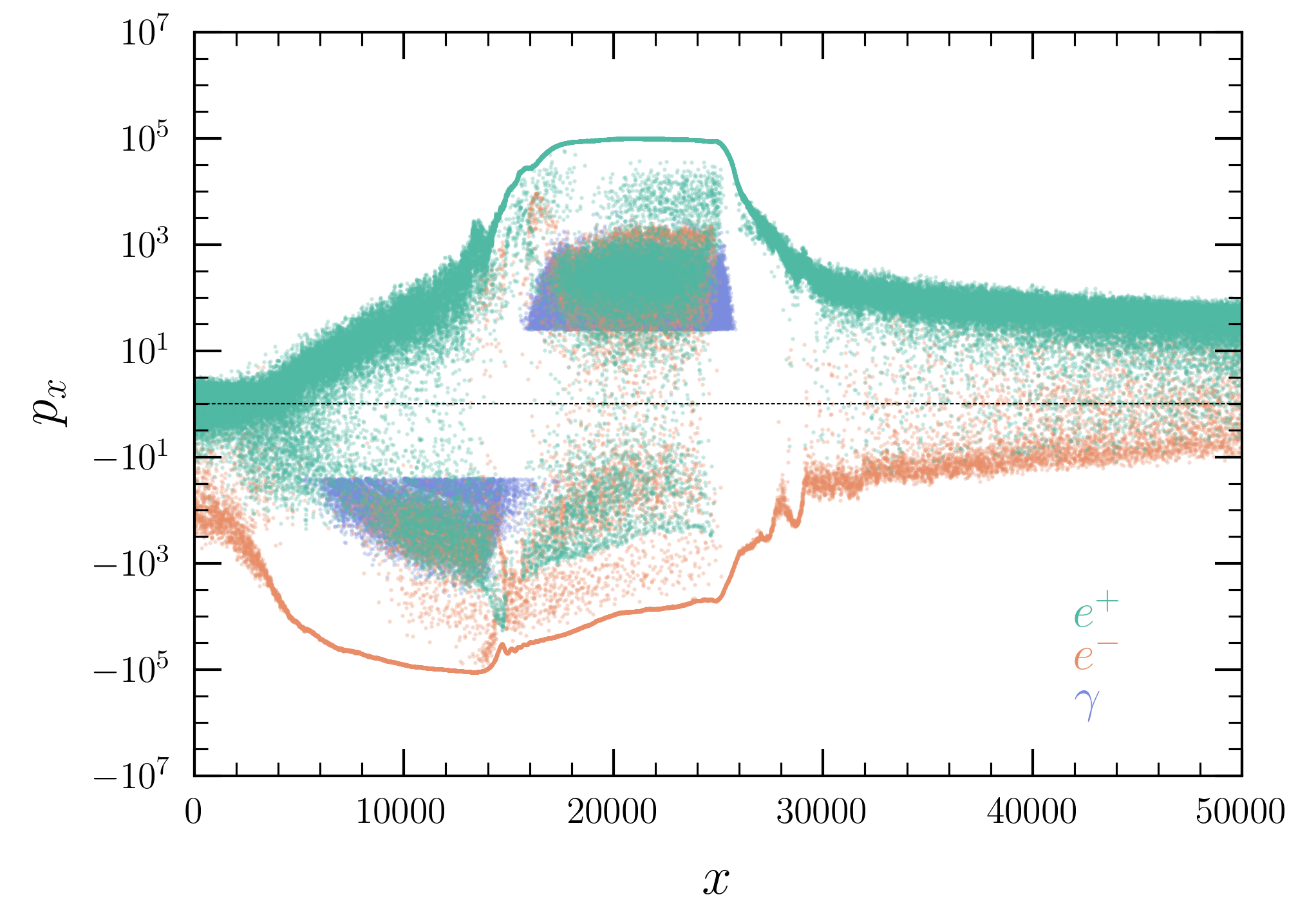} \\
            
            \includegraphics[trim={0cm 1.7cm 0cm 0cm},clip]{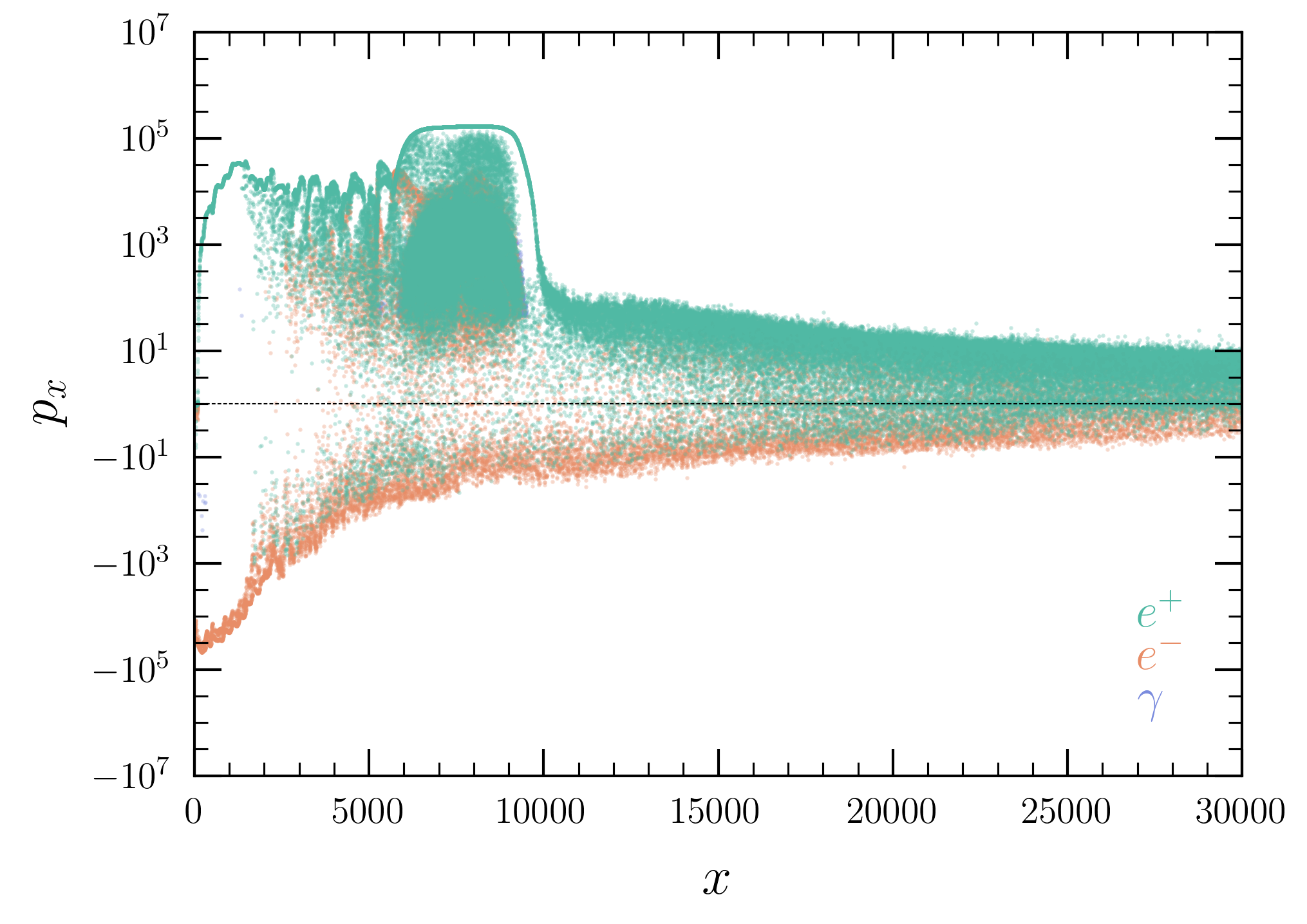} &
            \includegraphics[trim={2.7cm 1.7cm 0cm 0cm},clip]{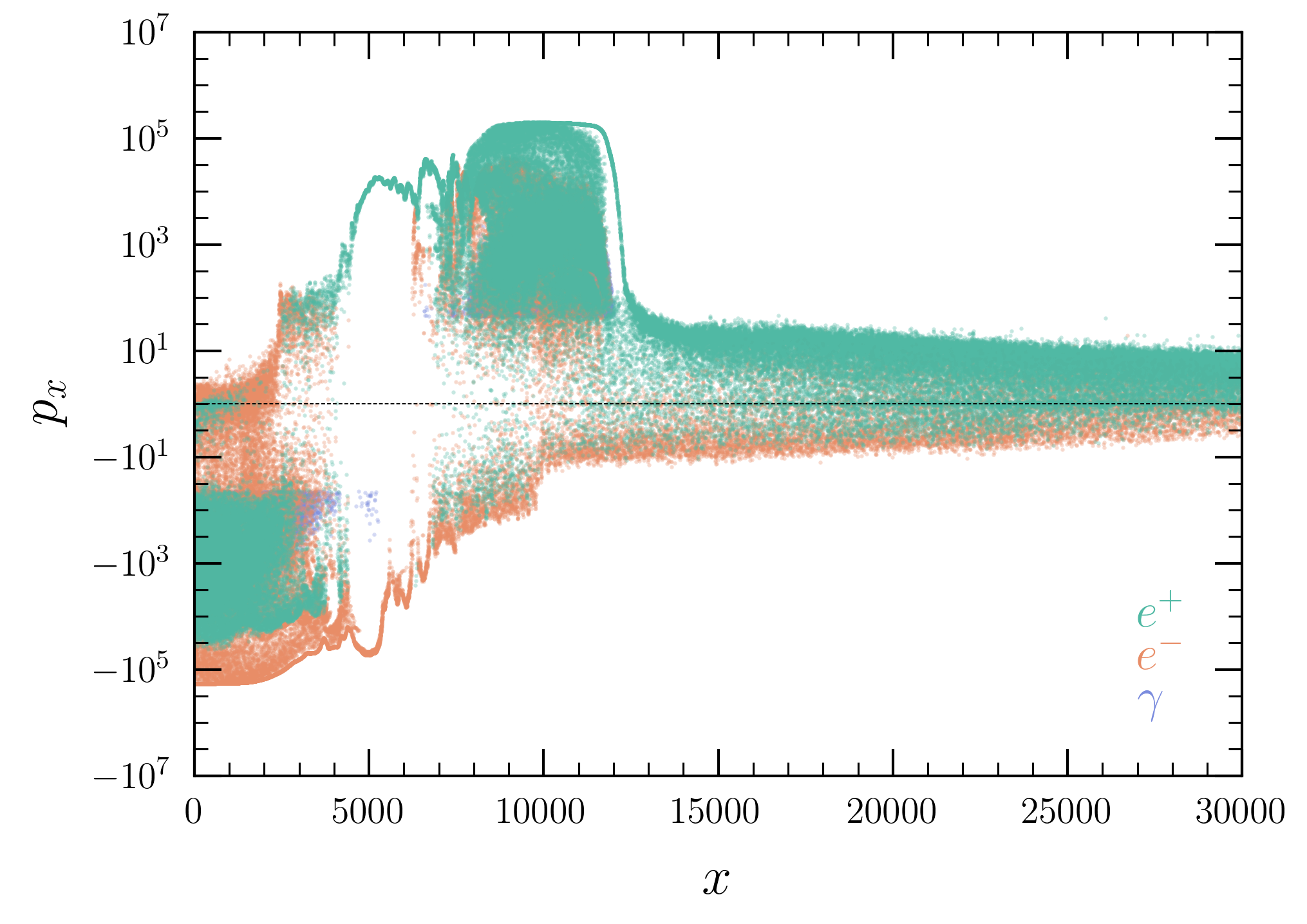} &
            \includegraphics[trim={2.7cm 1.7cm 0cm 0cm},clip]{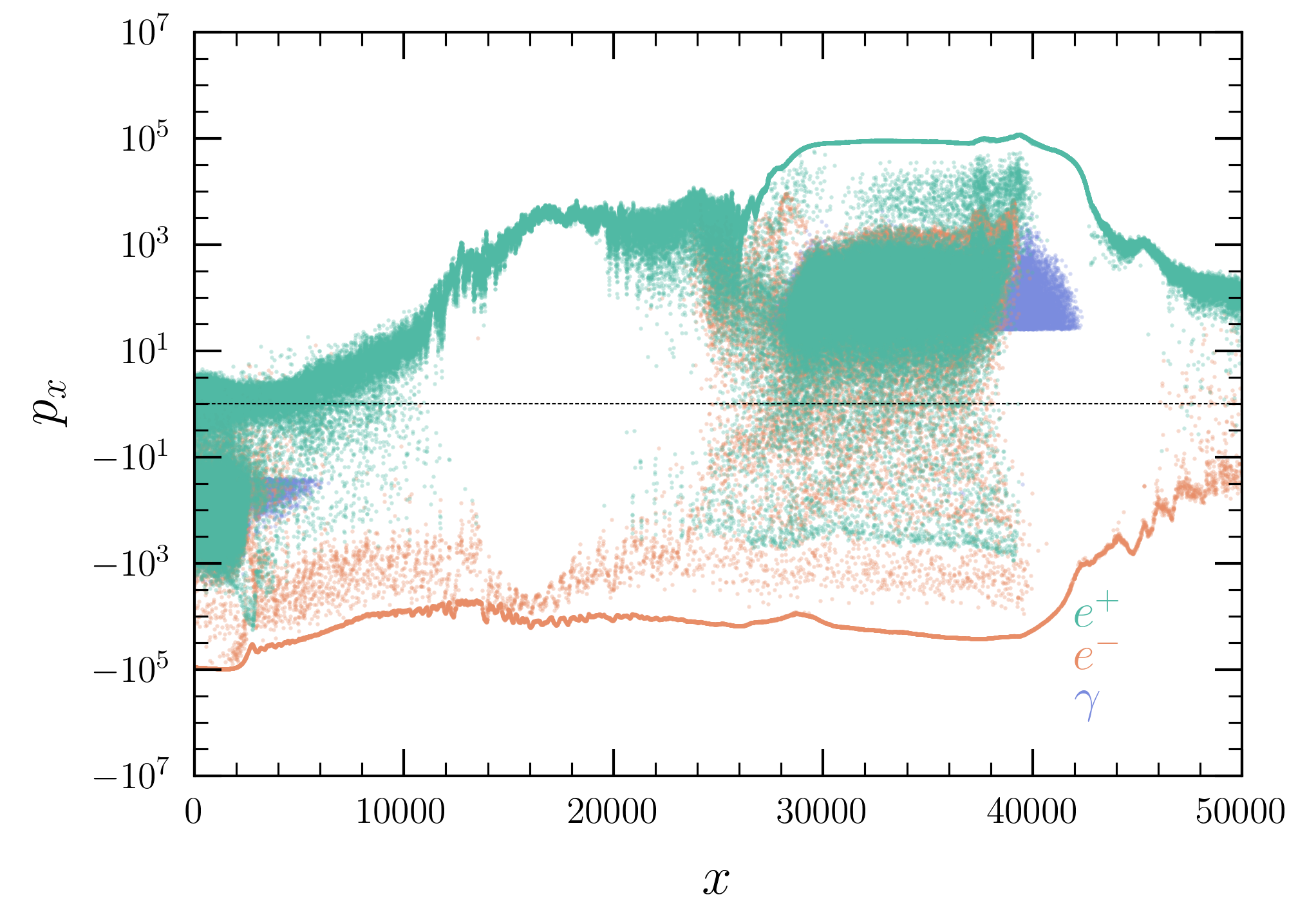} \\
            
            \includegraphics[trim={0cm 0cm 0cm 0cm},clip]{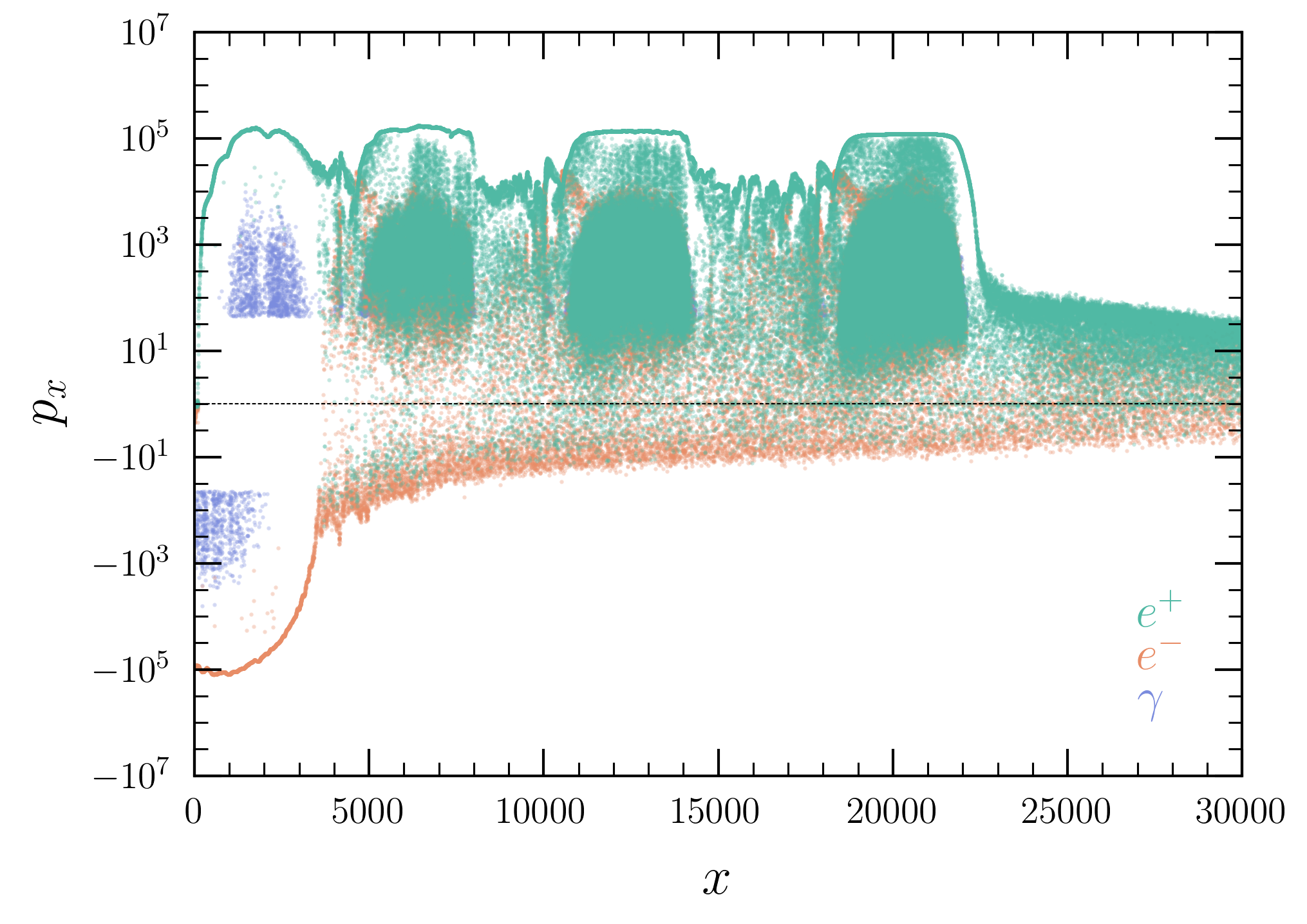} &
            \includegraphics[trim={2.7cm 0cm 0cm 0cm},clip]{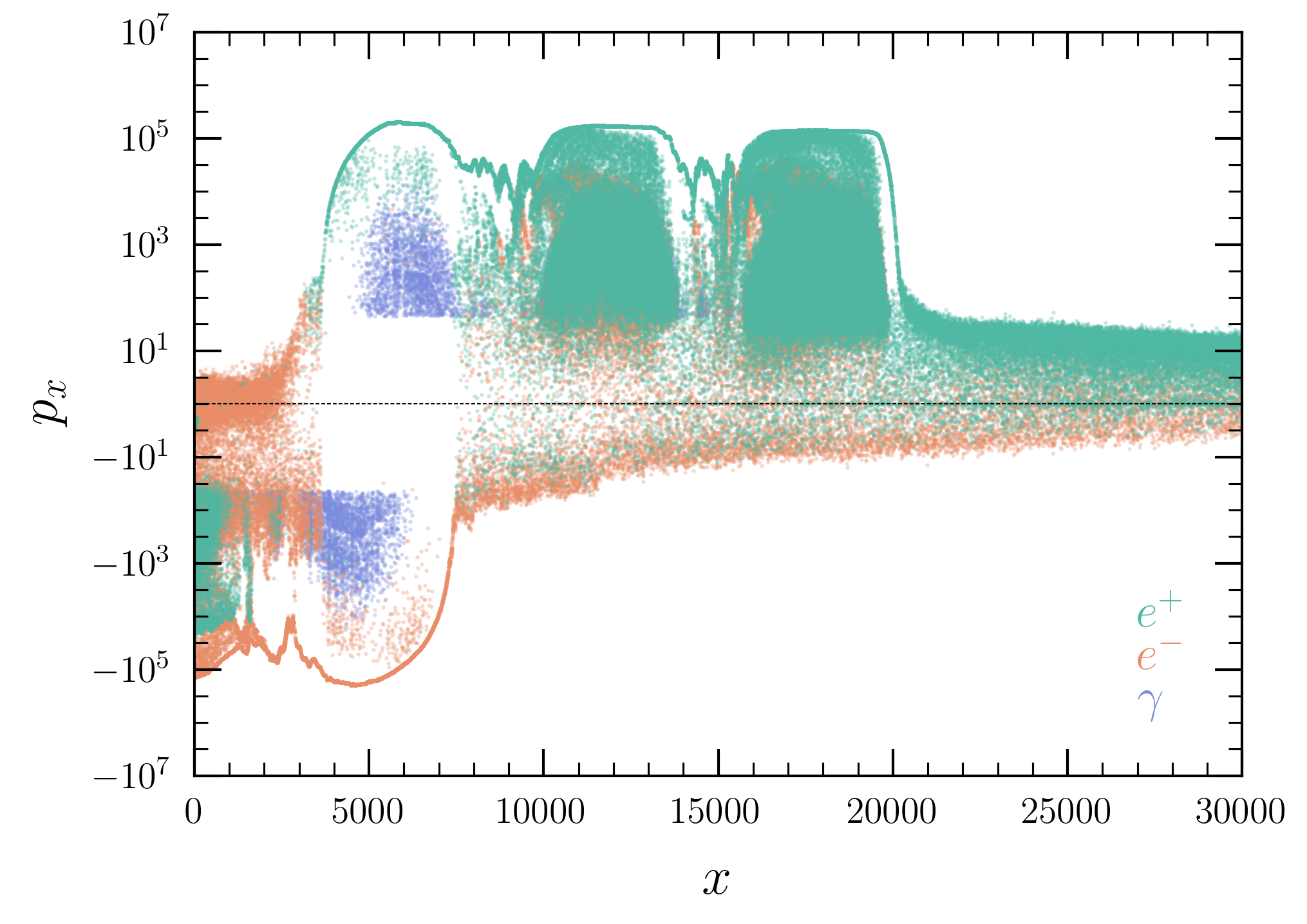} &
            \includegraphics[trim={2.7cm 0cm 0cm 0cm},clip]{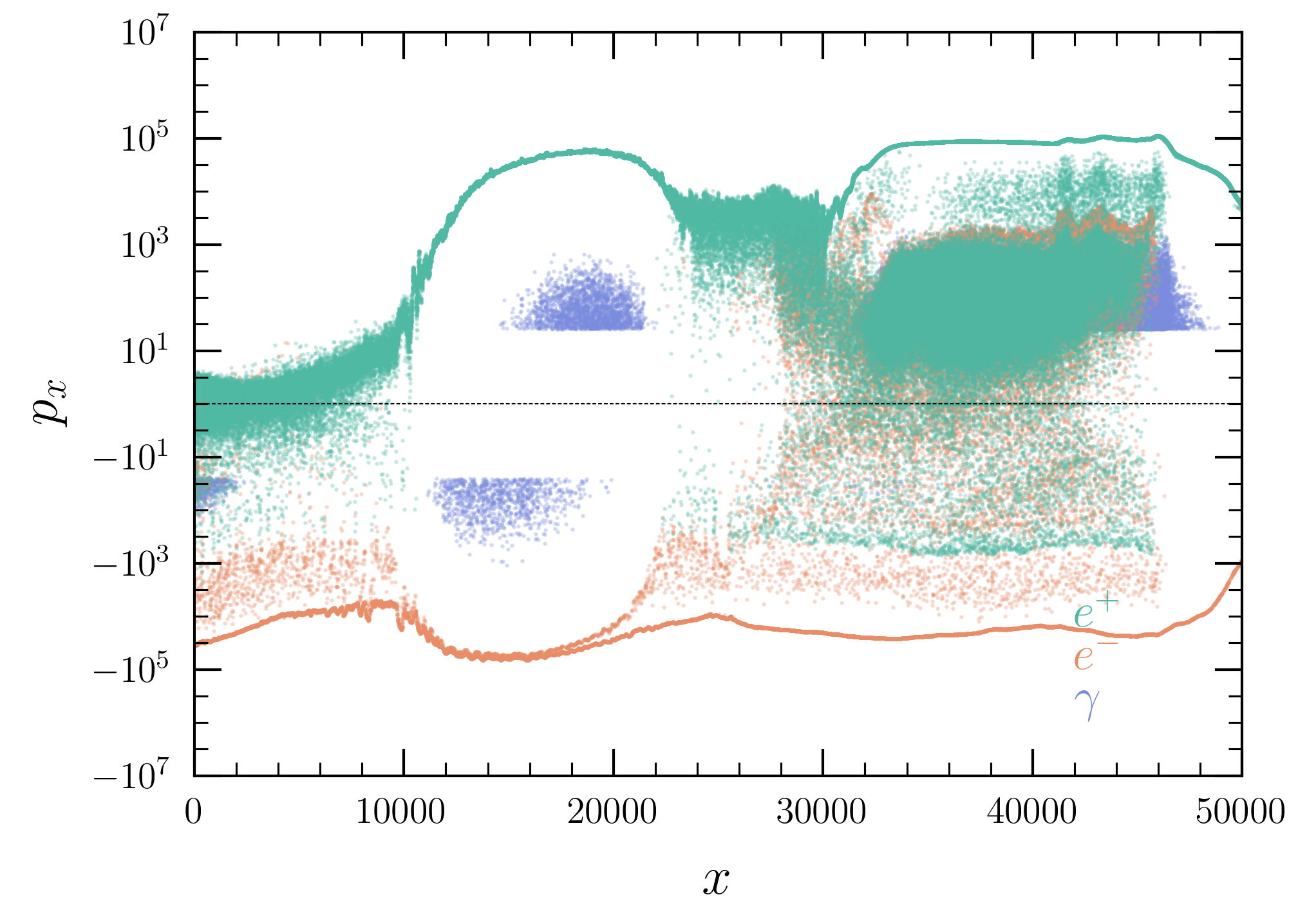} \\
        \end{tabular}
    \end{adjustbox}
    \caption{Same as Fig.~\ref{fig:picE}, but showing the evolution of the $e^-$ (orange dots), $e^+$ (green) and $\gamma$ (blue)  phase space distributions. Time snapshots correspond to the same moments as shown in Fig.~\ref{fig:picE}. }\label{fig:picP}
\end{figure*}

\begin{figure*}
    \includegraphics[width=0.32\linewidth]{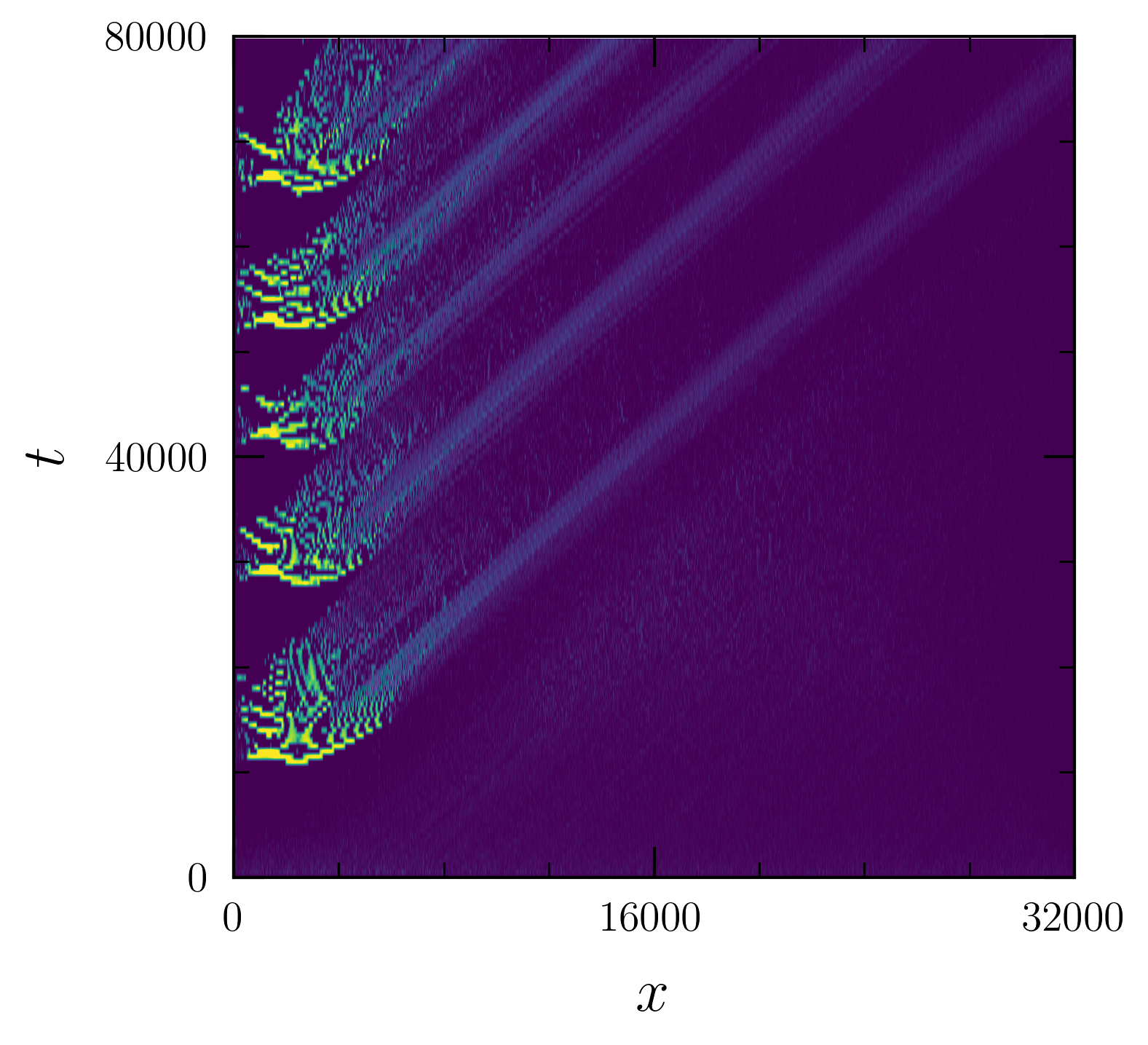}
    \includegraphics[width=0.32\linewidth]{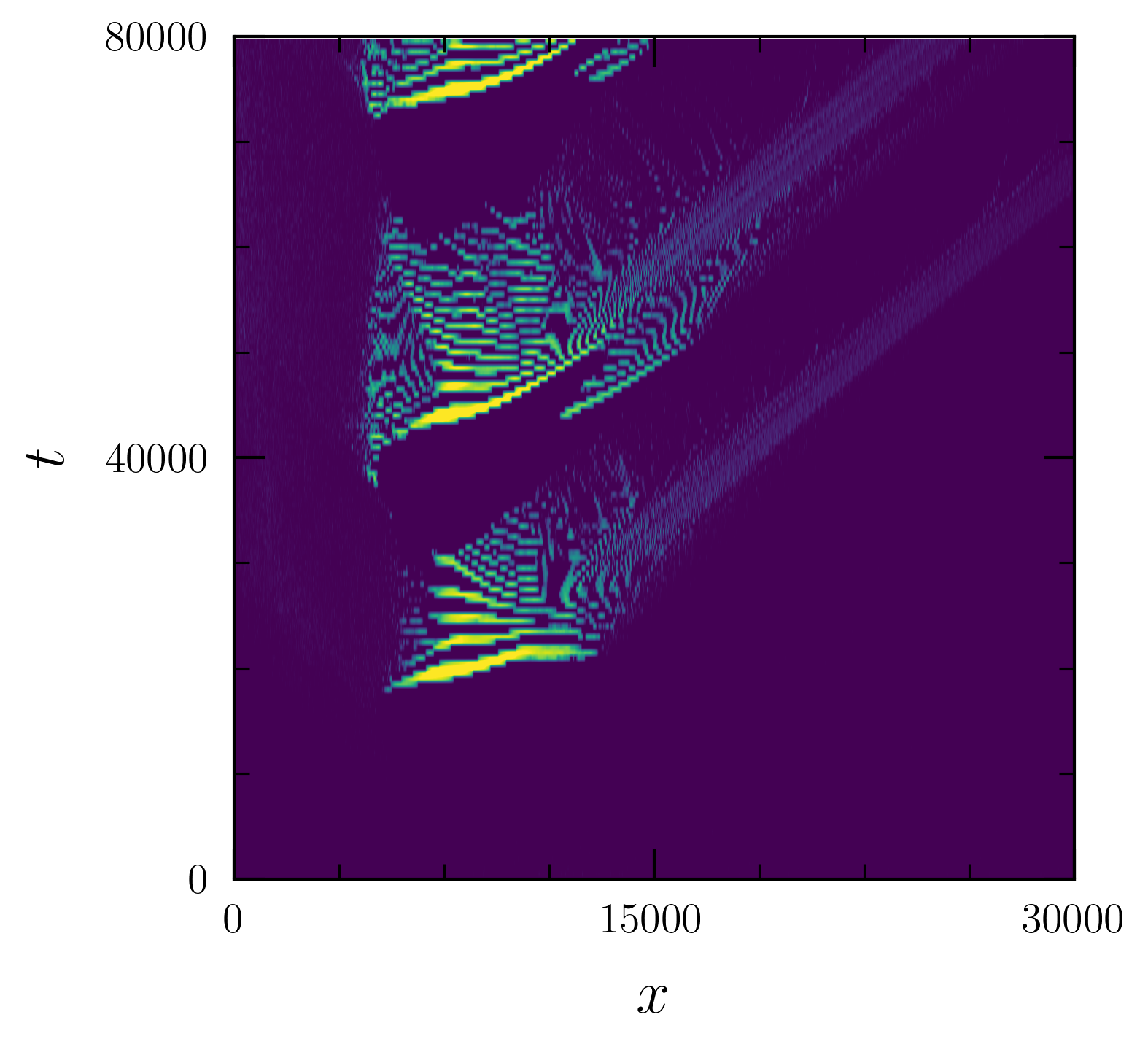}
    \includegraphics[width=0.32\linewidth]{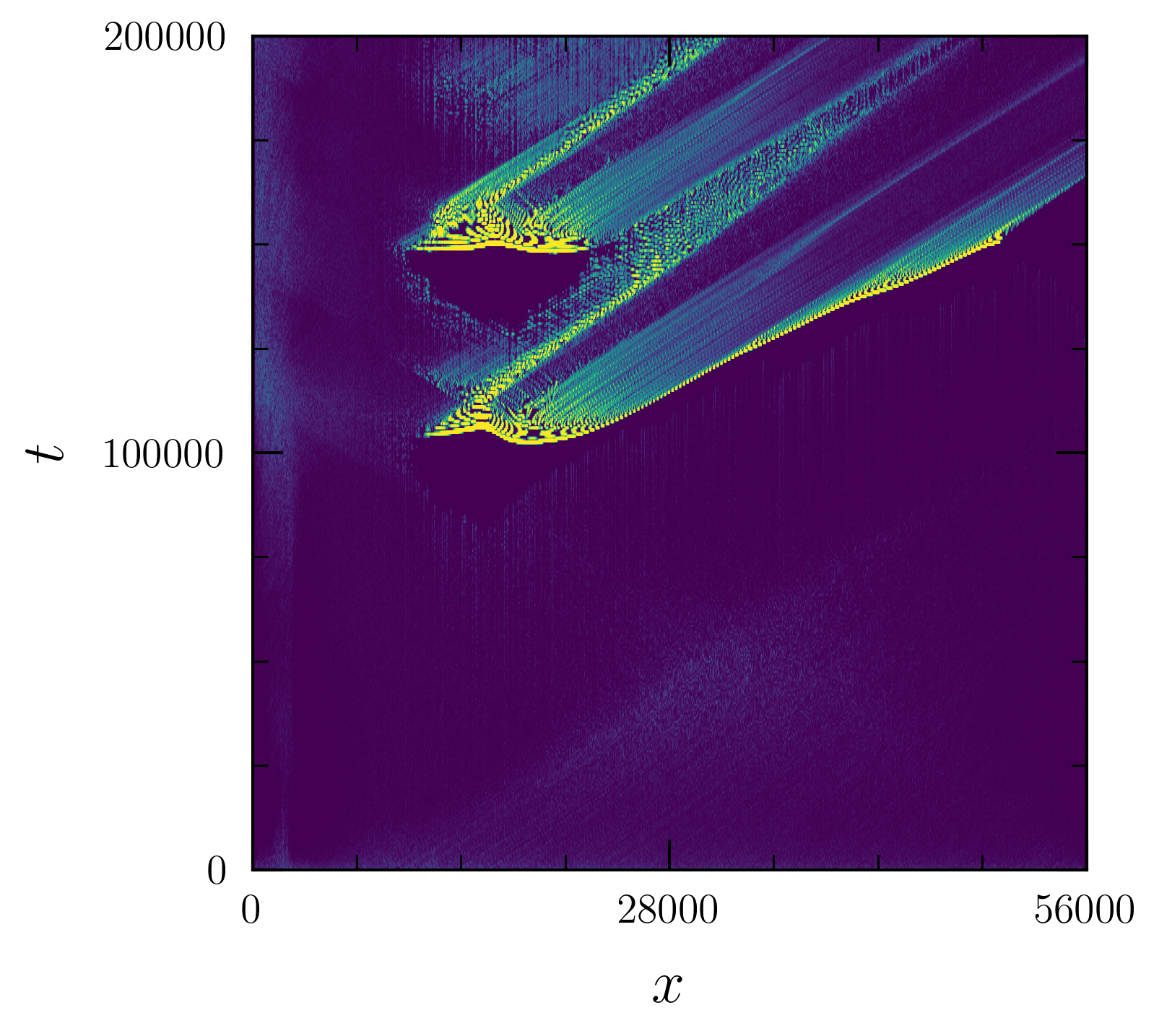}
    \caption{Spatiotemporal evolution of the amplitude of $(E_{||}/E_0)^2$ (truncated at a threshold near $\sim 10^{-8}$ to remove small scale noise, and plotted with a log-scale color scheme) for the default run in Fig.~\ref{fig:picE} (left) the anti-aligned axion gradient (center), and the aligned axion gradient (right). X- and y-axes denote spatial and temporal coordinates, respectively (note the right panel has a longer timescale and spatial scale as the gap size and periodicity is notably larger). This image provides an alternate view of how the gap collapse operates, clearly illustrating the evolution in the damped oscillations and the quasi-periodicity of the collapse process. In Fig.~\ref{fig:picE} and~\ref{fig:picP}, we show only one full discharge cycle and the beginning of the following cycle; this figure shows that the system is in an oscillating steady state with all discharges having similar behavior. }\label{fig:cntr_pic}
\end{figure*}

We start by performing three simulations which illustrate the impact of axion hair on a field line with active pair discharge. Among these simulations is one with $\xi = 0$ (corresponding to no axion), one with $\xi = 10$ (corresponding to an axion-induced electric field with the same sign at the stellar surface) and one with $\xi = -10$ (corresponding to an axion-induced electric field with the opposite sign at the stellar surface). Fixed-time snapshots showing the growth and collapse of $E_{||}$, as well as the $e^\pm$ and $\gamma$ distributions, are shown in Figs.~\ref{fig:picE} and ~\ref{fig:picP}; here, each column corresponds to a fixed $\xi$, with $\xi = 0$ being on the left, $\xi = -10$ being in the center, and $\xi = 10$ being on the right. We also include Fig.~\ref{fig:cntr_pic}, which shows the 2-dimensional evolution of $(E_{||})^2$ as a function of spatial and temporal coordinates, clearly illustrating the damped oscillations and the quasi periodicity of the gap collapse process (here, we have truncated small values of $E_{||}$ in order to cut-out excess noise, and plotted using a log-scale color scheme in order to highlight the damped oscillations).  In the absence of an axion gradient, particles are accelerated near the surface, reach high Lorentz factors $>\gamma_{\rm emit}$, emit pair producing gamma-rays, the gap collapses, and the process repeats. For $\xi = -10$, particles are not accelerated right at the surface, but rather remain non-relativistic until $\rho_a$ begins to fall; here, a gap forms, but it is displaced by a distance $\delta x \sim m_a^{-1}$. After pair production and gap collapse, one can see the phase space distribution of newly formed pairs differs significantly from the standard scenario -- namely the charges flowing back to the surface are not purely $e^-$, but instead also contain a high density of slowly moving $e^+$. For $\xi = 10$, the gap once again appears displaced from the stellar surface, showing a reduced amplitude but larger spatial extent (corresponding to similar voltage drops). Note that the larger displacement in the appearance of the gap observed in this simulation is due to the spatial dependence of $\rho_{\rm GJ}$, whose absolute value increases  with distance from the stellar surface.  

\begin{figure*}
    \includegraphics[width=0.32\linewidth, trim={0cm, 1.7cm, 0cm, 0cm }, clip]{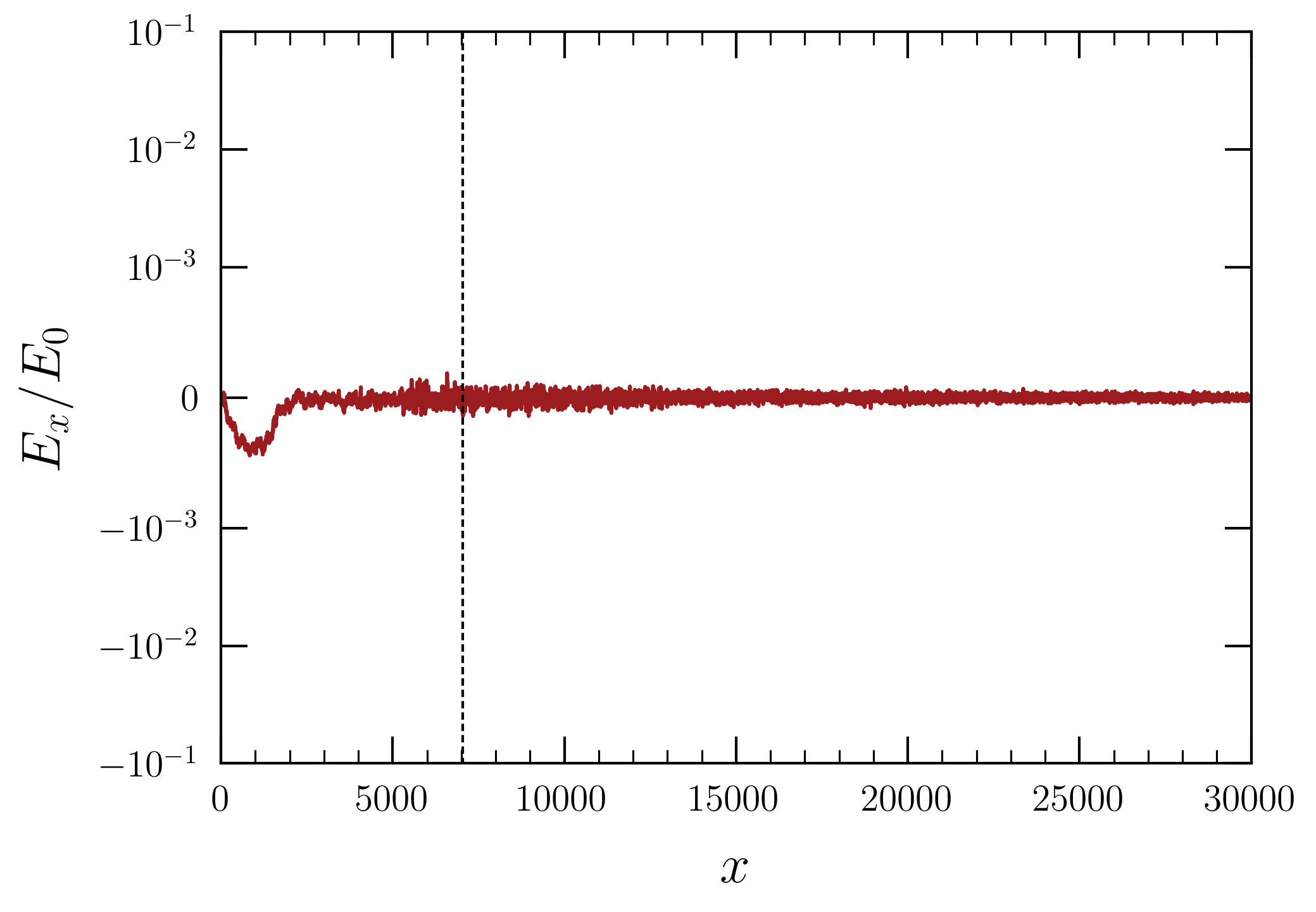}
    \includegraphics[width=0.27\linewidth,trim={2.7cm, 1.7cm, 0cm, 0cm }, clip]{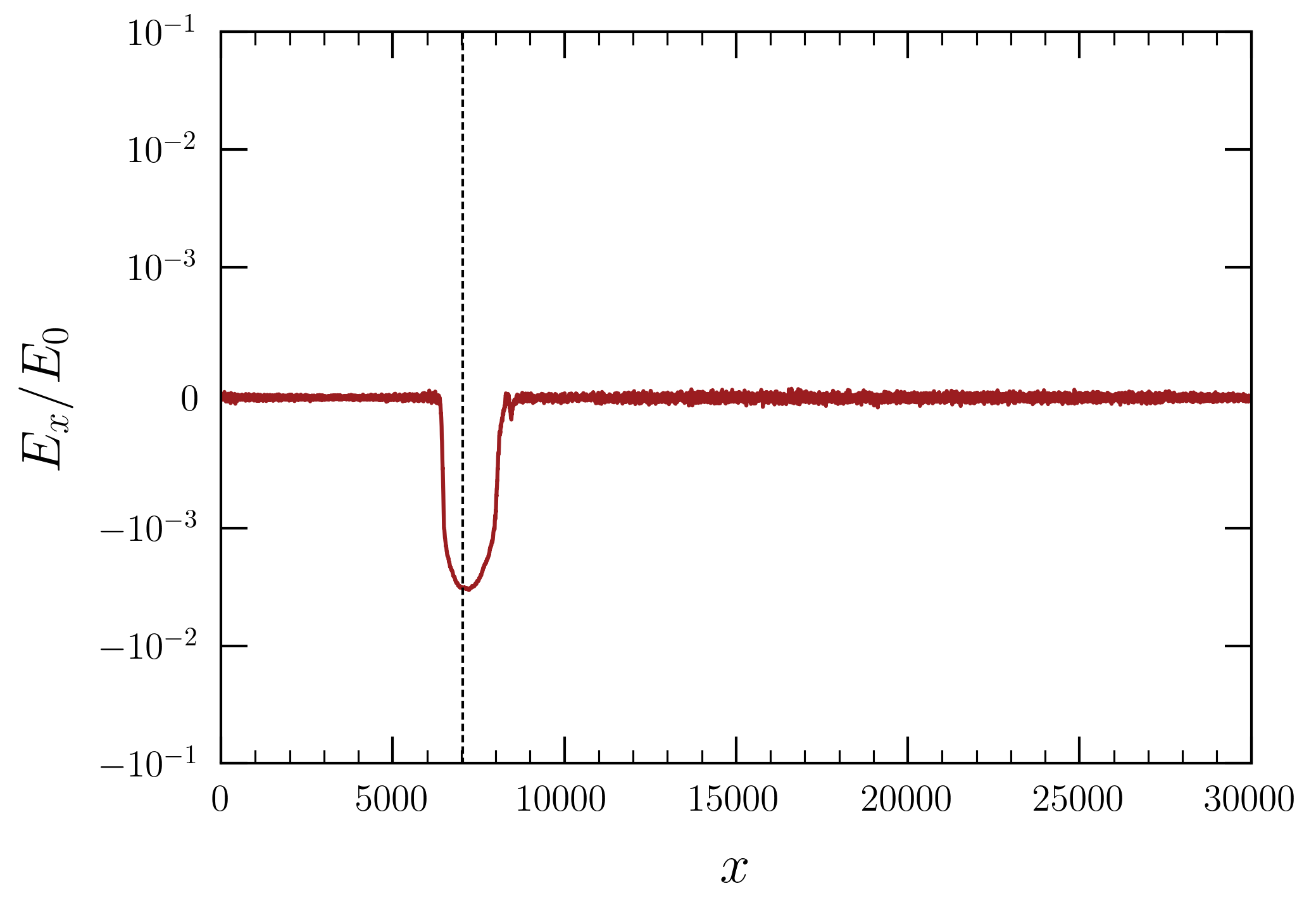}
    \includegraphics[width=0.27\linewidth,trim={2.7cm, 1.7cm, 0cm, 0cm }, clip]{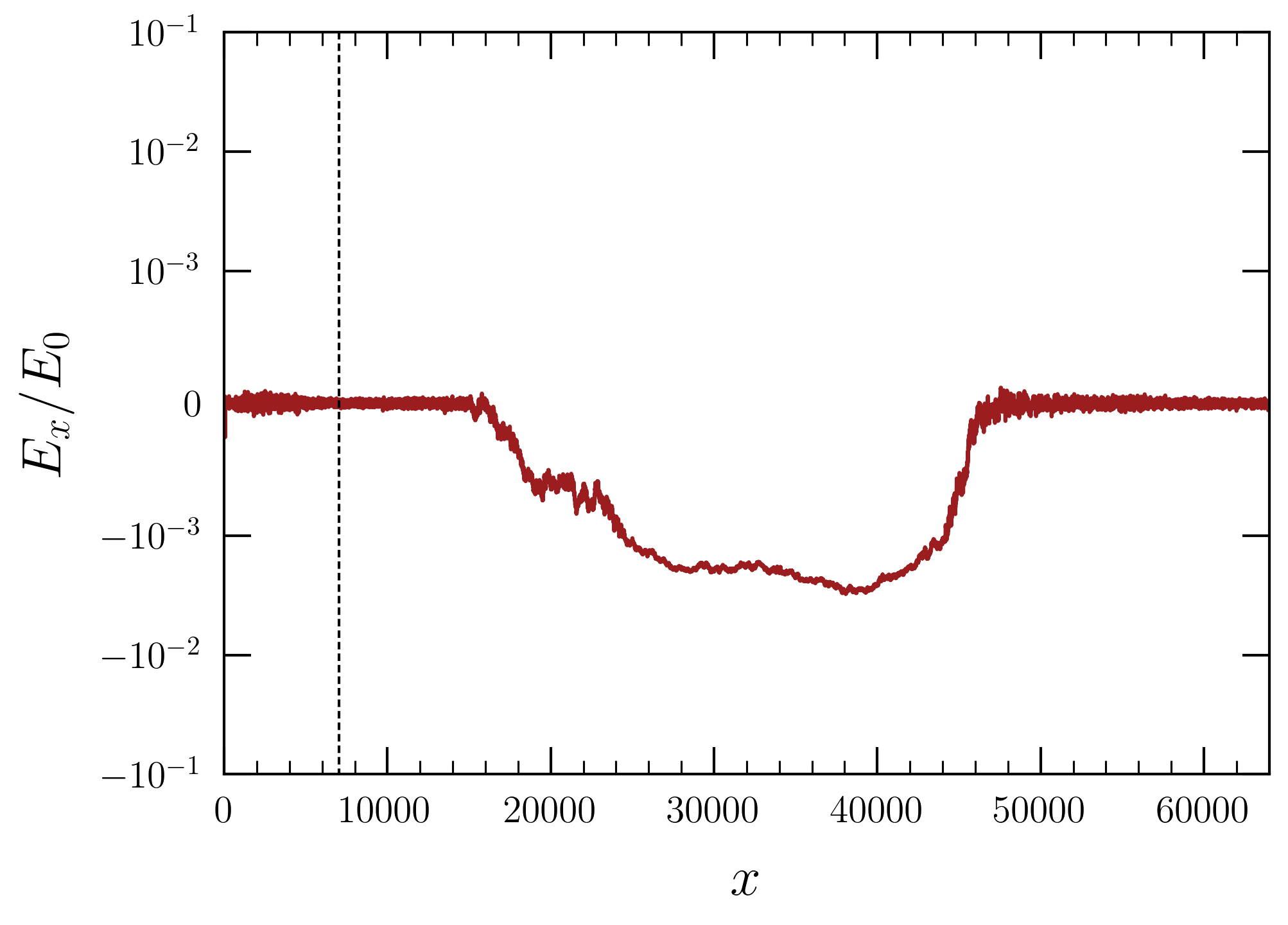}

    \includegraphics[width=0.32\linewidth, trim={0cm, 1.7cm, 0cm, 0cm }, clip]{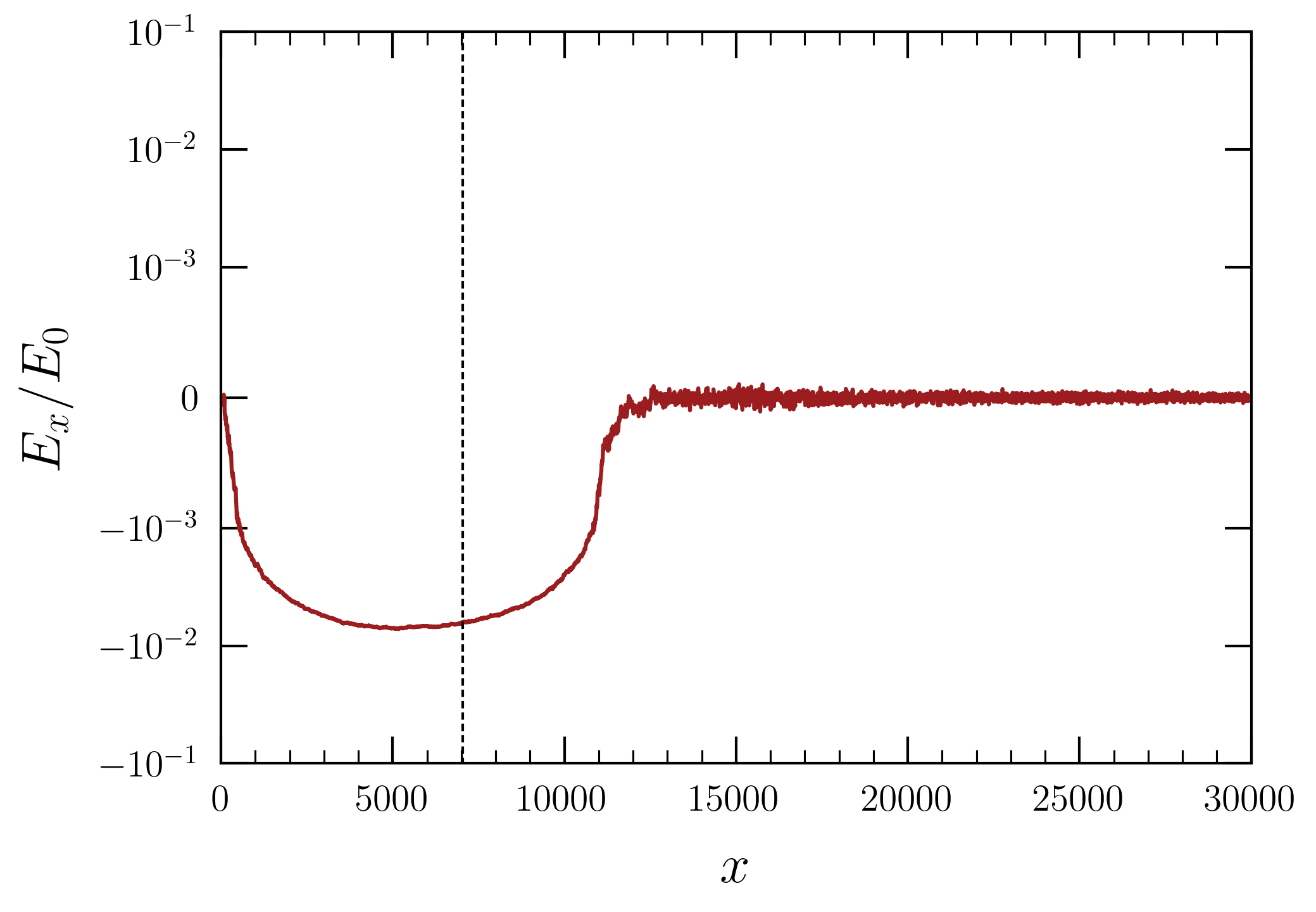}
    \includegraphics[width=0.27\linewidth,trim={2.7cm, 1.7cm, 0cm, 0cm }, clip]{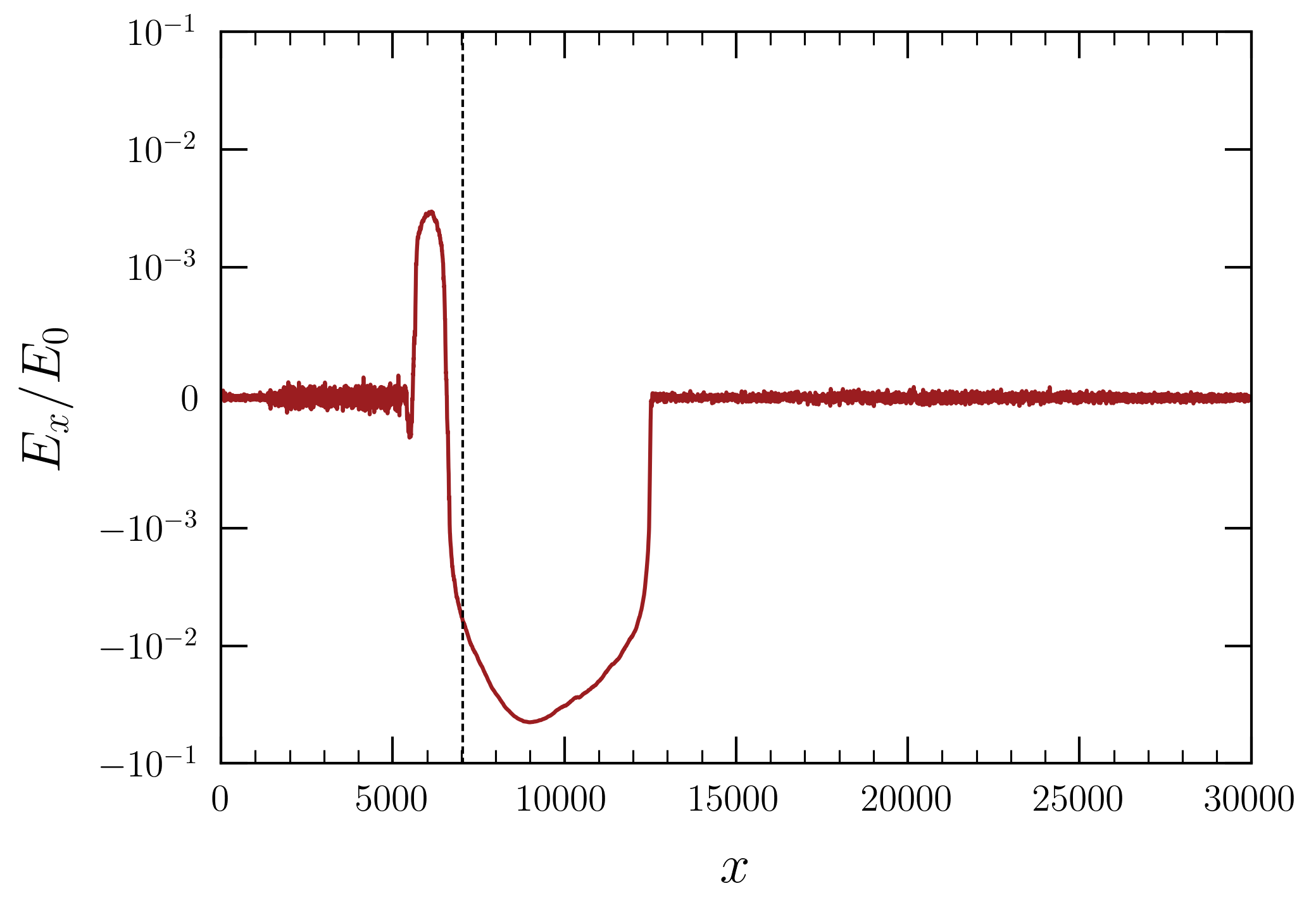}
    \includegraphics[width=0.27\linewidth,trim={2.7cm, 1.7cm, 0cm, 0cm }, clip]{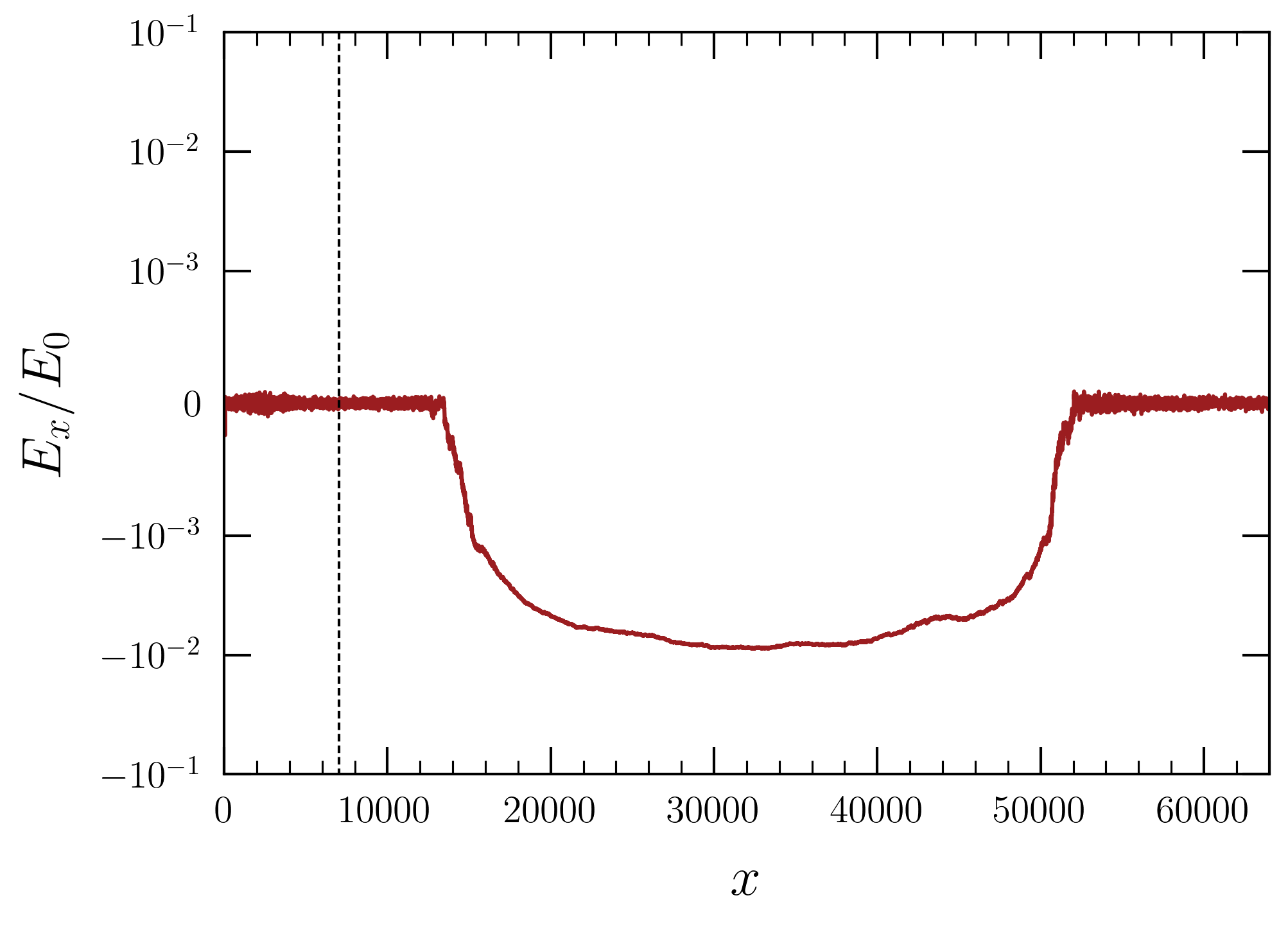}

    \includegraphics[width=0.32\linewidth, trim={0cm, 1.7cm, 0cm, 0cm }, clip]{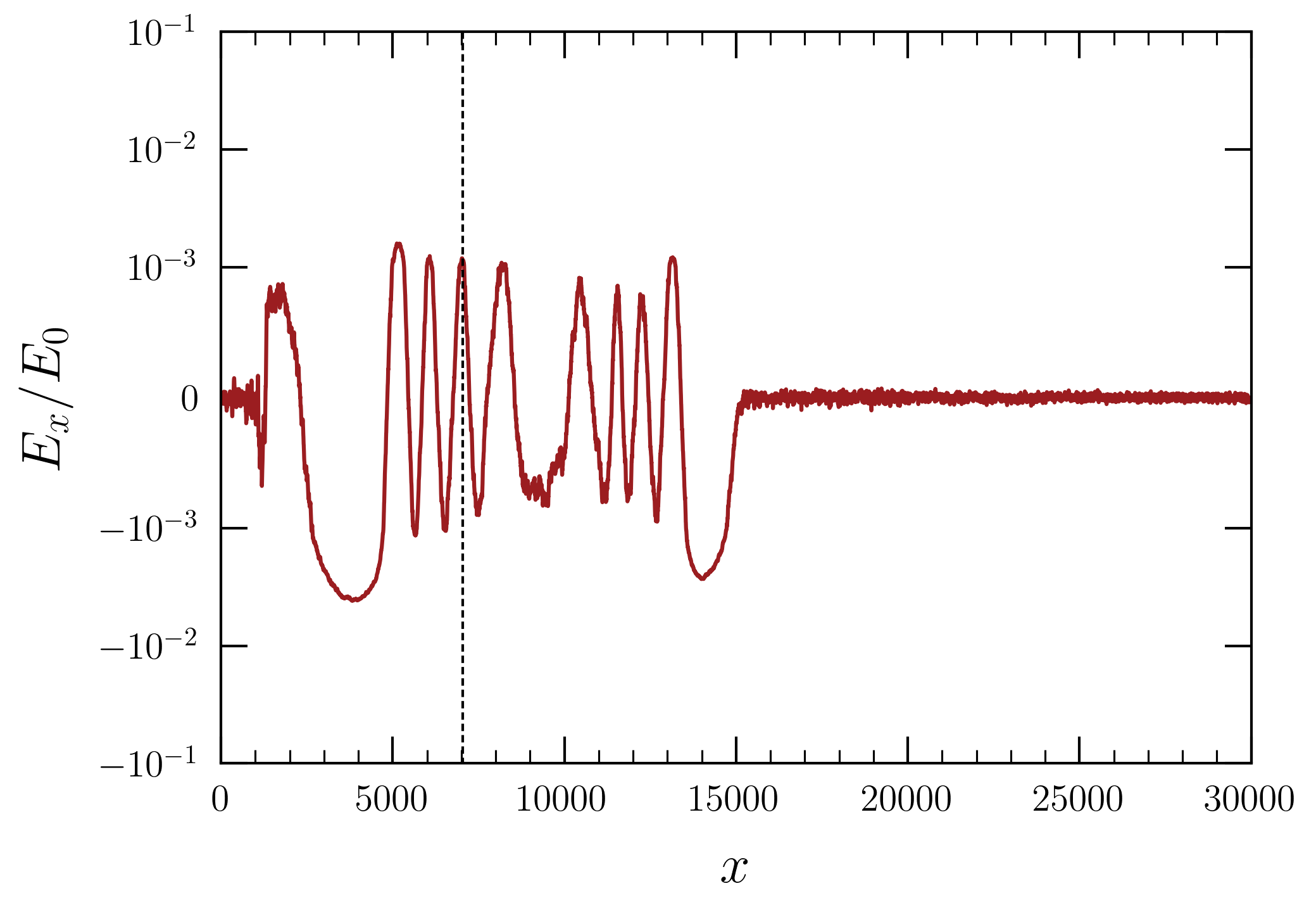}
    \includegraphics[width=0.27\linewidth,trim={2.7cm, 1.7cm, 0cm, 0cm }, clip]{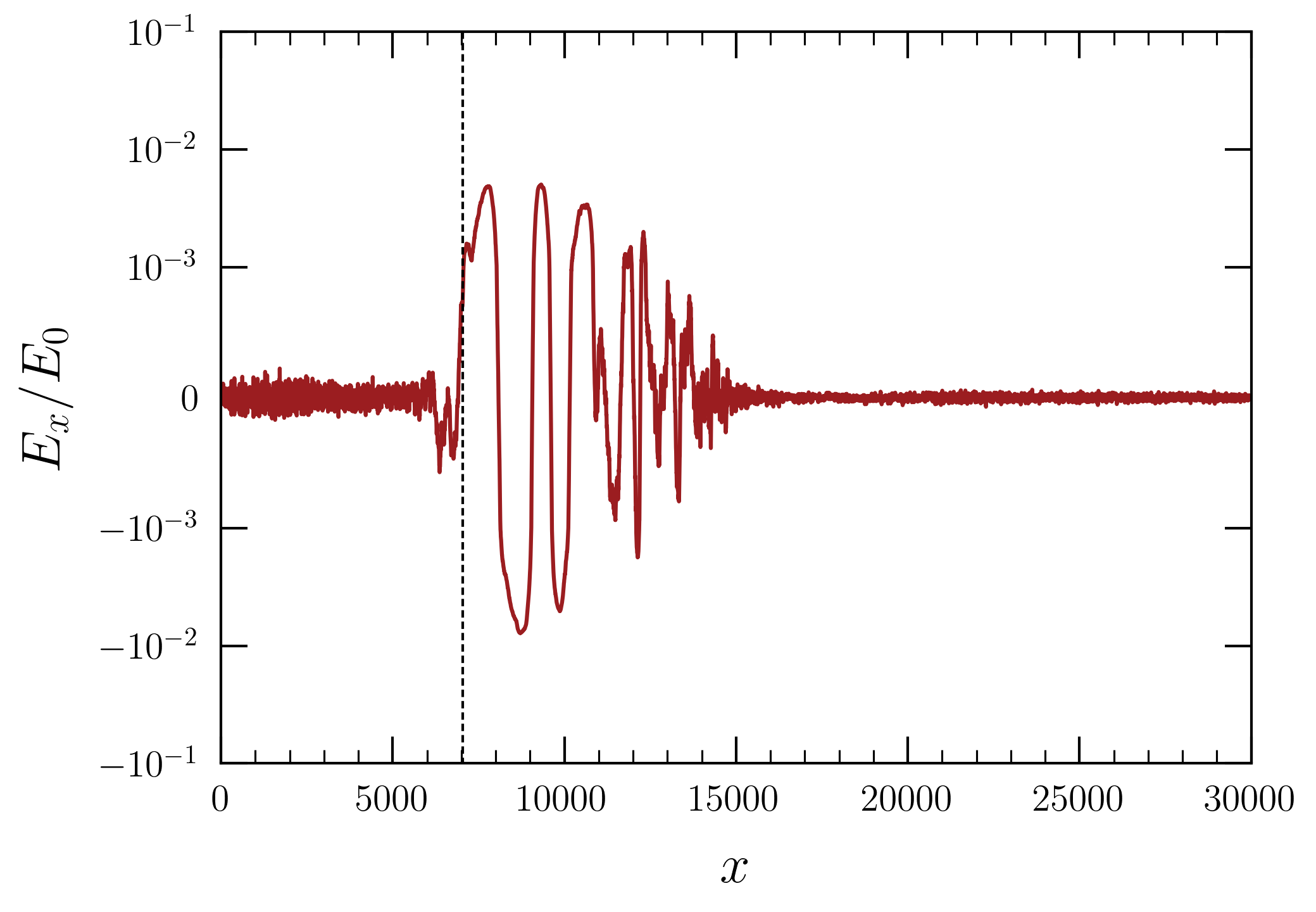}
    \includegraphics[width=0.27\linewidth,trim={2.7cm, 1.7cm, 0cm, 0cm }, clip]{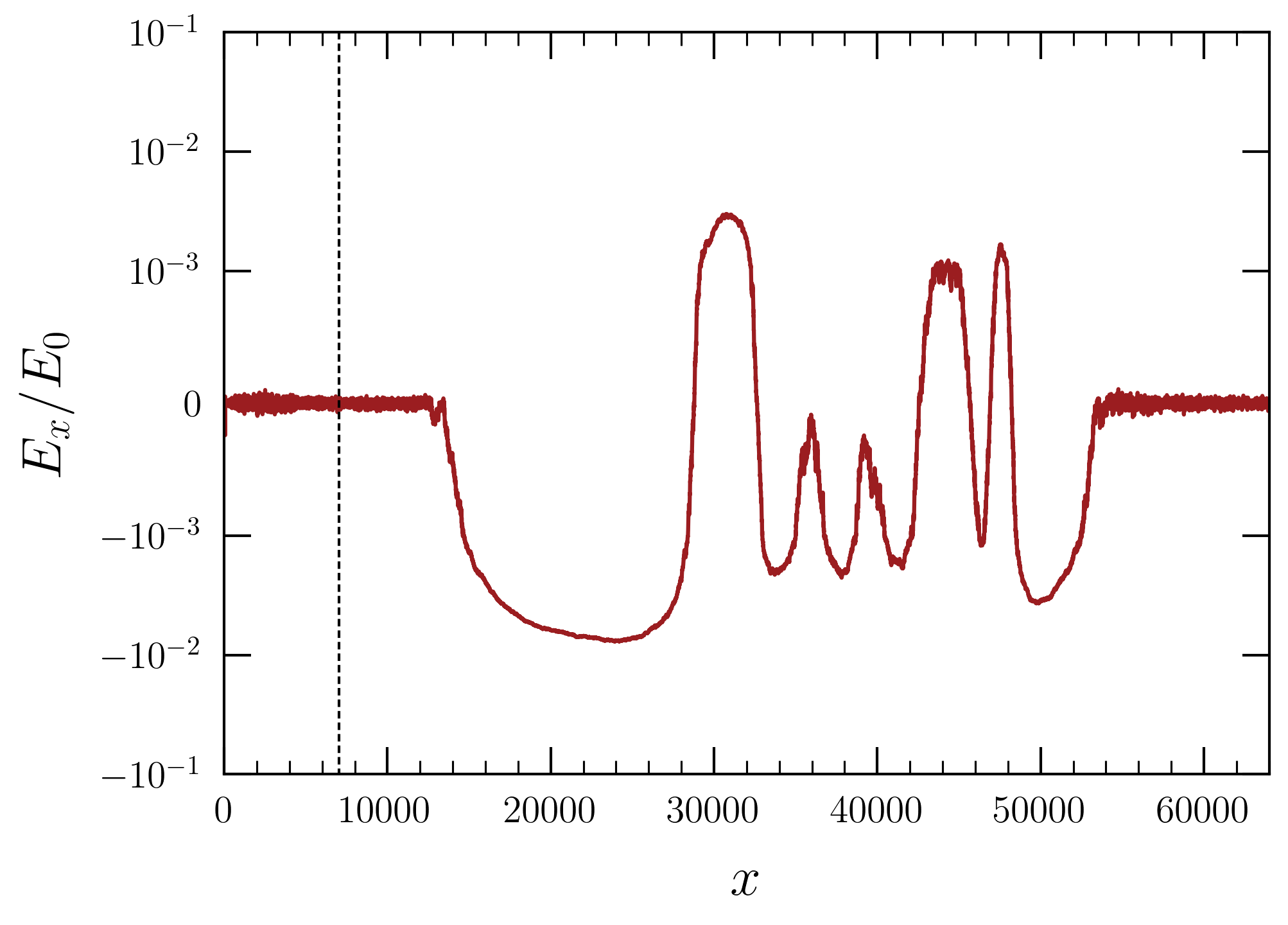}

    \includegraphics[width=0.32\linewidth, trim={0cm, 1.7cm, 0cm, 0cm }, clip]{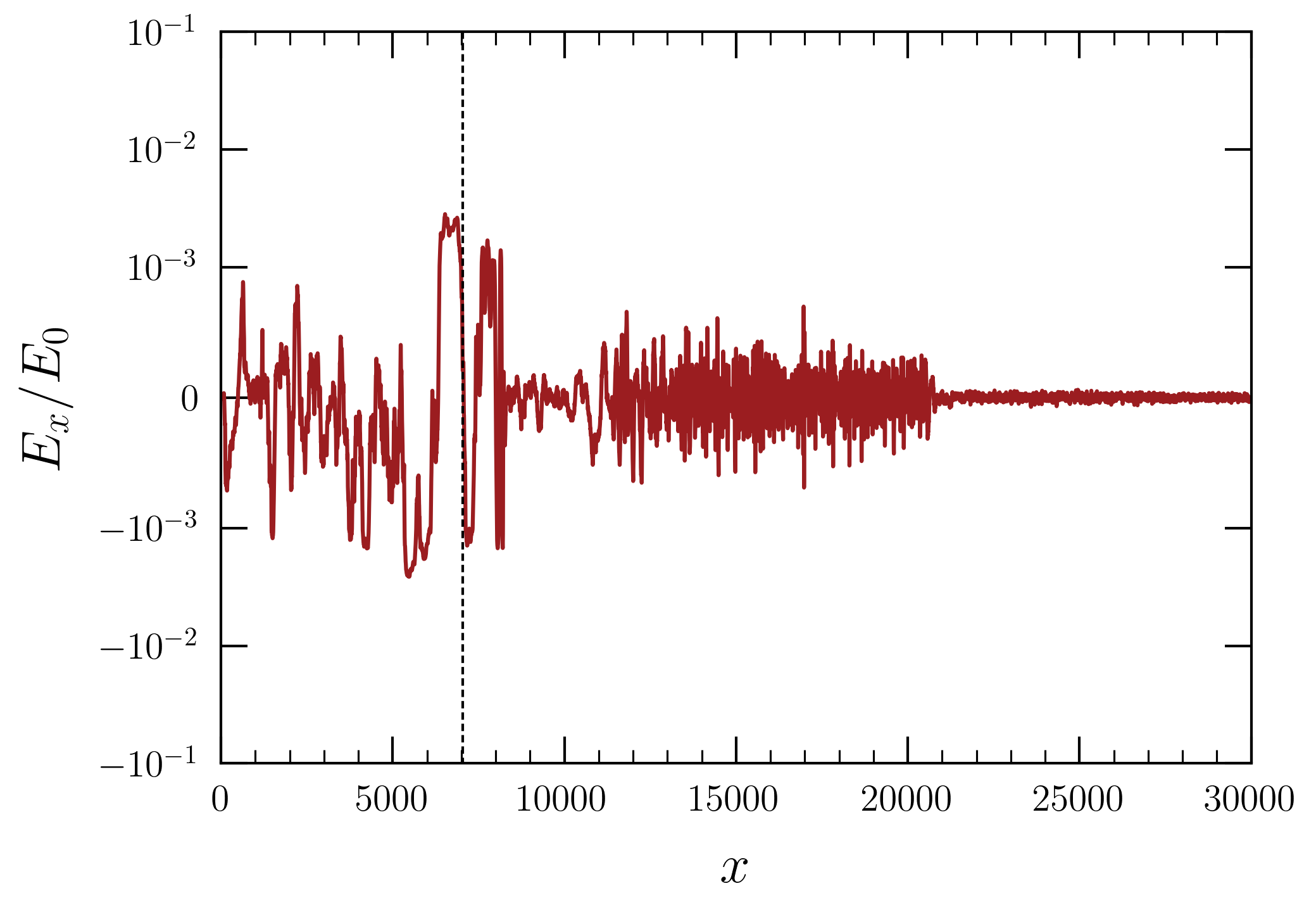}
    \includegraphics[width=0.27\linewidth,trim={2.7cm, 1.7cm, 0cm, 0cm }, clip]{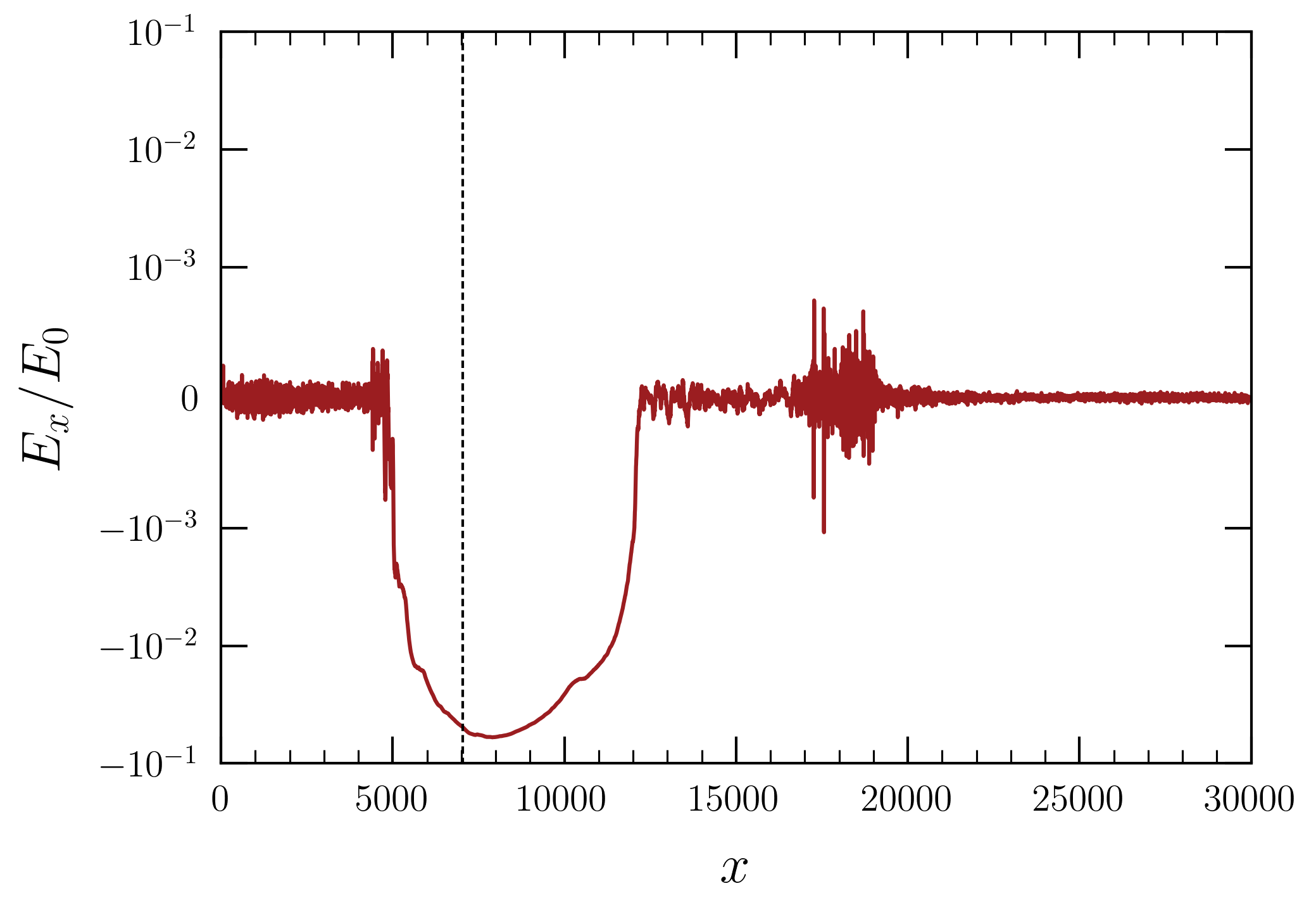}
    \includegraphics[width=0.27\linewidth,trim={2.7cm, 1.7cm, 0cm, 0cm }, clip]{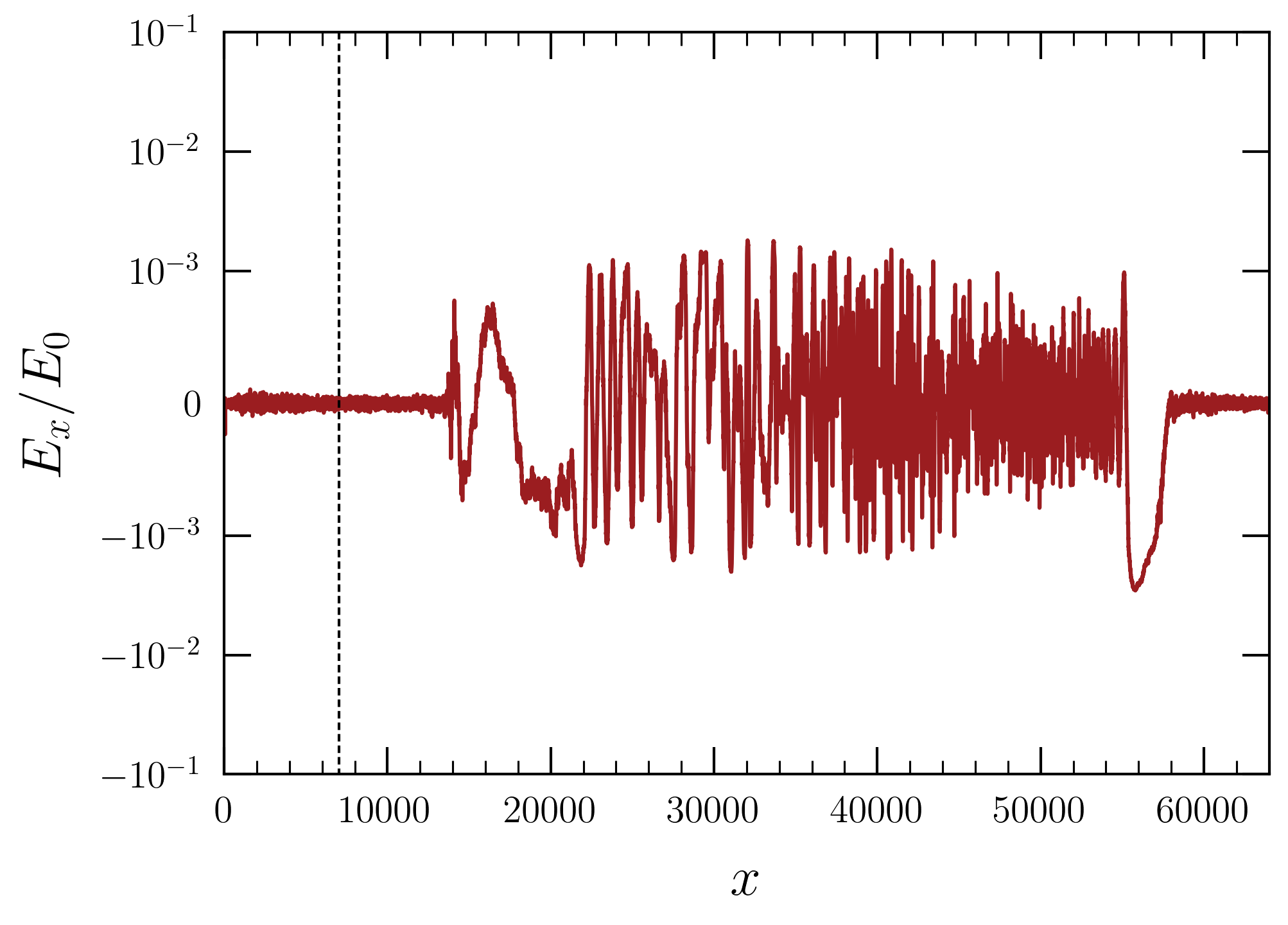}

    \includegraphics[width=0.32\linewidth, trim={0cm, 0cm, 0cm, 0cm }, clip]{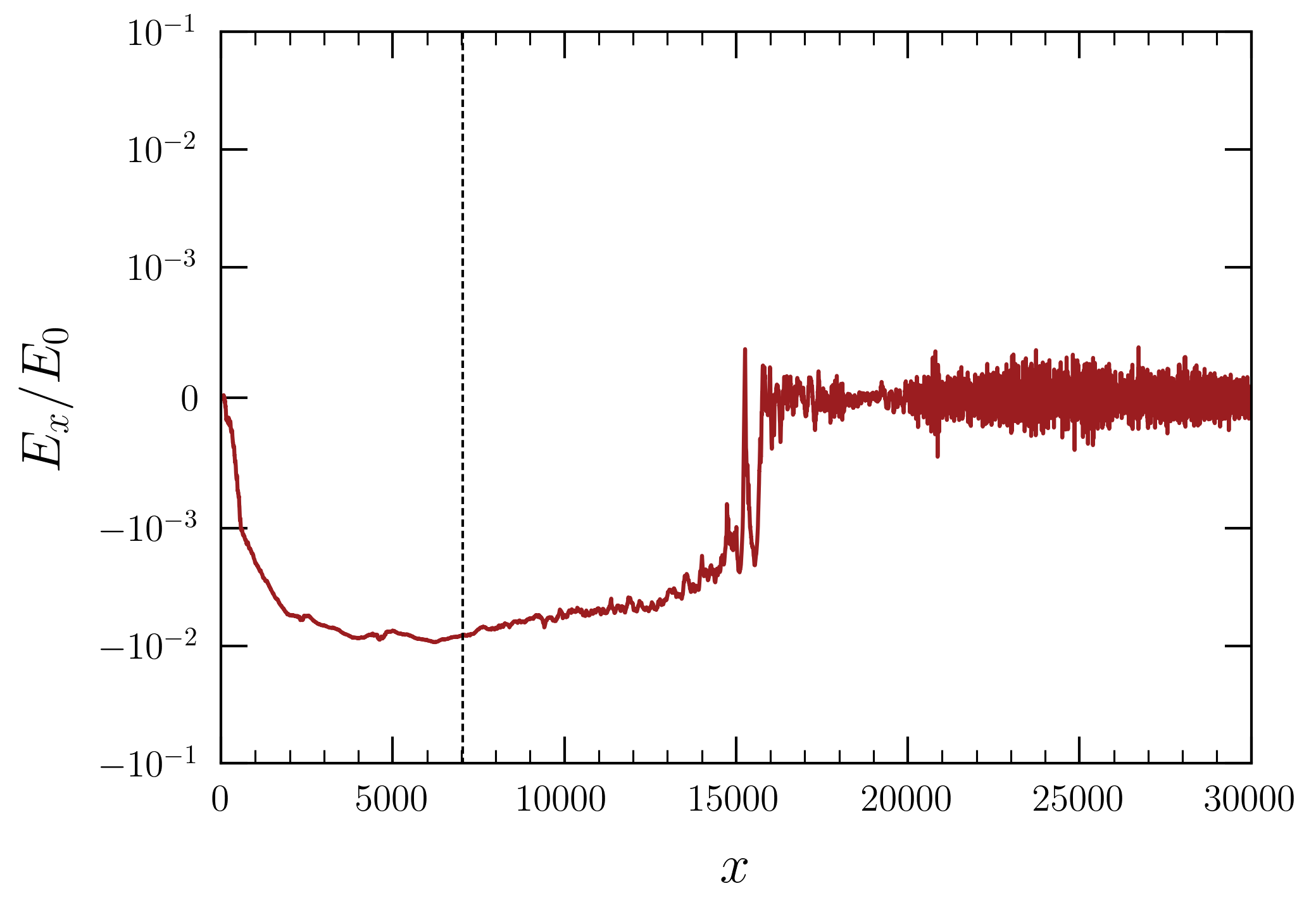}
    \includegraphics[width=0.27\linewidth,trim={2.7cm, 0cm, 0cm, 0cm }, clip]{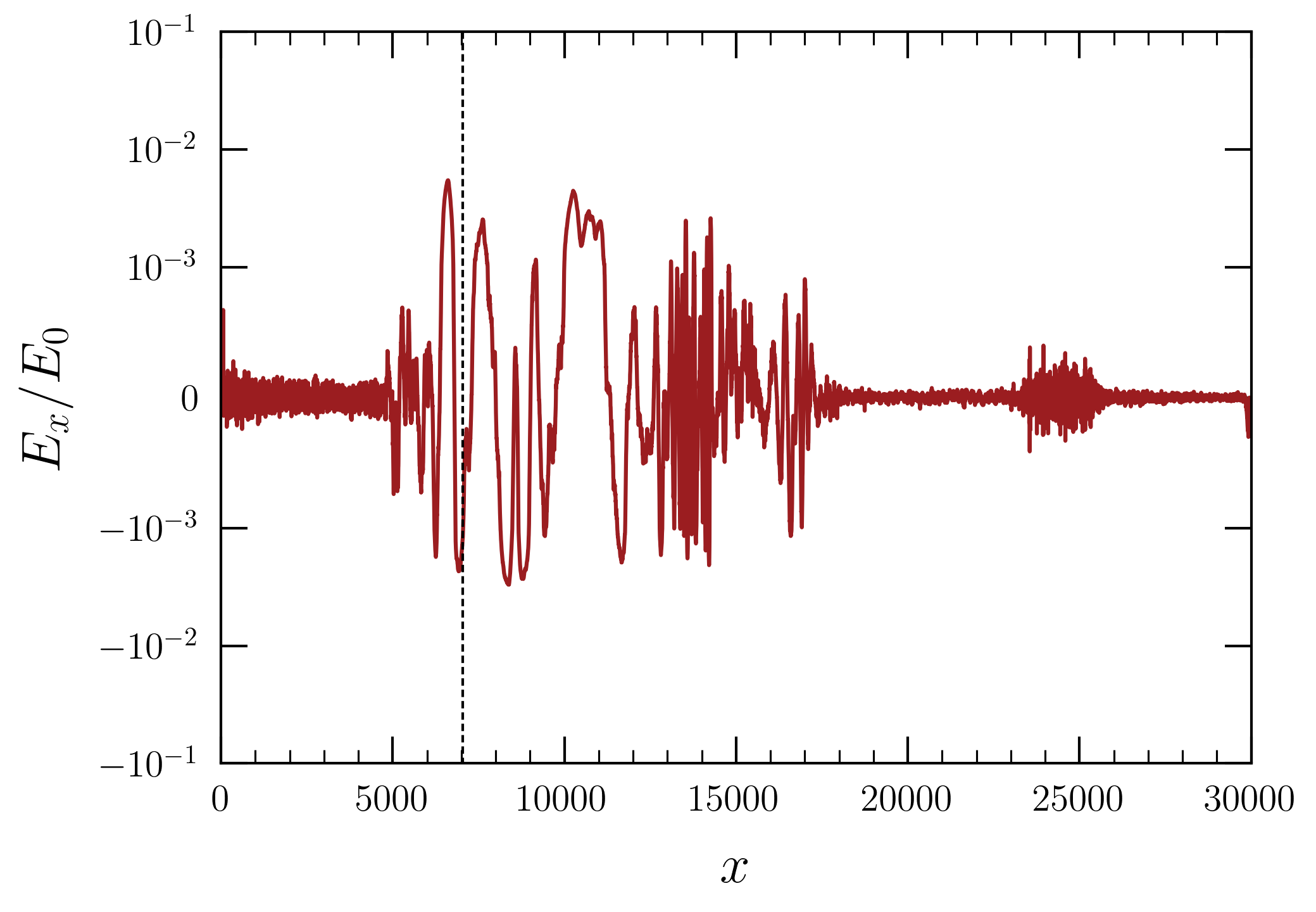}
    \includegraphics[width=0.27\linewidth,trim={2.7cm, 0cm, 0cm, 0cm }, clip]{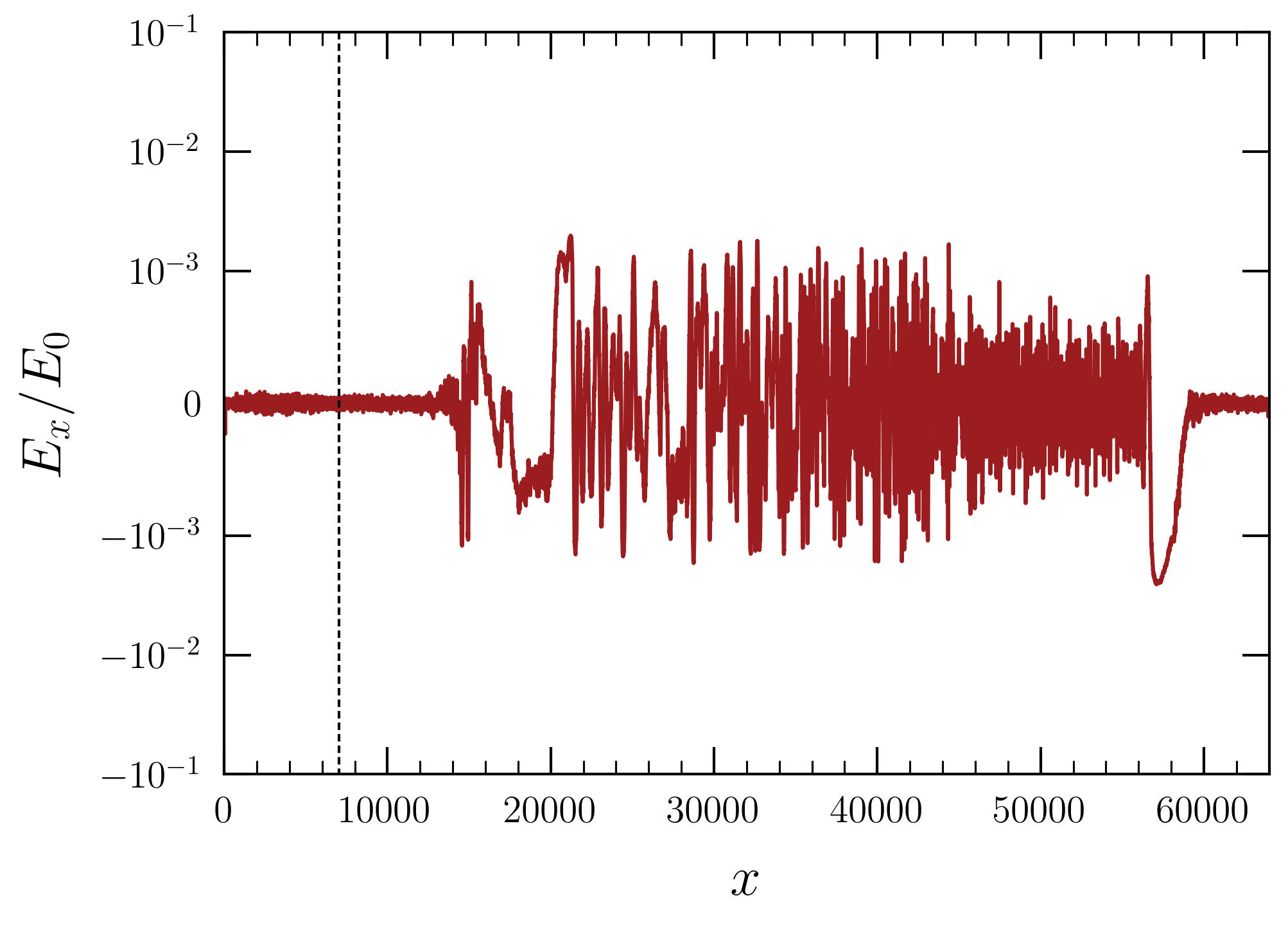}

    \caption{Same as Fig.~\ref{fig:picE}, but for the rotational frequency of the star reduced by a factor of four, while keeping the axion amplitude fixed. A vertical dashed line has been placed at the position $x$ where $|\rho_a(x)| = |\rho_{\rm GJ}(x)|$, taking $|\rho_a(x=0)| = 40 |\rho_{\rm GJ}(x=0)| $.  Note the extended x-axis in the right panel. The snapshots in the left panel correspond to time step: 12000, 33500, 42000, 56000, 78000; in the center panel to time step: 7500, 18500, 27500, 40500, 55000; and in the right panel to time step: 111500, 123000, 127000, 136500, 139000. }\label{fig:picE_v2}
\end{figure*}

\begin{figure*}
    \includegraphics[width=0.32\linewidth, trim={0cm, 1.7cm, 0cm, 0cm }, clip]{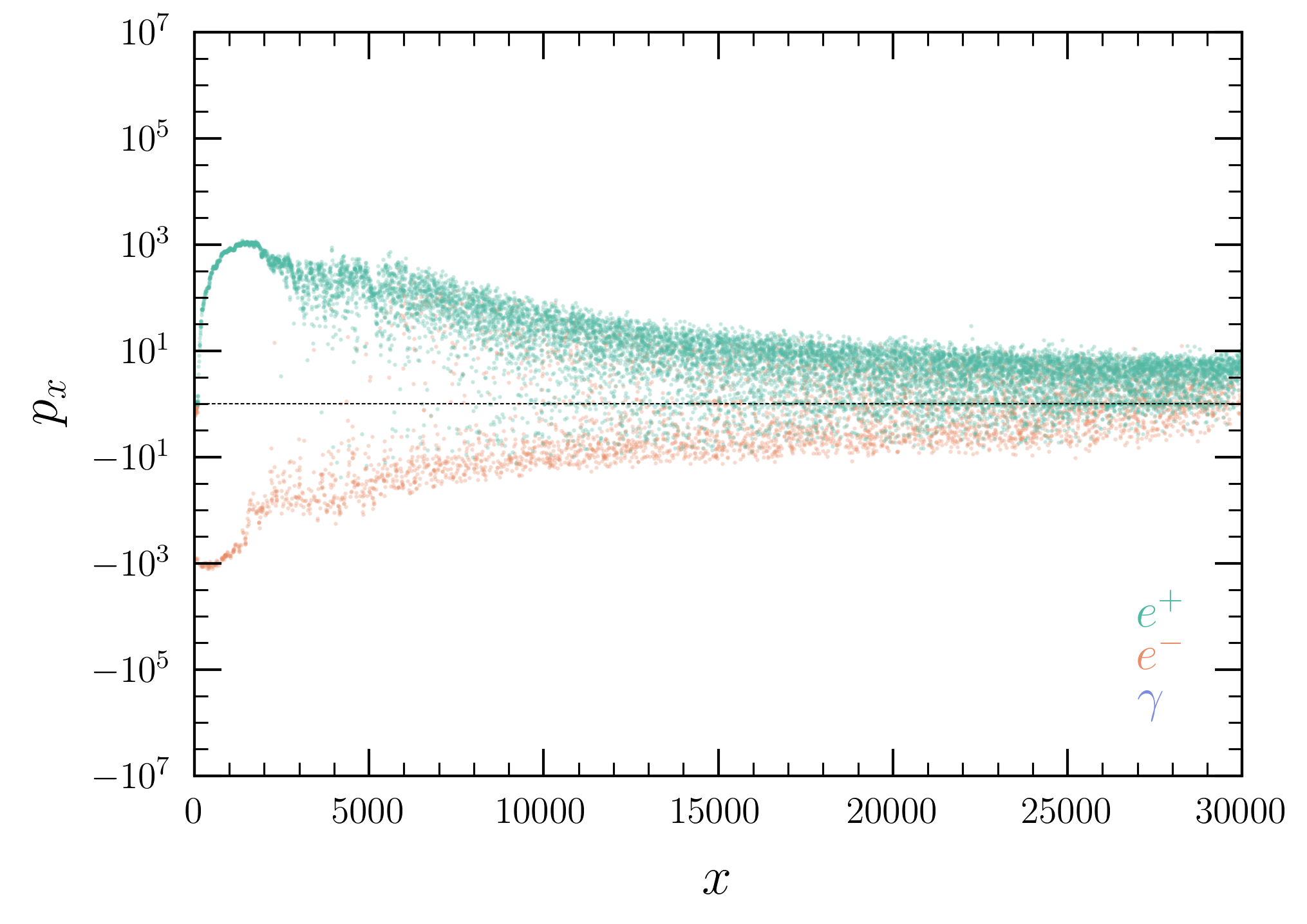}
    \includegraphics[width=0.275\linewidth,trim={2.4cm, 1.7cm, 0cm, 0cm }, clip]{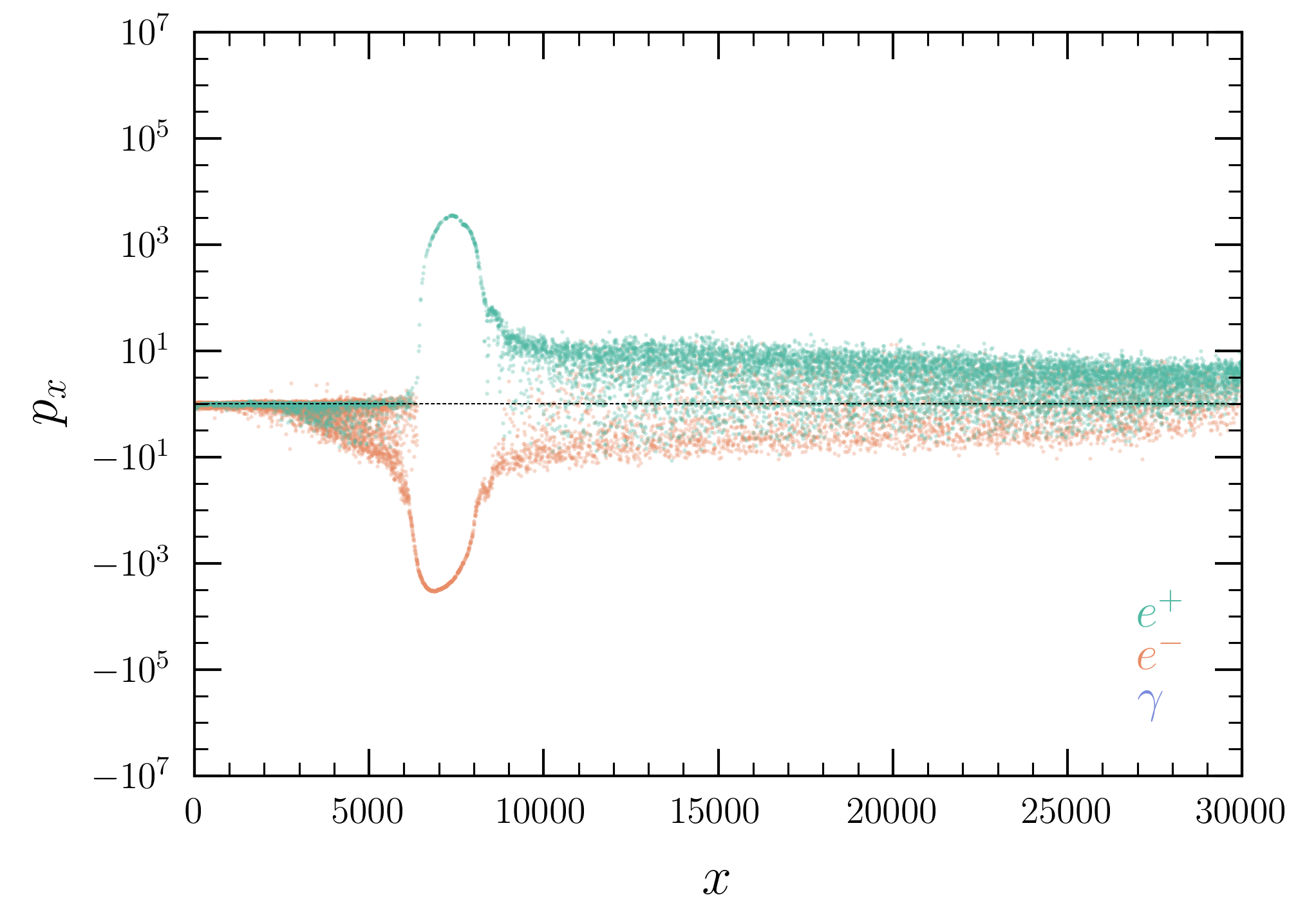}
    \includegraphics[width=0.275\linewidth,trim={2.4cm, 1.7cm, 0cm, 0cm }, clip]{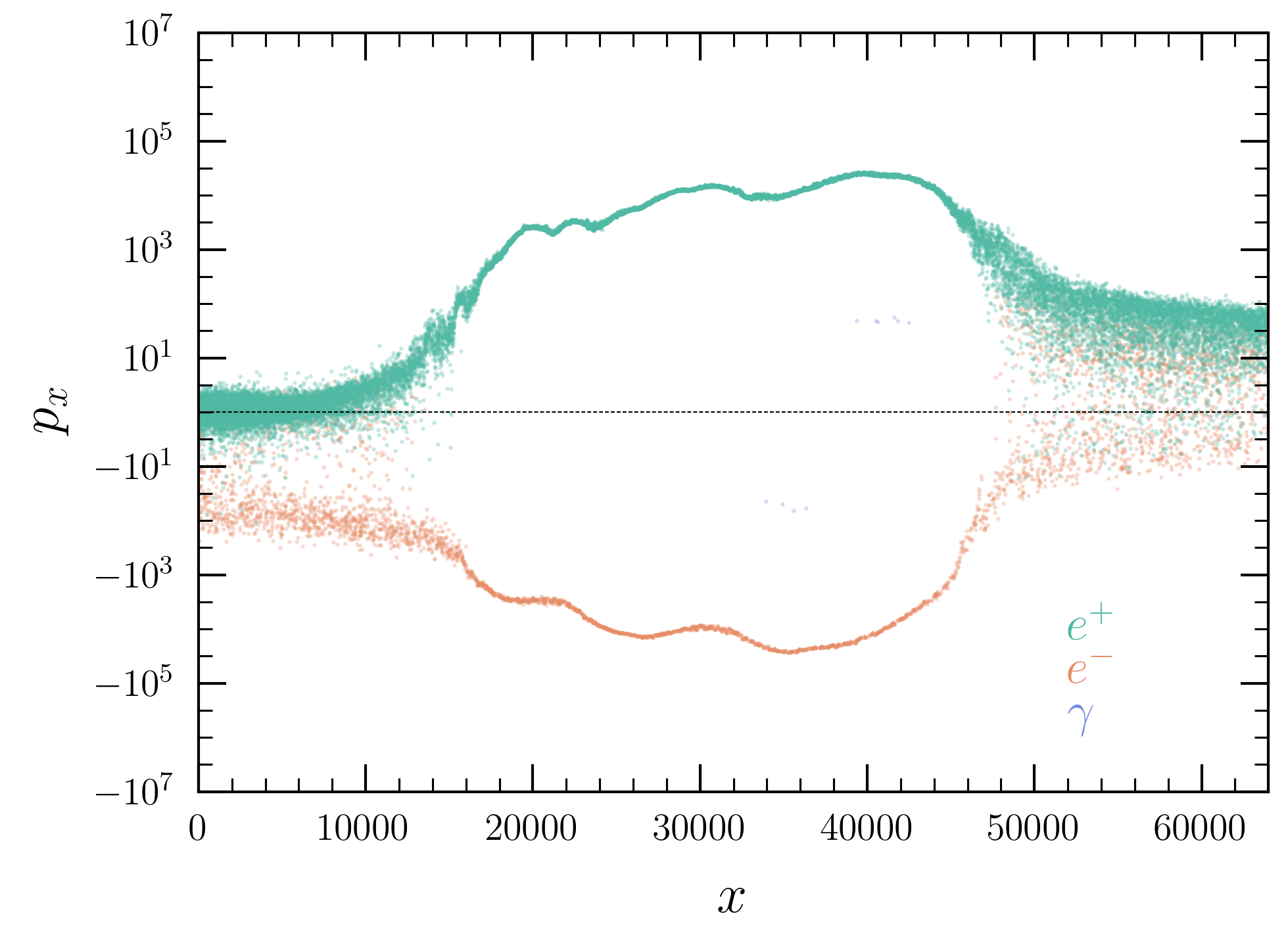}

    \includegraphics[width=0.32\linewidth, trim={0cm, 1.7cm, 0cm, 0cm }, clip]{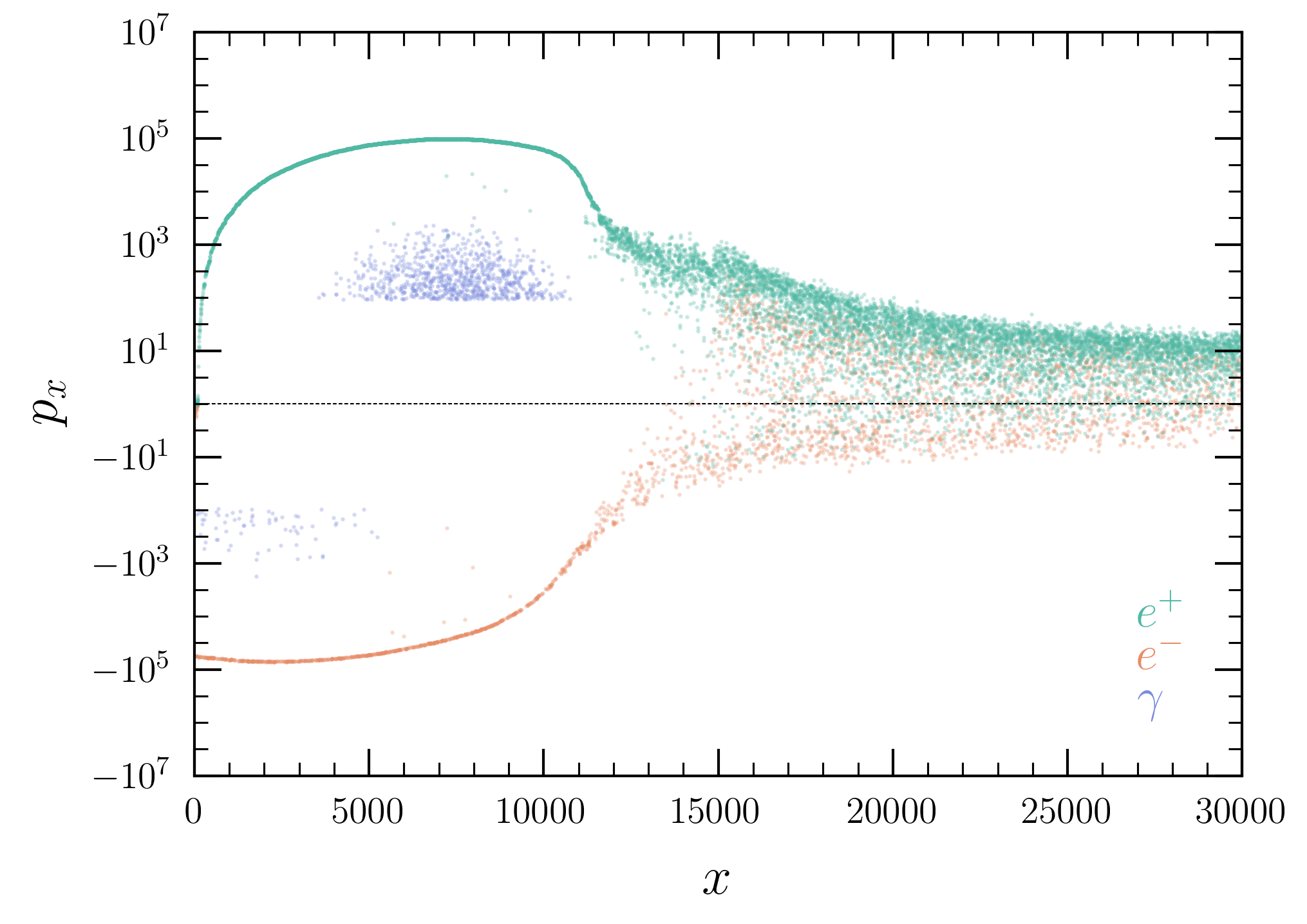}
    \includegraphics[width=0.275\linewidth,trim={2.4cm, 1.7cm, 0cm, 0cm }, clip]{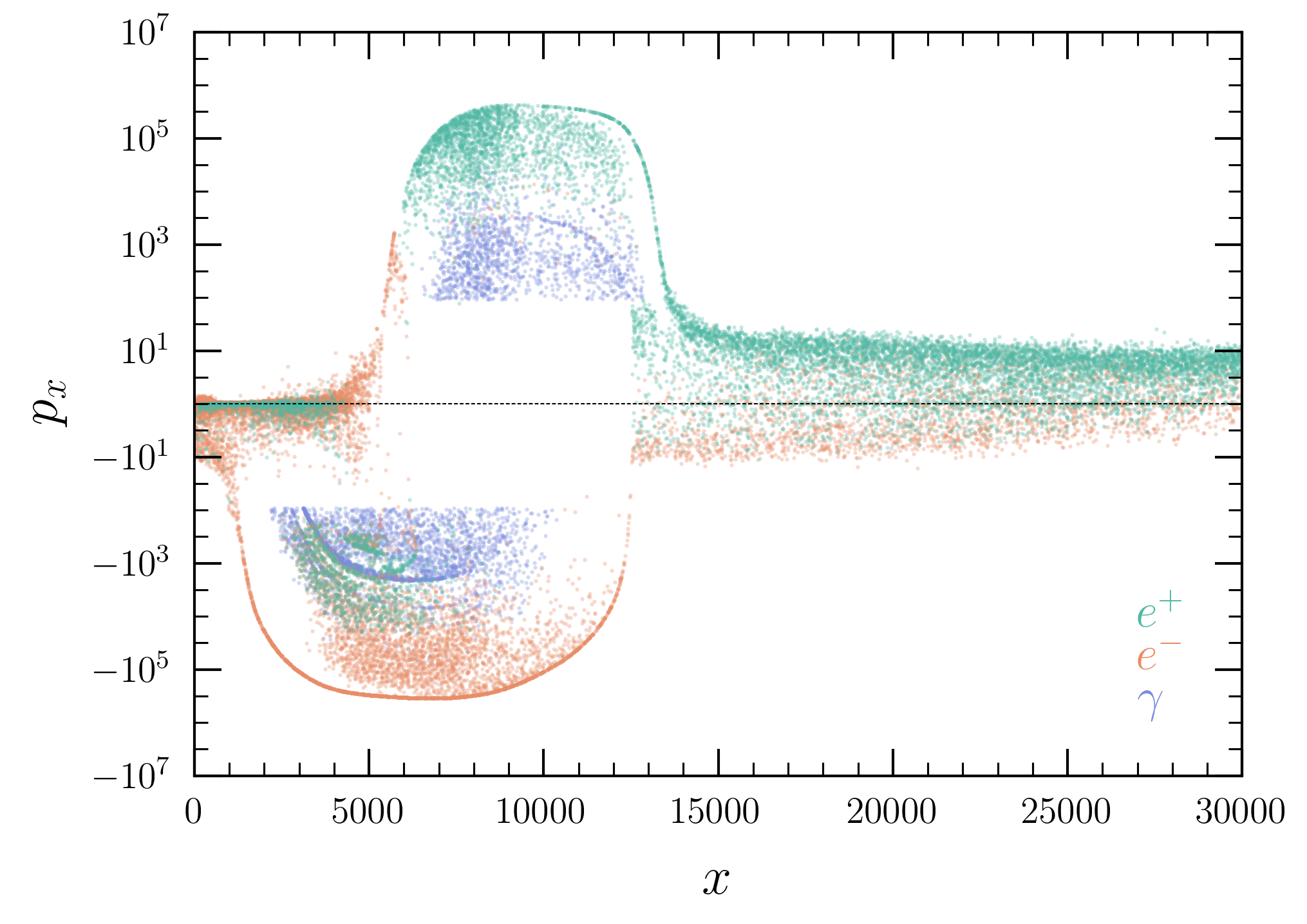}
    \includegraphics[width=0.275\linewidth,trim={2.4cm, 1.7cm, 0cm, 0cm }, clip]{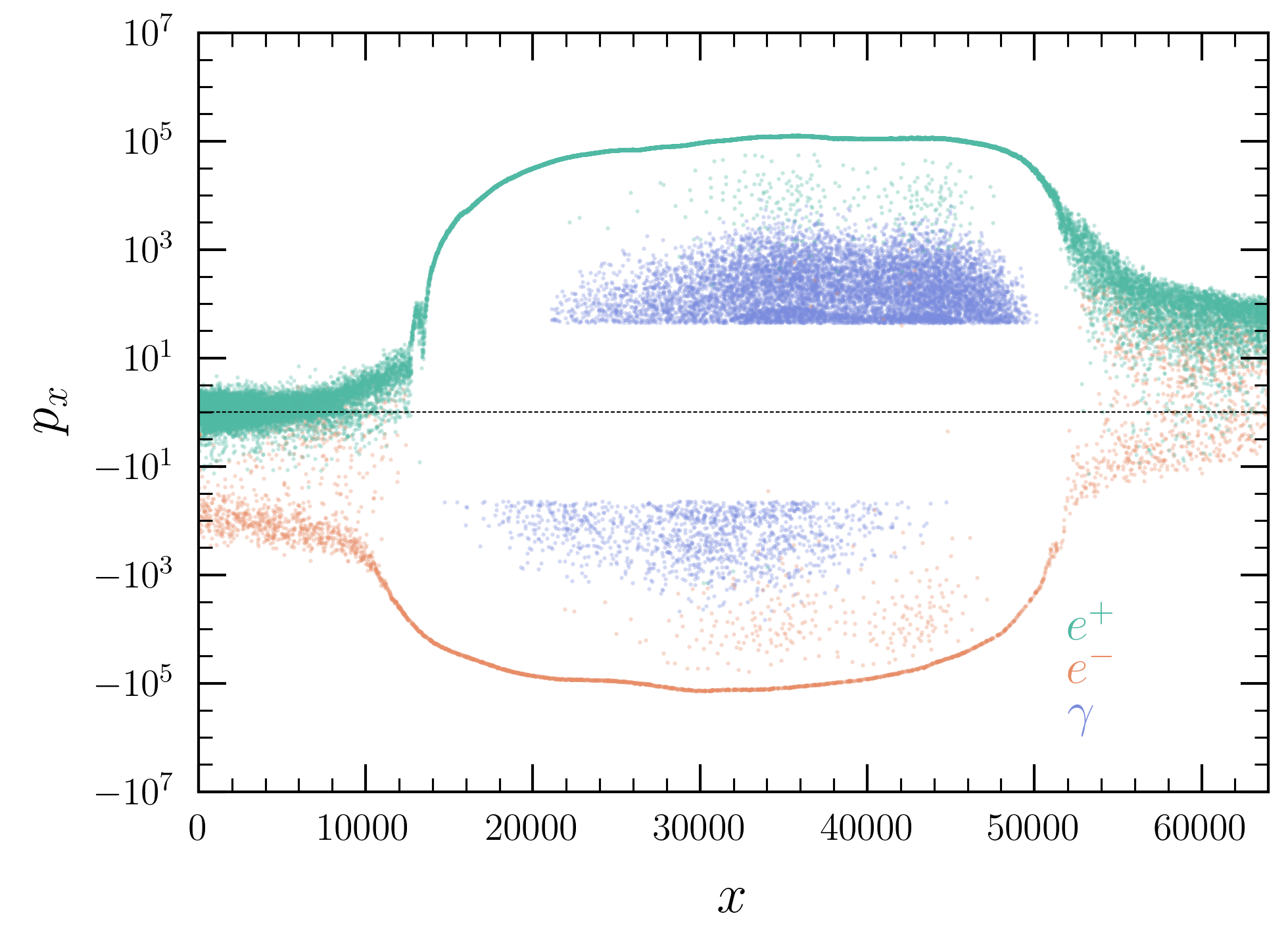}

    \includegraphics[width=0.32\linewidth, trim={0cm, 1.7cm, 0cm, 0cm }, clip]{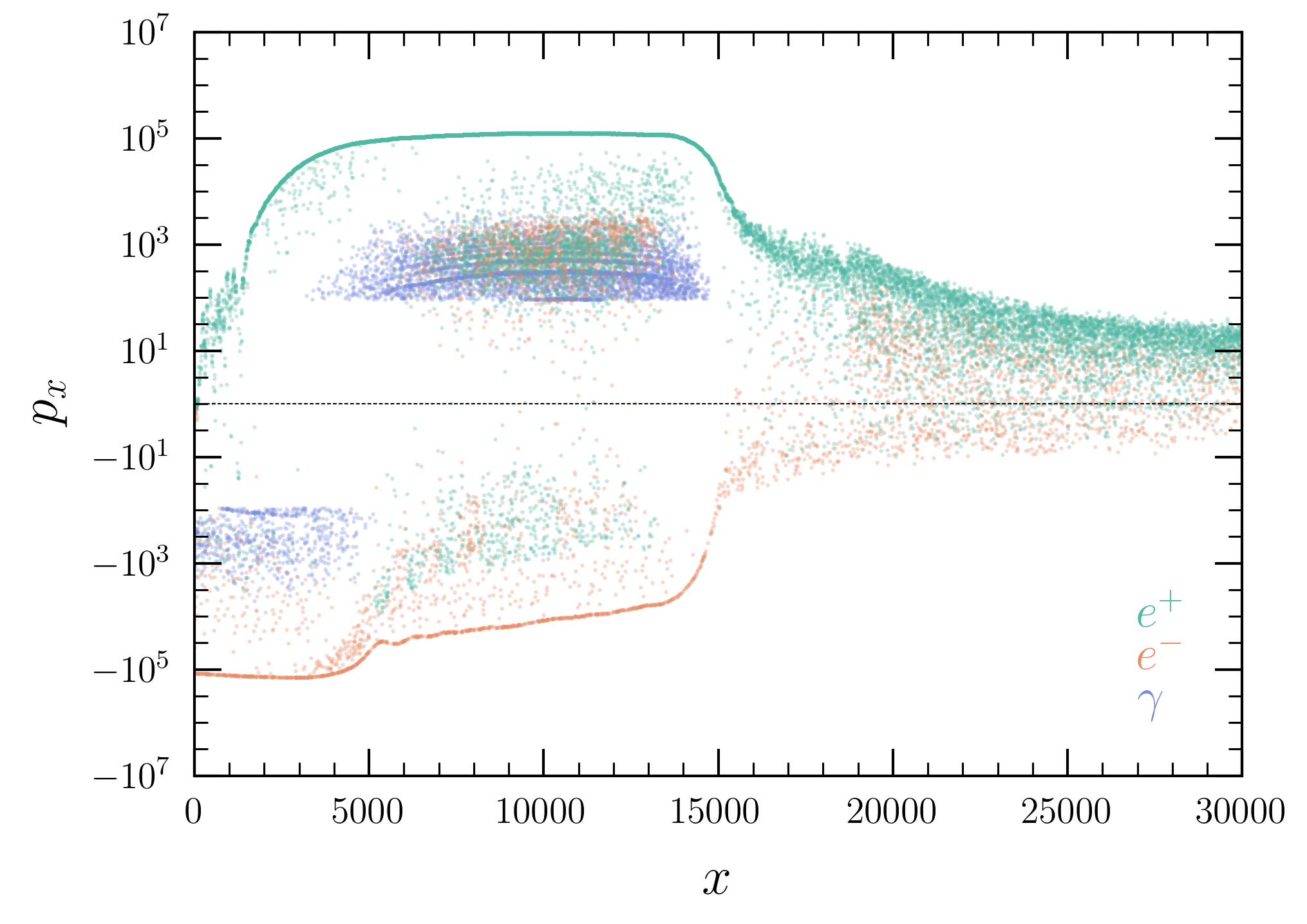}
    \includegraphics[width=0.275\linewidth,trim={2.4cm, 1.7cm, 0cm, 0cm }, clip]{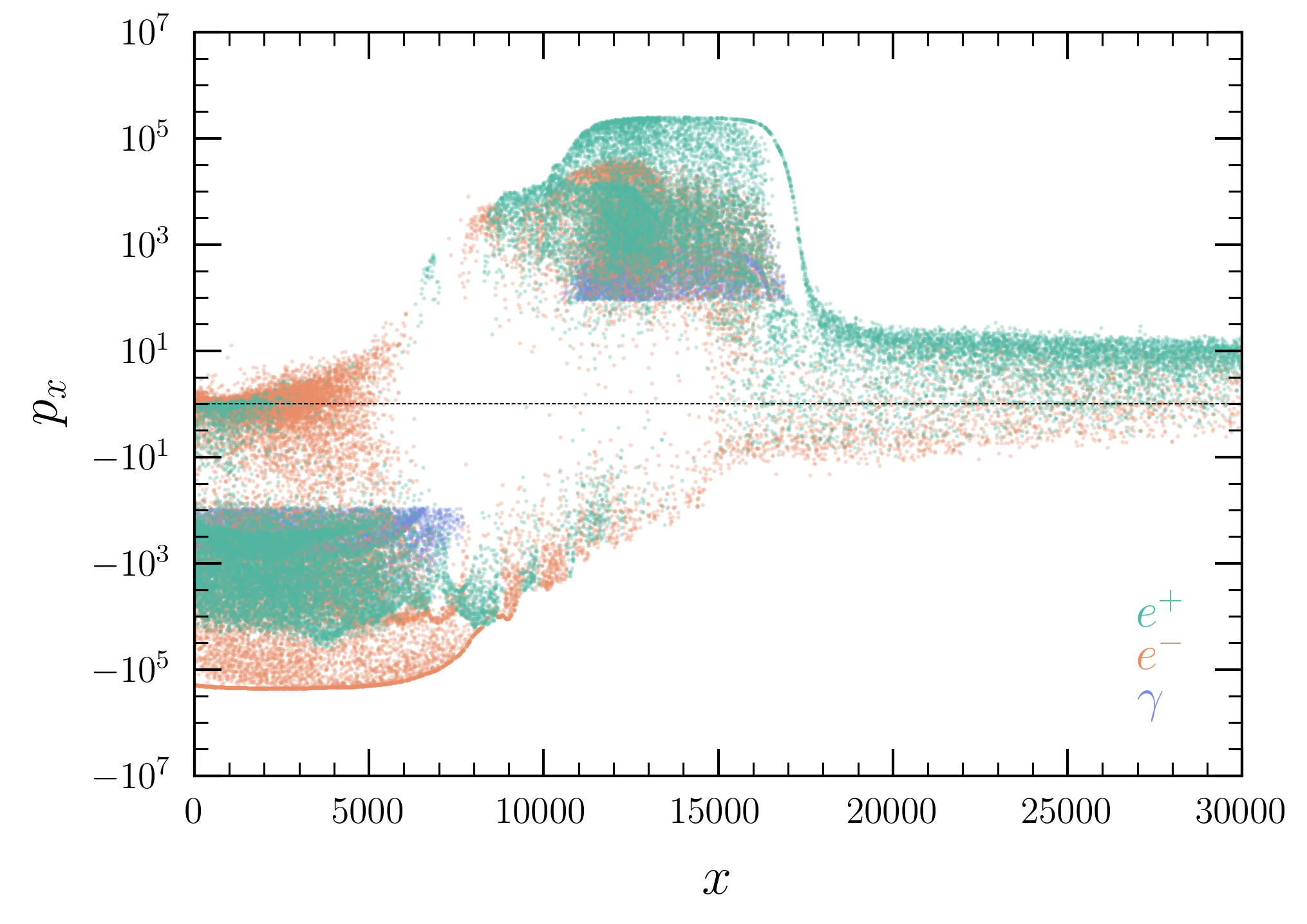}
    \includegraphics[width=0.275\linewidth,trim={2.4cm, 1.7cm, 0cm, 0cm }, clip]{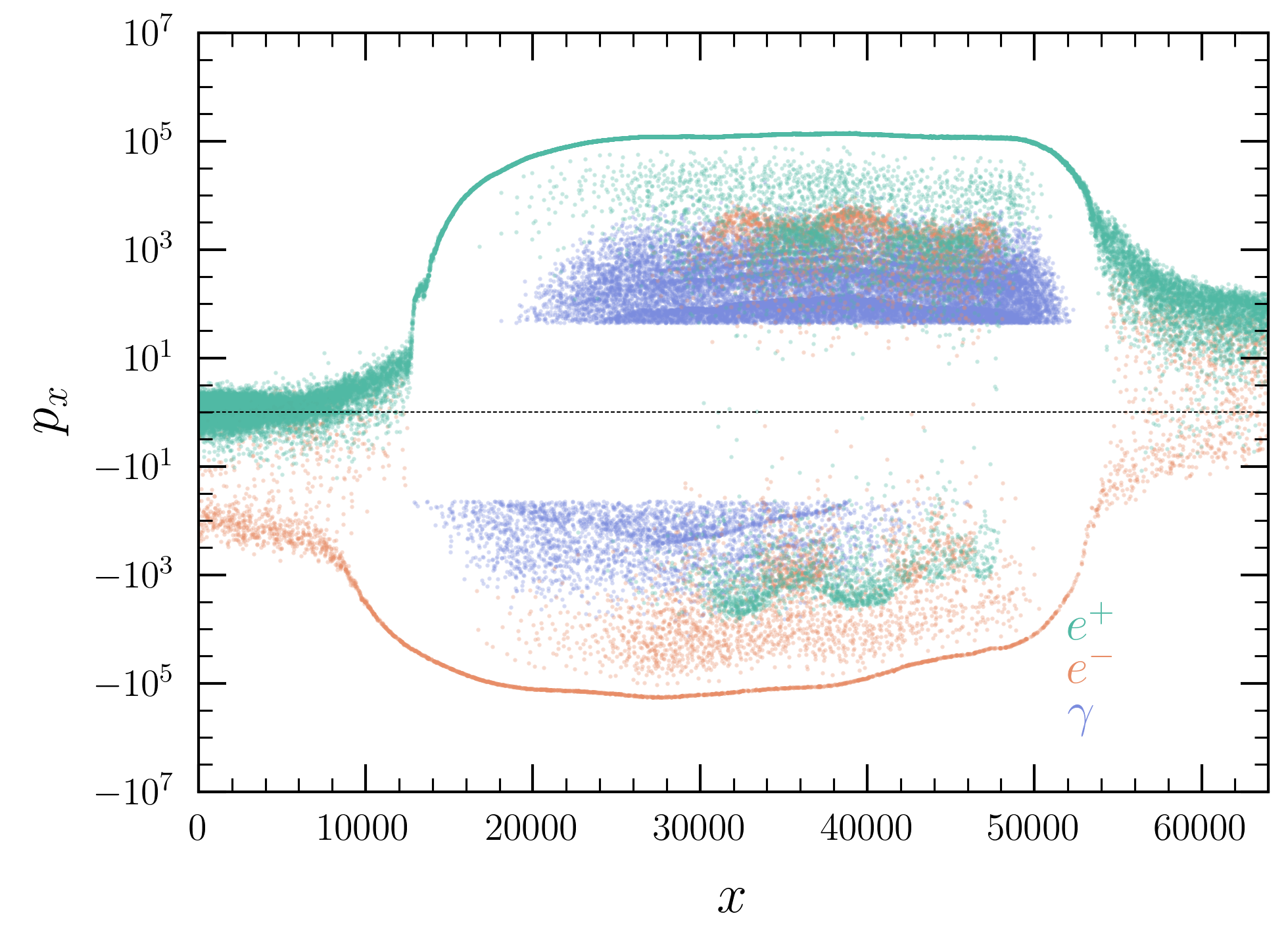}

    \includegraphics[width=0.32\linewidth, trim={0cm, 1.7cm, 0cm, 0cm }, clip]{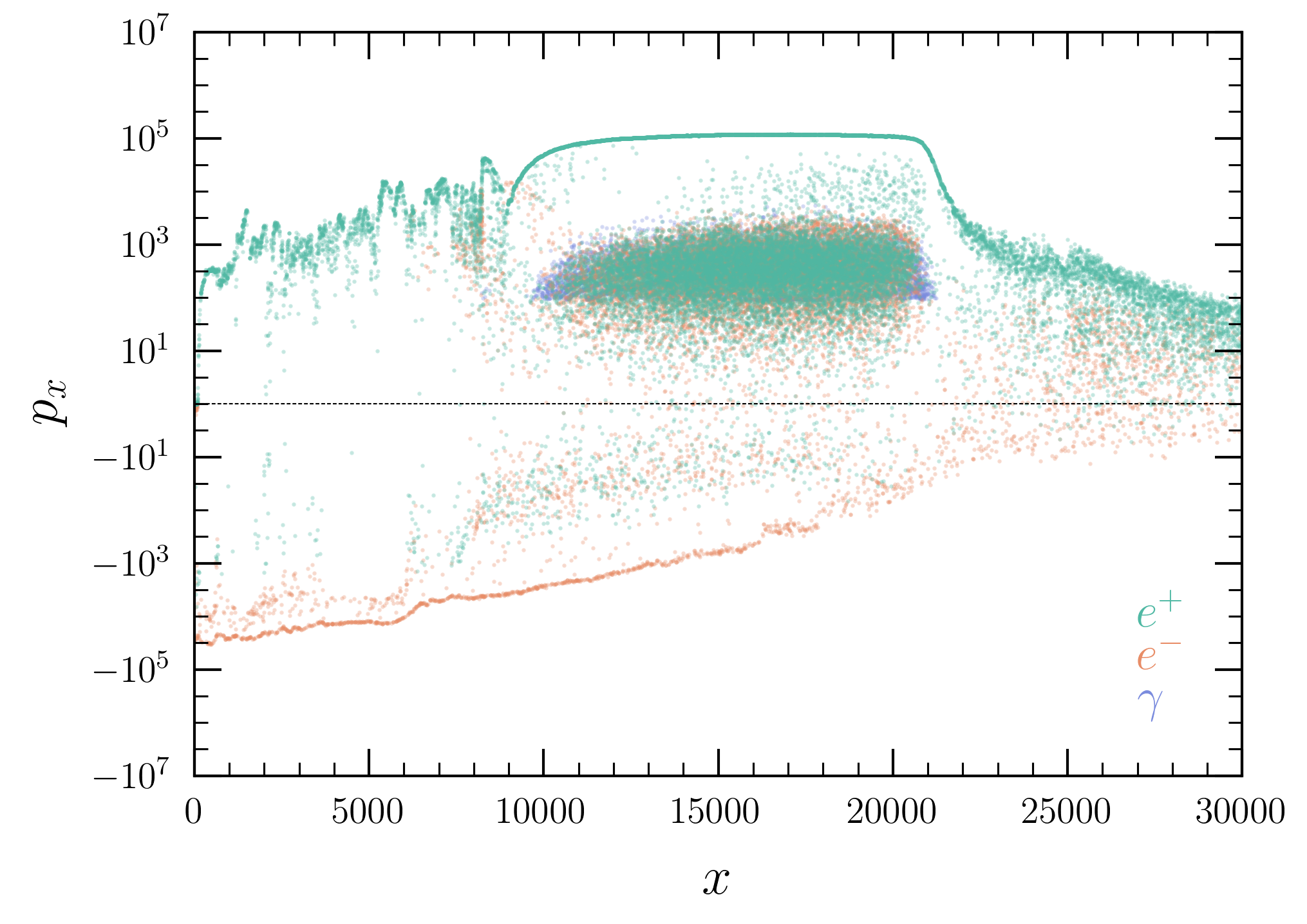}
    \includegraphics[width=0.275\linewidth,trim={2.4cm, 1.7cm, 0cm, 0cm }, clip]{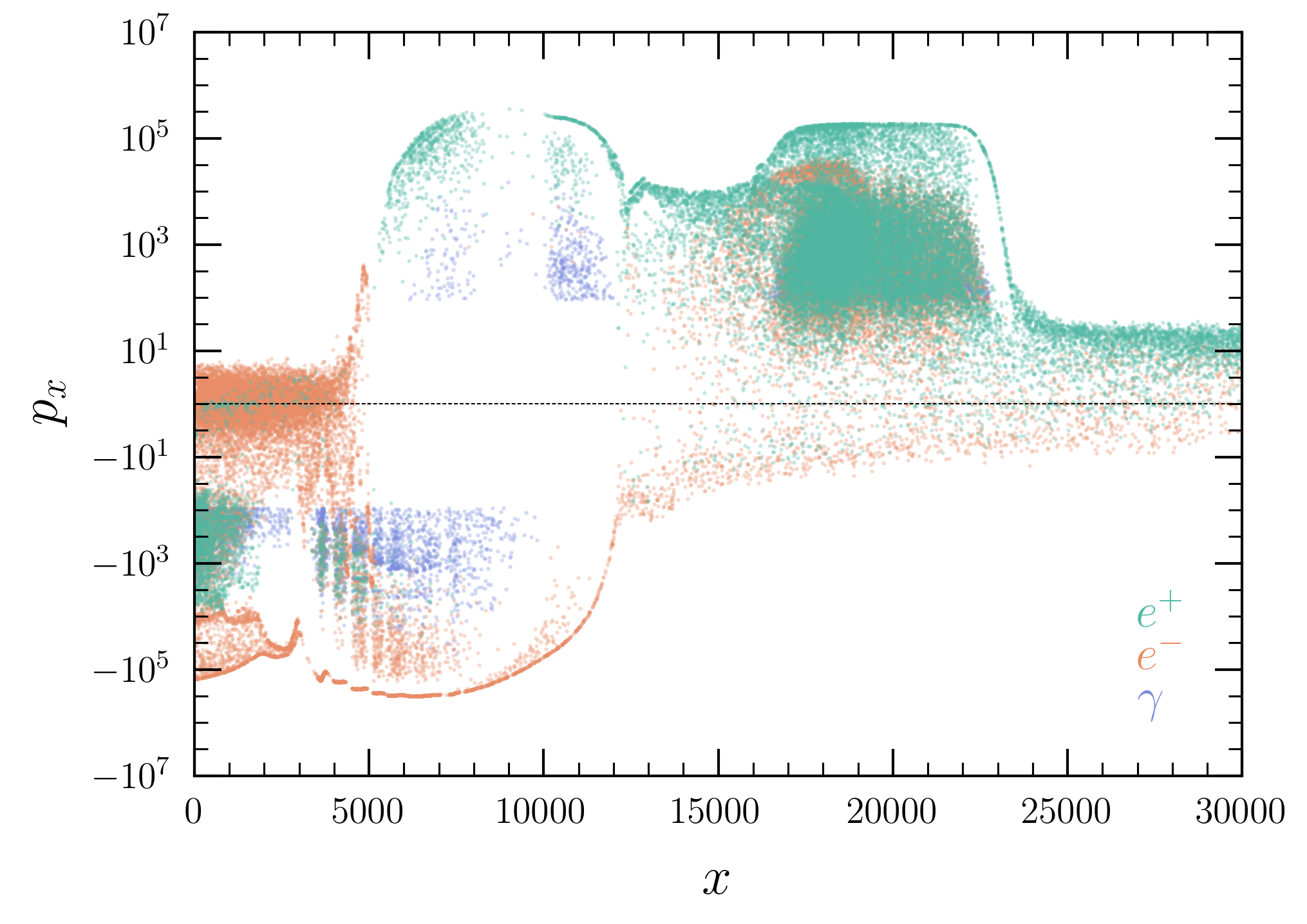}
    \includegraphics[width=0.275\linewidth,trim={2.4cm, 1.7cm, 0cm, 0cm }, clip]{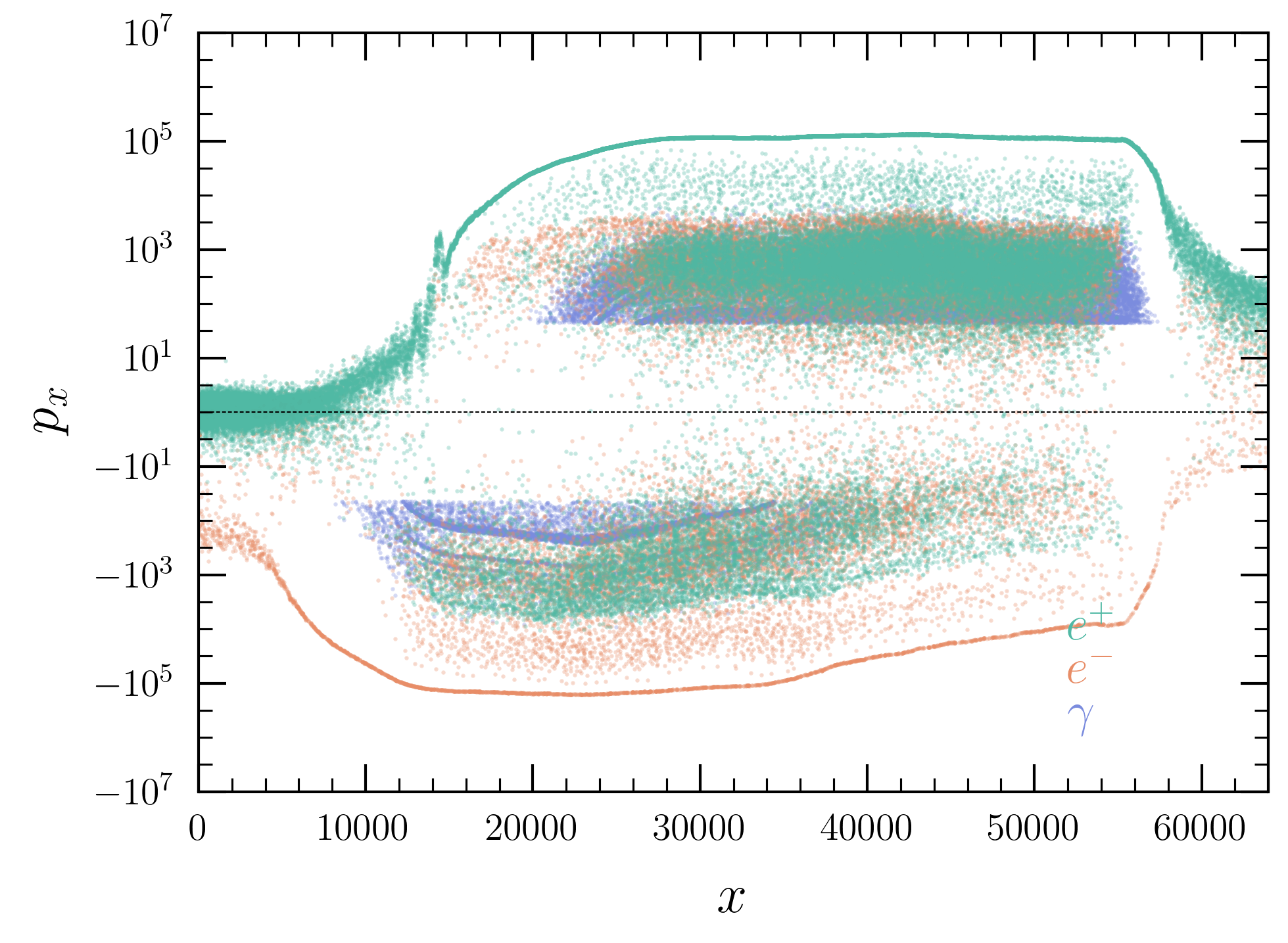}

    \includegraphics[width=0.32\linewidth, trim={0cm, 0cm, 0cm, 0cm }, clip]{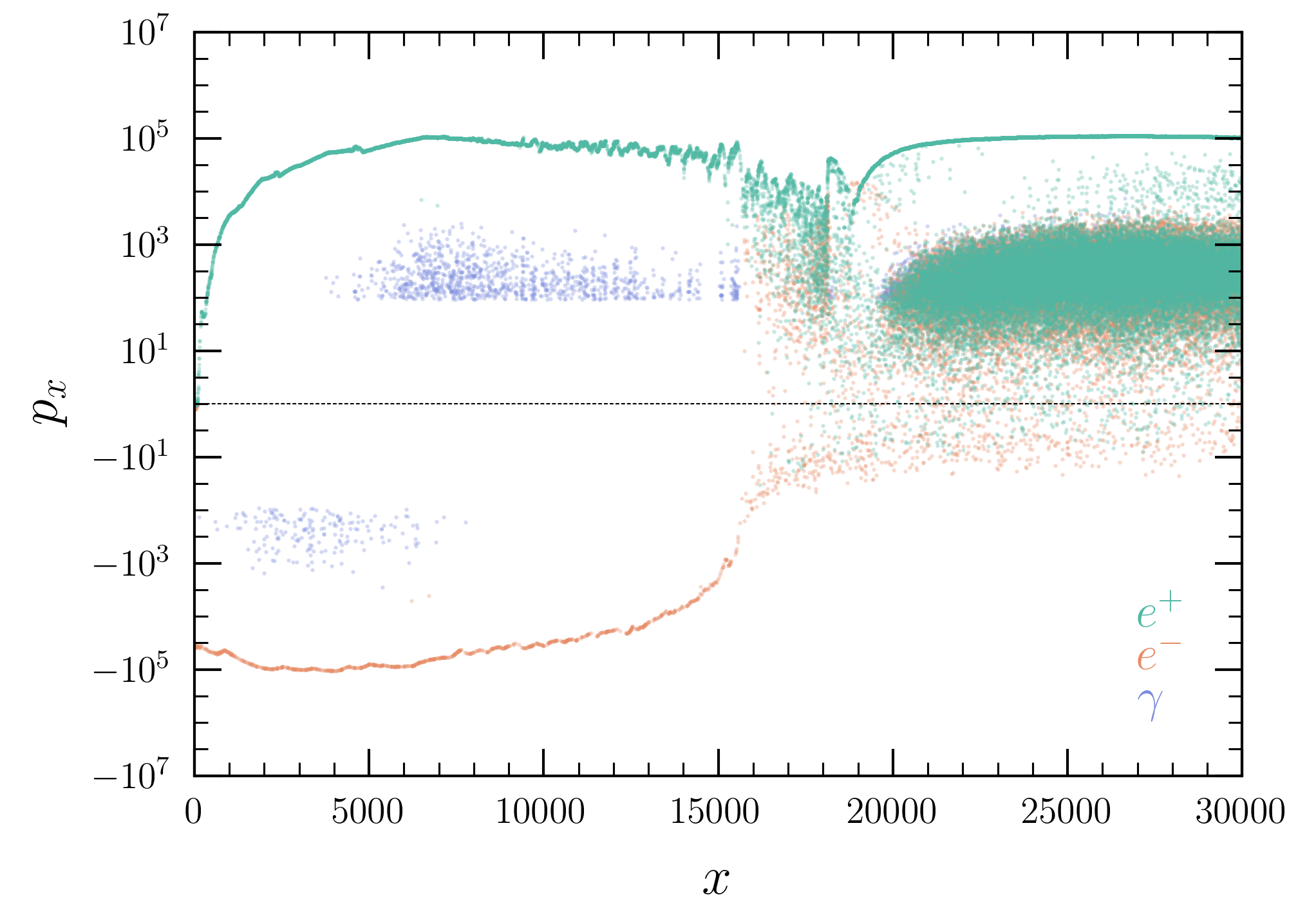}
    \includegraphics[width=0.275\linewidth,trim={2.4cm, 0cm, 0cm, 0cm }, clip]{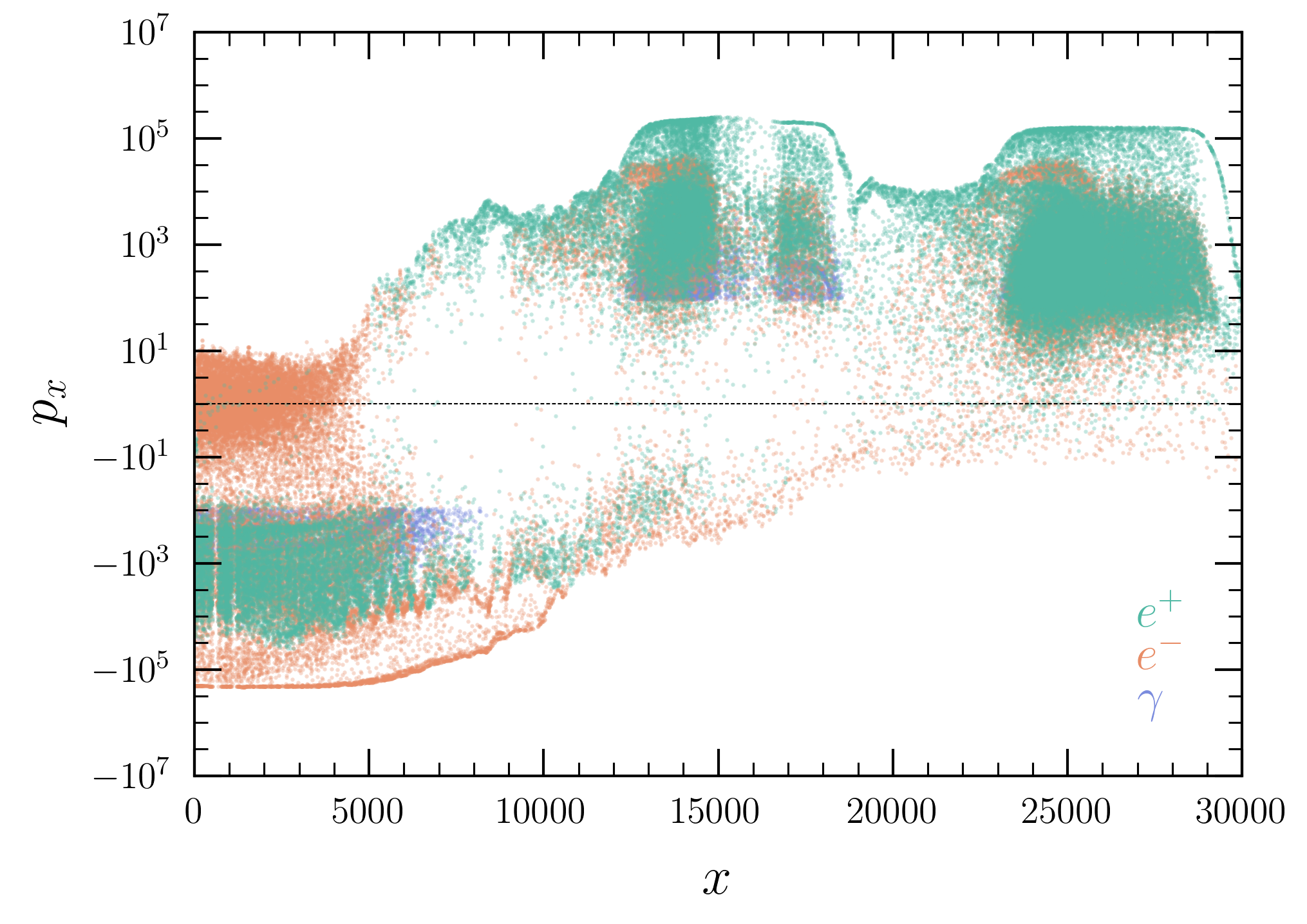}
    \includegraphics[width=0.275\linewidth,trim={2.4cm, 0cm, 0cm, 0cm }, clip]{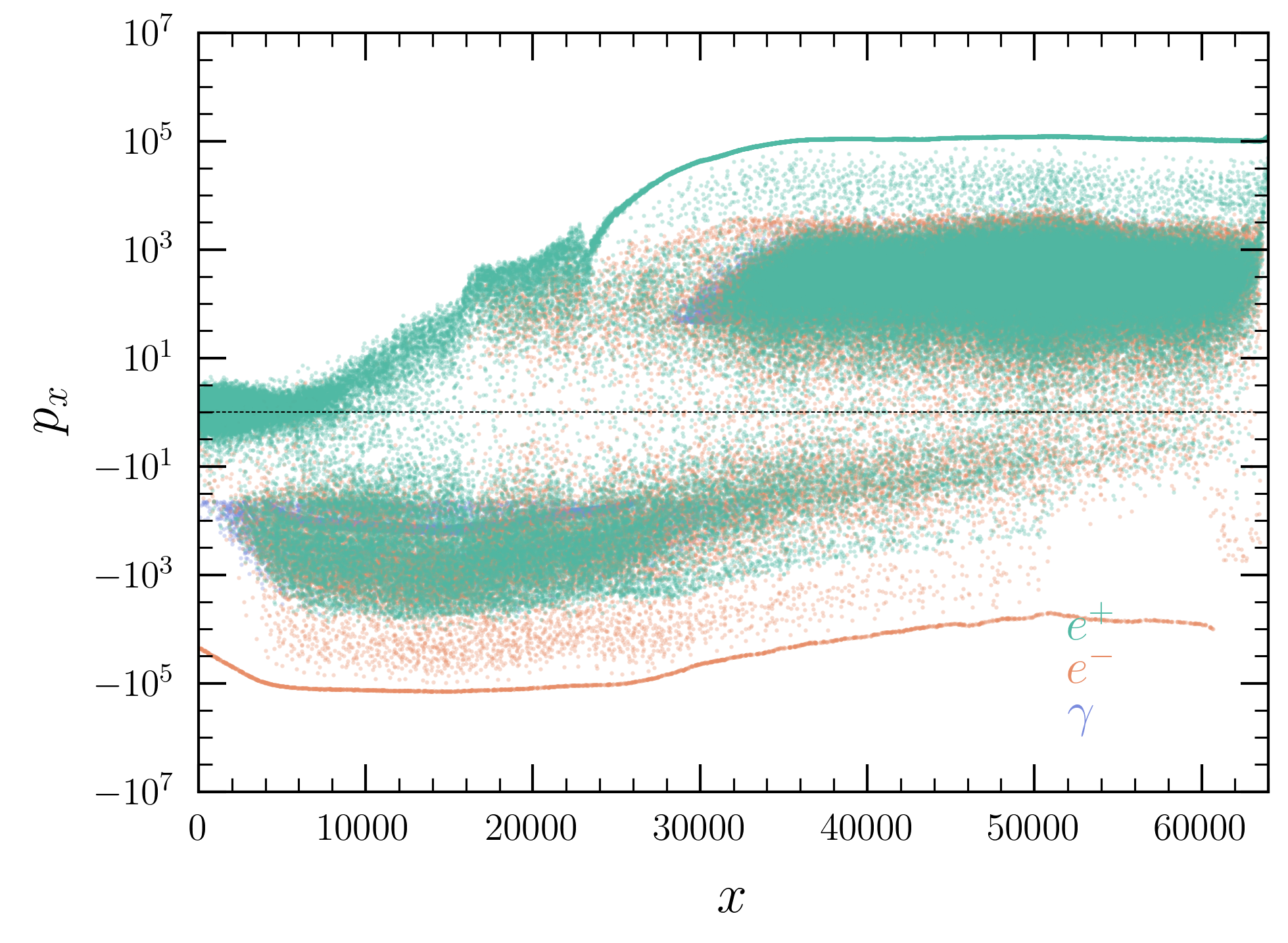}

    \caption{Same as Fig.~\ref{fig:picP}, but for the rotational frequency of the star reduced by a factor of four, while keeping the axion amplitude fixed.}\label{fig:picP_v2}
\end{figure*}

Fig.~\ref{fig:picE} illustrates that while the presence of axion hair modifies the pair discharge process around active neutron stars, it nevertheless proceeds despite these changes in dynamics. In order to illustrate the effect highlighted in this work, namely the decoupling of pair production from the rotational frequency of the pulsar, we now alter the simulations performed in Figs.~\ref{fig:picE} by reducing the rotationally induced electric field while maintaining the amplitude of the axion-induced electric field. In order to avoid increases in computational cost, we achieve this by reducing the characteristic plasma skin depth by a factor of two, corresponding to a factor of four in the rotational frequency, while increasing $\xi$ by a factor of four. Our expectation is that the $\xi = 0$ case leads to a slower particle acceleration, and a more spatially extended gap, while the $\xi = -40$ case yields a much more compact gap (with more efficient acceleration), once again manifesting near $r \sim m_a^{-1}$. Snapshots of the evolution of the electric field and phase space are shown in Fig.~\ref{fig:picE_v2} and \ref{fig:picP_v2}, which confirm the calculations and arguments laid out here, and in the main part of the text. Note that we have also verified that the axion-induced voltage drop cannot be screened in one of the polar caps. We show this by turning off pair production in the simulations and identifying the steady state solutions (which, along one pole, always involve amplified voltage drops).  In future work it would be interesting to determine whether there exist secondary, more subtle, observables induced by the axion cloud, \eg on the spectrum, amplitude, or polarization of the signal itself. We leave such endeavors to future work. 

\begin{figure*}
    \includegraphics[width=0.32\linewidth, trim={0cm, 1.7cm, 0cm, 0cm }, clip]{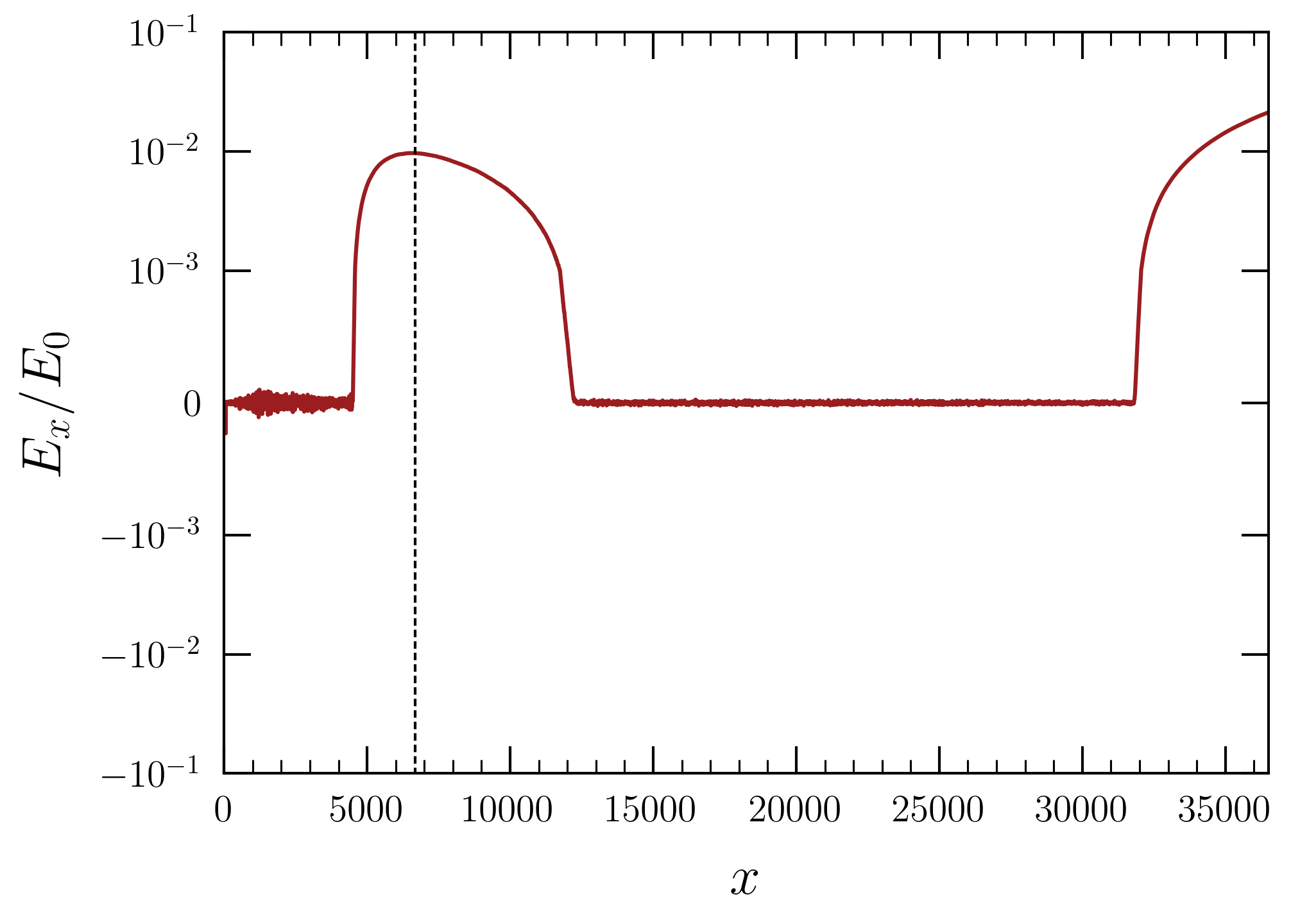}
    \includegraphics[width=0.32\linewidth,trim={0cm, 1.7cm, 0cm, 0cm }, clip]{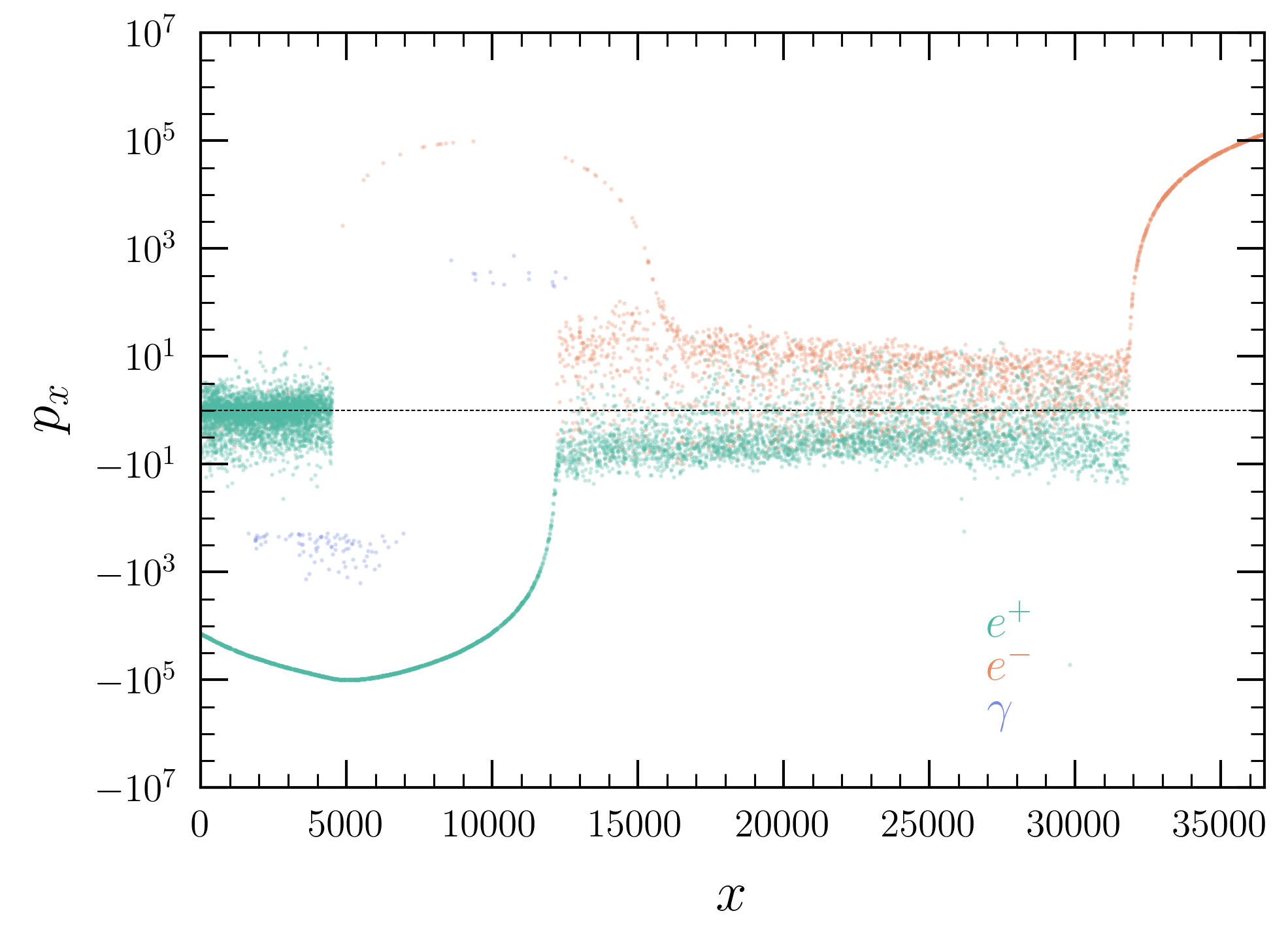}

   \includegraphics[width=0.32\linewidth, trim={0cm, 1.7cm, 0cm, 0cm }, clip]{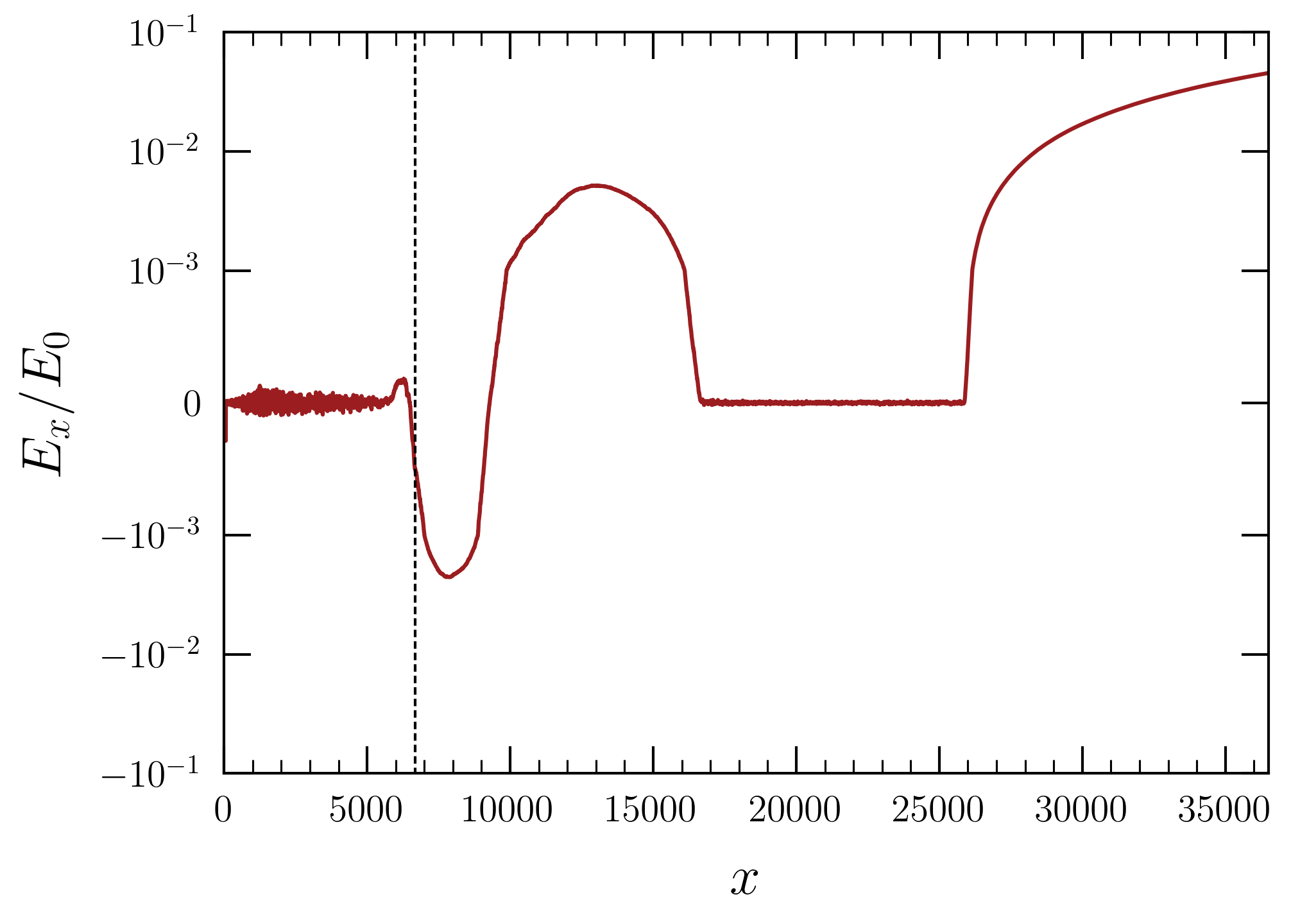}
    \includegraphics[width=0.32\linewidth,trim={0cm, 1.7cm, 0cm, 0cm }, clip]{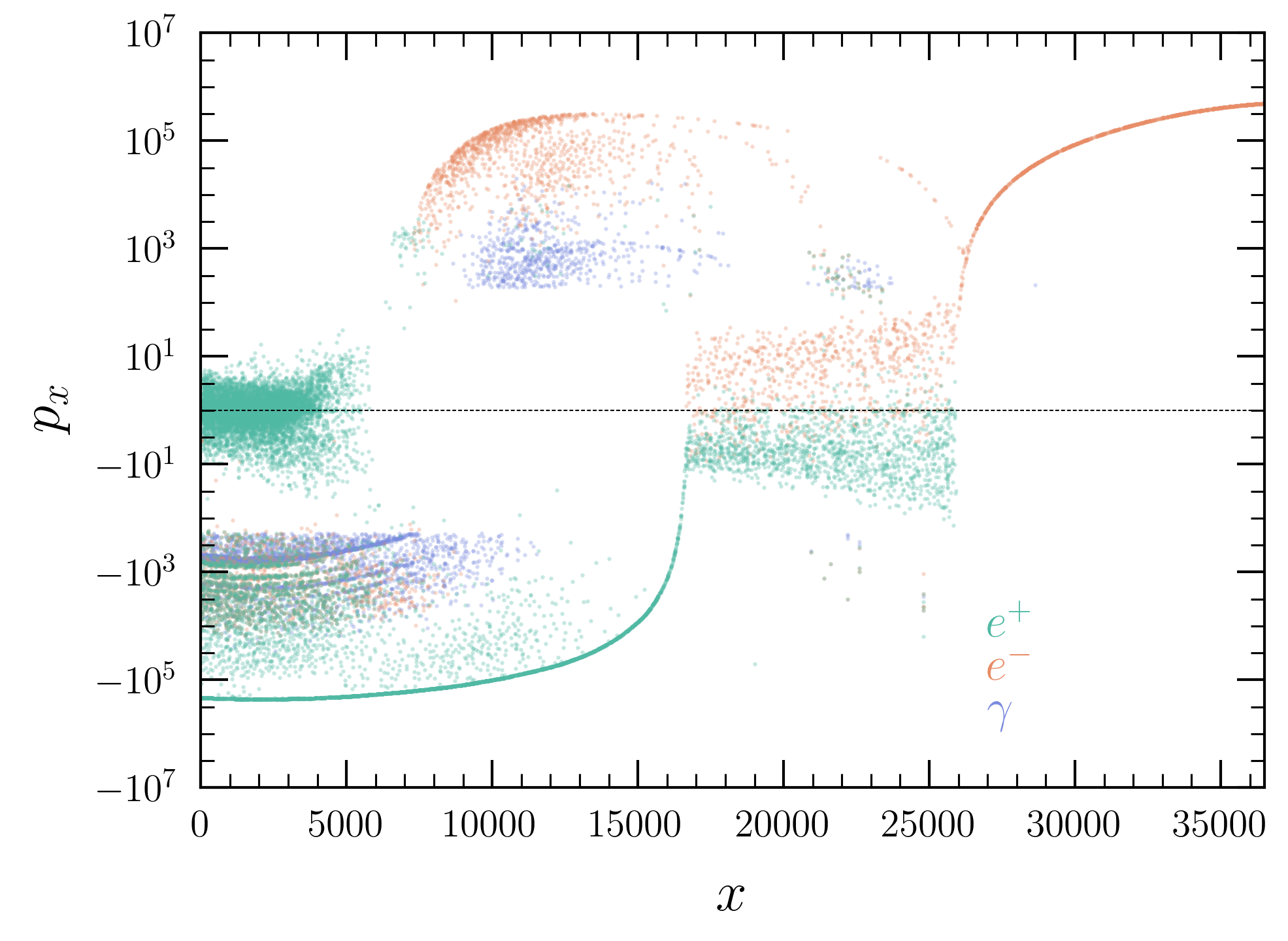}

    \includegraphics[width=0.32\linewidth, trim={0cm, 1.7cm, 0cm, 0cm }, clip]{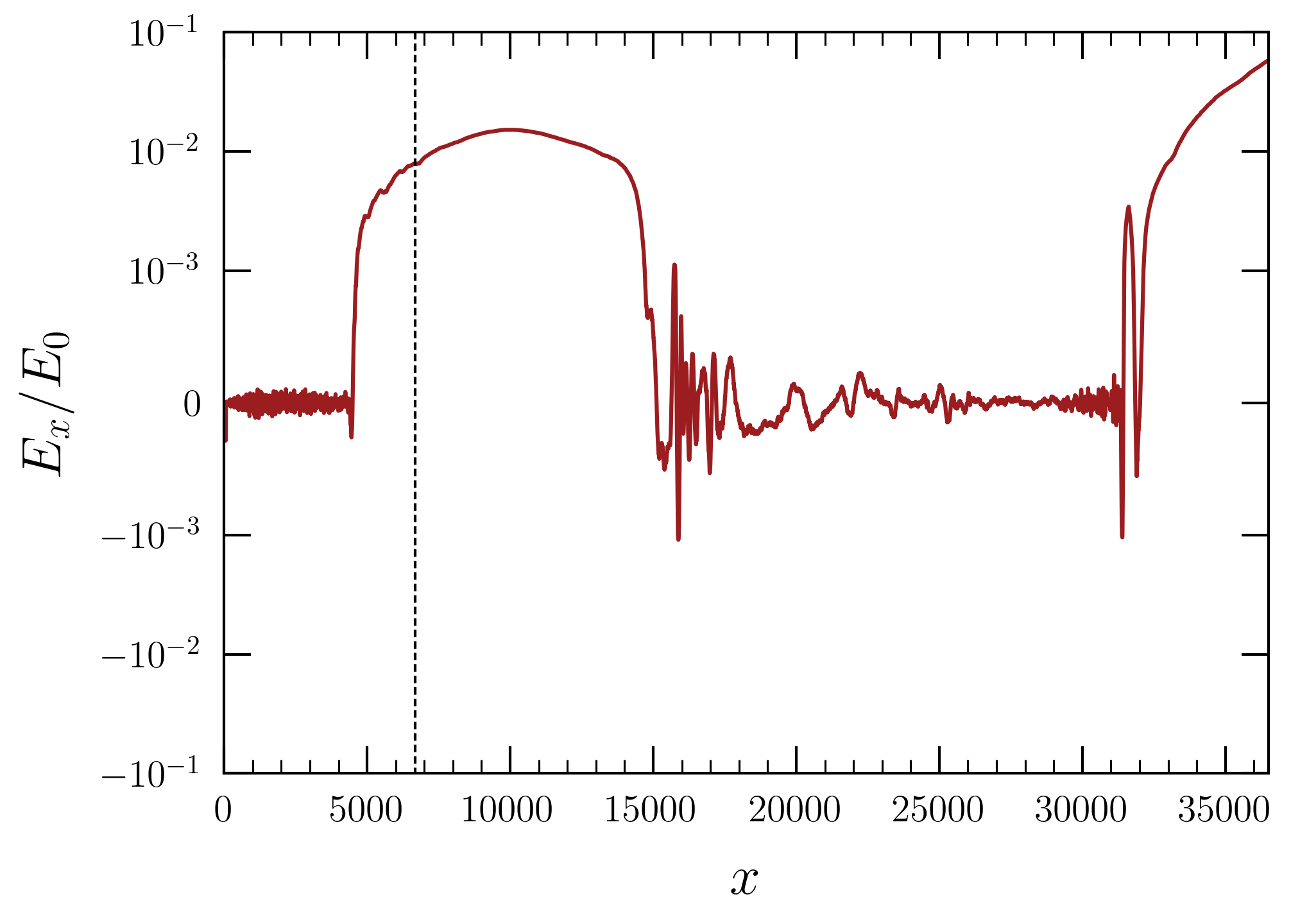}
    \includegraphics[width=0.32\linewidth,trim={0cm, 1.7cm, 0cm, 0cm }, clip]{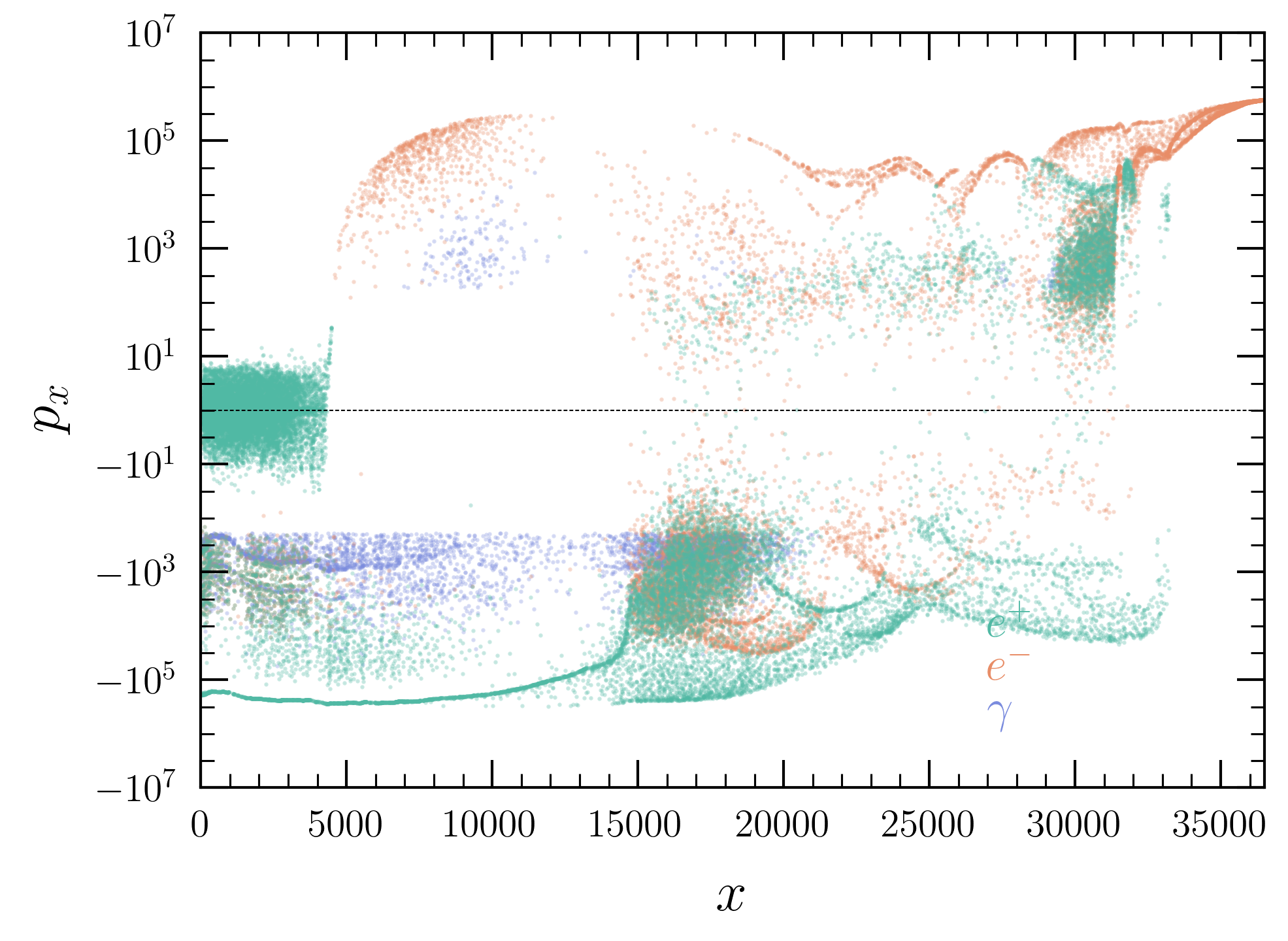}

    \includegraphics[width=0.32\linewidth, trim={0cm, 1.7cm, 0cm, 0cm }, clip]{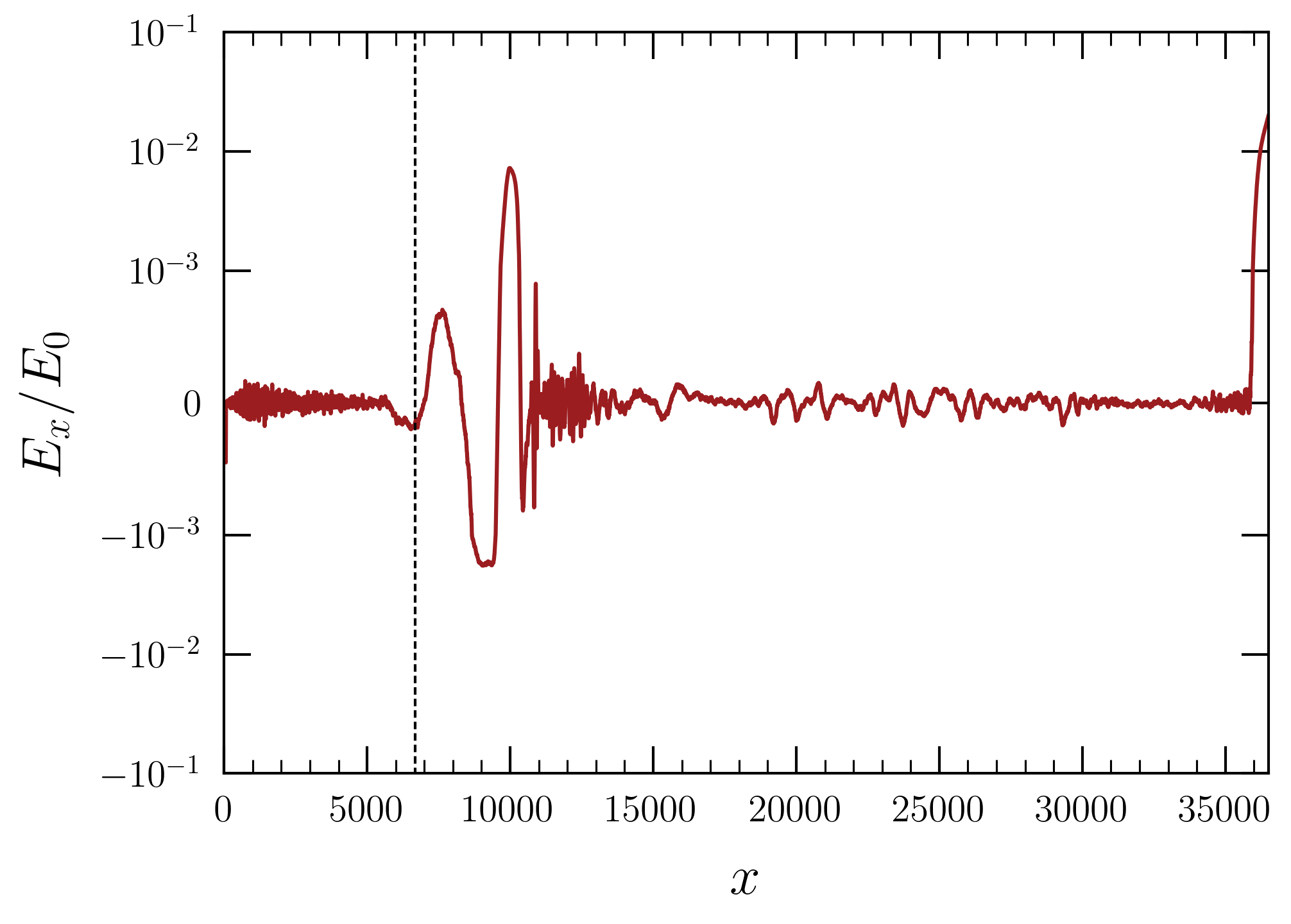}
    \includegraphics[width=0.32\linewidth,trim={0cm, 1.7cm, 0cm, 0cm }, clip]{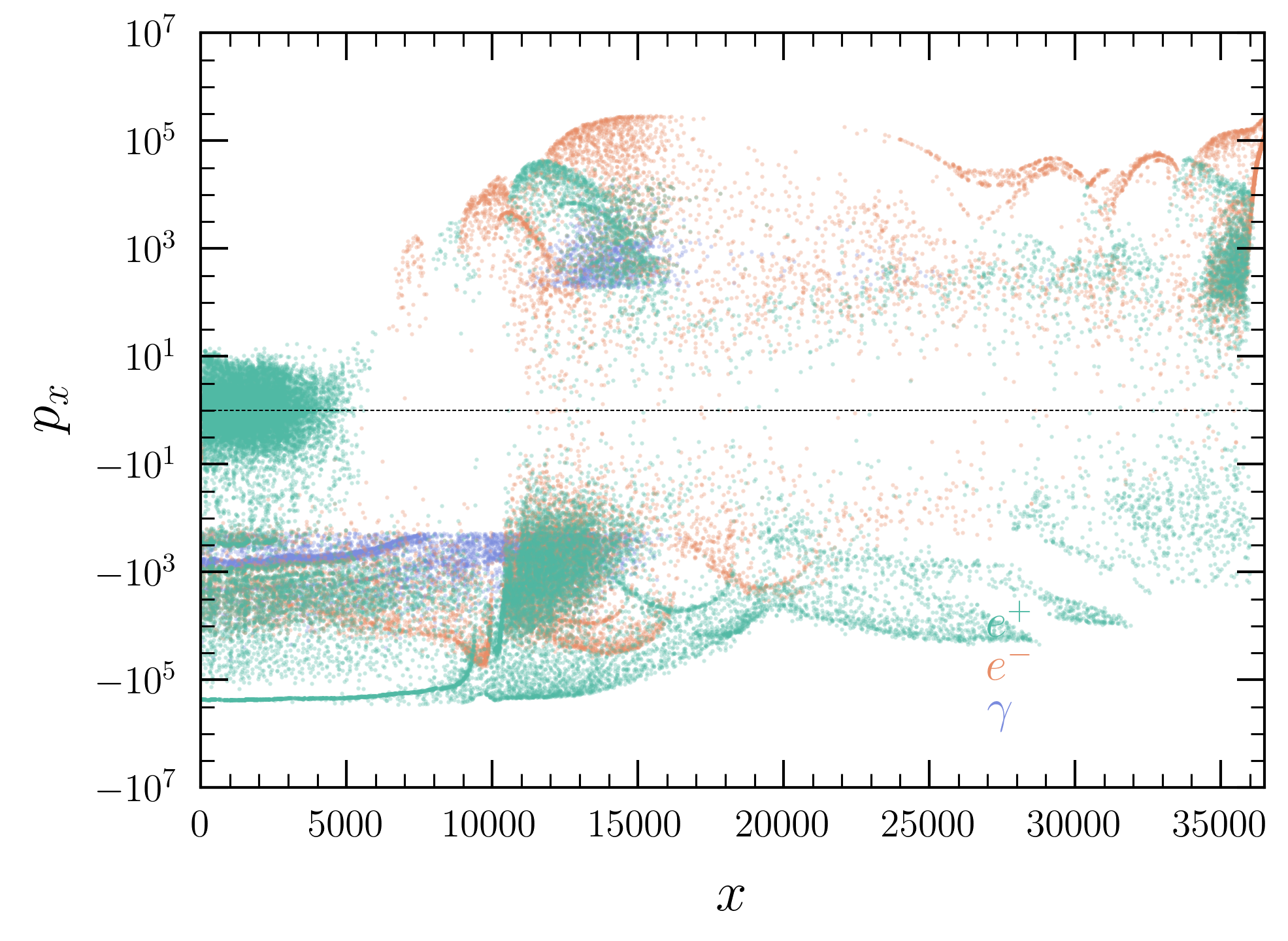}

    \includegraphics[width=0.32\linewidth, trim={0cm, 0cm, 0cm, 0cm }, clip]{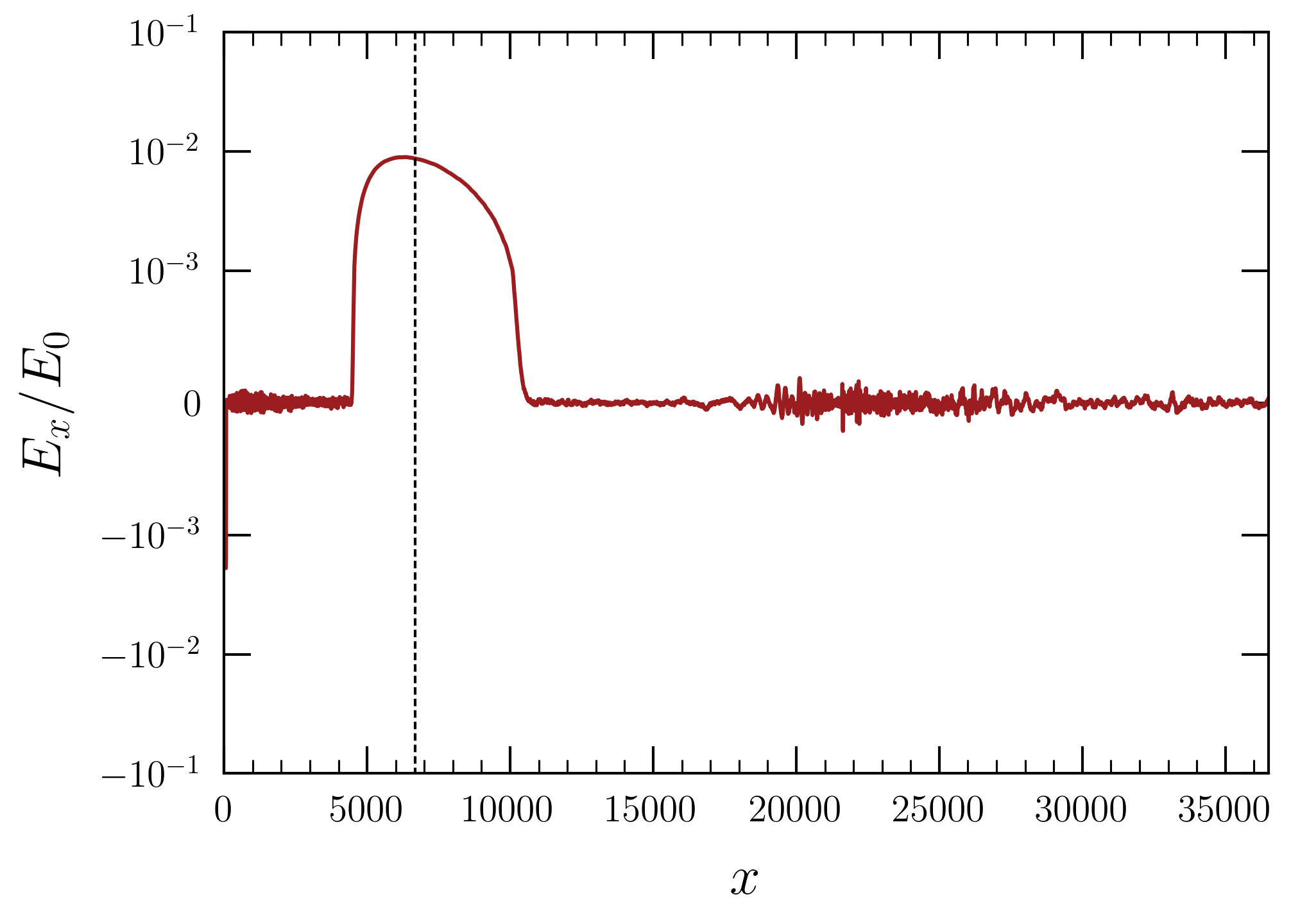}
    \includegraphics[width=0.32\linewidth,trim={0cm, 0cm, 0cm, 0cm }, clip]{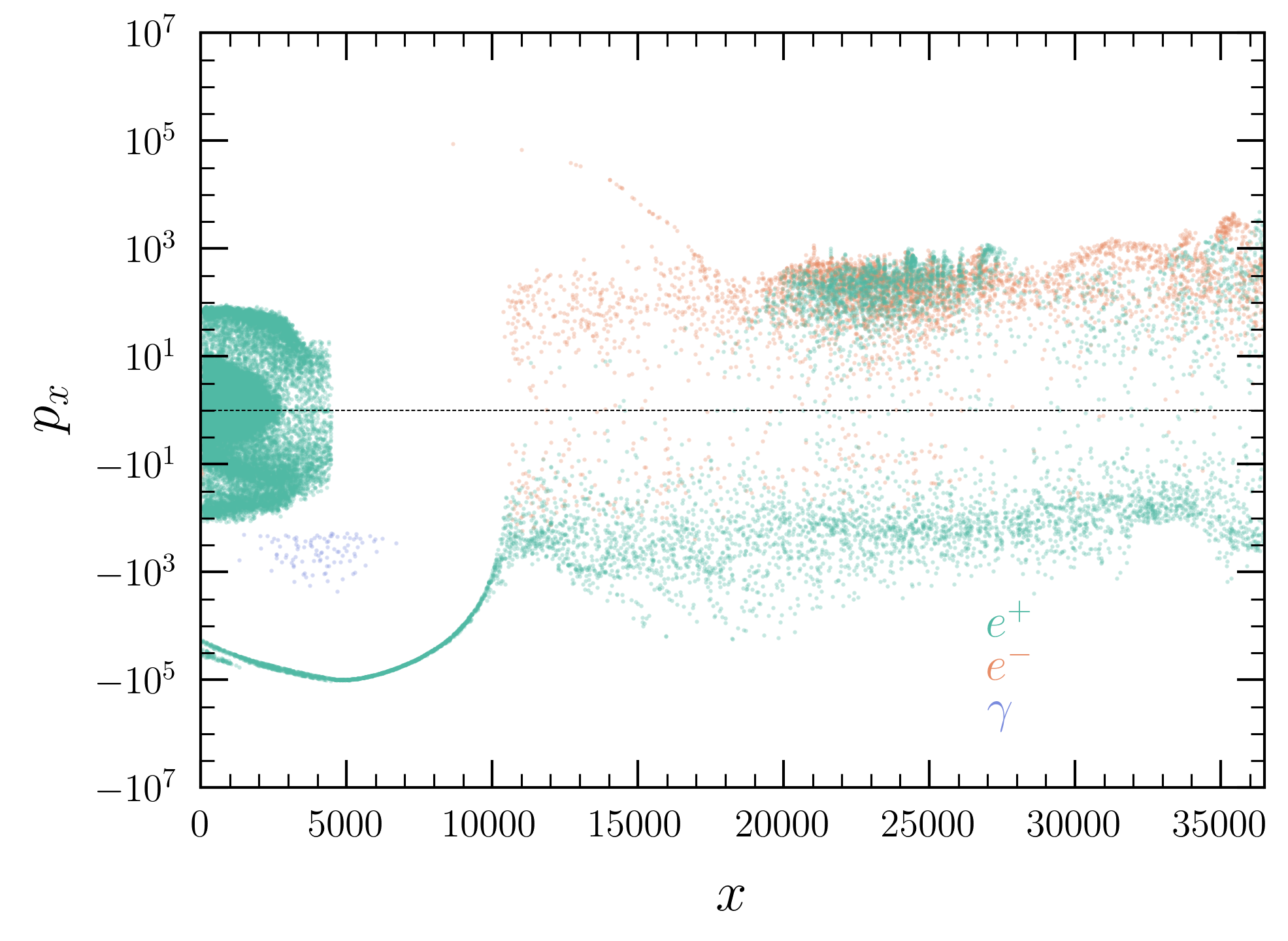}

    \caption{Same as Fig.~\ref{fig:picE} and ~\ref{fig:picP}, but showing the evolution of the electric field (left) and phase space (right) for the return current simulation, in which the axion induces pair production near the stellar surface. Here, one can see that the gap opens both at the right edge of the box (as expected for a return current simulation) and at a distance $r \sim m_a^{-1}$. Pair discharges close to the star eventually lead to full screening of the gap at larger distances. Simulations correspond to time steps: 143500, 155500, 162000, 197000, 216500. }\label{fig:pic_return}
\end{figure*}

Finally, before continuing, we briefly comment on the case of the return current. As discussed in the companion paper, active pair production along the outward flowing current is not sufficient by itself to ensure the pulsar remains active, one must also ensure that plasma can be efficiently supplied by the return current (otherwise the global configuration will untwist, invalidating the assumption of using the force-free current densities). We had also argued~\cite{Witte:CompanionPRL} that particle production should be easily sustainable along the return current in the near-field regime (by the axion potential drop) for ${\rm sgn}(\rho_a) = {\rm sgn}(\rho_{\rm GJ})$. In order to illustrate this explicitly, we perform an additional simulation, taking $j_{||} = -2 \rho_{\rm GJ}$, and adopt a strong axion field, setting $|\rho_{a}(x=0)| = 800 |\rho_{\rm GJ}(x=0)|$ (this choice enables the axion-potential drop to over-power the far-field drop, the latter occurring on the right-edge of the simulation domain). In order to consider the regime in which the conventional gap on the return field lines, at distances comparable to the stellar radius, is not capable of igniting pair production, we further forbid photon production near the right-most boundary $x \geq 0.85 \times L$ and remove the gradient of $\rho_{\rm GJ}$. We adopt an axion mass of $m_a = 0.001$ (in normalized units). The resulting evolution of $E_{||}$ and the particle phase space is shown in Fig.~\ref{fig:pic_return}. Here, one can see that the return current can easily be sustained by pair discharges operating at the near-field null surface, indicating the intuition outlined in the above is indeed valid. 

\begin{figure*}
    \includegraphics[width=0.32\linewidth, trim={0cm, 1.7cm, 0cm, 0cm }, clip]{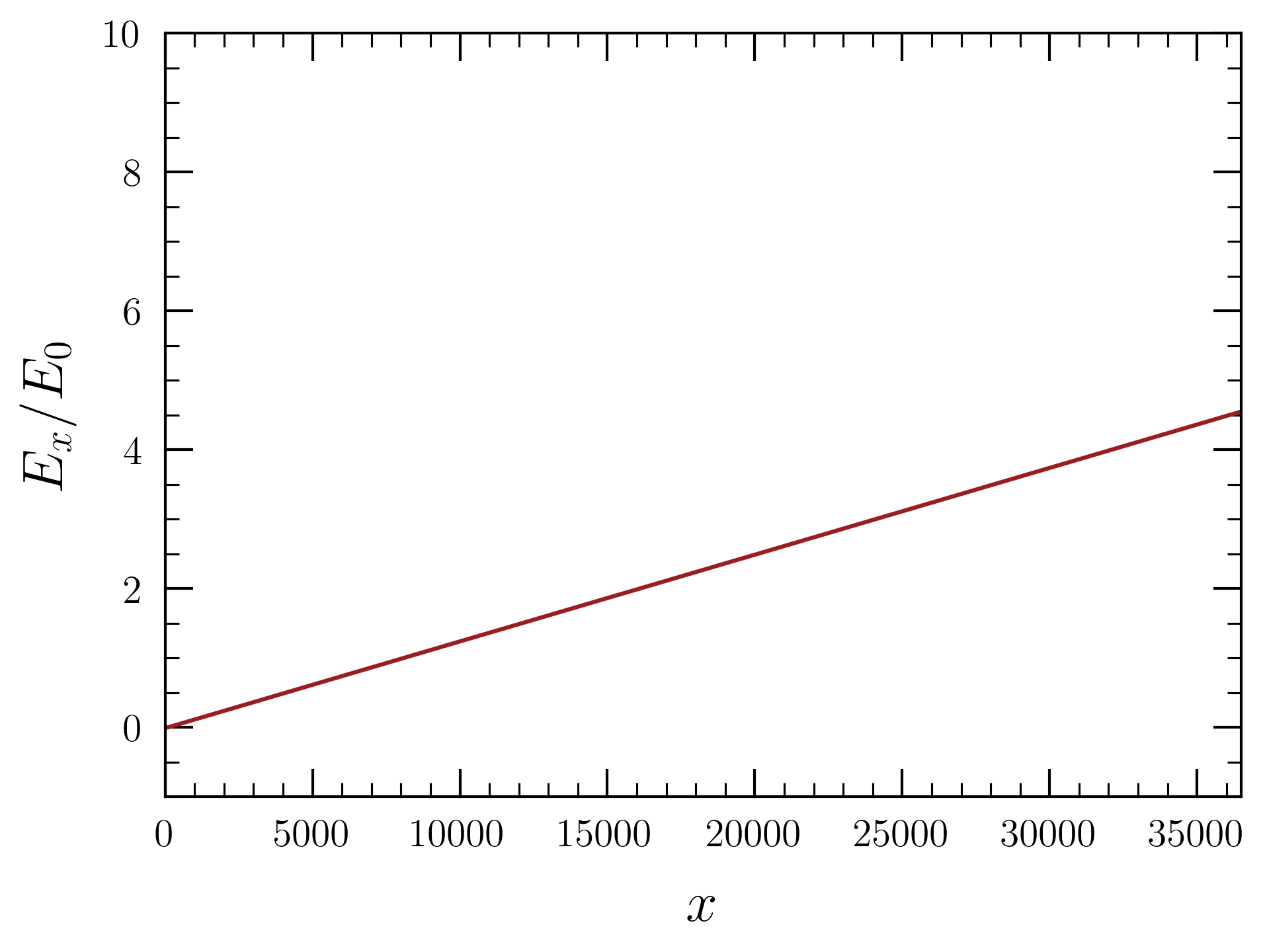}
    \includegraphics[width=0.32\linewidth,trim={0cm, 1.7cm, 0cm, 0cm }, clip]{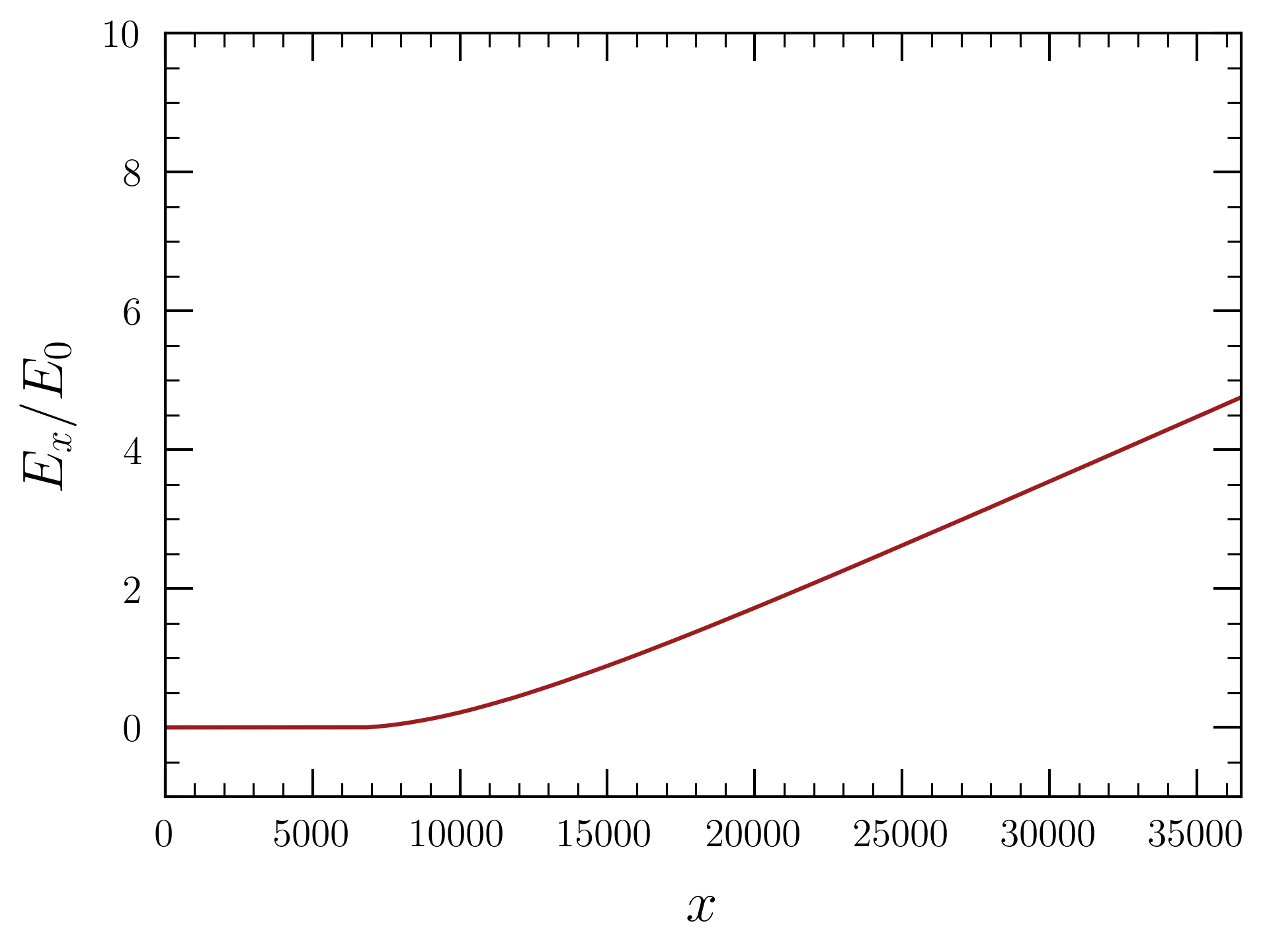}
    \includegraphics[width=0.32\linewidth,trim={0cm, 1.7cm, 0cm, 0cm }, clip]{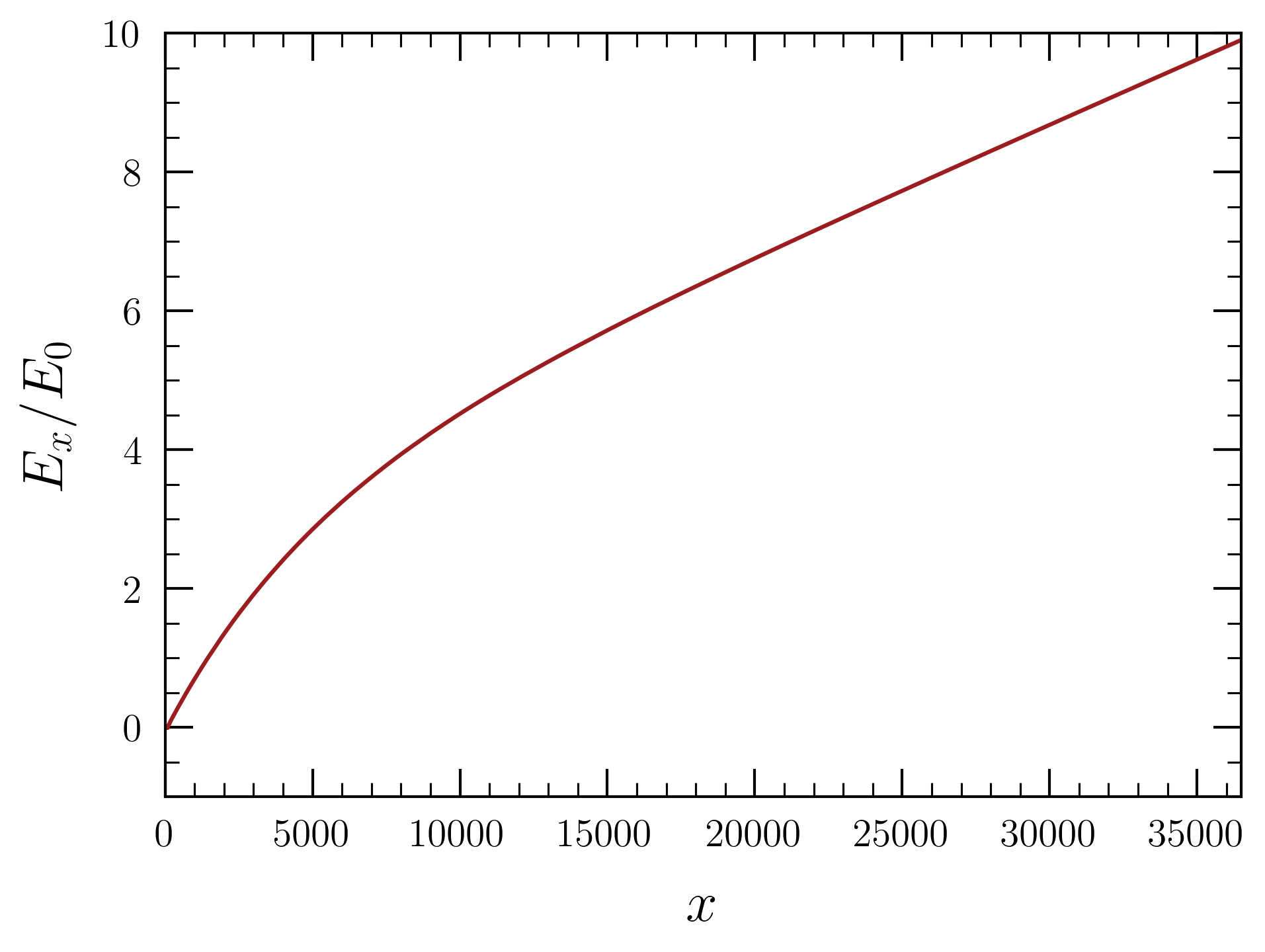}

    \includegraphics[width=0.32\linewidth, trim={0cm, 0cm, 0cm, 0cm }, clip]{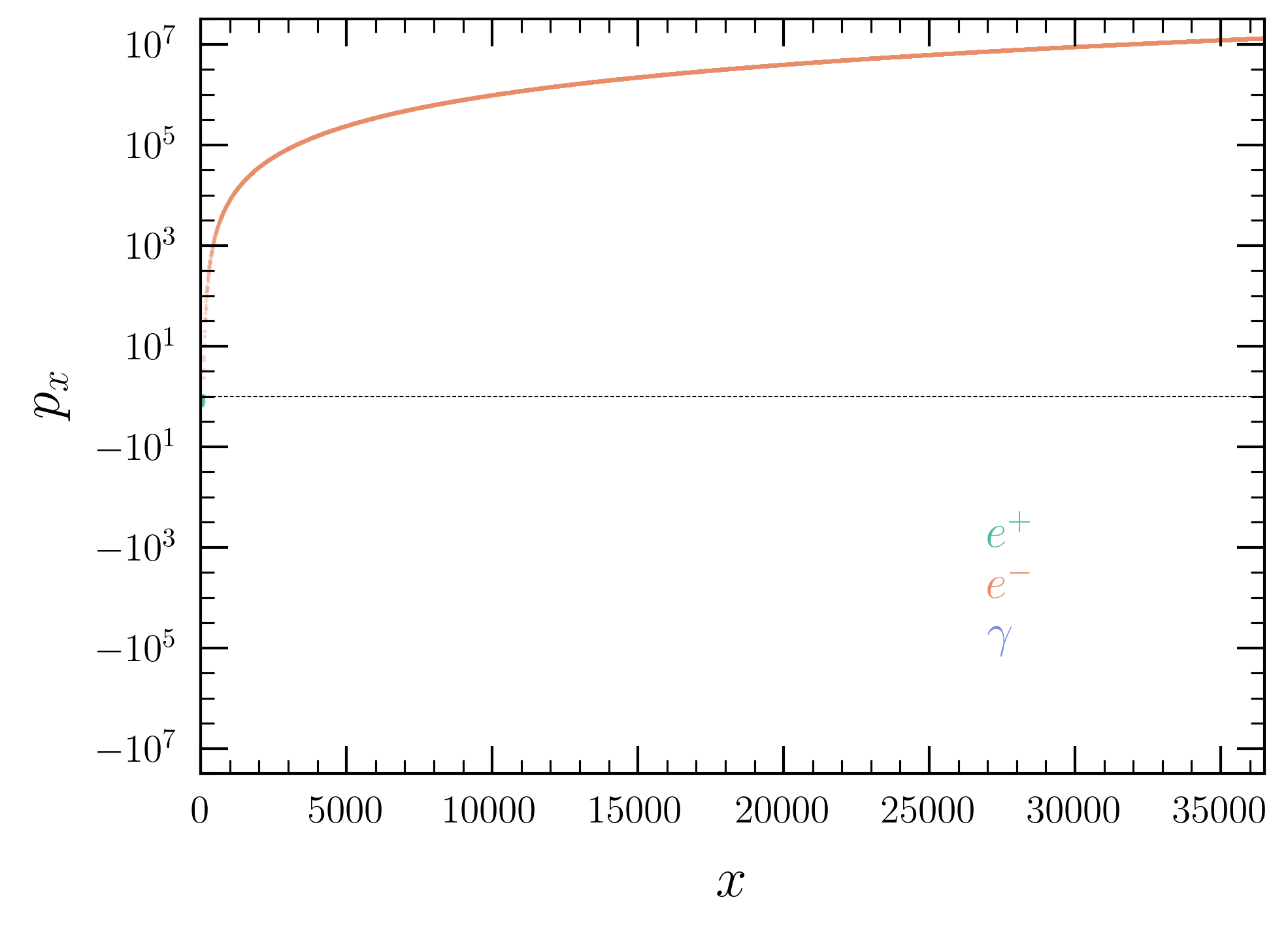}
    \includegraphics[width=0.32\linewidth,trim={0cm, 0cm, 0cm, 0cm }, clip]{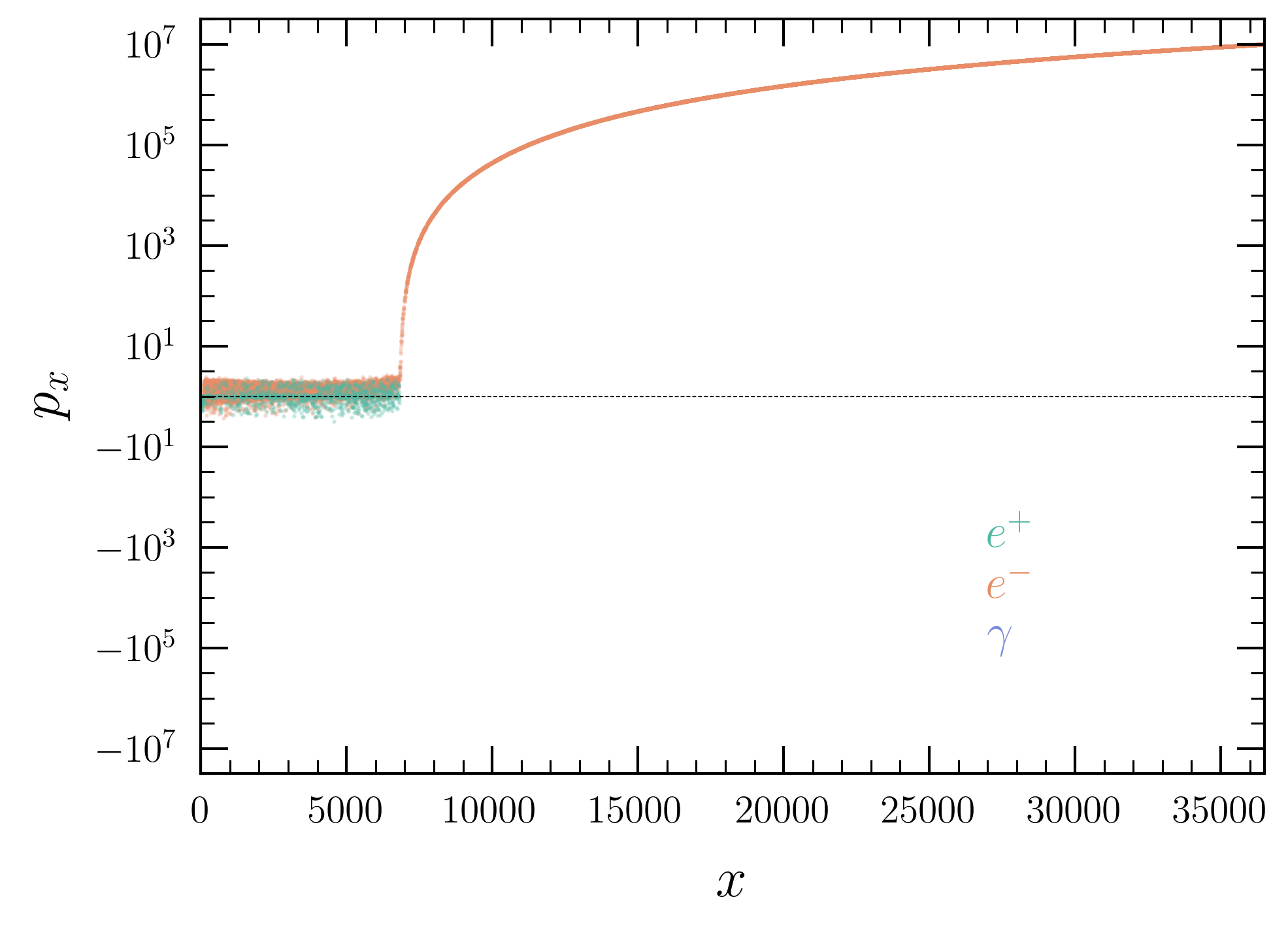}
     \includegraphics[width=0.32\linewidth,trim={0cm, 0cm, 0cm, 0cm }, clip]{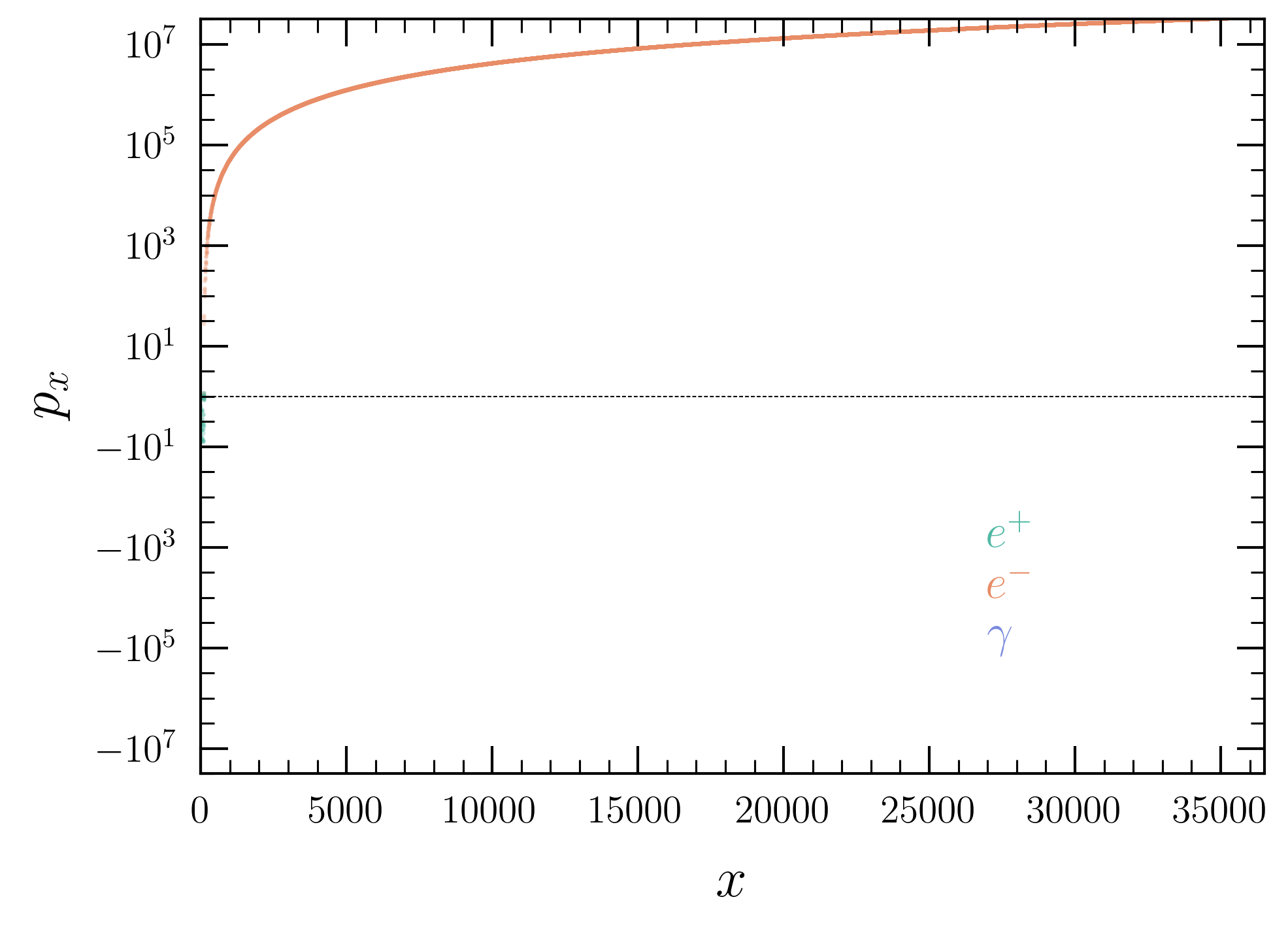}

     \caption{Simulation of return current in which pair production is turned off: without axion (left), $\xi = -10$ (center), and $\xi = 10$ (right). For $\xi < 0$, one can see that the axion-induced electric field at distances below the new null surface can be fully screened while supporting the magnetospheric current (the residual trapped $e^-$ population near the stellar surface is expected to vanish at later times, while the electric field will remain fully screened), while for $\xi > 0$ the voltage drop is substantially enhanced. }\label{fig:pic_return_blockpp}
\end{figure*}

For the sake of completeness, we also illustrate the steady state achieved by the return current with ${\rm sgn}(\rho_a) = -{\rm sgn}(\rho_{\rm GJ})$, in which a stable plasma configuration screens the axion-induced field (here, particles only experience the rotationally-induced voltage drop, implying pair production only occurs for sufficiently rapidly rotating stars). We do this by explicitly forbidding pair production, and looking at late-time distribution of particles and electric fields in the system. The particle phase space and voltage drop is illustrated for $\rho_a = 0$, $\rho_a = -10 \rho_{\rm GJ}$, and $\rho_a = 10 \rho_{\rm GJ}$, in Fig.~\ref{fig:pic_return_blockpp}.

\section{Consideration on energetics}

A natural question to ask at this point is: what is actually responsible for supplying the additional energy used to accelerate the primary particles? The axion gradient is being supported by the shift in the new ground state of nuclear matter, but itself is static in time, sitting at the minimum of its potential. In that sense, it cannot be either the shift in the nuclear energy density, nor the axion field, which supplies the energy, as both quantities are sitting a fixed equilibrium state. It is also clear that it is not the magnetic field which contributes to the energy drain -- the magnetic field is being sourced by currents internal to the star, and the discussion of particle acceleration can be had without any discussion about the micro-physical energy dissipation mechanisms which may alter these currents. We are therefore left with only one possibility: it must be that the axion is indirectly siphoning rotational energy, and using this to drive the acceleration. Let us provide intuitive arguments below for why this must be the case.

Perhaps the easiest way to see that this acceleration is inherently tied to the rotational energy of the star is to notice that the rate at which energy is transferred from the axion-induced electric field to the plasma 
is given by $\partial_t \mathcal{E} \sim 2 \pi  \rho r_{\rm pc}^2 \Delta V$, where $\Delta V$ is the pair production-limited voltage drop (which in the standard scenario is positively correlated with the magnetic field and the rotational frequency of the pulsar). When axion hair is present, the size of the voltage drop remains comparable (this follows from the fact that it is limited in size by the onset of pair production), but the functional dependence is altered such that it now depends on the axion field profile and the magnetic field. Nevertheless, $r_{\rm pc}^2 \propto \Omega$, and thus in the limit where the neutron star stops rotating, the energy transfer goes to zero.

There is another way in which this can be easily seen. A plasma filled magnetosphere induces a larger spin-down rate than a vacuum dipole. We have argued the effect of axion hair is to continue to replenish the supply of plasma to the magnetosphere when the rotationally induced electric field can no longer do so. Since this effect naturally comes with an enhanced loss of rotational energy, it seems clear that it is the rotational energy which is indirectly responsible for the production itself. 

This can also be inferred by looking at the solution for a non-rotating star. Here, one can understand that there exists a stable electrostatic solution in which free charges are lifted from the surface of the star, and placed in a configuration which directly cancels the axion induced electric field, implying there are no energy losses in this limit. As before, one sees that it is the rotation itself which prevents stable screening configurations of charges.

Given that the axion is directly sourcing an electric field, another natural question which arises at this point is whether one can neglect the energy cost associated with producing the electric field in the calculation of the sourcing of the axion field gradient. In general, one could imagine that this could perturb the axion potential in such a way so as to shift its minimum back toward the vacuum expectation. Here, we demonstrate explicitly that this is not he case, and that neglecting the electromagnetic energy is clearly a valid approximation for all systems.

In order to determine whether the presence of the magnetic field alters the formation of the cloud, we can compute the characteristic energy density being stored in the axion induced electric fields, and compare this directly with the energy stored in the axion gradient.  Taking the neutron star in vacuum, we can estimate the ratio of the energy densities as
\begin{eqnarray}
    \frac{E_{\rm EM}}{E_{\nabla a}} \sim \frac{\int dV \, |E_{\rm ax}|^2}{\int dV \, (\nabla a)^2}
\end{eqnarray}
where $E_{\rm ax}$ is the axion induced electric field, which can be inferred from Gauss' law. Using a rough estimator of $E_{\rm ax} \sim g_{a\gamma} a(r) \, B_0 \, (r_{\rm NS} / r)^3 \, \hat{r}$, and integrating only for $r \geq R_{\rm NS}$ (this is done to avoid unphysical divergences at the origin arising from the adopted fitting formulae, and matter effects inside the star), we obtain a ratio of 
\begin{eqnarray}
E_{\rm EM} / E_{\nabla a} &\sim& B_0^2 \, R_{\rm NS}^2 \alpha_{\rm EM}^2 / (f_a^2 \pi^2) \\ &\sim& 8 \times 10^{-12}\Big(\frac{B_0}{10^{12}\, \text{G}}\Big)^2\Big(\frac{R_{\rm NS}}{12\, \rm km}\Big)^2\Big(\frac{10^{15}\, \rm GeV}{f_a}\Big)^2, \nonumber
\end{eqnarray}
in the $m_a \rightarrow 0$ limit, and
\begin{eqnarray}
E_{\rm EM} / E_{\nabla a} &\sim& B_0^2 \alpha_{\rm EM}^2 / (f_a^2 \pi^2 m_a^2) \\ &\sim& 2 \times 10^{-17}\Big(\frac{B_0}{10^{12}\, \text{G}}\Big)^2\Big(\frac{10^{-8}\text{eV}}{m_a}\Big)^2\Big(\frac{10^{15}\, \rm GeV}{f_a}\Big)^2, \nonumber
\end{eqnarray}
in the $m_a \rightarrow \infty$ limit. Evaluating for typical parameters tends to yield values in the ball park of $E_{\rm EM}/ E_{\nabla a} \sim \mathcal{O}(10^{-16})$, although for magnetar magnetic field strengths we note that this could be as large as $E_{\rm EM}/ E_{\nabla a} \sim \mathcal{O}(10^{-8}) $. While the precise value depends on the properties of the neutron star and the location in parameter space, we find that this energy density is always sufficiently small that it can be neglected.

\section{Summary}
\label{sec:summary}
This work described all the machinery needed to quantify how axion hair modifies particle acceleration and pair discharge in the inner magnetosphere of rotation-powered pulsars.
The companion paper~\cite{Witte:CompanionPRL} uses these ingredients to construct death-valley curves for axion benchmarks, confront them with the observed pulsar distribution in the $P$--$\dot P$ plane, and derive constraints on light QCD axions and on linearly-coupled (monopole--dipole) axions. Population synthesis and the associated statistical treatment are presented in the Appendices of the companion paper, and are not repeated here so as to keep the focus of this article on the modeling and simulation infrastructure. 

In future work, it would be valuable to extend the PIC simulations described here to include \textit{dynamical} fields and investigate their impact on nonlinear regimes, which are analytically intractable~\cite{Caputo:2023cpv}. Results from these simulations will be presented in a forthcoming publication.

\section{Acknowledgments}%
We thank Anson Hook, Konstantin Springmann and Kai Bartnick for enlightening discussions, and Ani Prabhu for their useful comments on the manuscript. SJW acknowledges support from a Royal Society University Research Fellowship (URF-R1-231065). This work is also supported by the Deutsche Forschungsgemeinschaft under Germany's Excellence Strategy EXC 2121 Quantum Universe 390833306. AC is supported by an ERC STG grant (``AstroDarkLS'', grant No. 101117510). AC acknowledges the Weizmann Institute of Science for hospitality at different stages of this project and the support from the Benoziyo Endowment Fund for the Advancement of
Science. 
SS acknowledges financial support from the Spanish Ministry of Science and Innovation (MICINN) through the Spanish State Research Agency, under Severo Ochoa Centres of Excellence Programme 2025-2029 (CEX2024001442-S).
This work is also part of the R\&D\&i project PID2023-146686NB-C31, funded by MICIU/AEI/10.13039/501100011033/ and by ERDF/EU. IFAE is partially funded by the CERCA program of the Generalitat de Catalunya.
A.Ch. is supported by Martin A. and Helen Chooljian Member Fund and the Fund for Memberships in Natural Sciences.  This article/publication is based upon work from COST Action COSMIC WISPers CA21106, supported by COST (European Cooperation in Science and Technology). This work was supported by a grant from the Simons Foundation (MP-SCMPS-00001470), as well as an Alfred P. Sloan Fellowship and a Packard Foundation Fellowship in Science and Engineering to AP. This work was supported by the U.S.~Department of Energy~(DOE), Office of Science, National Quantum Information Science Research Centers, Superconducting Quantum Materials and Systems Center~(SQMS) under Contract No.~DE-AC02-07CH11359.  S.R.\ is supported in part by the U.S.~National Science Foundation~(NSF) under Grant No.~PHY-2412361.The work of S.R.\  was also supported by the Simons Investigator Award No.~827042.

\bibliographystyle{bibi}
\bibliography{biblio}

\end{document}